\documentclass{pasa}%

\usepackage{graphicx}
\usepackage{amsmath}	
\usepackage{amssymb}	
\usepackage{aas_macros}
\usepackage{pdflscape}
\usepackage{array}
\usepackage{multirow}
\usepackage[figuresright]{rotating}

\def\degr{\hbox{$^\circ$}}
\def\fdg{\hbox{$.\!\!^\circ$}}
\def\farcm{\hbox{$.\mkern-4mu^\prime$}}
\def\farcs{\hbox{$.\!\!^{\prime\prime}$}}
\def\arcmin{\hbox{$^\prime$}}
\def\arcsec{\hbox{$^{\prime\prime}$}}

\DeclareMathOperator{\sgn}{sgn}

\title[HzRG candidates from GLEAM and VIKING]{The GLEAMing of the first supermassive black holes: \newline II. A new sample of high-redshift radio galaxy candidates}

\author[J.~W. Broderick et al.]{J.~W.~Broderick$^{1}$\thanks{E-mail: jess.broderick@curtin.edu.au}, G.~Drouart$^{1}$, N.~Seymour$^{1}$, T.~J.~Galvin$^{1}$, N.~Wright$^{1}$, A.~Carnero~Rosell$^{2,3}$, R.~Chhetri$^{1,4}$, H.~Dannerbauer$^{2,3}$, S.~P.~Driver$^{5}$, J.~S.~Morgan$^{1}$, V.~A.~Moss$^{6,7}$, S.~Prabu$^{1,6}$, J.~M.~Afonso$^{8,9}$, C.~De~Breuck$^{10}$, B.~H.~C.~Emonts$^{11}$, T.~M.~O.~Franzen$^{12}$, C.~M.~Guti\'{e}rrez$^{2,3}$, P.~J.~Hancock$^{1,13}$, G.~H.~Heald$^{4}$, N.~Hurley-Walker$^{1}$, R.~J.~Ivison$^{10}$, M.~D.~Lehnert$^{14}$, G.~Noirot$^{15}$, M.~Read$^{16}$, S.~S.~Shabala$^{17}$, D.~Stern$^{18}$, W.~J.~Sutherland$^{19}$, E.~Sutorius$^{16}$, R.~J.~Turner$^{17}$ and J.~Vernet$^{10}$ 
\affil{$^{1}$International Centre for Radio Astronomy Research, Curtin University, GPO Box U1987, Bentley, WA 6845, Australia}
\affil{$^{2}$Instituto de Astrof\'{i}sica de Canarias (IAC), E-38205 La Laguna, Tenerife, Spain}
\affil{$^{3}$Universidad de La Laguna, Dpto. Astrof\'{i}sica, E-38206 La Laguna, Tenerife, Spain}
\affil{$^{4}$CSIRO Space and Astronomy, PO Box 1130, Bentley, WA 6102, Australia}
\affil{$^{5}$International Centre for Radio Astronomy Research, The University of Western Australia, 35 Stirling Highway, Crawley, WA 6009, Australia}
\affil{$^{6}$CSIRO Space and Astronomy, PO Box 76, Epping, NSW 1710, Australia}
\affil{$^{7}$Sydney Institute for Astronomy, School of Physics, The University of Sydney, NSW 2006, Australia}
\affil{$^{8}$Instituto de Astrof\'{i}sica e Ci\^{e}ncias do Espa\c co, Universidade de Lisboa, OAL, Tapada da Ajuda, PT1349-018 Lisboa, Portugal}
\affil{$^{9}$Departamento de F\'{i}sica, Faculdade de Ci\^{e}ncias, Universidade de Lisboa, Edif\'{i}cio C8, Campo Grande, PT1749-016 Lisbon, Portugal}
\affil{$^{10}$European Southern Observatory, Karl-Schwarzschild-Stra{\ss}e 2, D-85748 Garching bei M\"{u}nchen, Germany}
\affil{$^{11}$National Radio Astronomy Observatory, 520 Edgemont Road, Charlottesville, VA 22903, USA}
\affil{$^{12}$ASTRON, the Netherlands Institute for Radio Astronomy, Oude Hoogeveensedijk 4, NL-7991 PD Dwingeloo, The Netherlands}
\affil{$^{13}$Curtin Institute for Computation, Curtin University, GPO Box U1987, Perth, WA 6845}
\affil{$^{14}$Universit\'e Lyon 1, ENS de Lyon, CNRS UMR5574, Centre de Recherche Astrophysique de Lyon, F-69230 Saint-Genis-Laval, France}
\affil{$^{15}$Institute for Computational Astrophysics and Department of Astronomy \& Physics, Saint Mary’s University, 923 Robie Street, Halifax, NS B3H 3C3, Canada}
\affil{$^{16}$Institute for Astronomy, University of Edinburgh, Royal Observatory, Blackford Hill, Edinburgh, EH9 3HJ, UK}
\affil{$^{17}$School of Natural Sciences, University of Tasmania, Private Bag 37, Hobart, TAS 7001, Australia}
\affil{$^{18}$Jet Propulsion Laboratory, California Institute of Technology, 4800 Oak Grove Drive, Pasadena, CA 91109, USA}
\affil{$^{19}$Astronomy Unit, School of Physical and Chemical Sciences, Queen Mary University of London, Mile End Road, London, E1 4NS, UK}
}

\jid{PASA}
\doi{10.1017/pas.\the\year.xxx}
\jyear{\the\year}

\usepackage{hyperref} 
\hypersetup{colorlinks,citecolor=blue,linkcolor=blue,urlcolor=blue}

\begin{document}

\begin{frontmatter}
\maketitle

\begin{abstract}
While unobscured and radio-quiet active galactic nuclei are regularly being found at redshifts $z > 6$, their obscured and radio-loud counterparts remain elusive. We build upon our successful pilot study, presenting a new sample of low-frequency-selected candidate high-redshift radio galaxies (HzRGs) over a sky area twenty times larger. We have refined our selection technique, in which we select sources with curved radio spectra between 72--231 MHz from the GaLactic and Extragalactic All-sky Murchison Widefield Array (GLEAM) survey. In combination with the requirements that our GLEAM-selected HzRG candidates have compact radio morphologies and be undetected in near-infrared $K_{\rm s}$-band imaging from the Visible and Infrared Survey Telescope for Astronomy Kilo-degree Infrared Galaxy (VIKING) survey, we find 51 new candidate HzRGs over a sky area of approximately 1200 deg$^2$. Our sample also includes two sources from the pilot study: the second-most distant radio galaxy currently known, at $z=5.55$, with another source potentially at $z \sim 8$. We present our refined selection technique and analyse the properties of the sample. We model the broadband radio spectra between 74 MHz and 9 GHz by supplementing the GLEAM data with both publicly available data and new observations from the Australia Telescope Compact Array at 5.5 and 9 GHz. In addition, deep $K_{\rm s}$-band imaging from the High-Acuity Widefield $K$-band Imager (HAWK-I) on the Very Large Telescope and from the Southern {\it Herschel} Astrophysical Terahertz Large Area Survey Regions $K_{\rm s}$-band Survey (SHARKS) is presented for five sources. We discuss the prospects of finding very distant radio galaxies in our sample, potentially within the epoch of reionisation at $z \gtrsim 6.5$.     
\end{abstract}

\begin{keywords}
galaxies: high-redshift -- galaxies: active -- radio continuum: galaxies -- infrared: galaxies
\end{keywords}
\end{frontmatter}

\section{Introduction}\label{section:intro}
The formation and evolution of the most extreme supermassive black holes (SMBHs; mass $\gtrsim 10^{8} M_\odot$) in the early Universe is a topic for which theoretical models must take into account very challenging observational constraints (e.g. reviews by \citealt{volonteri12}; \citealt*{smith17}; \citealt{smith19}). With the recent discovery of the ultra-distant quasar J0313$-$1806 by \citet[][]{wang21}, active galactic nuclei (AGN) have now been found at redshifts as high as $z = 7.64$, well within the epoch of reionisation \citep[EoR; e.g. review by][]{koopmans15}. The SMBH in J0313$-$1806 has a mass of $(1.6 \pm 0.4) \times 10^9 M_\odot$ \citep[][]{wang21}, only 0.68 Gyr\footnote{In this paper, we assume a flat Lambda cold dark matter ($\Lambda$CDM) cosmology with Hubble constant $H_0=67.7$ km s$^{-1}$ Mpc$^{-1}$, matter density parameter $\Omega_{\rm M}=0.31$ and vacuum density parameter $\Omega_{\Lambda}=0.69$ \citep{planck20}.} after the Big Bang. As discussed in \citet[][]{wang21}, this SMBH may have been able to grow so fast if the seed was a direct-collapse black hole of mass $10^4$--$10^5 M_\odot$. Finding more AGN at very high redshift is vital to further build our knowledge of the efficiency of accretion and the properties of black hole seeds at early cosmic epochs. 

Radio emission is a crucial tracer of the accretion onto the central SMBH. The search for ultra-high-redshift radio-loud AGN has now progressed beyond $z=6$. Radio-loud quasars have recently been identified near the end of the EoR ($z \sim 6.5$): VIK~J2318$-$3113 at $z=6.44$ \citep[][]{ighina21,ighina22} and P172$+$18 at $z=6.82$ \citep[][]{banados21,momjian21}. In addition, \citet[][]{belladitta20} discovered PSO~J0309$+$27 at $z=6.10$; this radio-loud source is the most distant known blazar. However, the most radio-powerful known AGN in the distant Universe remain the high-redshift radio galaxies \citep[HzRGs; $z \gtrsim 2$; review by][]{miley08}, which, as shown by the Hubble $K$--$z$ relation, are also among the most massive galaxies \citep[where $K$ is the near-infrared 2.2-\textmu m apparent magnitude; e.g.][]{roccavolmerange04}. The luminous radio emission from HzRGs allows us to efficiently pinpoint these rare systems. Their host galaxies can also be more easily studied than for quasars because the AGN emission is obscured along our line of sight, reducing AGN contamination in the rest-frame ultraviolet through near-infrared wavebands \citep[e.g.][]{seymour07,debreuck10,drouart16,podigachoski16}. HzRGs are therefore of great importance for studying the co-evolution of massive galaxies and their central SMBHs in the early Universe.  

For nearly two decades, the most distant HzRG known was TN~J0924$-$2201 at $z=5.19$ \citep[][]{vanbreugel99}. However, with the advent of a number of deep, low-frequency radio surveys, momentum has been regained in the search for even more distant HzRGs. \citet[][]{saxena18a} cross-correlated the 147.5-MHz 
Tata Institute of Fundamental Research (TIFR) Giant Metrewave Radio Telescope \citep[GMRT;][]{swarup91} Sky Survey \citep[TGSS;][]{intema17} with both the 1.4-GHz Faint Images of the Radio Sky at Twenty Centimetres Karl G. Jansky Very Large Array \citep[VLA;][]{thompson80} survey \citep*[FIRST;][]{becker95,helfand15} and the 1.4-GHz National Radio Astronomy Observatory (NRAO) VLA Sky Survey \citep[NVSS;][]{condon98}. HzRG candidates were selected on the basis of their ultra-steep radio spectra (USS; radio spectral index\footnote{In this paper, we use the radio spectral index convention $S_{\nu} \propto \nu^{\alpha}$, where $S_{\nu}$ is the flux density at frequency $\nu$. Our $S_{\nu}$ notation assumes that the frequency is in MHz. We also denote a two-point spectral index between $\nu_1$ and $\nu_2$ MHz as $\alpha^{\nu_2}_{\nu_1}$.} $\alpha \leq -1.3$, although in the literature $\alpha \lesssim -1.0$ is often used as a USS cutoff) between the two widely spaced frequencies, a selection technique that has been used for many decades to increase the efficiency of an HzRG search (e.g. \citealt*{tielens79}; \citealt{blumenthal79}; \citealt{roettgering94}; \citealt{debreuck00,debreuck04}; \citealt{cohen04}; \citealt{cruz06}; \citealt{broderick07}; \citealt{afonso11}). The correlation between redshift and spectral index (such that steeper sources are at higher redshifts) has been the subject of extensive analysis in the literature \citep[e.g.][]{athreya98,blundell99b,klamer06,ker12,morabito18} and has been postulated to arise from either (i) selection effects combined with large inverse-Compton losses at high redshift due to the energy density of the cosmic microwave background, (ii) a correlation between spectral index and radio luminosity (such that more luminous sources have steeper spectral indices) coupled with Malmquist bias, or (iii) sources at high redshift residing in denser environments on average.  

From the sample of 32 USS HzRG candidates presented in \citet[][]{saxena18a}, TGSS~J1530$+$1049 was discovered at $z=5.72$, which is currently the most distant known radio galaxy \citep[][]{saxena18b}. An additional four HzRGs in this sample have redshifts in the range $4.01 \leq z \leq 4.86$ \citep[][]{saxena19}. Furthermore, the ongoing 144-MHz Low-Frequency Array \citep[LOFAR;][]{vanhaarlem13} Two-metre Sky Survey \citep[LoTSS;][]{shimwell17,shimwell19,shimwell22} has considerable potential for finding many HzRGs \citep*[e.g. see predictions in][]{saxena17}, as does the ongoing 54-MHz LOFAR Low-band antenna Sky Survey \citep[LoLSS;][]{degasperin21}. Other relevant studies of interest are USS searches that made use of deep 150-MHz GMRT data \citep[][]{ishwarachandra10,ishwarachandra11,bisoi11} and the TGSS--NVSS spectral index map that covers 80 per cent of the celestial sphere \citep*[][]{degasperin18}.   

As an alternative to USS-selected HzRG samples, the wide frequency coverage of the Murchison Widefield Array \citep[MWA;][]{tingay13} allows one to use broadband low-frequency radio spectral properties to search for HzRGs. In a pilot study centred on one of the Galaxy And Mass Assembly survey \citep[GAMA;][]{driver09,driver11} equatorial fields, GAMA-09 (60 deg$^2$), we used data from the 72--231 MHz GaLactic and Extragalactic All-sky MWA survey \citep[GLEAM;][]{wayth15} to conduct an HzRG search using a new radio selection technique that takes into account both spectral steepness and curvature (\citealt[][]{drouart20}; henceforth D20). From just four HzRG candidates, we discovered the radio galaxy GLEAM~J0856$+$0223 at $z=5.55$, which is the second-most distant radio galaxy currently known. The Atacama Large Millimeter/submillimeter Array \citep[ALMA;][]{wootten09} was used to determine the redshift of J0856$+$0223 via the detection of two CO molecular emission lines near 100 GHz, as opposed to the detection of redshifted Lyman alpha and/or other emission lines using classical optical/near-infrared spectroscopy. Additionally, we compiled multiwavelength data for a second source from the D20 pilot project, GLEAM~J0917$-$0012, which may be at $z \sim 2$ or $\sim 8$ \citep[][]{drouart21,seymour22}. J0856$+$0223 and J0917$-$0012 have spectral indices within the GLEAM band of $-1.01 \pm 0.04$ and $-1.00 \pm 0.06$, respectively; our technique does not require a USS spectrum (see Section~\ref{section:selection criteria} for further details).

With the caveat of small number statistics, our pilot sample selection technique has demonstrated the potential to efficiently select very distant radio galaxies. In this paper, we build on the success of the D20 pilot study by applying a refined radio/near-infrared selection technique to define a larger sample of 53 sources: 51 new HzRG candidates as well as J0856$+$0223 and J0917$-$0012 from D20. In Section~\ref{section:sample definition}, we describe how we used GLEAM and the Visible and Infrared Survey Telescope for Astronomy (VISTA; \citealt[][]{dalton06}; \citealt*[][]{emerson06}) Kilo-degree Infrared Galaxy survey \citep[VIKING;][]{edge13} to construct our sample. The 2.15-\textmu m near-infrared $K_{\rm s}$-band properties from VIKING are presented in Section~\ref{section: K-band}, along with deeper $K_{\rm s}$-band imaging for five sources from the High-Acuity Widefield $K$-band Imager \citep[HAWK-I;][] {kissler08} on the Very Large Telescope \citep[VLT;][]{eso98} and from the Southern {\it Herschel} Astrophysical Terahertz Large Area Survey \citep[H-ATLAS;][]{eales10,valiante16,bourne16} Regions $K_{\rm s}$-band Survey (SHARKS\footnote{\url{http://research.iac.es/proyecto/sharks/pages/en/home.php}}; Dannerbauer et al. in prep.). Australia Telescope Compact Array \citep*[ATCA;][]{frater92} 5.5- and 9-GHz follow-up observations are presented in Section \ref{section:ATCA}. An analysis of the radio data, including radio/near-infrared overlay plots and modelling of broadband radio spectra, can be found in Section~\ref{section:overlays}. A discussion of the sample properties then follows in Section~\ref{section:discussion}. Lastly, we present our conclusions and plans for future work in Section~\ref{section:conclusions}. 

Unless noted otherwise, all uncertainties in this paper are given as $\pm 1\sigma$. All near-infrared magnitudes are given in the AB system \citep[][]{oke74}. Throughout the paper, $\log$ refers to the decimal logarithm (base 10) and radio synthesised beam position angles (BPA) are measured north through east.  

\section{Sample definition}\label{section:sample definition}

\begin{table}
\setlength{\tabcolsep}{2.00pt}
 \centering
  \caption{Summary of our HzRG candidate sample selection. SGP and EQU refer to the two strips in the VIKING survey footprint: south Galactic pole and equatorial. Selection criteria were applied in the order specified in the table, although the steps are commutative. See Section~\ref{section:sample definition} for further details.}
  \begin{tabular}{lcc}
  \hline\hline
   \multicolumn{1}{c}{Criterion} & \multicolumn{2}{c}{No. sources} \\ 
     & SGP & EQU \\ 
   \hline
1. GLEAM source in VIKING & 15\,393 & 7684 \\
2. Single NVSS < 50\arcsec\, from GLEAM & 14\,344 & 7250 \\
3. Unresolved in NVSS & 10\,607 & 4760 \\
4. Single TGSS < 40\arcsec\:from NVSS & 6311 & 4397 \\
5. Single FIRST < 10\arcsec\:from NVSS & $\cdots$\rlap{$^{\rm a}$} & 3583 \\
6. Curved spectrum; $S_{0} \geq 40$ mJy & 2008 & 1319 \\
7. $\alpha \leq -0.7 \cap \beta \leq -0.2$ & 643 & 544 \\
8. No ALLWISE < 2\arcsec\:from TGSS/FIRST & 444 & 309 \\
9. Visual inspection of multi-wavelength & 20 & 26\rlap{$^{\rm b}$} \\
data for subset; VIKING non-detection & & \\
($K_{\rm s} \gtrsim 21.2$); radio LAS $\leq 5\arcsec$ & & \\
11. Final (extended) sample$^{\rm c}$ & 24 & 29\rlap{$^{\rm b}$} \\ 
\hline\hline
\multicolumn{3}{p{85mm}}{Notes. $^{\rm a}$Not covered by the FIRST survey. $^{\rm b}$Including J0856$+$0223 and J0917$-$0012 from the pilot study (D20). $^{\rm c}$We added back in seven sources that do not fully meet our selection criteria; see discussion in Section~\ref{section: HzRG sample}.}\\
\end{tabular}
\label{table:sample selection}
\end{table}

Table~\ref{table:sample selection} summarises how we defined our HzRG candidate sample. This selection process was very similar to the one presented in our pilot study (see D20), but using more extensive and refined criteria. We now describe the catalogues that we used and our selection criteria.    

\subsection{Input catalogues}\label{section:GLEAM VIKING selection}

\subsubsection{GLEAM}

As in D20, GLEAM was the basis catalogue for defining our sample. We used the first GLEAM extragalactic data release \citep[GLEAM Exgal;][]{hurleywalker17} as well as a deeper catalogue centered on the south Galactic pole generated from both years of GLEAM data combined \citep[GLEAM SGP;][]{franzen21}. GLEAM has an angular resolution of approximately 2\arcmin\:at 200 MHz, with flux density measurements from $20 \times 7.68$-MHz sub-bands centred at 76, 84, 92, 99, 107, 115, 122, 130, 143, 151, 158, 166, 174, 181, 189, 197, 204, 212, 220 and 227 MHz. GLEAM Exgal covers 24\,831 deg$^2$ at declination  $\delta < +30\degr$; the $5\sigma$ root-mean-square (RMS) detection threshold is $\approx 50$ mJy beam$^{-1}$ in the 170--231 MHz wideband images. GLEAM SGP covers 5113 deg$^2$ to a $5\sigma$ detection threshold $\approx 25$ mJy beam$^{-1}$ in the 200--231 MHz wideband images. Note that where data were available from both GLEAM Exgal and SGP, we used the latter catalogue only, including the source names (which can be slightly different from the Exgal release). 

\subsubsection{VIKING}\label{section:VIKING}

Given the well-known $K$--$z$ relation (see Section~\ref{section:intro}), it is well established in the literature that the efficiency of an HzRG search can be significantly improved by only selecting those sources that have $K_{\rm s}$-band magnitudes fainter than a given threshold (e.g. \citealt{ker12} and references therein). For example, the $K_{\rm s}$-band magnitudes of J0924$-$2201, J0856$+$0223 and J1530$+$1049 are $23.2 \pm 0.3$, $23.2 \pm 0.1$ and $>23.7$ $(5\sigma)$, respectively (applying $K_{\rm s, AB} \approx K_{\rm s, Vega} + 1.85$ as given in e.g. \citealt[][]{blanton07} to the reported magnitudes of J0924$-$2201 and J1530$+$1049 in \citealt{vanbreugel99} and \citealt{saxena18b}; also see D20). 

The VIKING survey was carried out in two distinct regions: an equatorial strip (EQU) and a SGP strip centred at $\delta \approx -31\fdg5$ (see further descriptions in \citealt[][]{edge13} and \citealt[][]{driver16}). The total area surveyed was $\approx 1200$ deg$^{2}$. The nominal $5\sigma$ magnitude limit is 21.2 in $K_{\rm s}$-band. VIKING overlaps with the fields from GAMA. We used reprocessed images from the GAMA collaboration (see \citealt[][]{bellstedt21} for further details).   

A total of 23\,077 GLEAM sources with complete flux density information are within the VIKING survey footprint: 15\,393 SGP sources and 7684 EQU sources (step 1 in Table~\ref{table:sample selection}). The next steps were to then narrow this list down to the best HzRG candidates.   

\subsection{Selection criteria for HzRG candidates}\label{section:selection criteria}

\subsubsection{Isolated and compact radio sources} 

In the cosmology assumed in this paper, J0924$-$2201, J0856$+$0223 and J1530$+$1049 have projected linear sizes of 7.6, 30 and 3.6 kpc, respectively (\citealt[][]{vanbreugel99}; \citealt[][]{saxena18b}; D20). These relatively small radio sources are consistent with a scenario where HzRGs are youthful, luminous radio sources that will subsequently rapidly fade away as they age and expand as a result of significant inverse-Compton losses \citep[e.g.][]{blundell99,saxena17}. However, the extent to which HzRGs expand and remain detectable may be larger than previously expected (\citealt[][]{turner18a}; Turner et al. in prep.).   

In this study, we made an assumption that is relatively common in the literature: the efficiency of an HzRG search can be improved by removing large radio sources that are most likely low-redshift interlopers \citep[e.g.][]{ker12}. Making use of data from NVSS, TGSS and FIRST, we therefore applied a number of criteria to select isolated radio sources that are also unresolved in NVSS (steps 2--5 in Table~\ref{table:sample selection}). The cross-matching radii used between the various pairs of catalogues were conservative choices based on the angular resolution of the higher-resolution catalogue in a given pair. The criteria also removed GLEAM sources with multiple matches in NVSS, TGSS and/or FIRST, i.e. the possible multiple components of extended radio sources at lower redshift. Steps 2--5 reduced the number of sources from 23\,077 to 9894.    

\subsubsection{GLEAM spectral properties: steepness and curvature}\label{section: alpha_beta selection} 

Using a similar approach to our pilot study, we then fitted a model to the GLEAM flux density data for each of the remaining sources. In the pilot project, a second-order polynomial was fitted in $\log{(S_{\nu}})$--$\log{(\nu)}$ space:  
\begin{equation}\label{eqn:loglog}
\log(S_{\nu}) = \alpha\log\left(\frac{\nu}{\nu_{0}}\right) + \beta\log^2\left(\frac{\nu}{\nu_{0}}\right) + \log(S_{0}),
\end{equation}
where $\alpha$ is the spectral index at reference frequency $\nu_{0} = 151$ MHz, i.e. at the centre of the GLEAM band, $\beta$ the curvature term, and $S_{0}$ the flux density at $\nu_{0}$. In this study, however, the fitting was done in linear space to preserve the Gaussian characteristics of the flux density errors, which is especially important for the fainter sources that we considered (see below). Equation~\ref{eqn:loglog} is then equivalent to  
\begin{equation}\label{eqn:curved}
S_{\nu} = S_{0}\left(\frac{\nu}{\nu_{0}}\right)^{\alpha}  \times 10^{\beta\log^2\left(\frac{\nu}{\nu_{0}}\right)}.     
\end{equation}
For each sub-band flux density, the uncertainty was calculated by combining the fitting uncertainty from the catalogue and the internal GLEAM flux density calibration uncertainty, the latter being 2 per cent for the targets of interest in this study \citep[][]{hurleywalker17,franzen21}. Correlations between the sub-band flux densities \citep[e.g. see][]{hurleywalker17} were not modelled; each sub-band flux density was assumed to be an independent measurement. To first order, this is not expected to affect the accuracy of the spectral steepness/curvature selection technique. We did, however, take the correlations into account when modelling the broadband radio spectra (Section~\ref{section:SED modelling}).  

The next step was to isolate those sources with significantly curved spectra (step 6 in Table~\ref{table:sample selection}). The scientific rationale for this step follows the same argument presented in D20, that is many well-studied lower-redshift radio galaxies have observed-frame radio spectra that begin to flatten or turn over at low frequencies, and by `shifting' these sources to larger distances (higher redshifts) we can predict the optimal region within the $\alpha$--$\beta$ parameter space to search for HzRGs. To carry out step 6, we used fitting criteria of (i) a reduced chi-squared goodness-of-fit statistic $\chi^2_{\rm red} < 2$ (probability of obtaining a more extreme $\chi^2_{\rm red}$ by chance is $p \approx 0.008$) and (ii) $\lvert\beta\rvert > \sigma_{\beta}$, where $\sigma_{\beta}$ is the uncertainty for $\beta$. The latter criterion generally filters out those sources that are better fitted with a single power law, as $\beta$ will be close to zero for these cases. However, for three sources in our sample, J0053$-$3256, J1037$-$0325 and J2311$-$3359, $\chi^2_{\rm red}$ is slightly larger for a curved fit compared with a single-power-law fit (i.e. the first part of Equation~\ref{eqn:curved}: $S_{\nu} = S_{0}(\nu/\nu_{0})^{\alpha}$) across the GLEAM band: $\Delta \chi^2_{\rm red}$ is in the range $2.6 \times 10^{-4}$ -- 0.036. These sources do not fully meet all of our selection criteria and we provide further details in Section~\ref{section: HzRG sample}. More generally, a comparison of the reduced chi-squared values for curved and single-power-law fits suggests that we may have overfitted about 6 per cent of the sources classified as having curved GLEAM spectra.    

In addition, we used a fitted flux density cutoff $S_0 \geq 40$ mJy, i.e. an order of magnitude fainter than in the D20 pilot project. Such a cutoff enables the discovery of less luminous sources such as J1530$+$1049, in addition to powerful radio galaxies such as J0856$+$0223 and J0924$-$2201 (and possibly J0917$-$0012). We also removed the $S_{151 \rm MHz} < 1.0$ Jy upper flux density cutoff that was used in the pilot project. For example, some radio galaxies with $z > 4$ have $S_{150} > 1$ Jy (see e.g. Table~4 in \citealt[][]{saxena18b}). 

\begin{figure*}
\begin{minipage}{1.0\textwidth}
\centering
\includegraphics[height=13.5cm]{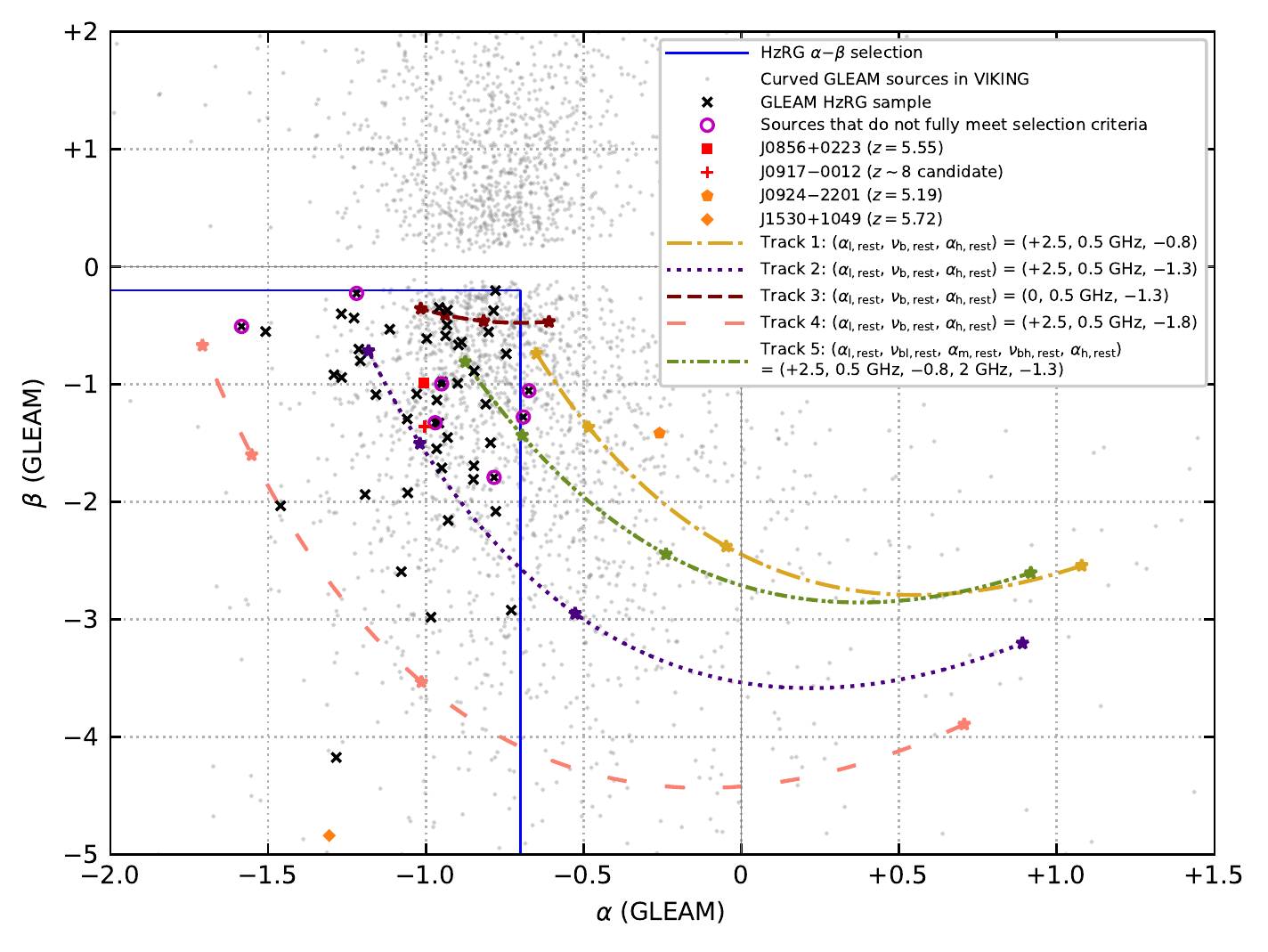}
\end{minipage}
\caption{The $\alpha$--$\beta$ parameter space for all GLEAM sources in the VIKING survey area that have curved low-frequency radio spectra (see Section~\ref{section: alpha_beta selection}). Error bars on individual data points are not shown for the sake of clarity. The paucity of sources near $\beta=0$ is due to our $\lvert\beta\rvert > \sigma_{\beta}$ selection criterion. HzRG candidates were mostly selected in the region of the plot delineated by the blue solid lines ($\alpha \leq -0.7 \cap \beta \leq -0.2$). The seven sources that do not fully meet our selection criteria are marked (see discussion in Section~\ref{section: HzRG sample}). We do not show the full range of $\alpha$ and $\beta$ values in the plot; indeed, extreme values often correspond to much poorer fits. One source in our sample, J1141$-$0158, is not visible in the figure as it has a very large curvature term ($\beta = -6.5 \pm 2.7$); in this case, the fit is likely significantly affected by the S/N in the lower part of the GLEAM band (see Figure~\ref{fig:overlays_spectra}). J0856$+$0223 and J0917$-$0012 from the pilot study, also in our larger sample, are marked separately. Furthermore, we show the HzRGs J0924$-$2201 ($\alpha = -0.26 \pm 0.04$; $\beta = -1.42 \pm 0.27$) and J1530$+$1049 ($\alpha = -1.31 \pm 0.28$; $\beta = -4.8 \pm 2.2$), both of which are detected in GLEAM in other parts of the sky not covered by VIKING. We can see that J0924$-$2201 would be missed by our selection criteria; this is because the source turns over in the middle of the GLEAM band (Figure~11 in \citealt[][]{callingham17}; also see  Section~\ref{section:discussion spectra2} and Figure~\ref{fig:all_sed} in this paper). J1530$+$1049 has a similar USS spectral index in the GLEAM band compared with its spectrum between TGSS and NVSS/FIRST ($\alpha^{1400}_{147.5} = -1.4 \pm 0.1$; \citealt[][]{saxena18a,saxena18b}), and $\lvert\beta\rvert$, while large, also has a large uncertainty due to low S/N measurements at 92, 204 and 212 MHz. We analyse the broadband spectrum of J1530$+$1049 in Section~\ref{section:discussion spectra2} and Figure~\ref{fig:all_sed}. Lastly, we plot five tracks corresponding to the predicted observed-frame values of $\alpha$ and $\beta$ of a model source at a given redshift with particular rest-frame spectral indices and break frequencies. Along each track, redshift increases from right to left, and the star markers indicate $z=$ 2, 4, 6 and 8. The tracks shown are not an exhaustive range of possibilities, but instead illustrate how the track trajectories can change depending on the underlying rest-frame spectrum of the source. The track model parameters correspond to the smoothly varying double- and triple-power-law fits that we use in this paper to model the broadband spectra of our targets (Equations~\ref{fit:dpl} and \ref{fit:tpl}). Tracks 1, 2, 3, 4 and 5 move into our selection region at redshifts of 9.3 (not shown in the panel), 4.5, 2.7, 3.4 and 6.0, respectively. The track trajectories would continue to the upper left if the redshift is increased. If the (lower) rest-frame break frequency is below/above 500 MHz, then a particular track trajectory is still the same as plotted, but the position corresponding to a given redshift is further along the track to the left/right.} 
\label{fig:alpha_beta}
\end{figure*}

The distribution of the remaining 3327 sources in the $\alpha$--$\beta$ parameter space is shown in Figure~\ref{fig:alpha_beta}. Using the same argument as in the pilot study for the tracks that sources follow in this parameter space as they are progressively redshifted (Figure~1 in D20; additional examples shown in Figure~\ref{fig:alpha_beta}), our next selection criterion (step 7 in Table~\ref{table:sample selection}) was $\alpha \leq -0.7 \cap \beta \leq -0.2$. The first part of this expression is almost identical to the spectral steepness criterion used in D20 ($\alpha < -0.7$), while the second part was relaxed from $-1.0 < \beta < -0.4$ in the pilot project to a wider range, given the potential trajectories of the tracks mentioned above. The total number of sources that remained after this step was 1187.  

Although we applied a number of selection criteria above to restrict the list of sources to those that are compact and isolated, source blending remains a potential issue that must be considered carefully given the relatively low angular resolution of the GLEAM data. This is particularly relevant regarding the reliability of the $\alpha$ and $\beta$ measurements. We discuss this potential issue further in Section~\ref{section:blending}. 

\subsubsection{Further selection criteria}\label{section: other selection criteria} 

To further reduce the fraction of low-redshift interlopers in our sample, step 8 in Table~\ref{table:sample selection} was to remove those sources with mid-infrared detections in data from the {\it Widefield Infrared Survey Explorer} \citep[{\it WISE};][]{wright10}, in particular the AllWISE \citep[][]{cutri14} data release. We searched for mid-infrared counterparts within 2\arcsec\:from the radio position in FIRST for EQU sources and TGSS for SGP sources. This left a reasonable number of sources (753 in total) that could potentially be examined in more detail, particularly visual inspection of multi-wavelength data (step 9 in Table~\ref{table:sample selection}). An important caveat for step 9, however, is that while our sample selection technique is designed to be efficient, {\em it is not complete}. In particular, there were considerations regarding follow-up observing campaigns, for example ensuring an adequate typical signal-to-noise ratio (S/N) of the ATCA data that we present in Section~\ref{section:ATCA}. In practice, we visually inspected about half of the 753 sources. 

For the visual inspection, the primary check was to overlay radio contours on the VIKING $K_{\rm s}$-band images. The goal was to identify those sources that were sufficiently compact in the radio and not detected in VIKING at the $5\sigma$ level (the latter confirmed from analysis of the reprocessed images and not from the latest VIKING catalogue available in the literature, i.e. \citealt[][]{edge16}); these sources were then deemed to be the best HzRG candidates for further follow-up and analysis.  We used the radio data from TGSS, NVSS and FIRST that had been considered in the previous steps as well as higher-resolution radio data that became available during the course of our analysis: 887.5-MHz images from the Rapid Australian Square Kilometre Array Pathfinder \citep[ASKAP;][]{johnston07,hotan21} Continuum Survey \citep[RACS;][]{mcconnell20,hale21} and 3-GHz `quick-look' images from the first epoch of the VLA Sky Survey \citep[VLASS;][]{lacy20}. Furthermore, for the $K_{\rm s}$-band non-detections, we also inspected the corresponding reprocessed $J$-band ($1.25$ \textmu m) and $H$-band ($1.64$ \textmu m) VIKING images from the GAMA collaboration (nominal $5\sigma$ magnitude limits of 22.1 and 21.5, respectively) to check that each host galaxy was not detected in these bands either at the $5\sigma$ level. 

After confirming the radio morphology in each case, we removed all sources with a largest angular size\footnote{In this paper, for a single-component source, the LAS is considered to be the deconvolved major axis full width at half maximum (FWHM) of the source. For a resolved multi-component source, the LAS is considered to be the maximum angular distance between the centroids of two components.} (LAS) $>5\arcsec$ in FIRST and/or VLASS (e.g. projected linear size $< 32.1$ kpc at $z > 5$). If there was sufficient uncertainty regarding the angular extent of the radio emission, particularly if a potential HzRG candidate was instead possibly a single component of a larger radio source, we erred on the side of caution and removed the candidate in question from the sample. Our LAS cutoff is a somewhat arbitrary choice, but such a cutoff can improve the efficiency of an HzRG search \citep[e.g.][]{ker12}. From the ALMA data presented in D20, LAS $= 4\farcs9$ for J0856$+$0223, which is the largest of the three currently known radio galaxies at $z > 5$. Assuming the modelling of \citet[][]{saxena17}, a selection criterion of LAS $\leq 5\arcsec$ would be sensitive to radio galaxies at redshifts beyond that of J1530$+$1049 at $z=5.72$ (see Figure~15 in \citealt[][]{saxena19}). One caveat is that we would then filter out larger HzRGs that may exist in the early Universe if the jets grow on shorter time-scales than predicted in the \citet[][]{saxena17} framework, as is suggested from modelling based on different assumptions (\citealt[][]{turner18a}; Turner et al. in prep.). 

Apart from searching for VIKING non-detections, we also inspected images, where available, from the second data release of the Hyper Suprime-Cam \citep[][]{miyazaki18} Subaru Strategic Program \citep[][]{aihara18a,aihara19}, removing any sources from our candidate list with deep $Y$-band ($0.98$ {\textmu}m) detections ($5\sigma$ magnitude limit $= 24.2$ in a 2\arcsec-diameter aperture). We also searched the ATNF pulsar database \citep[][]{manchester05}\footnote{\url{https://www.atnf.csiro.au/research/pulsar/psrcat/}}, finding no known pulsars at the positions of the HzRG candidates presented in this paper.   

Lastly, when we had nearly finalised our sample, some data became available from SHARKS, which is a deep survey of $\sim$300 deg$^2$ conducted with the VISTA telescope, covering several H-ATLAS fields, with a target $5\sigma$ magnitude limit of $K_{\rm s} \sim 22.7$. Details on the survey strategy and data reduction will be presented in Dannerbauer et al. in prep. The data that we used were from observations conducted between 2017 March and 2019 January. We removed sources with SHARKS detections and noted those sources with SHARKS non-detections, which remained in the sample (discussed further in Section~\ref{section: K-band}). 

\subsection{The GLEAM--VIKING HzRG candidate sample}\label{section: HzRG sample}

After applying the criteria described in Section~\ref{section: other selection criteria}, we were left with a sample of 53 sources within the VIKING survey region (step 10 in Table~\ref{table:sample selection}). The sample comprises 51 new HzRG candidates and both J0856$+$0223 and J0917$-$0012 from our pilot study (Section~\ref{section:intro}), which satisfy the refined selection criteria used in this paper.\footnote{However, the two low-redshift interlopers from the D20 study are filtered out by our selection criteria.} Note that the sample includes six sources that do not fully meet the $\alpha$ and $\beta$ selection criteria: J0034$-$3112, J0053$-$3256, J0129$-$3109, J1037$-$0325, J1246$-$0017 and J2311$-$3359. For all of these sources bar one, this is because (i) they marginally fall outside of our selection region in $\alpha$--$\beta$ parameter space (Figure~\ref{fig:alpha_beta}), yet within the $1\sigma$ uncertainties they are consistent with the selection criteria, or (ii) $\lvert\beta\rvert$ is marginally smaller than $\sigma_{\beta}$. Our original approach for the $\alpha$--$\beta$ fitting had been to use Equation~\ref{eqn:loglog}, and due to the propagation of the flux density uncertainties in $\log{(S_{\nu}})$--$\log{(\nu)}$ space, these sources initially fully satisfied our selection criteria and had been followed up with the ATCA (Section~\ref{section:ATCA}). The remaining source, J2311$-$3359 from the GAMA-23 field, was removed in step 2 in Table~\ref{table:sample selection}, as it is too faint to be detected in NVSS; moreover, it does not satisfy our $\lvert\beta\rvert > \sigma_{\beta}$ criterion, and ASKAP early science data suggests an LAS of 5\farcs6 (see Section~\ref{section:notes sources}). J2311$-$3359 was identified as an HzRG candidate of particular interest given its USS nature ($\alpha = -1.58 \pm 0.19$ in GLEAM) and followed up with the ATCA before the selection criteria for this project were fully defined. 

In addition to these six sources, our sample includes the source J1335$+$0112. This source meets all of our selection criteria except that it has an AllWISE identification (which we discuss later in Section~\ref{section:notes sources}). Unfortunately, this source was followed up with the ATCA before our 2\arcsec AllWISE criteria was introduced (it had been 1\arcsec). As we outline in Section~\ref{section:notes sources}, J1335$+$0112 may still be a high-redshift target, and therefore we decided that there was sufficient scientific interest to include the source in our sample.   

The sample is presented in Table~\ref{table:fluxes}, including the $\alpha$ and $\beta$ values for each source and the fitted 151-MHz GLEAM flux density. The fitted 151-MHz flux densities span the range 47.6--2439 mJy, with a median of 260.6 mJy. We have marked the locations of the sources in our sample in the $\alpha$--$\beta$ parameter space in Figure~\ref{fig:alpha_beta}. The median spectral index of the sample at 151 MHz is $-0.96$, and only three sources would be traditionally classified as USS with $\alpha \leq -1.3$: J0007$-$3040, J2311$-$3359 (discussed above) and J2314$-$3517. 

\begin{sidewaystable*}
\footnotesize
\setlength{\tabcolsep}{3.5pt}
\begin{minipage}{1.0\textwidth}
  \caption{Radio properties of our sample listed in order of right ascension. We list whether the source is in the equatorial VIKING strip (i.e. covered by GLEAM Exgal) or the south Galactic pole strip (i.e. covered by GLEAM SGP). Here we report the fitted $151$-MHz GLEAM flux density from Section~\ref{section: alpha_beta selection} only; the full set of catalogued GLEAM flux densities can be found in \citet[][]{hurleywalker17} and \citet[][]{franzen21}. TGSS flux densities at 147.5 MHz are from a rescaled version of this catalogue \citep[][]{hurleywalker17b}. FIRST and NVSS 1400-MHz flux densities are reported (denoted by the subscripts F and N, respectively). The SUMSS and NVSS flux density upper limits for J2311$-$3359 are at the $3\sigma$ level. Various table entries with a horizontal ellipsis ($\cdots$) indicate that observations are not available at a particular frequency because either the source is not located within the sky coverage of the survey (SUMSS and FIRST) or observations were not taken (ATCA). For the SGP sources, we report the ATCA and VLASS LAS values (in this order); similarly, for the equatorial sources, we report the FIRST and VLASS LAS values (again in this order). For the ATCA LAS measurements, where we have two observing frequencies, we took into account both angular resolution and S/N so as to choose the best LAS measurement from a given 5.5- and 9-GHz pair; this was generally the value at 9 GHz for sufficiently high S/N. FIRST and VLASS LAS measurements are from \citet[][]{helfand15} and \citet[][]{gordon20}, respectively. LAS upper limits were calculated using a $5\sigma$ upper bound on the deconvolved major axis FWHM (from the {\sc pybdsf} output, or, for the FIRST LAS upper limit for J1246$-$0017, following e.g. \citealt[][]{fomalont99}), except for J2311$-$3559 where the upper limit is taken as the 9-GHz synthesised beam major axis FWHM.  See Sections~\ref{section:sample definition}, \ref{section:ATCA} and \ref{section:overlays} for further details.}
  \begin{tabular}{ccrrrrrrrrrrrr}
  \hline\hline
  Source & Region &  \multicolumn{1}{c}{$\alpha$} & \multicolumn{1}{c}{$\beta$} & \multicolumn{1}{c}{$S_{147.5}$} & \multicolumn{1}{c}{$S_{151}$} & \multicolumn{1}{c}{$S_{843}$} & \multicolumn{1}{c}{$S_{887.5}$} &  \multicolumn{1}{c}{$S_{1400{\rm,\,F}}$} & \multicolumn{1}{c}{$S_{1400{\rm,\,N}}$} & \multicolumn{1}{c}{$S_{3000}$} & \multicolumn{1}{c}{$S_{5500}$} & \multicolumn{1}{c}{$S_{8800/9000}$} & \multicolumn{1}{c}{LAS} \\ 
 \multicolumn{1}{c}{(GLEAM)} & & \multicolumn{2}{c}{(GLEAM)} & \multicolumn{1}{c}{(mJy)} &  \multicolumn{1}{c}{(mJy)} &  \multicolumn{1}{c}{(mJy)} &  \multicolumn{1}{c}{(mJy)} &  \multicolumn{1}{c}{(mJy)$^{\rm a}$}  &  \multicolumn{1}{c}{(mJy)}  &  \multicolumn{1}{c}{(mJy)}  &  \multicolumn{1}{c}{(mJy)}  & \multicolumn{1}{c}{(mJy)} & \multicolumn{1}{c}{(\arcsec)} \\ 
   \hline
J000216$-$351433 & SGP &  $ -1.21 \pm 0.09 $ & $ -0.80 \pm 0.63 $ & $140 \pm 15$ & $139.7 \pm 3.6$ & $30.3 \pm 2.0$ & $24.1 \pm 2.3$ & $\cdots$ & $13.4 \pm 0.6$ & $4.7 \pm 0.6$ & $2.9 \pm 0.2$ & $1.62 \pm 0.13$ & $1.4,1.5$ \\
J000614$-$294640 & SGP & $ -1.23 \pm 0.05 $ & $ -0.44 \pm 0.39 $ & $177 \pm 19$ & $175.7 \pm 3.0$ & $\cdots$ & $33.3 \pm 2.9$ & $\cdots$ & $19.4 \pm 0.7$ & $9.9 \pm 1.1$ & $4.4 \pm 0.3$ & $2.30 \pm 0.17$ & $0.4,1.5$ \\
J000737$-$304030 & SGP & $ -1.51 \pm 0.02 $ & $ -0.55 \pm 0.16 $ & $633 \pm 65$ & $599.0 \pm 4.9$ & $46.3 \pm 2.2$ & $43.8 \pm 3.6$ & $\cdots$ & $15.9 \pm 0.7$ & $4.5 \pm 0.7$ & $1.04 \pm 0.12$ & $0.20 \pm 0.03$\rlap{$^{\rm b}$} & $3.0,3.0$ \\
J000845$-$300731 & SGP & $ -1.16 \pm 0.03 $ & $ -1.09 \pm 0.27 $ &  $292 \pm 31$ & $289.5	\pm	3.4$ & $30.2 \pm 1.8$ & $32.4 \pm 2.8$ & $\cdots$ & $14.0 \pm 0.7$ & $4.8 \pm 0.6$ & $1.37 \pm 0.12$ & $0.53 \pm 0.08$  & $<0.6,2.1$ \\
J003402$-$311210\rlap{$^{\rm \dag}$} & SGP & $ -0.97 \pm 0.19 $ & $ -1.3 \pm 1.4 $ &  $46.4 \pm 6.4$ & $47.6	\pm	2.7$ & $9.0 \pm 3.3$\rlap{$^{\rm b}$} & $11.7 \pm 1.4$ & $\cdots$ & $5.6 \pm 0.5$ & $3.5 \pm 0.5$ & $1.69 \pm 0.14$ &  $1.09 \pm 0.15$ & $1.3,2.3$ \\
J004219$-$351516 & SGP & $ -1.19 \pm 0.16 $ & $ -1.9 \pm 1.2 $ &  $71.2 \pm 8.4$ & $74.9	\pm	3.8$ & $11.3 \pm 1.7$ & $14.4 \pm 2.1$ & $\cdots$ & $8.0 \pm 0.5$ & $4.0 \pm 0.5$ & $1.96 \pm 0.16$ & $0.90 \pm 0.11$ & $<0.3,2.8$ \\
J004828$-$354005 & SGP & $ -1.27 \pm 0.04 $ & $ -0.40 \pm 0.33 $ &  $226 \pm 23$ & $216.2	\pm	3.2$ & $23.2 \pm 1.4$ & $22.2 \pm 2.2$ & $\cdots$ & $11.0 \pm 0.6$ & $5.5 \pm 0.8$ & $1.95 \pm 0.15$ & $1.04 \pm 0.12$ & $1.5,4.1$ \\
J005332$-$325630\rlap{$^{\rm \dag}$} & SGP & $ -1.22 \pm 0.03 $ & $ -0.23 \pm 0.25 $ &  $260 \pm 26$ & $260.6	\pm	3.1$ & $30.3 \pm 1.7$ & $29.3 \pm 2.6$ & $\cdots$ & $14.7 \pm 0.6$ & $5.2 \pm 0.6$ & $1.76 \pm 0.14$ & $0.66 \pm 0.11$ & $0.9,2.4$ \\
J010826$-$350157 & SGP & $ -0.79 \pm 0.04 $ & $ -0.37 \pm 0.28 $ &  $364 \pm 37$ & $325.5	\pm	4.2$ & $85.9 \pm 3.0$ & $80.7 \pm 6.2$ & $\cdots$ & $47.0 \pm 1.5$ & $21.5 \pm 2.2$ & $9.4 \pm 0.7$ & $4.5 \pm 0.3$ & $1.0,2.2$ \\
J012929$-$310915\rlap{$^{\rm \dag}$} & SGP & $ -0.67 \pm 0.07 $ & $ -1.05 \pm 0.56 $ &  $113 \pm 13$ & $115.0	\pm	2.4$ & $36.2 \pm 2.0$ & $37.8 \pm 3.2$ & $\cdots$ & $27.2 \pm 0.9$ & $15.8 \pm 1.6$ & $9.2 \pm 0.7$ & $5.7 \pm 0.4$ & $0.2,1.5$ \\
J013340$-$305638 & SGP & $ -0.89 \pm 0.03 $ & $ -0.64 \pm 0.20 $ & $ 366 \pm	37 $ & $423.4	\pm	3.9$ & $ 81.7 \pm 3.3 $ & $ 73.3 \pm 5.7 $ & $\cdots$ & $ 43.3 \pm 1.4 $ & $ 19.9 \pm 2.1 $ & $\cdots$ & $\cdots$ & $ 4.3$\rlap{$^{\rm c}$} \\ 
J020118$-$344100 & SGP & $ -1.06 \pm 0.10 $ & $ -1.29 \pm 0.75 $ &  $98 \pm 11$ & $95.6	\pm	2.9$ & $23.8 \pm 1.6$ & $19.4 \pm 1.9$ & $\cdots$ & $10.0 \pm 0.6$ & $6.0 \pm 0.9$ & $2.7 \pm 0.2$ & $1.38 \pm 0.15$ & $2.1,3.9$ \\
J021618$-$330148 & SGP & $ -0.81 \pm 0.14 $ & $ -1.2 \pm 1.0 $ &  $63.5 \pm 8.8$ & $79.7	\pm	3.3$ & $22.9 \pm 1.8$ & $21.0 \pm 2.0$ & $\cdots$ & $12.5 \pm 0.6$ & $7.1 \pm 0.8$ & $3.9 \pm 0.3$ & $2.16 \pm 0.17$	& $1.8,2.4$ \\
J023937$-$304337 & SGP & $ -0.95 \pm 0.06 $ & $ -0.98 \pm 0.43 $ &  $268 \pm 27$ & $223.4	\pm	3.9$ & $37.7 \pm 2.2$ & $37.6 \pm 3.2$ & $\cdots$ & $19.9 \pm 0.8$ & $8.1 \pm 0.9$ & $3.2 \pm 0.2$ & $1.53 \pm 0.12$ & $1.1,1.6$ \\
J024019$-$320659 & SGP & $ -0.85 \pm 0.17 $ & $ -1.8 \pm 1.4 $ &  $35.6 \pm 7.8$ & $58.8	\pm	2.8$ & $13.1 \pm 1.5$ & $12.3 \pm 1.4$ & $\cdots$ & $5.9 \pm 0.5$ & $4.0 \pm 0.5$ & $2.10 \pm 0.16$ & $1.29 \pm 0.10$ & $0.9,2.3$ \\
J030108$-$313211 & SGP & $ -0.75 \pm 0.05 $ & $ -0.74 \pm 0.35 $ &  $228 \pm 24$ & $243.7	\pm	3.5$ & $69.4 \pm 2.8$ & $68.9 \pm 5.4$ & $\cdots$ & $45.1 \pm 1.4$ & $23.1 \pm 2.4$ & $11.1 \pm 0.8$ & $6.0 \pm 0.4$ & $1.0^{\rm b},2.0$ \\
J030931$-$352623 & SGP & $ -0.93 \pm 0.12 $ & $ -2.16 \pm 0.97 $ &  $80.8 \pm 9.8$ & $93.5	\pm	3.2$ & $13.8 \pm 1.3$ & $13.2 \pm 1.5$ & $\cdots$ & $6.0 \pm 0.5$ & $3.1 \pm 1.2$\rlap{$^{\rm b}$} & $0.8 \pm 0.2$\rlap{$^{\rm b}$} & $0.2 \pm 0.1$\rlap{$^{\rm b}$} & $5^{\rm b},5$\rlap{$^{\rm b}$} \\
J032634$-$301359 & SGP & $ -0.78 \pm 0.02 $ & $ -0.20 \pm 0.17 $ &  $750 \pm 75$ & $744.0	\pm	6.1$ & $188.2 \pm 5.9$ & $171 \pm 12$ & $\cdots$ & $98.9 \pm 3.0$ & $38.5 \pm 3.9$  & $17.8 \pm 1.3$ & $8.6 \pm 0.6$ & $<0.1,<0.1$ \\
J084256$-$015722 & EQU & $-0.94 \pm	0.06$ & $-0.59 \pm	0.43$  & $382 \pm 39$ & $435.9	\pm	9.5$ & $\cdots$  & $ 83.6 \pm 6.4 $ & $ 54.2 $ & $ 51.7 \pm	1.6 $ & $21.5 \pm 2.2$ & $\cdots$ & $\cdots$ & $1.7,1.1$ \\
J085614$+$022359\rlap{$^{\rm d}$} & EQU & $-1.01 \pm	0.04$ &	$-0.99 \pm	0.25$ & $852	\pm	85$  & $844	\pm	12$ & $\cdots$ & $ 148 \pm 11 $ & $87.7$ & $86.5 \pm 2.6$ & $ 30.3 \pm 3.1 $ & $15.5 \pm 1.6$ & $7.6 \pm 0.8$ & $ 2.8,3.4 $ \\
J090942$-$015409 & EQU & $ -0.93 \pm 0.04 $ & $ -1.45 \pm 0.31 $ &  $848 \pm 85$ & $807	\pm	12$ & $\cdots$ & $65.3 \pm 5.1$ & $31.5$ & $30.1 \pm 1.0$ & $9.0 \pm 1.0$ & $3.5 \pm  0.3$ & $1.53 \pm 0.12$ & $1.6,1.0$ \\
J091734$-$001243\rlap{$^{\rm d}$} & EQU & $-1.00 \pm	0.06$ & $-1.36 \pm	0.44$ & $459	\pm	46$  & $477	\pm	11$ & $\cdots$ & $84.8 \pm 6.5$  & $49.0$ &	$46.6 \pm 1.5$ & $ 18.5 \pm 1.9 $ & $7.7 \pm 0.8$ & $3.5 \pm 0.4$ & $ 1.2,0.9 $ \\
J103055$+$013519 & EQU & $ -0.78 \pm 0.12 $ & $ -2.1 \pm 1.0 $ &  $207 \pm 21$ & $213	\pm	10$ & $\cdots$ & $73.9 \pm 5.7$ & $55.8$ & $53.6 \pm 1.7$ & $32.7 \pm 3.3$ & $20.2 \pm  1.4$ & $12.9 \pm 0.9$ & $1.8,0.5$ \\
J103223$+$033933 & EQU & $ -1.11 \pm 0.03 $ & $ -0.53 \pm 0.25 $ &  $829 \pm 83$ & $863	\pm	12$ & $\cdots$ & $145 \pm 11$ & $77.7$ & $76.4 \pm 2.3$ & $30.1 \pm 3.0$ & $13.6 \pm  1.0$ & $6.9 \pm 0.5$ & $3.1,3.2$ \\
J103340$+$010725 & EQU & $ -0.90 \pm 0.07 $ & $ -0.99 \pm 0.51 $ &  $339 \pm 34$ & $386	\pm	10$ & $\cdots$ & $78.0 \pm 6.0$ & $48.7$ & $47.2 \pm 1.5$ & $22.3 \pm 2.3$ & $11.0 \pm  0.8$ & $6.1 \pm 0.4$ & $1.3,0.9$ \\
J103747$-$032519\rlap{$^{\rm \dag}$} & EQU & $ -0.78 \pm 0.25 $ & $ -1.8 \pm 1.9 $ &  $78 \pm 10$ & $101.4	\pm	8.4$ & $\cdots$ & $26.9 \pm 2.5$ & $12.3$ & $16.3 \pm 0.6$ & $6.0 \pm 0.7$ & $4.4 \pm  0.4$ & $2.5 \pm 0.2$ & $3.4,2.6$ \\
J104041$+$015003 & EQU & $ -1.27 \pm 0.06 $ & $ -0.94 \pm 0.44 $ & $486 \pm 49$ & $434	\pm	10$ & $\cdots$ & $46.7 \pm 3.9$ & $22.7$ & $18.7 \pm 0.7$ & $7.4 \pm 0.8$ & $2.5 \pm  0.3$ & $1.05 \pm 0.11$ & $1.8,1.6$ \\
J105232$-$031808 & EQU & $ -1.06 \pm 0.16 $ & $ -1.9 \pm 1.2 $ &  $124 \pm 14$ & $159.9	\pm	8.4$ & $\cdots$ & $21.1 \pm 2.0$ & $9.4$ & $11.0 \pm 0.5$ & $4.4 \pm 0.6$ & $2.12 \pm  0.19$ & $0.97 \pm 0.12$ & $1.5,1.8$ \\
J111211$+$005607 & EQU & $ -0.85 \pm 0.16 $ & $ -1.7 \pm 1.3 $ &  $139 \pm 14$ & $177	\pm	11$ & $\cdots$ & $30.3 \pm 2.7$ & $19.5$ & $17.1 \pm 0.7$ & $9.5 \pm 1.0$ & $4.8 \pm  0.4$ & $2.9 \pm 0.2$ & $2.0,1.9$ \\
J112557$-$034203 & EQU & $ -0.93 \pm 0.05 $ & $ -0.37 \pm 0.35 $ &  $507 \pm 51$ & $503.7	\pm	8.7$ & $\cdots$ & $102.5 \pm 7.7$ & $57.4$ & $58.4 \pm 1.8$ & $26.8 \pm 2.7$ & $14.3 \pm  1.0$ & $8.4 \pm 0.6$ & $2.8,2.9$ \\
J112706$-$033210 & EQU & $ -1.21 \pm 0.08 $ & $ -0.70 \pm 0.53 $ &  $293 \pm 30$ & $313.2	\pm	8.3$ & $\cdots$ & $48.7 \pm 4.0$ & $24.3$ & $25.9 \pm 0.9$ & $10.2 \pm 1.1$ & $5.8 \pm  0.4$ & $3.1 \pm 0.2$ & $2.7,2.5$ \\
J113601$-$035122 & EQU & $ -0.79 \pm 0.14 $ & $ -1.5 \pm 1.0 $ &  $214 \pm 22$ & $185.6	\pm	8.8$ & $\cdots$ & $31.5 \pm 2.8$ & $17.1$ & $15.1 \pm 0.6$ & $7.3 \pm 0.8$ & $3.5 \pm  0.3$ & $2.01 \pm 0.17$ & $1.8,1.6$ \\
J114103$-$015846 & EQU & $ -1.10 \pm 0.33 $ & $ -6.5 \pm 2.7 $ &  $66.5 \pm 9.3$ & $109.6	\pm	9.9$ & $\cdots$ & $26.6 \pm 2.6$ & $16.4$ & $15.8 \pm 0.6$ & $7.1 \pm 0.8$ & $3.7 \pm  0.3$ & $2.09 \pm 0.15$ & $1.7,1.2$ \\
J121103$-$025603 & EQU & $ -0.96 \pm 0.13 $ & $ -1.33 \pm 0.89 $ &  $223 \pm 23$ & $293	\pm	11$ & $\cdots$ & $45.4 \pm 3.7$ & $22.3$ & $22.9 \pm 0.8$ & $10.1 \pm 1.1$ & $3.9 \pm  0.3$ & $1.69 \pm 0.14$\rlap{$^{\rm e}$} & $3.6,5.0$ \\
\hline\hline
\end{tabular}
\label{table:fluxes}
\end{minipage}
\end{sidewaystable*}

\setcounter{table}{1} 
\begin{sidewaystable*}
\footnotesize
\setlength{\tabcolsep}{3.5pt}
\begin{minipage}{1.0\textwidth}
\caption{{\em - continued.}}
\begin{tabular}{ccrrrrrrrrrrrr}
  \hline\hline
  Source & Region & \multicolumn{1}{c}{$\alpha$} & \multicolumn{1}{c}{$\beta$} & \multicolumn{1}{c}{$S_{147.5}$} & \multicolumn{1}{c}{$S_{151}$} & \multicolumn{1}{c}{$S_{843}$} & \multicolumn{1}{c}{$S_{887.5}$} & \multicolumn{1}{c}{$S_{1400{\rm,\,F}}$} & \multicolumn{1}{c}{$S_{1400{\rm,\,N}}$} & \multicolumn{1}{c}{$S_{3000}$} & \multicolumn{1}{c}{$S_{5500}$} & \multicolumn{1}{c}{$S_{8800/9000}$} & \multicolumn{1}{c}{LAS} \\
 \multicolumn{1}{c}{(GLEAM)} & & \multicolumn{2}{c}{(GLEAM)} & \multicolumn{1}{c}{(mJy)} &  \multicolumn{1}{c}{(mJy)} &  \multicolumn{1}{c}{(mJy)}  & \multicolumn{1}{c}{(mJy)}  &  \multicolumn{1}{c}{(mJy)$^{\rm a}$}  &  \multicolumn{1}{c}{(mJy)}  &  \multicolumn{1}{c}{(mJy)}  &  \multicolumn{1}{c}{(mJy)}  & \multicolumn{1}{c}{(mJy)} & \multicolumn{1}{c}{(\arcsec)} \\
   \hline
J124617$-$001741\rlap{$^{\rm \dag}$} & EQU & $ -0.69 \pm 0.15 $ & $ -1.3 \pm 1.0 $ &  $235 \pm 24$ & $279	\pm	13$ & $\cdots$ & $82.5 \pm 6.3$ & $51.0$ & $49.7 \pm 1.5$ & $26.4 \pm 2.7$ & $14.5 \pm  1.1$ & $8.0 \pm 0.6$\rlap{$^{\rm e}$} & $<0.09,0.6$ \\   
J131748$+$033906 & EQU & $ -1.29 \pm 0.11 $ & $ -0.92 \pm 0.66 $ &  $359 \pm 36$ & $358	\pm	12$ & $\cdots$ & $67.5 \pm 5.3$ & $39.2$ & $37.8 \pm 1.2$ & $17.9 \pm 1.9$ & $8.3 \pm  0.6$ & $4.4 \pm 0.3$\rlap{$^{\rm e}$} & $2.2,2.6$ \\
J132918$+$013341 & EQU & $ -0.90 \pm 0.06 $ & $ -0.67 \pm 0.40$ &  $695 \pm 70$ & $666	\pm	14$ & $\cdots$ & $161 \pm 12$ & $102.6$ & $100.9 \pm 3.1$ & $51.1 \pm 5.1$ & $26.1 \pm  1.9$ & $15.3 \pm 1.1$\rlap{$^{\rm e}$} & $3.4,3.0$ \\
J133531$+$011219\rlap{$^{\rm \dag}$} & EQU & $ -0.95 \pm 0.08 $ & $ -1.00 \pm 0.54 $ &  $451 \pm 45$ & $453	\pm	12$ & $\cdots$ & $111.1 \pm 8.3$ & $64.6$ & $64.5 \pm 2.0$ & $31.2 \pm 3.1$ & $15.4 \pm  1.1$ & $8.2 \pm 0.6$\rlap{$^{\rm e}$} & $2.0,2.2$ \\
J133744$+$032813 & EQU & $ -0.97 \pm 0.12 $ & $ -1.55 \pm 0.89 $ &  $253 \pm 26$ & $274	\pm	12$ & $\cdots$ & $75.7 \pm 5.8$ & $50.5$ & $49.5 \pm 1.5$ & $24.8 \pm 2.5$ & $13.2 \pm  1.0$ & $7.1 \pm 0.5$\rlap{$^{\rm e}$} & $1.7,0.9$ \\
J134030$+$000953 & EQU & $ -1.08 \pm 0.11 $ & $ -2.59 \pm 0.81 $ &  $322 \pm 33$ & $328	\pm	12$ & $\cdots$ & $79.2 \pm 6.1$ & $49.2$ & $48.0 \pm 1.5$ & $25.4 \pm 2.6$ & $14.1 \pm  1.0$ & $8.3 \pm 0.6$\rlap{$^{\rm e}$} & $4.8,4.6$ \\
J134747$+$001243 & EQU & $ -0.93 \pm 0.07 $ & $ -0.50 \pm 0.48 $  &  $511 \pm 51$ & $495	\pm	12$ & $\cdots$ & $105.3 \pm 7.9$ & $63.5$ & $59.3 \pm 1.8$ & $25.5 \pm 2.6$ & $10.9 \pm  0.8$ & $5.4 \pm 0.4$\rlap{$^{\rm e}$} & $1.8,1.1$ \\
J134912$+$022200 & EQU & $ -0.95 \pm 0.14 $ & $ -1.7 \pm 1.1$ &  $233 \pm 24$ & $236	\pm	12$ & $\cdots$ & $55.6 \pm 4.4$ & $32.8$ & $32.5 \pm 1.1$ & $14.8 \pm 1.5$ & $6.6 \pm  0.5$ & $3.0 \pm 0.2$\rlap{$^{\rm e}$} & $0.9,0.9$ \\
J135158$-$020956 & EQU & $ -0.96 \pm 0.02 $ & $ -0.35 \pm 0.14 $ &  $2420 \pm 240$ & $2439	\pm	19$ & $\cdots$ & $482 \pm 34$ & $277.4$ & $266.6 \pm 8.0$ & $118 \pm 12$ & $50.5 \pm  3.6$ & $25.6 \pm 1.8$\rlap{$^{\rm e}$} & $2.8,2.8$ \\
J140214$+$031753 & EQU & $ -0.96 \pm 0.05 $ & $ -1.13 \pm 0.33 $ &  $708 \pm 71$ & $714	\pm	12$ & $\cdots$ & $165 \pm 12$ & $102.9$ & $99.3 \pm 3.0$ & $50.4 \pm 5.1$ & $25.6 \pm  1.8$ & $13.9 \pm 1.0$\rlap{$^{\rm e}$} & $2.2,1.5$ \\
J141023$+$025958 & EQU & $ -0.98 \pm 0.19 $ & $ -3.0 \pm 1.4 $ &  $168 \pm 18$ & $197	\pm	13$ & $\cdots$ & $46.2 \pm 4.0$ & $26.3$ & $28.4 \pm 0.9$ & $15.8 \pm 1.8$ & $7.9 \pm  0.6$ & $4.4 \pm 0.3$\rlap{$^{\rm e}$} & $3.9,4.1$ \\
J144305$+$022940 & EQU & $ -0.73 \pm 0.26 $ & $ -2.9 \pm 1.8 $ &  $284 \pm 29$ & $266	\pm	20$ & $\cdots$ & $38.0 \pm 3.2$ & $18.3$ & $18.9 \pm 0.7$ & $7.0 \pm 0.8$ & $2.9 \pm  0.2$ & $1.44 \pm 0.15$\rlap{$^{\rm e}$} & $3.3,2.6$ \\
J152154$-$010413 & EQU & $ -1.28 \pm 0.41 $ & $ -4.2 \pm 3.1 $ &  $62 \pm 10$ & $93	\pm	11$ & $\cdots$ & $13.9 \pm 1.7$ & $4.8$ & $5.0 \pm 0.5$ & $1.4 \pm 0.3$\rlap{$^{\rm b}$} & $0.44 \pm 0.09$\rlap{$^{\rm b}$} & $0.14 \pm 0.06$\rlap{$^{\rm be}$} & $0.9$\rlap{$^{\rm c}$} \\
J221921$-$331206 & SGP & $ -0.85 \pm 0.07 $ & $ -0.89 \pm 0.55 $ &  $149 \pm 16$ & $179.0	\pm	3.4$ & $38.3 \pm 1.6$ & $38.9 \pm 3.3$ & $\cdots$ &  $23.3 \pm 0.8$ & $11.1 \pm 1.2$ & $6.2 \pm 0.4$  & $3.5 \pm 0.3$ & $0.9,<0.4$ \\
J231148$-$335918\rlap{$^{\rm \dag}$} & SGP & $ -1.58 \pm 0.19 $ & $ -0.5 \pm 1.4 $ &  $42.5 \pm 6.9$ & $51.2	\pm	3.3$ & $<6.3$\rlap{$^{\rm b}$} & $2.5 \pm 0.9$ & $\cdots$ & $<1.6$\rlap{$^{\rm b}$} & $0.18 \pm 0.13$\rlap{$^{\rm b}$} & $0.047 \pm 0.012$\rlap{$^{\rm b}$} & $0.044 \pm 0.010$\rlap{$^{\rm b}$} & $<18$\rlap{$^{\rm c}$} \\
J231456$-$351721 & SGP & $ -1.46 \pm 0.08 $ & $ -2.03 \pm 0.65 $ &  $124 \pm 13$ & $126.1	\pm	3.1$ & $19.5 \pm 1.7$ & $20.5 \pm 2.0$ & $\cdots$ & $10.6 \pm 0.6$ & $3.9 \pm 0.5$ & $3.4 \pm 0.3$ & $2.02 \pm 0.15$ & $0.8,<0.7$ \\
J232614$-$302839 & SGP & $ -0.80 \pm 0.03 $ & $ -0.55 \pm 0.20 $ &  $472 \pm 48$ & $507.0	\pm	4.9$ & $155.6 \pm 5.0$ & $146 \pm 11$ & $\cdots$ & $90.1 \pm 2.7$ & $46.3 \pm 4.8$ & $19.8 \pm 1.4$ & $10.2 \pm 0.7$ & $0.9,2.1$ \\
J233020$-$323729 & SGP & $ -1.03 \pm 0.08 $ & $ -1.08 \pm 0.65 $ &  $101 \pm 13$ & $104.6	\pm	2.4$ & $18.3 \pm 1.7$ & $25.7 \pm 2.3$ & $\cdots$ & $15.0 \pm 0.7$ & $6.3 \pm 0.7$ & $4.0 \pm 0.3$ & $2.11 \pm 0.16$ & $4.2,1.4$ \\
J234019$-$323059 & SGP & $ -1.00 \pm 0.05 $ & $ -0.61 \pm 0.36 $ &  $223 \pm 26$ & $218.8	\pm	3.2$ & $57.6 \pm 2.4$ & $55.1 \pm 4.4$ & $\cdots$ & $33.9 \pm 1.1$ & $13.1 \pm 1.3$ & $7.9 \pm 0.6$ & $4.6 \pm 0.3$ & $1.5,<0.3$ \\
\hline
\hline
\multicolumn{14}{p{230mm}}{Notes. $^{\rm \dag}$These sources do not satisfy all of our selection criteria (see Section~\ref{section: HzRG sample}). $^{\rm a}$Uncertainties are not given in the catalogue, but see \citet{becker95}, \citet{helfand15} and \url{https://sundog.stsci.edu/first/catalogs/readme.html} for further details. $^{\rm b}$Estimated from visual inspection and subsequent analysis of the image of this source. $^{\rm c}$One LAS measurement only for J0133$-$3056 (VLASS; no ATCA observations taken), J1521$-$0104 (FIRST; insufficient S/N in VLASS) and J2311$-$3359 (ATCA; insufficient S/N in VLASS). The LAS for J2311$-$3359 may be larger than our 5\arcsec\:cutoff (see Section~\ref{section:notes sources}). $^{\rm d}$Source is from the pilot study; see Section~\ref{section:intro} and D20 for further details. Note that the 5.5- and 9-GHz flux densities are from the ATCA observations presented in D20. $^{\rm e}$8.8-GHz flux densities; other flux density values in this column without a footnote are 9-GHz values.} \\
\end{tabular}
\end{minipage}
\end{sidewaystable*}

\section{Near-infrared $K_{\rm s}$-band data}\label{section: K-band}

In this section, we summarise the $K_{\rm s}$-band data for our sample. All sources are not detected in VIKING at the $5\sigma$ level, but deeper limits or detections are available in some cases.   

Firstly, as presented in D20, J0856$+$0223 and J0917$-$0012 were observed with VLT/HAWK-I. The host galaxy was detected in both cases; the magnitudes are $23.2 \pm 0.1$ (J0856$+$0223) and $23.01 \pm 0.04$ (J0917$-$0012; see the most recent analysis in \citealt[][]{seymour22}).

J0133$-$3056 and J0842$-$0157 were also observed with HAWK-I as part of the ESO service-mode `filler' programme 0104.A-0599(A). The eight targets for which data were obtained were a combination of USS sources and those selected from an earlier version of the curved-spectrum technique described in Section~\ref{section:selection criteria}. For both J0133$-$3056 and J0842$-$0157, the exposure time was 35 min using a standard jitter pattern. We ran the source-finding code {\sc sextractor} \citep{bertin96} on the pipeline-reduced images. We measured magnitudes within an aperture of diameter 2\arcsec\: (one of several standard choices for HzRG $K_{\rm s}$-band measurements in the literature; e.g. \citealt[][]{debreuck02}) and applied an aperture correction derived from a curve of growth of $-$0.16~mag. To confirm the photometric scale, we cross-matched the {\sc sextractor} source catalogues from the pipeline-produced images with the Two-Micron All Sky Survey \citep[2MASS;][]{skrutskie06}, using a maximum search radius of 1\farcs5. In both cases, the difference was $\leq 0.1$ mag; therefore, an associated correction was not applied. The final $K_{\rm s}$-band aperture magnitudes, converted from Vega to AB, are $22.96 \pm 0.12$ (J0842$-$0157) and $22.23 \pm 0.08$ (J0133$-$3056); further discussion can be found in Section~\ref{section:notes sources}. The uncertainties are the combination of the measurement uncertainty and a 10 per cent calibration uncertainty.

Three of the sources, J0007$-$3040, J0008$-$3007 and J2340$-$3230, were not detected in SHARKS. For these sources and the remaining 46 sources in the sample with VIKING $K_{\rm s}$-band non-detections, we calculated the $5\sigma$ magnitude lower limit as follows:

\begin{equation}\label{eqn:viking limit} 
K_{\rm{s,\,5\sigma}} = m_{\rm ZP} -  2.5\log(5\sigma_{\rm ADU} \times 1.13309 \times \theta^2).
\end{equation}
In Equation~\ref{eqn:viking limit}, $\sigma_{\rm ADU}$ is the root-mean-square (RMS) standard deviation in the vicinity of the radio position in analogue-to-digital units (ADU), $m_{\rm ZP}$ is the zero-point magnitude (30.0), $\theta$ is the seeing disc FWHM in pixels (median seeing 0\farcs84; pixel size 0\farcs339), and the factor 1.13309 is from the standard formula for a two-dimensional Gaussian function. The $5\sigma$ VIKING limits span the range 21.2--22.3, with a median of 21.7, i.e. deeper than the nominal 5$\sigma$ limit of 21.2. The three SHARKS limits are consistent with the nominal $5\sigma$ threshold of 22.7.  

The $K_{\rm s}$-band images for the 51 new HzRG candidates are presented in Section~\ref{section:overlays} (Figure~\ref{fig:overlays_spectra}).

\section{ATCA data}\label{section:ATCA}

\begin{table*}
\begin{minipage}{1.0\textwidth}
 \centering
  \caption{ATCA observing log for our 5.5- and 9-GHz observations. On 2020 December 2/3, the effective frequency of the upper band was 8.8 rather than 9 GHz. Noise levels are those measured near the target; for the C3377 data, we give the median noise levels at 5.5 and 8.8/9 GHz. Furthermore, for the C3377 data, the reported angular resolutions were determined by taking the separate medians of the major and minor axis FWHMs as well as the BPAs. PKS B1934$-$638 was the primary calibrator in all of the observing runs. Further details on the ATCA observations can be found in Section~\ref{section:ATCA}.} 
  \begin{tabular}{lc}
  \hline\hline
  \multicolumn{2}{c}{Project CX437} \\
  \hline
   Dates (UTC) & 2019 July 4/5 \\
   Configuration & 750C  \\
   No. targets observed & 1 \\
   Integration time on source & 6.57 hr \\ 
   Angular resolution (5.5 GHz) & $29\farcs03 \times 9\farcs95$ (BPA $= 18\fdg4$) \\ 
   Angular resolution (9 GHz) & $18\farcs03 \times 6\farcs16$ (BPA $= 18\fdg3$) \\     
   Noise level (5.5 GHz) & 11.5 \textmu Jy beam$^{-1}$  \\
   Noise level (9 GHz) & 9 \textmu Jy beam$^{-1}$ \\  
   Secondary calibrator & B2254$-$367 \\ 
   \hline
   \\
   \multicolumn{2}{c}{Project C3377 (part 1)}  \\
   \hline
   Dates (UTC) & 2020 April 30/May 1, 2020 May 2/3 \\
   Configuration & 6A  \\
   No. targets observed & 22 \\
   Integration time on source & 20--40 min (median 24 min) \\ 
   Angular resolution (5.5 GHz) & $3\farcs1 \times 1\farcs9$ (BPA $= 4\fdg5$)  \\
   Noise level (5.5 GHz) & 37 \textmu Jy beam$^{-1}$  \\
   Angular resolution (9 GHz) & $1\farcs9 \times 1\farcs2$ (BPA $= 4\fdg2$)  \\
   Noise level (9 GHz) & 31 \textmu Jy beam$^{-1}$  \\
   Secondary calibrators & B0104$-$408, B0150$-$334, B0220$-$349, B0400$-$319 \\ 
   & B2245$-$328, B2337$-$334, B2357$-$318 \\
   \hline
   \\
   \multicolumn{2}{c}{Project C3377 (part 2)}  \\
   \hline
   Dates (UTC) & 2020 December 2/3, 2020 December 8/9 \\
   Configuration & H168  \\
   No. targets observed & 26 \\
   Integration time on source & 20--68 min (median 27 min) \\ 
   Angular resolution (5.5 GHz) & $51\arcsec \times 31\arcsec$ (BPA $= 74\degr$)  \\
   Noise level (5.5 GHz) & 62 \textmu Jy beam$^{-1}$ \\
   Angular resolution (8.8/9 GHz) & $32\arcsec \times 19\arcsec$ (BPA $= 73\degr$)  \\
   Noise level (8.8/9 GHz) & 45 \textmu Jy beam$^{-1}$ \\
   Secondary calibrators & B0906$+$015, B1021$-$006, B1055$+$018, B1145$-$071 \\ 
   & B1222$+$037, B1351$-$018, B1502$+$036 \\ 
\hline\hline
\end{tabular}
\label{table:atca}
\end{minipage}
\end{table*}

To facilitate modelling of each radio spectrum and, where possible, to obtain further high-resolution radio data to help confirm the compact radio morphologies and $K_{\rm s}$-band non-detections, we observed 49 of the 53 sources in our sample with the ATCA as part of projects CX437 and C3377. Of the remaining four sources, J0856$+$0223 and J0917$-$0012 were observed with the ATCA as part of the D20 pilot study. J0133$-$3056 and J0842$-$0157 were not observed as these sources were originally considered to be part of another closely related HzRG project (i.e. the $K_{\rm s}$-band `filler' targets discussed in Section~\ref{section: K-band}), before being added to the sample presented in this paper. ATCA data are not available for these sources in the Australia Telescope Online Archive\footnote{\url{https://atoa.atnf.csiro.au/}}. 

\subsection{Projects CX437 and C3377}

An observing log for our ATCA observations is presented in Table~\ref{table:atca}. Observations were carried out simultaneously at 5.5 and 9 GHz using the Compact Array Broadband Backend \citep[CABB;][]{wilson11}, with a nominal bandwidth of 2.048 GHz at each frequency comprising $2048 \times 1$-MHz channels. We used a general observational strategy of target snapshots interleaved with scans of phase calibrators. When more than one target was observed, we ensured that the individual snapshots were sufficiently well spread in hour angle to give sufficient $(u,v)$ coverage for imaging. The standard primary calibrator PKS B1934$-$638 was observed in each run, with the flux density scale reported in \citet[][]{reynolds94}. 

In project CX437, we used available Director's Discretionary Time to observe J2311$-$3359. The array was in the 750C configuration. 20-min target scans were interleaved with 2-min phase calibrator observations. Note that these observations were set up differently to what we describe below for C3377; this was due to the fact that the expected very faint flux densities for J2311$-$3359 at 5.5 and 9 GHz required a significantly longer on-source integration time to increase the likelihood of a detection.   

In project C3377, we observed the remaining 48 sources. Observations were carried out with the 6A array configuration for the SGP targets and with the hybrid H168 array configuration for the EQU sources. The H168 configuration was needed for the EQU sources to ensure adequate $(u,v)$ coverage. The individual target scans were mostly $3$ min in duration, although with some variations. Phase calibrators were observed every $\sim$ 10--25 min and the scans were either 1 or 1.5 min in duration. 

\subsection{Data reduction and imaging}

The data were reduced and imaged using standard procedures in {\sc miriad} \citep*[][]{sault95}. Radio-frequency interference (RFI) was particularly problematic in our observing run on 2020 December 2/3; hence, a significant amount of primary calibrator data in the upper part of the 9-GHz band had to be flagged. To make the subsequent flux density calibration of the phase calibrator and target data as straightforward as possible, we also flagged the corresponding channels in these data, resulting in a nominal sensitivity penalty of about 10--15 per cent and a shift in the effective frequency from 9 to 8.8 GHz.    

When the S/N was sufficient, we used several iterations of imaging and phase-only self-calibration, the latter based on the {\sc clean} component models. We used multifrequency deconvolution \citep[][]{sault94} and the robust weighting parameter \citep[][]{briggs95} was set to 0.5. Note that we did not include baselines to antenna 6 in the imaging and self-calibration step for the data from project CX437 as well as the H168 observations from C3377. This was to remove the large gap in the $(u,v)$ coverage, with a nominal sensitivity penalty of about 20 per cent and poorer angular resolution. However, high-resolution data were already available from FIRST and VLASS for these targets. The maximum antenna spacings for our ATCA data sets were then 192, 750 and 5939 m for the H168, 750C and 6A observations, respectively. A primary beam correction was applied to all {\sc clean}ed maps, although this did not make a significant difference to the integrated flux densities as all of our targets were at or very close to the pointing centre. Angular resolutions and noise levels can be found in Table~\ref{table:atca}.  

After imaging the data, we then used {\sc pybdsf} \citep[][]{mohan15} for source finding and integrated flux density measurements. Each flux density uncertainty was determined by combining the fitting uncertainty from {\sc pybdsf}, the internal calibration uncertainty and the flux density scale uncertainty in quadrature. We estimate that the internal calibration uncertainty is at the $\sim$ 5 per cent level at both 5.5 and 8.8/9 GHz; the flux density scale uncertainty is estimated to be an additional $\sim$ 5 per cent using the information available in \citet[][]{reynolds94} and \citet[][]{perley17}.      

The 5.5- and 8.8/9-GHz flux densities are given in Table~\ref{table:fluxes}. In a handful of cases noted in this table as well as in Section~\ref{section:overview available data}, due to sufficiently low S/N it was necessary to estimate the flux densities from an additional analysis of the images in question, rather than using {\sc pybdsf}. Overlay plots showing the high-resolution ATCA radio contours from the 6A configuration observations and the medium-resolution contours from the 750C configuration observation of J2311$-$3359 are presented in Section~\ref{section: morphologies} (Figure~\ref{fig:overlays_spectra}).  

\section{Analysis of radio data and overlay plots}\label{section:overlays}

\subsection{Overview of available radio data}\label{section:overview available data}

Our radio data from GLEAM and the ATCA were supplemented by flux density measurements from other radio surveys (Tables~\ref{table:fluxes} and \ref{table:extra_fluxes}). We used the following catalogues: the 74-MHz VLA Low-Frequency Sky Survey Redux \citep[VLSSr;][]{cohen07,lane14}, TGSS at $147.5$ MHz, $325$-MHz GMRT survey data of H-ATLAS/GAMA fields from \citet[][]{mauch13}, the $365$-MHz Texas Survey \citep[TXS;][]{douglas96}, the $408$-MHz Molonglo Reference Catalogue (MRC; \citealt[][]{large81,large91}), the $843$-MHz Sydney University Molonglo Sky Survey (SUMSS; \citealt*[][]{bock99}; \citealt[][]{mauch03}; \citealt[][]{murphy07}), RACS at 887.5 MHz, FIRST at 1.4 GHz, NVSS at 1.4 GHz and VLASS at 3 GHz. Following \citet[][]{debreuck00}, we only considered those sources that are well modelled (`$+++$' flags) in the TXS catalogue. SUMSS data are only available for SGP sources and FIRST for EQU sources. Furthermore, we used the VLASS catalogue derived from first-epoch data presented and analysed in \citet[][]{gordon20,gordon21}, applying a multiplicative flux density scale correction of 1/0.87 to the catalogued flux densities and flux density fitting uncertainties (see discussion and analysis in \citealt[][]{gordon21}). For the VLASS data, we also assumed a conservative 10 per cent calibration uncertainty, adding this value in quadrature to the corrected fitting uncertainties from the catalogue to obtain the final uncertainties listed in Table~\ref{table:fluxes}.   

The radio data presented in Tables~\ref{table:fluxes} and \ref{table:extra_fluxes} are from surveys that are not always tied to the same flux density scale. We return to this topic in Section~\ref{section:SED modelling}, as it is an important consideration for the modelling of the broadband radio spectra of our sample. However, for the comparison between the 147.5-MHz TGSS and 151-MHz GLEAM flux densities described below in Section~\ref{section: flux comparison}, we note that we have used rescaled TGSS flux densities from \citet[][]{hurleywalker17b}; these measurements are reported in Table~\ref{table:fluxes}. The average multiplicative correction factor applied for the sources in our sample is 0.97, with minimum and maximum values of 0.79 and 1.20, respectively. The rescaling moves the TGSS flux densities onto the flux density scale of \citet{baars77} that was used to calibrate GLEAM, rather than the \citet[][]{scaife12} scale used by \citet[][]{intema17} for the main TGSS catalogue. It also corrects for position-dependent flux density scale variations in TGSS. 

There are three sources in the sample with one or more non-detections at the $3\sigma$ level: J0034$-$3112 (843 MHz), J1521$-$0104 (8.8 GHz) and J2311$-$3359 (843 MHz, 1.4 GHz and 3 GHz). For J0034$-$3112 and J1521$-$0104, we made use of a technique often used when studying the light curves of radio transients \citep[e.g.][]{swinbank15}: rather than including upper limits in the analysis described below, we instead measured the flux density from a forced point-source fit at the target location using {\sc imfit} in {\sc miriad}. This ensured the consistency of the characteristics of the data points, rather than a combination of detections and upper limits. The flux densities from these fits are $S_{843} = 9.0 \pm 3.3$ mJy for J0034$-$3112 and $S_{8800} = 0.14 \pm 0.06$ mJy for J1521$-$0104. Both of these flux densities are consistent with the formal $3\sigma$ upper limits ($<10$ and $<0.2$ mJy beam$^{-1}$, respectively). For J2311$-$3359, the SUMSS and NVSS upper limits are not significantly constraining and we therefore excluded them from our analysis. However, for the VLASS data point, we also measured the flux density from a forced fit: $S_{3000} = 0.18 \pm 0.13$ mJy. This measurement is consistent with the formal $3\sigma$ upper limit ($<0.38$ mJy beam$^{-1}$) to within the $2\sigma$ error on the forced-fit value.

For J0856$+$0223 and J0917$-$0012, these sources were analysed in detail in D20, \citet[][]{drouart21} and \citet[][]{seymour22} over a wider frequency range than is considered here. We report radio properties for these two sources in Tables~\ref{table:fluxes} and \ref{table:extra_fluxes}, but only over the same frequency range as for the other sources in the sample.

\begin{table}
 \centering
  \caption{Additional flux density measurements for sources in our sample. We have rescaled the values in the literature so as to place them on the \citet[][]{baars77} flux density scale; the multiplicative correction factors that we used are stated. VLSSr uncertainties are accurate to two significant figures. See Sections~\ref{section:overview available data} and \ref{section:SED modelling} for further details.}
  \begin{tabular}{cr}
  \hline\hline
  \multicolumn{2}{c}{74-MHz VLSSr ($\times 0.91^{\rm a}$)} \\
   \multicolumn{1}{c}{Source} & \multicolumn{1}{c}{$S_{74}$ (mJy)}\\
   \hline
J0006$-$2946 & $ 690 \pm 130 $ \\
J0007$-$3040 & $ 1660 \pm 290 $ \\
J0008$-$3007 & $ 500 \pm 110 $ \\
J0053$-$3256 & $ 610 \pm 140 $ \\
J0326$-$3013 & $ 1900 \pm 430 $ \\
J0842$-$0157 & $ 940 \pm 180 $ \\
J0856$+$0223 & $ 2140 \pm 380 $ \\
J0909$-$0154 & $ 1400 \pm 250 $ \\
J1032$+$0339 & $ 2330 \pm 470 $ \\
J1125$-$0342 & $ 1220 \pm 230 $ \\
J1317$+$0339 & $ 660 \pm 140 $ \\
J1329$+$0133 & $ 1150 \pm 210 $ \\
J1335$+$0112 & $ 990 \pm 200 $ \\
J1340$+$0009 & $ 860 \pm 180 $ \\
J1347$+$0012 & $ 800 \pm 150 $ \\
J1351$-$0209 & $ 4090 \pm 700 $ \\
J1402$+$0317 & $ 1400 \pm 250 $ \\
\hline
   \\ 
  \multicolumn{2}{c}{325-MHz GMRT ($\times 1.13$)} \\
   \multicolumn{1}{c}{Source} & \multicolumn{1}{c}{$S_{325}$ (mJy)}\\
   \hline
  J0856$+$0223 & $ 582 \pm 18$ \\
  J0917$-$0012 & $ 312.8 \pm 9.2$ \\
  J1141$-$0158 & $87.0 \pm 5.0$ \\
  J1443$+$0229 & $126.7 \pm 6.4$ \\
\hline
   \\ 
  \multicolumn{2}{c}{365-MHz TXS ($\times 1.04$)} \\
  \multicolumn{1}{c}{Source} & \multicolumn{1}{c}{$S_{365}$ (mJy)}\\
  \hline
  J0326$-$3013 & $402 \pm 26$ \\
  J0856$+$0223 & $ 419 \pm 29$ \\
  J1125$-$0342 & $368 \pm 52$ \\
  J1317$+$0339 & $415 \pm 62$ \\
\hline
   \\
  \multicolumn{2}{c}{408-MHz MRC ($\times 0.97$)} \\
  \multicolumn{1}{c}{Source} & \multicolumn{1}{c}{$S_{408}$ (mJy)}\\
  \hline
  J1351$-$0209 & $1000 \pm 40$ \\
\hline\hline
\multicolumn{2}{p{60mm}}{Notes. $^{\rm a}$The flux density calibration uncertainty was also increased by 5 per cent following the recommendation presented in Section 5.3 in \citet[][]{lane14}. References: \citet[][VLSSr]{lane14}, \citet[][GMRT]{mauch13}, \citet[][TXS]{douglas96} and \citet[][MRC]{large81,large91}.} \\
\end{tabular}
\label{table:extra_fluxes}
\end{table}

\subsection{Comparison of radio flux densities from data sets matched or closely spaced in observing frequency}\label{section: flux comparison}

To assess the reliability of the radio flux densities, we compared several of our data sets: FIRST and NVSS, SUMSS and RACS, and GLEAM and TGSS. Moreover, as our sample was selected using the criterion LAS $\leq 5\arcsec$, there was the possibility that one or more very compact components in a given source had resulted in significant variability between the epochs of the surveys used in the three comparisons listed above. Variability resulting from refractive interstellar scintillation (\citealt[][]{shapirovskaya78}; \citealt*[][]{rickett84}; \citealt[][]{rickett86,rickett90}) has been observed often at frequencies over the range that we are considering here (e.g. \citealt[][]{hunstead72}; \citealt[][]{gaensler00}; \citealt[][]{bannister11}; \citealt[][]{ofek11}; \citealt[][]{bell19}; \citealt[][]{hajela19}; \citealt[][]{ross21}; \citealt[][]{murphy21}; but see \citealt[][]{koay12} for the case of high-redshift AGN). Other possibilities include intrinsic variability from a compact core and/or jet component \citep[e.g.][]{mooley16,nyland20,ross21,wolowska21} or a combination of both refractive scintillation and intrinsic variability, the former occurring on shorter time-scales (e.g. \citealt*[][]{rickett06}; \citealt[][]{bhandari18}; \citealt[][]{sarbadhicary21}).

\subsubsection{Comparison at mid frequencies}\label{section: flux comparison mid} 

For the EQU sources, we compared the 1.4-GHz NVSS and FIRST flux densities. As our sources are selected to be compact in the radio, the difference in the angular resolutions of NVSS and FIRST for these particular targets ($45\arcsec \times 45\arcsec$ for NVSS and $6\farcs4 \times 5\farcs4$ with BPA $= 0\degr$ for FIRST) should not result in a significant difference between the two flux densities for a given source, i.e. the NVSS flux densities should not be systematically brighter as a result of extended emission being resolved out in FIRST. We determined that the NVSS/FIRST flux density ratio has a mean and standard error of the mean of 0.99 and 0.02, respectively; the minimum and maximum values are 0.82 (J1040$+$0150) and 1.33 (J1037$-$0325). Inspecting the uncertainties associated with individual NVSS and FIRST flux density measurements, J1037$-$0325 and J1040$+$0150 are the only sources where the difference from unity for the NVSS/FIRST flux density ratio is more than $3\sigma$. It is beyond the scope of this paper to thoroughly consider all possible systematic effects that may result in a statistically significant offset. Any significant difference could instead indicate mild variability. For J1037$-$0325, it is also possible that some extended emission has been resolved out in FIRST, which may in turn suggest that this source is not as compact as indicated by both the FIRST and VLASS data.  

We carried out a similar comparison for those sources with data in both SUMSS at 843 MHz (angular resolution $45\csc{\lvert \delta \rvert}\arcsec \times 45\arcsec$ with BPA $= 0\degr$) and RACS (angular resolution $25\arcsec \times 25\arcsec$) at 887.5 MHz. For a source with a canonical spectral index of $\alpha = -0.7$, the expected SUMSS/RACS flux density ratio is 1.037; similarly, the expected flux density ratio is 1.069 for $\alpha = -1.3$. We calculated a mean SUMSS/RACS flux density ratio of 1.01, with a standard error of the mean of 0.04. \citet[][]{hale21} also found an excellent agreement between SUMSS and RACS from a more general comparison of these two surveys. The minimum and maximum ratios from our comparison are 0.71 (J2330$-$3237) and 1.26 (J0002$-$3514). J2330$-$3237 is the only source for which there is evidence of a discrepancy at the $> 3\sigma$ level between the measured and expected SUMSS/RACS flux density ratios (note that J1037$-$0325 and J1040$+$0150 discussed above are not within the SUMSS survey footprint). Inspecting the broadband radio data for this source (Figure~\ref{fig:overlays_spectra}), there is a suggestion that the RACS flux density might be slightly overestimated. A comparison of the FIRST and NVSS flux densities was not possible for J2330$-$3237 as it is not within the FIRST survey footprint.  

\subsubsection{Comparison at low frequencies}\label{section: flux comparison low} 

We compared the rescaled TGSS data at 147.5 MHz with the fitted 151-MHz flux densities from GLEAM. For our sample, the angular resolutions of TGSS and GLEAM are $25\sec(\delta-19\degr)\arcsec \times 25\arcsec$ (BPA $= 0\degr$) and $2\farcm5 \times  2\farcm2\sec(\delta + 26\fdg7)$ (BPA $= 0\degr$ or $90\degr$), respectively, with the latter resolution at 154 MHz. For a source with $\alpha = -0.7$, the expected GLEAM/TGSS flux density ratio is 0.984 (0.970 for $\alpha = -1.3$). We found that the mean and standard error of the mean for the GLEAM/TGSS flux density ratio are 1.08 and 0.03, respectively; the minimum and maximum values are 0.83 (J0239$-$3043) and 1.65 (J0240$-$3206). Therefore, there is a suggestion that the fitted GLEAM flux densities are slightly overestimated on average or that the rescaled TGSS flux densities are slightly underestimated, with the caveat that this is not a fully like-to-like comparison (i.e. we are comparing a fitted 151-MHz flux density determined using the full bandwidth of GLEAM with a single 147.5-MHz measurement determined over a much narrower bandwidth). There are no sources where the measured and expected GLEAM/TGSS flux density ratios differ at the $> 3\sigma$ level.    

In addition, six of our sources were included in the low-frequency GLEAM spectral variability study by \citet[][]{ross21}: J0007$-$3040, J0008$-$3007, J0108$-$3501, J0301$-$3132, J0326$-$3013 and J2326$-$3028. However, none were identified as being variable between the two GLEAM epochs, separated by one year.  

\subsection{Overlay plots}\label{section: morphologies}

In Figure~\ref{fig:overlays_spectra}, we present $K_{\rm s}$-band/radio overlay plots for the sources in our sample. We show contours from our highest-resolution radio data sets, i.e. FIRST, VLASS (angular resolution $\approx$ 2\farcs5) and the ATCA 5.5- and 9-GHz data from the 6A array configuration observations. The deepest $K_{\rm s}$-band image available for a given source has been used (i.e. from VIKING, SHARKS or HAWK-I). Note, however, that we do not include J0856$+$0023 and J0917$-$0012, which have been analysed extensively elsewhere (D20; \citealt{drouart21}; \citealt{seymour22}). For the overlay plots, a summary of the $K_{\rm s}$-band host galaxy magnitudes or lower limits and the lowest radio contour levels used can be found in Table~\ref{table:overlay_contour_levels}.  

The host galaxies of J0133$-$3056 and J0842$-$0157 are detected in the HAWK-I $K_{\rm s}$-band images. Otherwise, any possible host galaxy detections in the VIKING or SHARKS images (e.g. for J1329$+$0133) are tentative at best and below a $5\sigma$ brightest pixel value, which indeed is why the sources are included in our sample (Section~\ref{section: other selection criteria}). 

Given our LAS $\leq 5\arcsec$ selection criterion and the angular resolutions of the VLASS and ATCA data (which are similar for the SGP sources), it is not particularly surprising that all but one of our sources have radio morphologies that can be classified as one of the following: single component, incipient double or resolved double. The one exception is the triple source J0309$-$3526. Offsets between the VLASS and ATCA centroids are within the astrometric uncertainties of the VLASS quick-look images (up to $\approx 1\arcsec$; \citealt[][]{lacy19}). Phase errors remaining in the VLASS maps can lead to erroneous extension visible in the overlay plots (e.g. for J0301$-$3132 and J0326$-$3013), and care must be taken interpreting the radio morphology at 3 GHz (see \citealt[][]{lacy19} for further details).  

Further information can be found in the notes on individual sources in Section~\ref{section:notes sources}. 

\begin{figure*}
\begin{minipage}{0.5\textwidth}
\vspace{0.2cm}
\includegraphics[width=7.5cm]{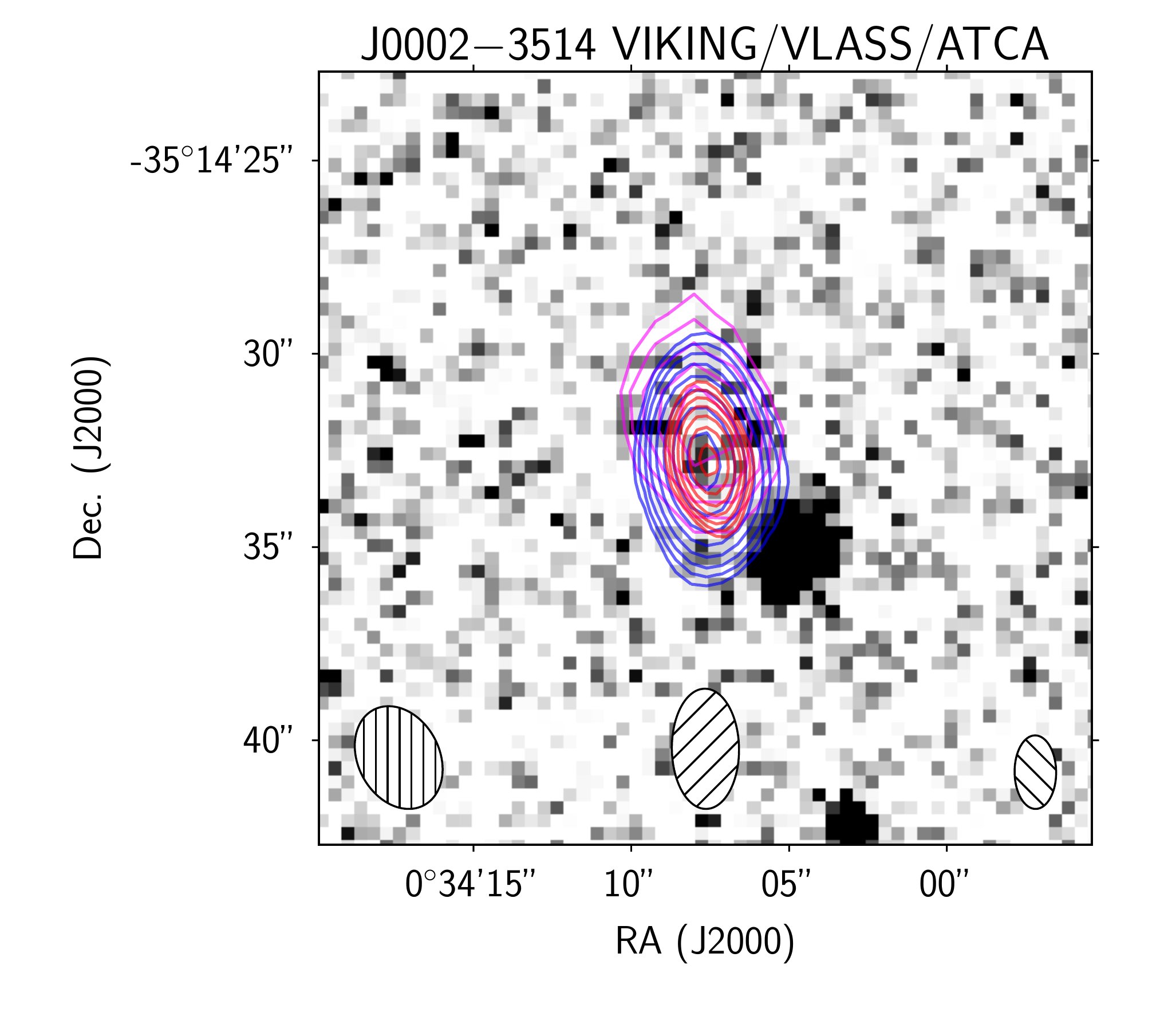}
\end{minipage}
\begin{minipage}{0.5\textwidth}
\includegraphics[width=8.5cm]{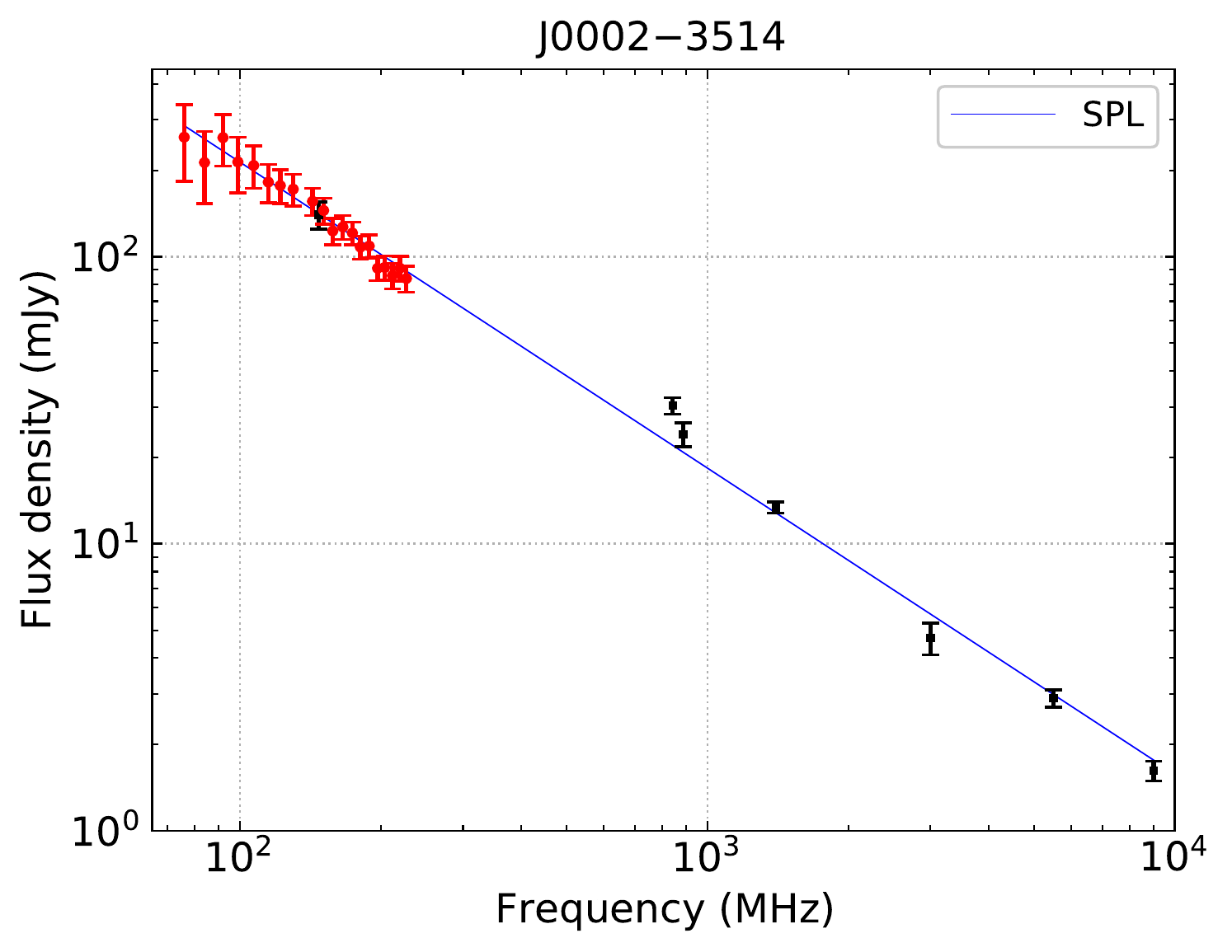}
\end{minipage}
\begin{minipage}{0.5\textwidth}
\vspace{0.2cm}
\includegraphics[width=7.5cm]{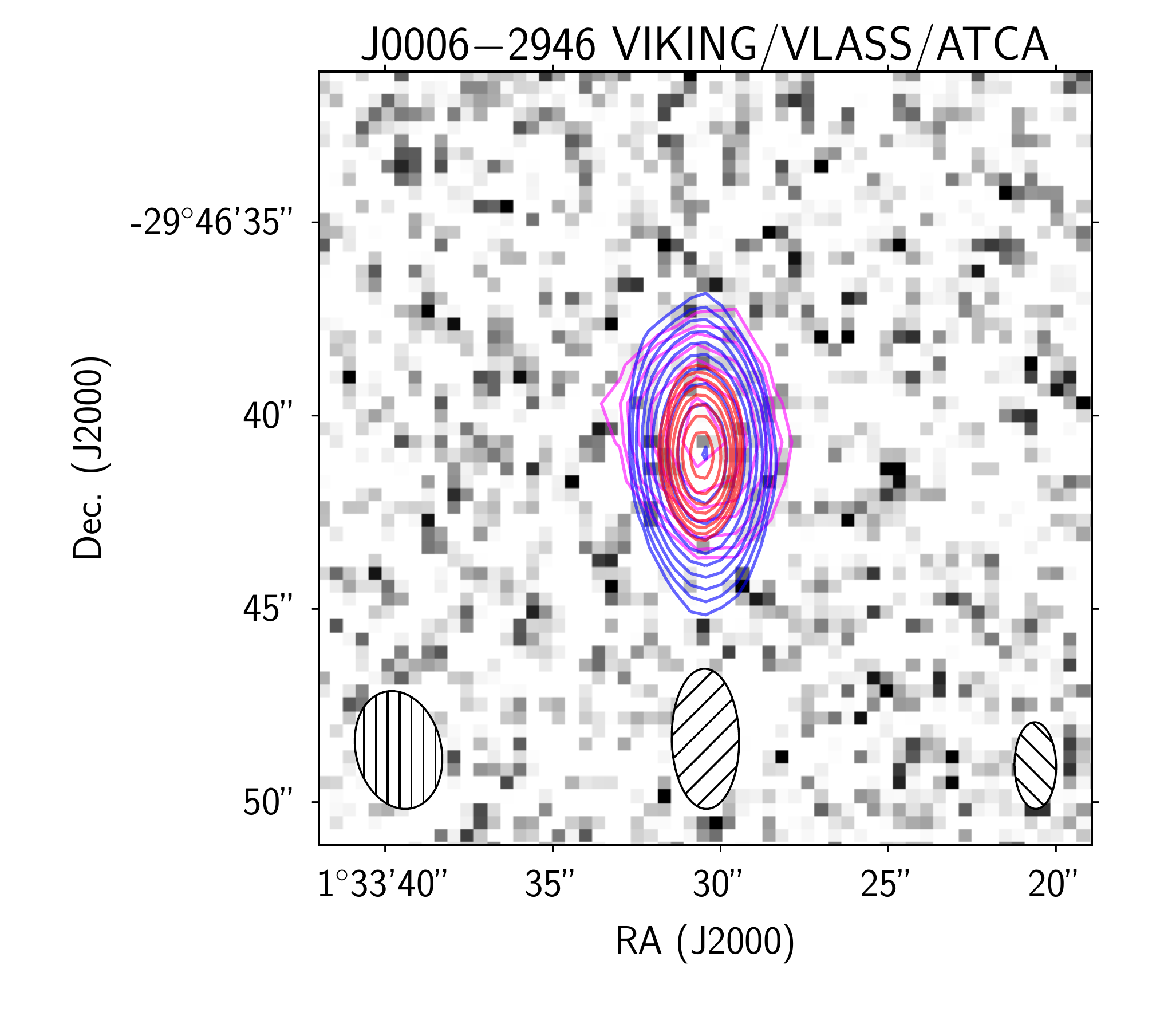}
\end{minipage}
\begin{minipage}{0.5\textwidth}
\includegraphics[width=8.5cm]{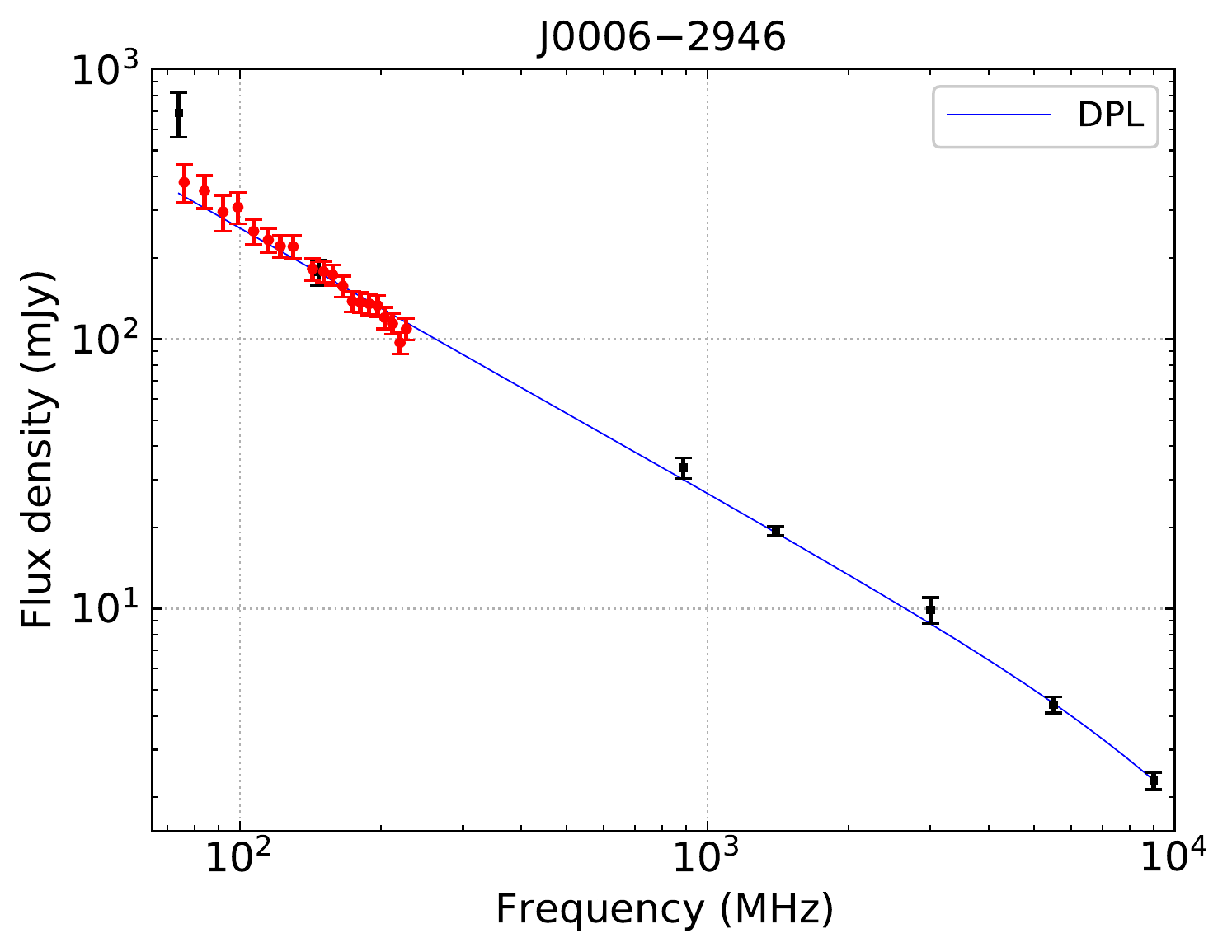}
\end{minipage}
\caption{Overlay plots (left) and observed-frame broadband radio spectra (right) for all of the sources in our sample apart from J0856$+$0223 and J0917$-$0012 from the pilot study. The $K_{\rm s}$-band image used in each overlay plot is listed. We plot FIRST (orange) and VLASS (magenta) contours for the EQU sources; similarly, we plot VLASS, ATCA 5.5-GHz (blue) and ATCA 9-GHz (red) contours for the SGP sources (note that J0133$-$3056 has VLASS data only). ATCA contours are not shown for the EQU sources given the poorer angular resolution compared to FIRST and VLASS. The contours are a geometric progression in $\sqrt{2}$, with the lowest contour usually at the $5\sigma$ level. A summary of the lowest contour levels used in each overlay plot can be found in Table~\ref{table:overlay_contour_levels}. Radio synthesised beams are shown in each overlay plot with different hatching styles (FIRST: horizontal; VLASS: vertical; ATCA 5.5 GHz: forward slash; ATCA 9 GHz: backslash). For the broadband radio spectra, we plot the data presented in Tables~\ref{table:fluxes} and \ref{table:extra_fluxes}, apart from the 151-MHz GLEAM fitted flux densities and the 1400-MHz FIRST flux densities in Table~\ref{table:fluxes} (in the latter case we show the NVSS flux density measurements only). Additionally, we plot catalogued GLEAM flux densities (20 measurements per source) from \citet[][EQU sources]{hurleywalker17} and \citet[][SGP sources]{franzen21}. In each panel, the GLEAM data points are shown as red circles and the remaining data points as black squares. In addition, the solid line represents the preferred model (either a single or double power law, indicated in each plot legend as SPL and DPL, respectively). In the panel showing the radio spectrum for J2311$-$3359, triangles represent $3\sigma$ flux density upper limits (Table~\ref{table:fluxes}), but we did not use these upper limits when modelling its radio spectrum. The 365-MHz data point for J1317$+$0339, a clear outlier, was also not used in the modelling. Error bars are $\pm1\sigma$. In a few of the panels, the relative GLEAM flux density uncertainties at certain frequencies are larger than 100 per cent, and therefore the full extent of the lower error bar cannot be shown on a log--log plot. See Section~\ref{section:SED modelling} for a description of the modelling and Table~\ref{table:fitting_results} for the fitted parameters from each preferred model.} 
\label{fig:overlays_spectra}
\end{figure*}

\setcounter{figure}{1} 
\begin{figure*}
\begin{minipage}{0.5\textwidth}
\vspace{0.2cm}
\includegraphics[width=7.5cm]{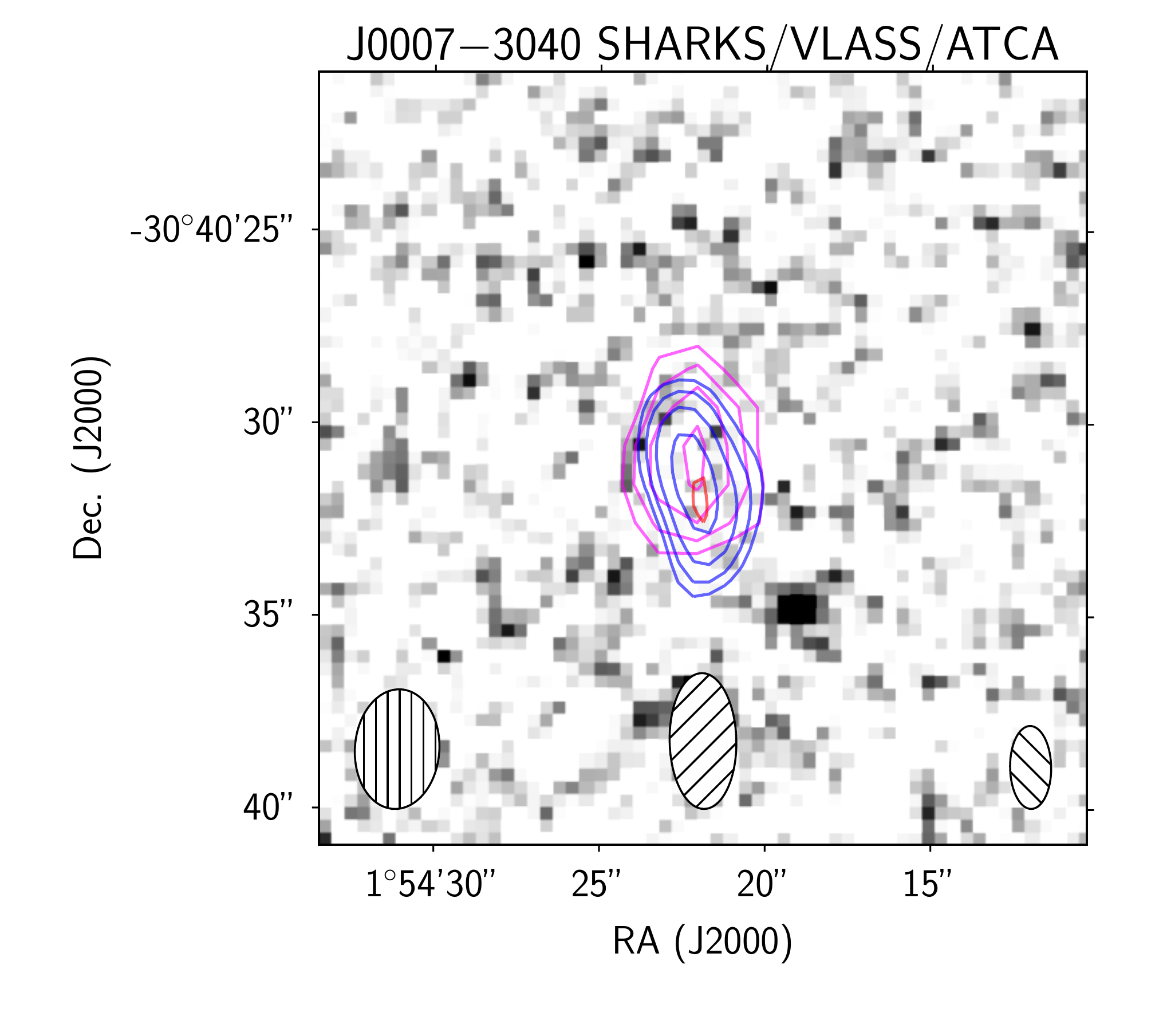}
\end{minipage}
\begin{minipage}{0.5\textwidth}
\includegraphics[width=8.5cm]{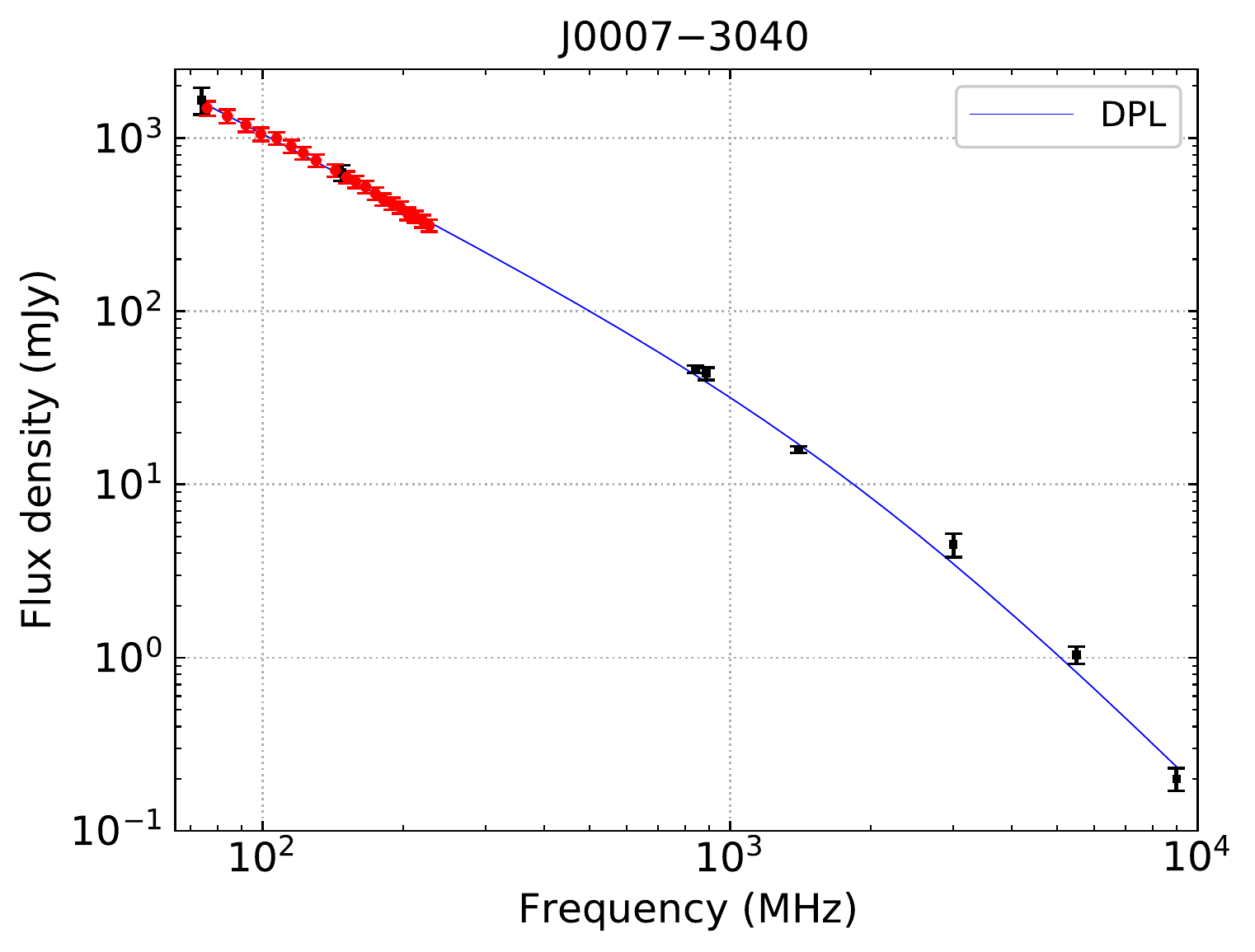}
\end{minipage}
\begin{minipage}{0.5\textwidth}
\vspace{0.2cm}
\includegraphics[width=7.5cm]{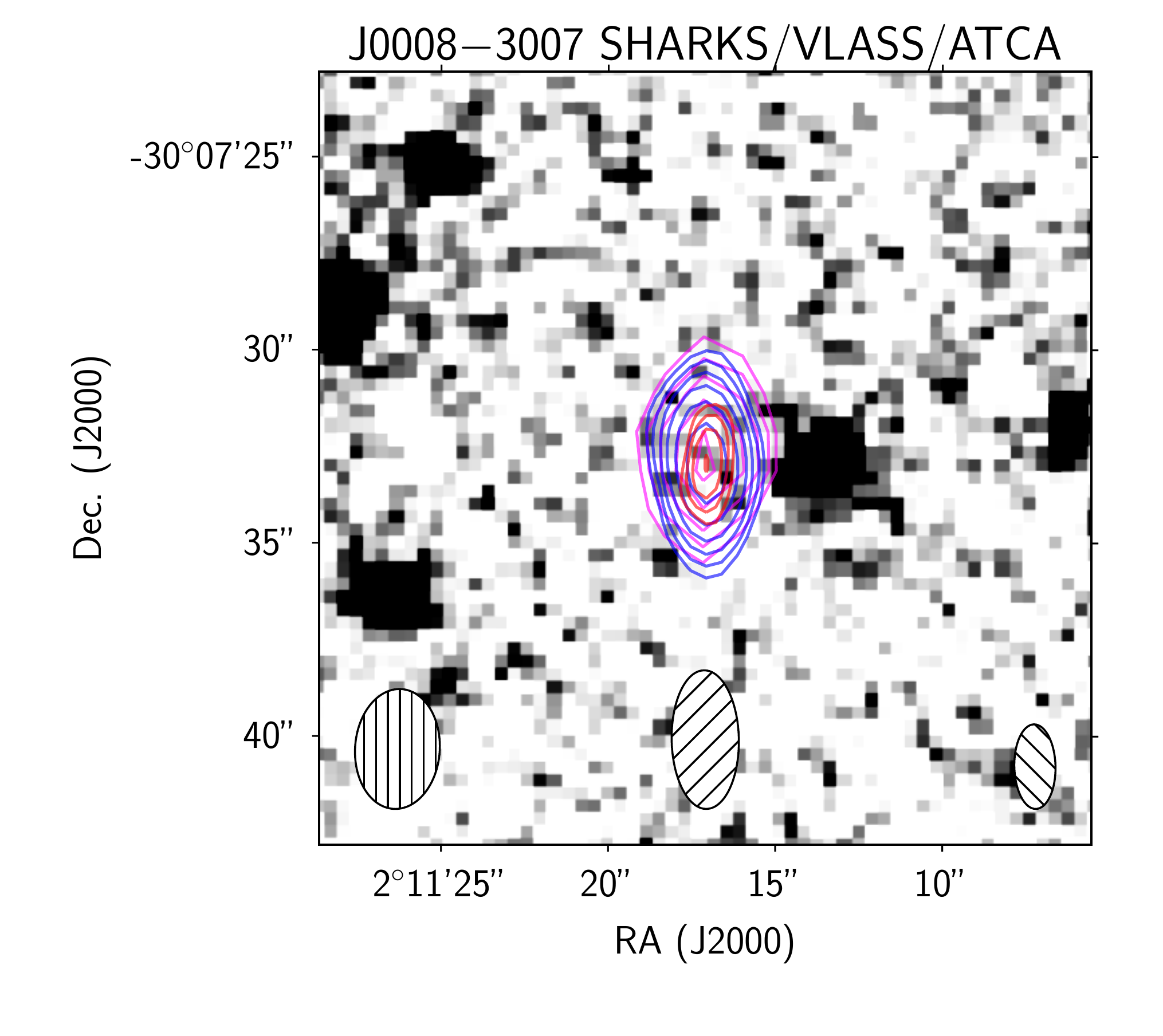}
\end{minipage}
\begin{minipage}{0.5\textwidth}
\includegraphics[width=8.5cm]{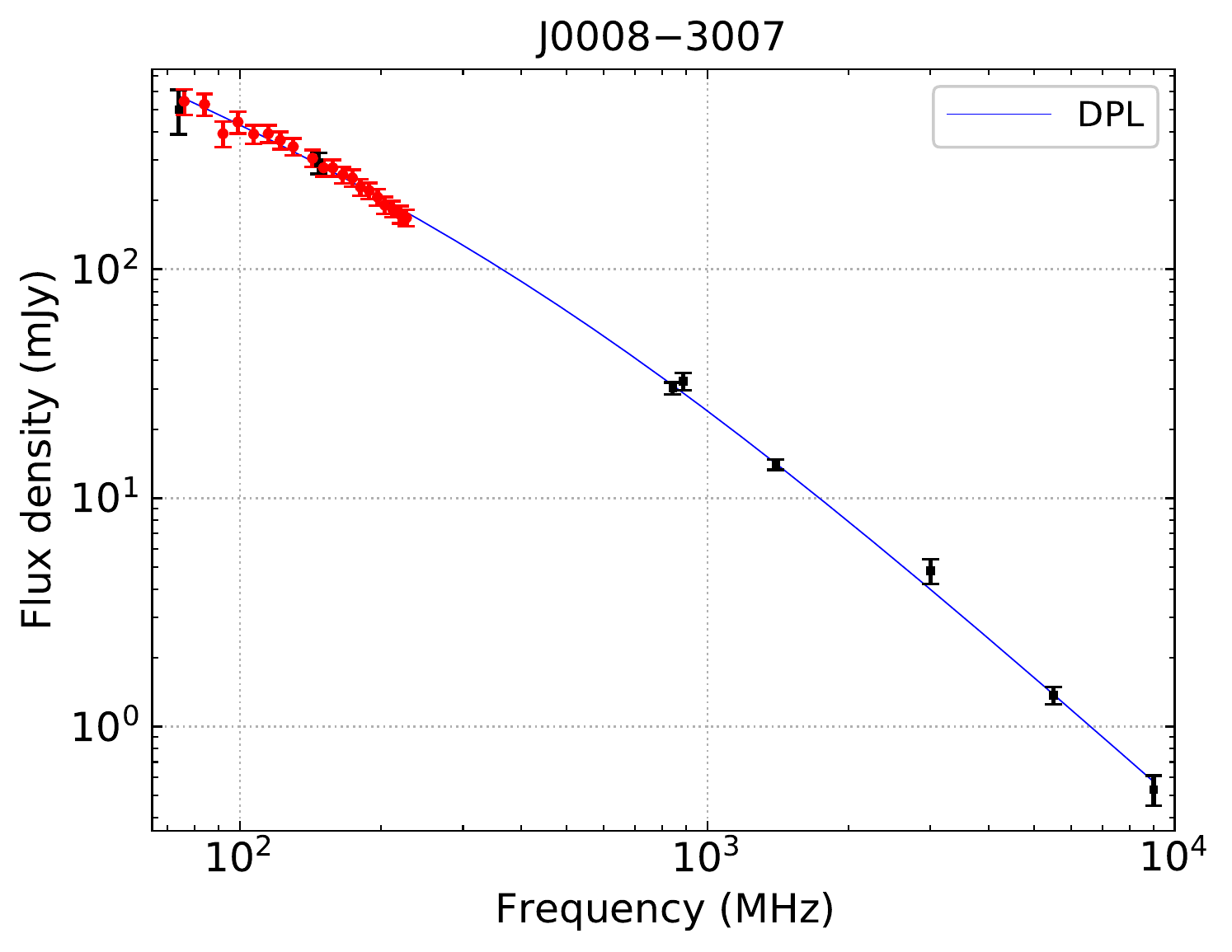}
\end{minipage}
\begin{minipage}{0.5\textwidth}
\vspace{0.2cm}
\includegraphics[width=7.5cm]{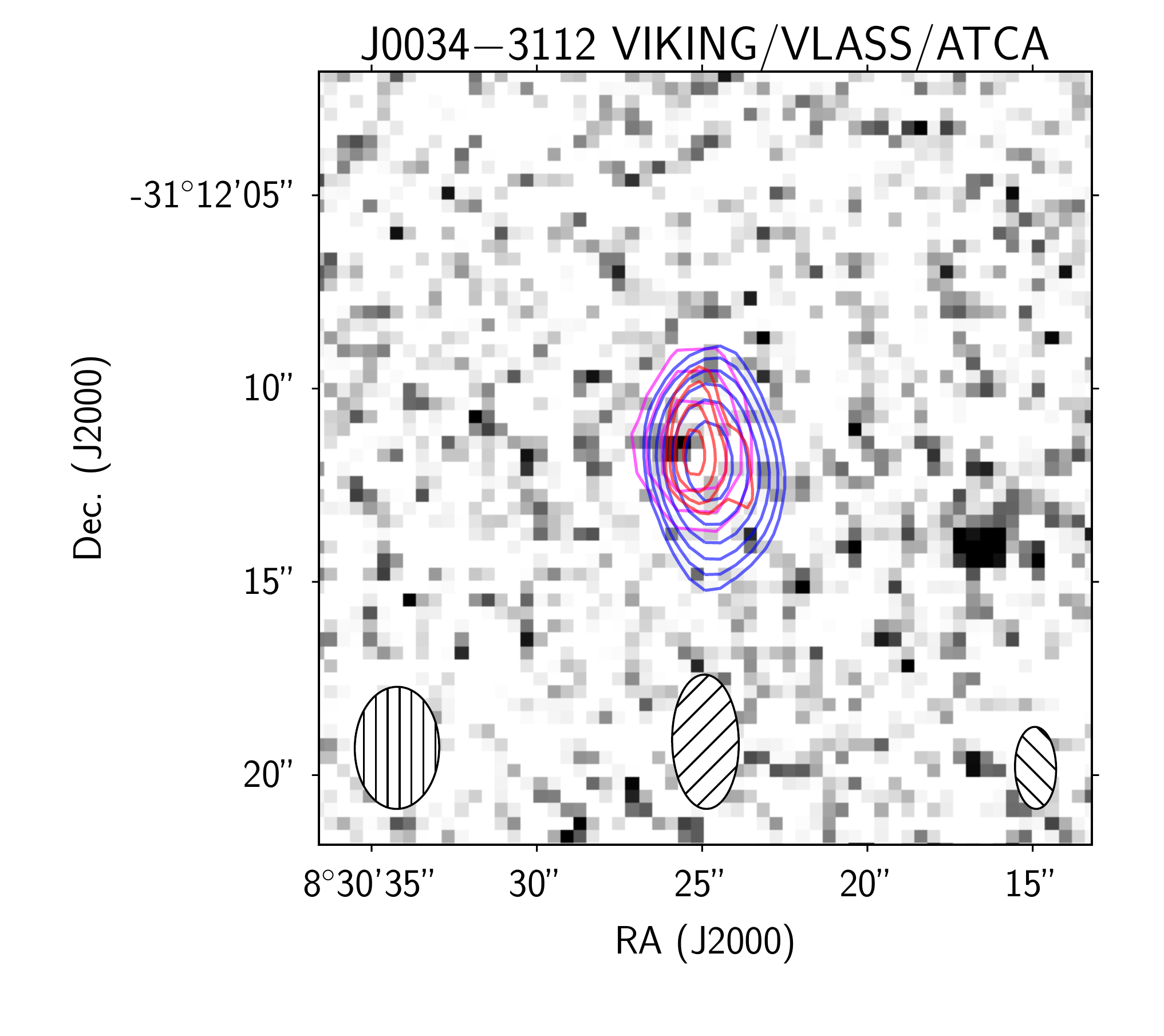}
\end{minipage}
\begin{minipage}{0.5\textwidth}
\includegraphics[width=8.5cm]{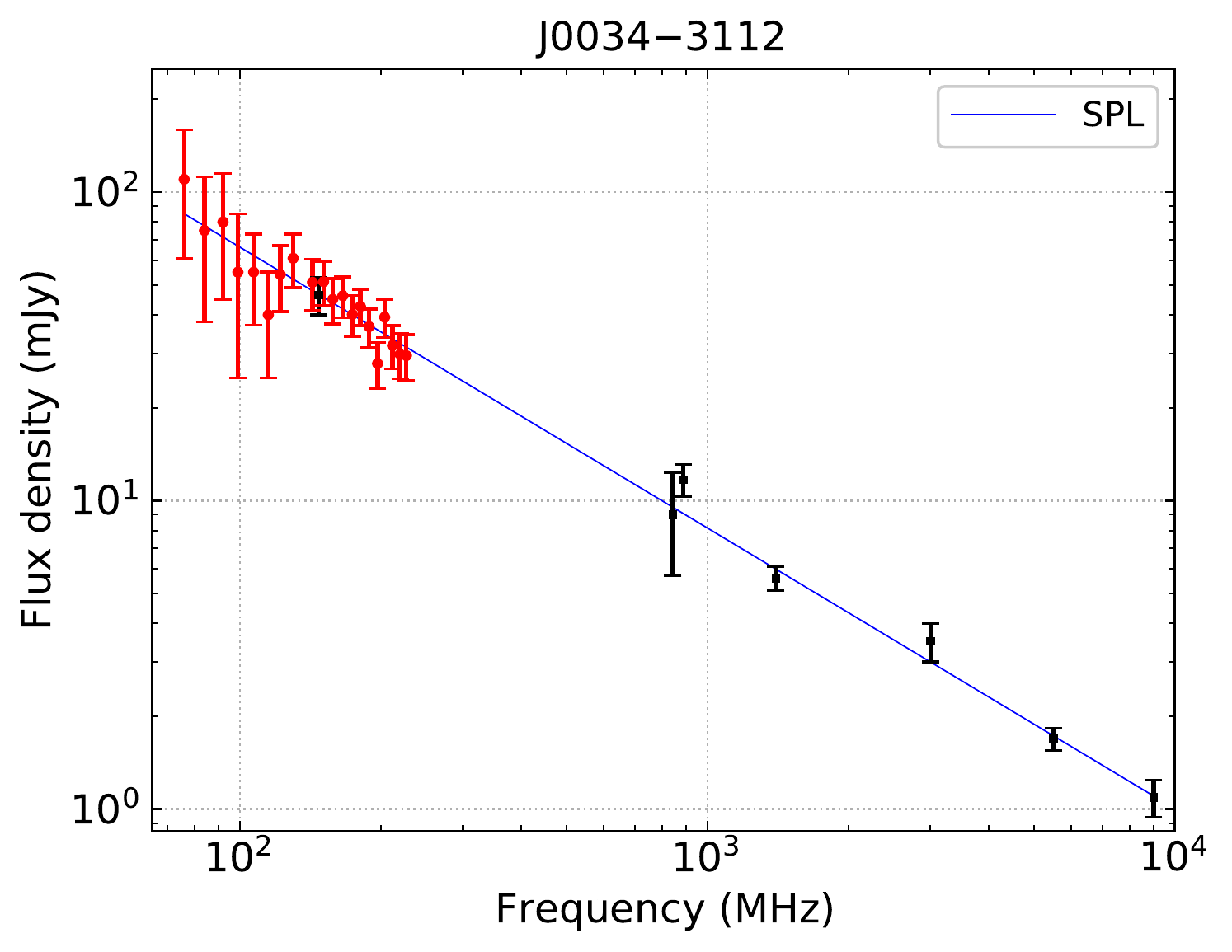}
\end{minipage}
\caption{{\em - continued.}}
\end{figure*}

\setcounter{figure}{1} 
\begin{figure*}
\vspace{-0.5cm}
\begin{minipage}{0.5\textwidth}
\vspace{0.2cm}
\includegraphics[width=7.5cm]{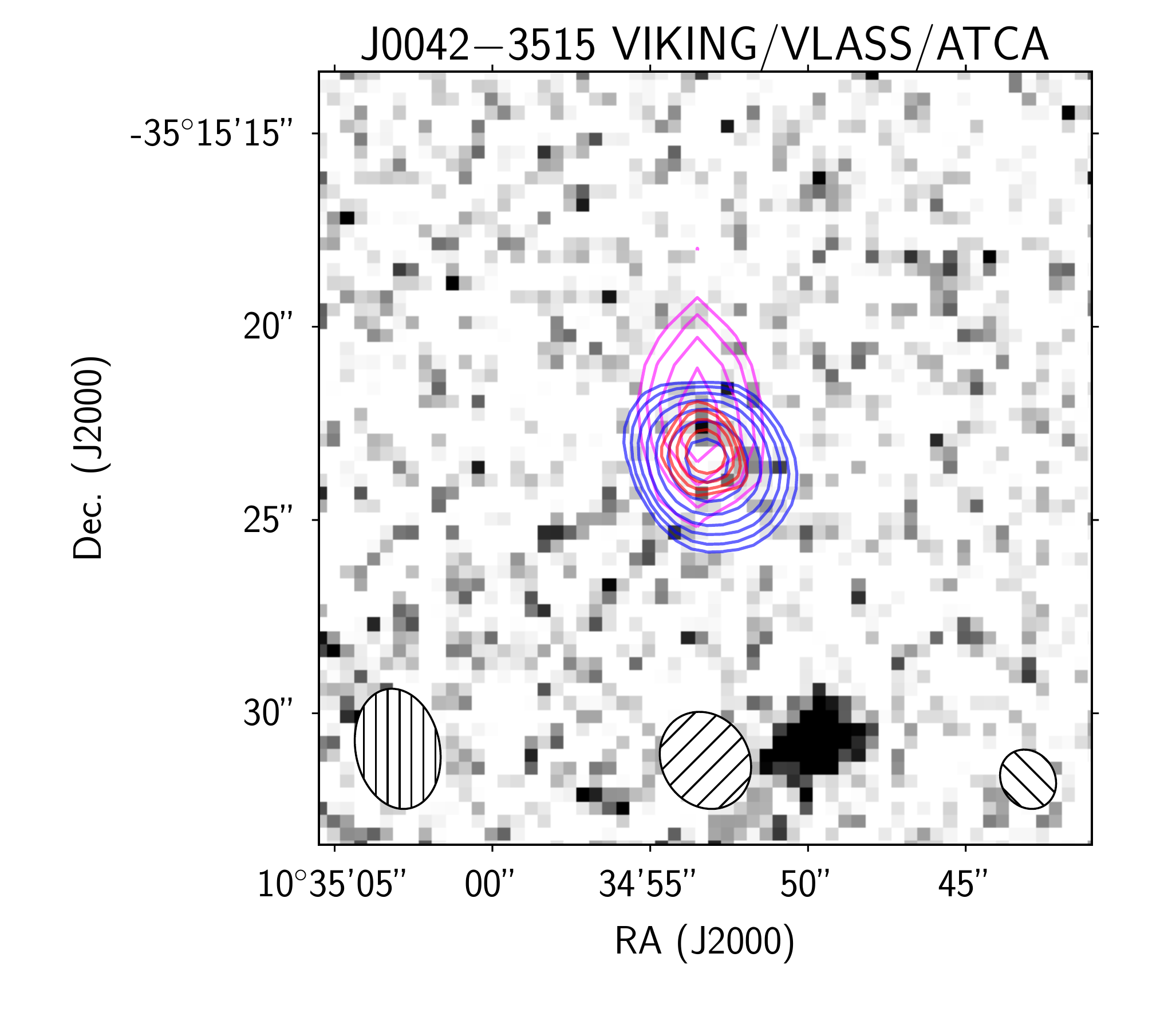}
\end{minipage}
\begin{minipage}{0.5\textwidth}
\includegraphics[width=8.5cm]{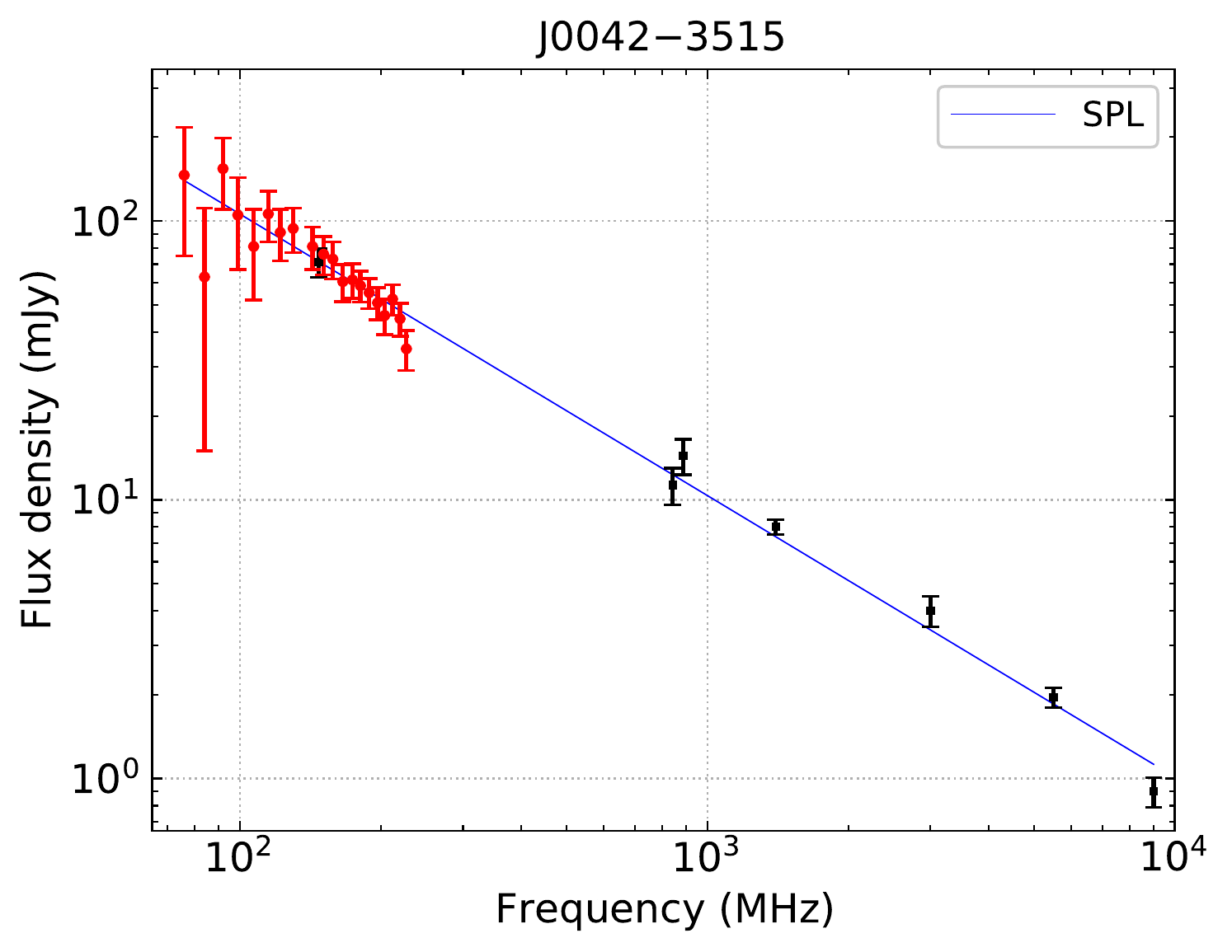}
\end{minipage}
\begin{minipage}{0.5\textwidth}
\vspace{0.2cm}
\includegraphics[width=7.5cm]{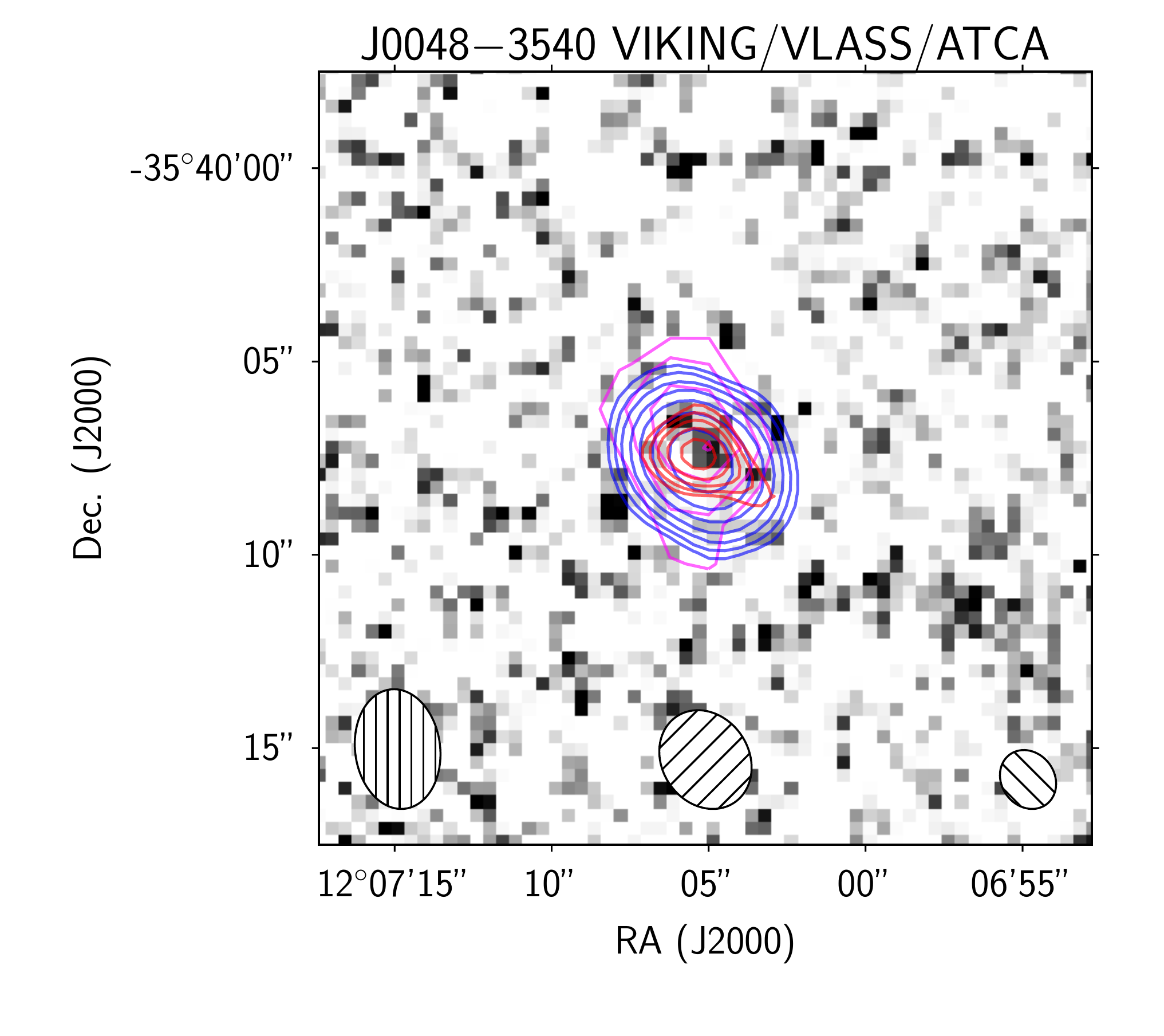}
\end{minipage}
\begin{minipage}{0.5\textwidth}
\includegraphics[width=8.5cm]{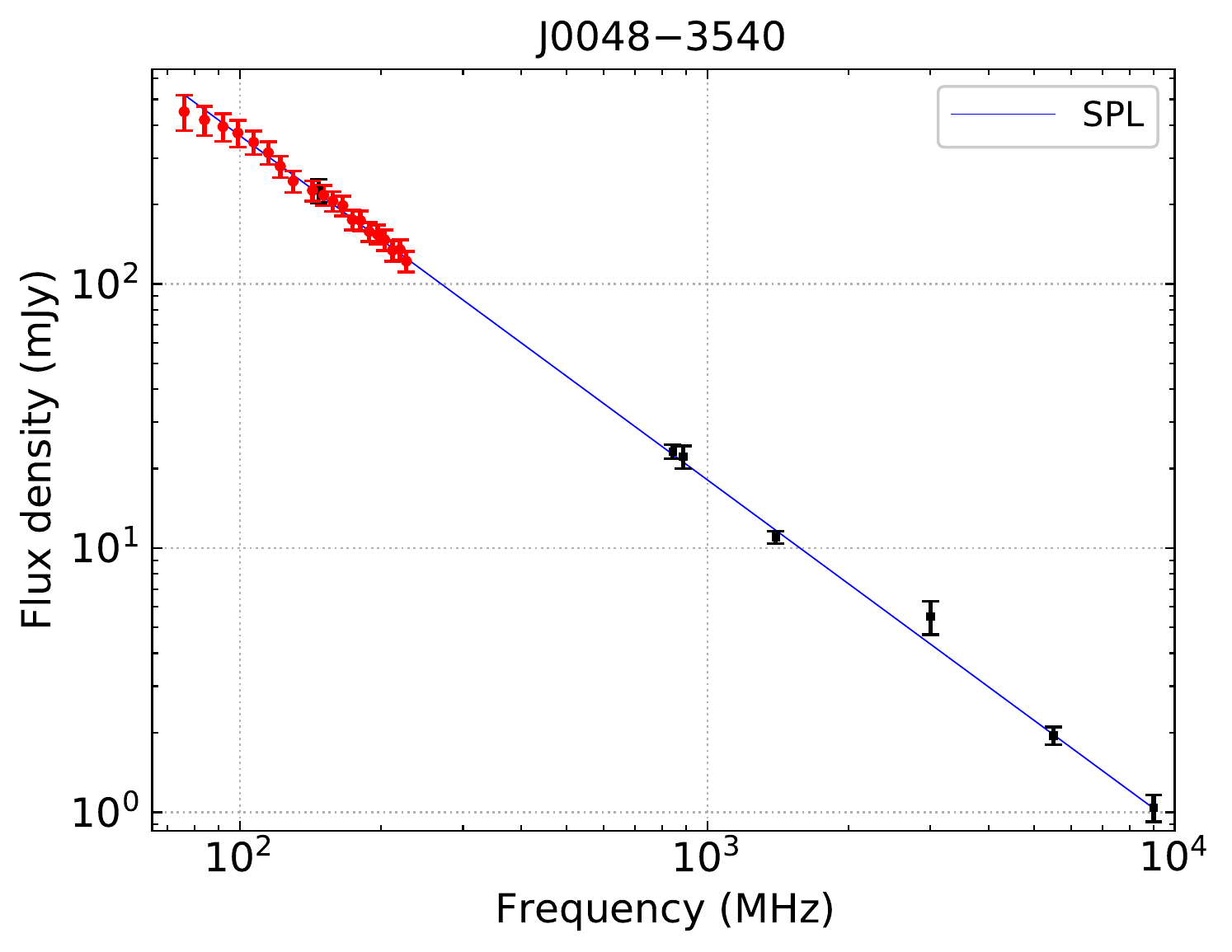}
\end{minipage}
\begin{minipage}{0.5\textwidth}
\vspace{0.2cm}
\includegraphics[width=7.5cm]{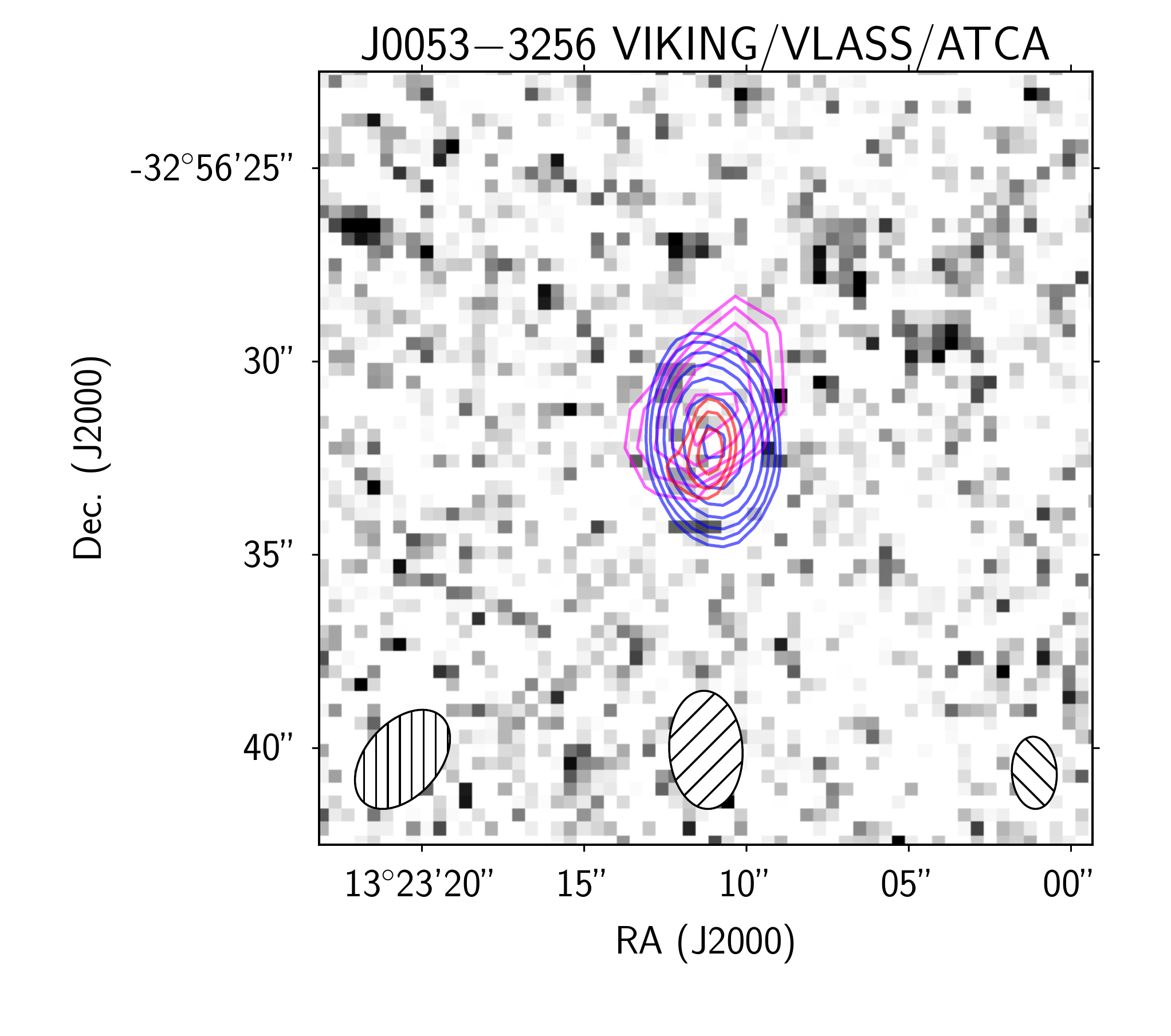}
\end{minipage}
\begin{minipage}{0.5\textwidth}
\includegraphics[width=8.5cm]{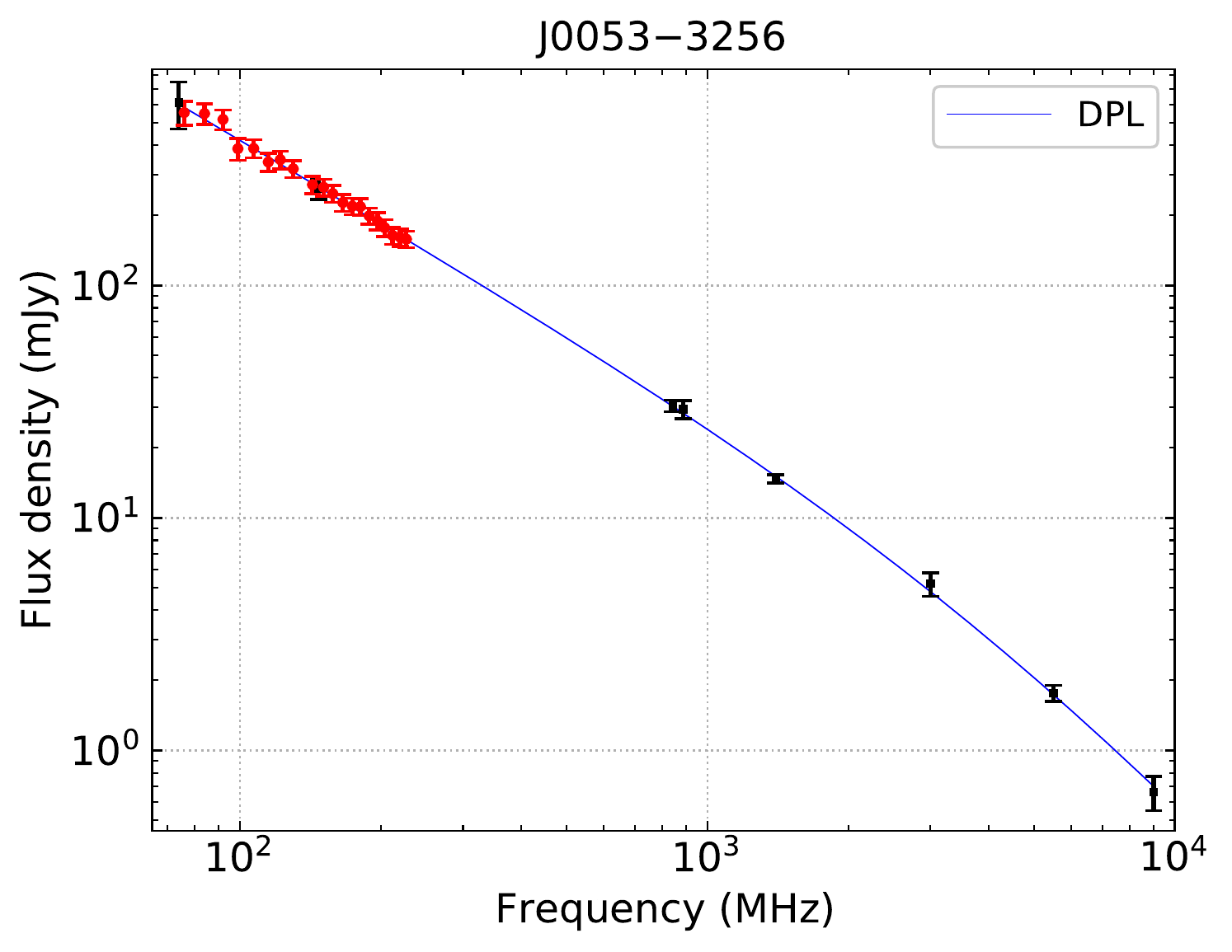}
\end{minipage}
\caption{{\em - continued.}} 
\end{figure*}

\setcounter{figure}{1} 
\begin{figure*}
\begin{minipage}{0.5\textwidth}
\vspace{0.2cm}
\includegraphics[width=7.5cm]{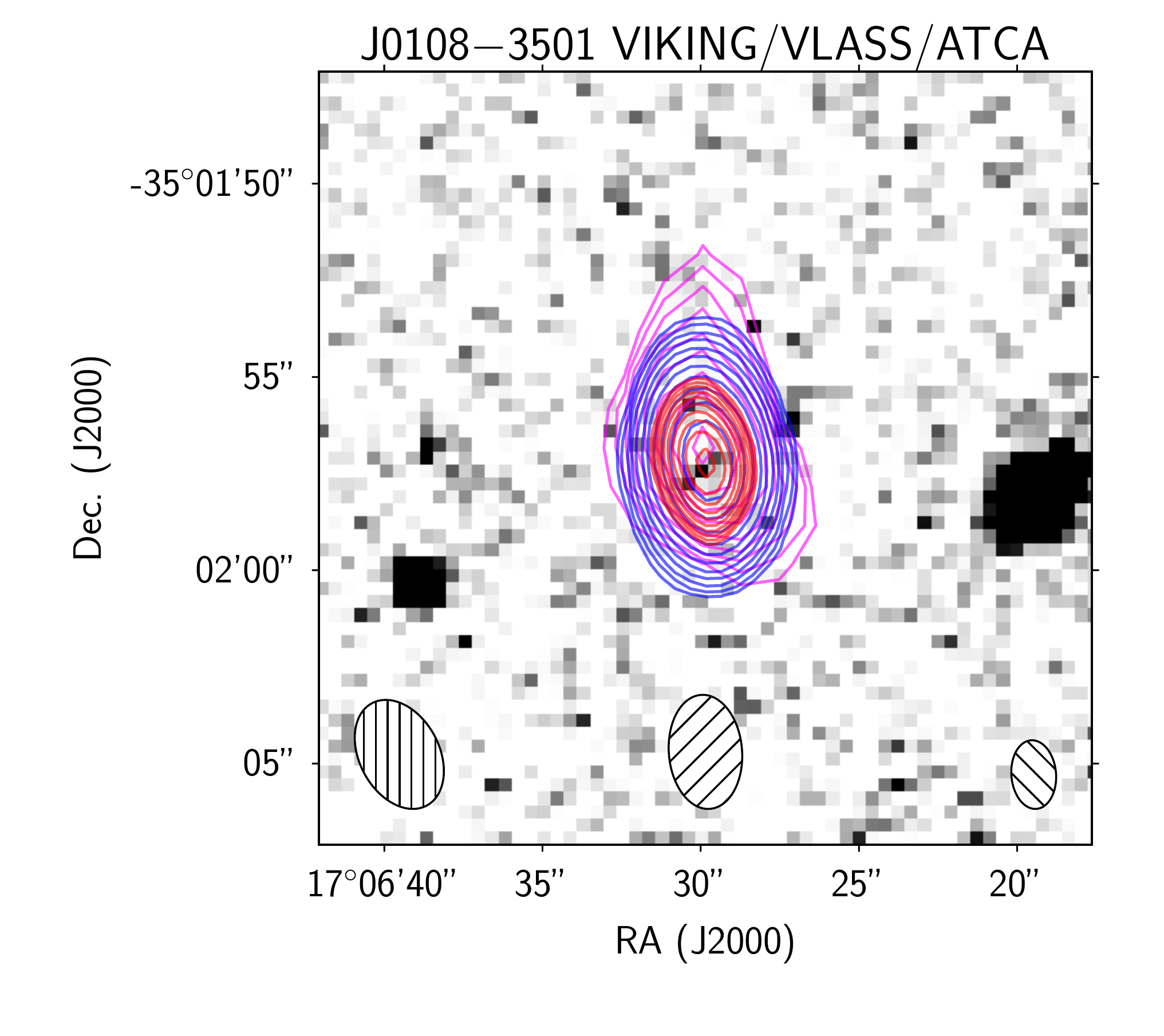}
\end{minipage}
\begin{minipage}{0.5\textwidth}
\includegraphics[width=8.5cm]{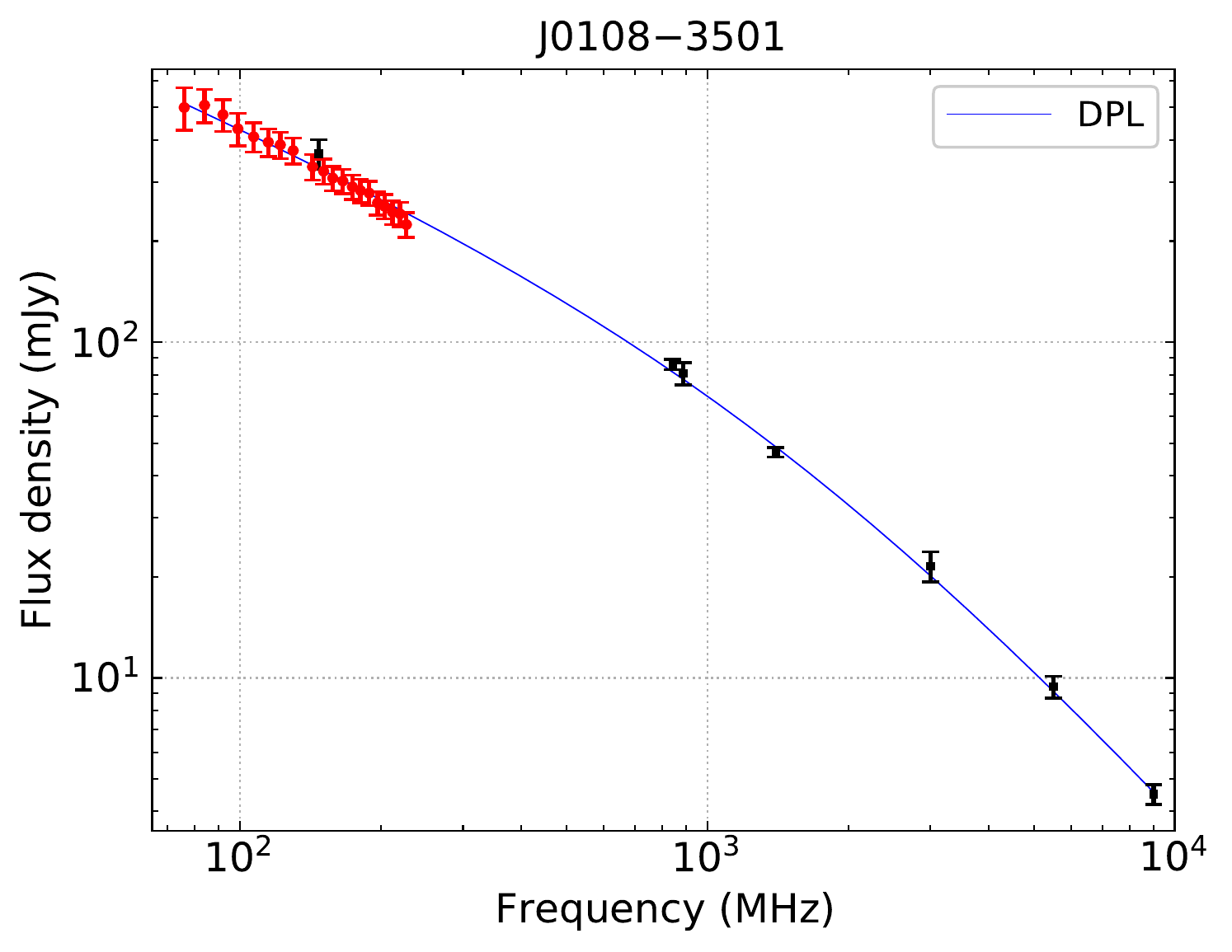}
\end{minipage}
\begin{minipage}{0.5\textwidth}
\vspace{0.2cm}
\includegraphics[width=7.5cm]{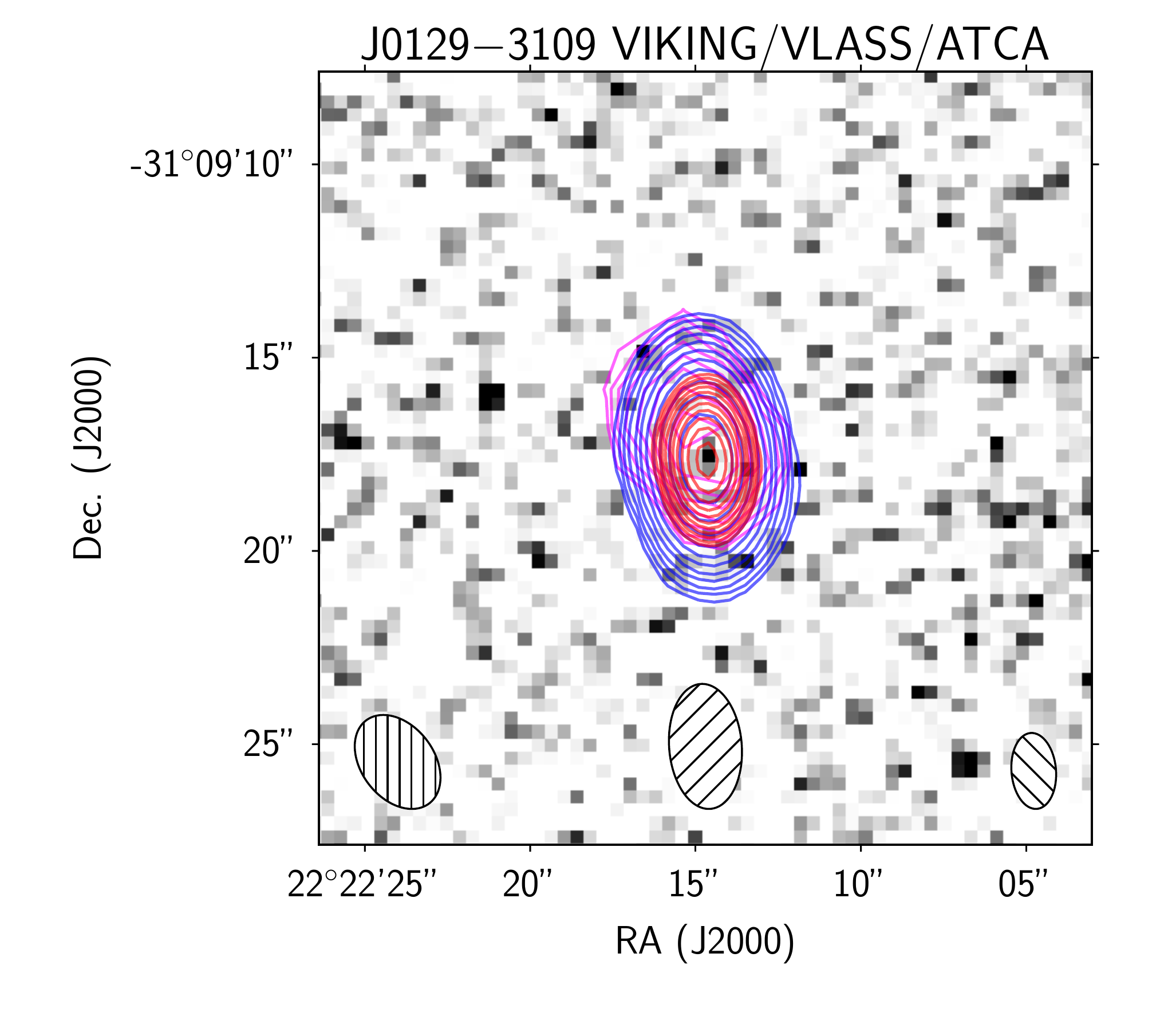}
\end{minipage}
\begin{minipage}{0.5\textwidth}
\includegraphics[width=8.5cm]{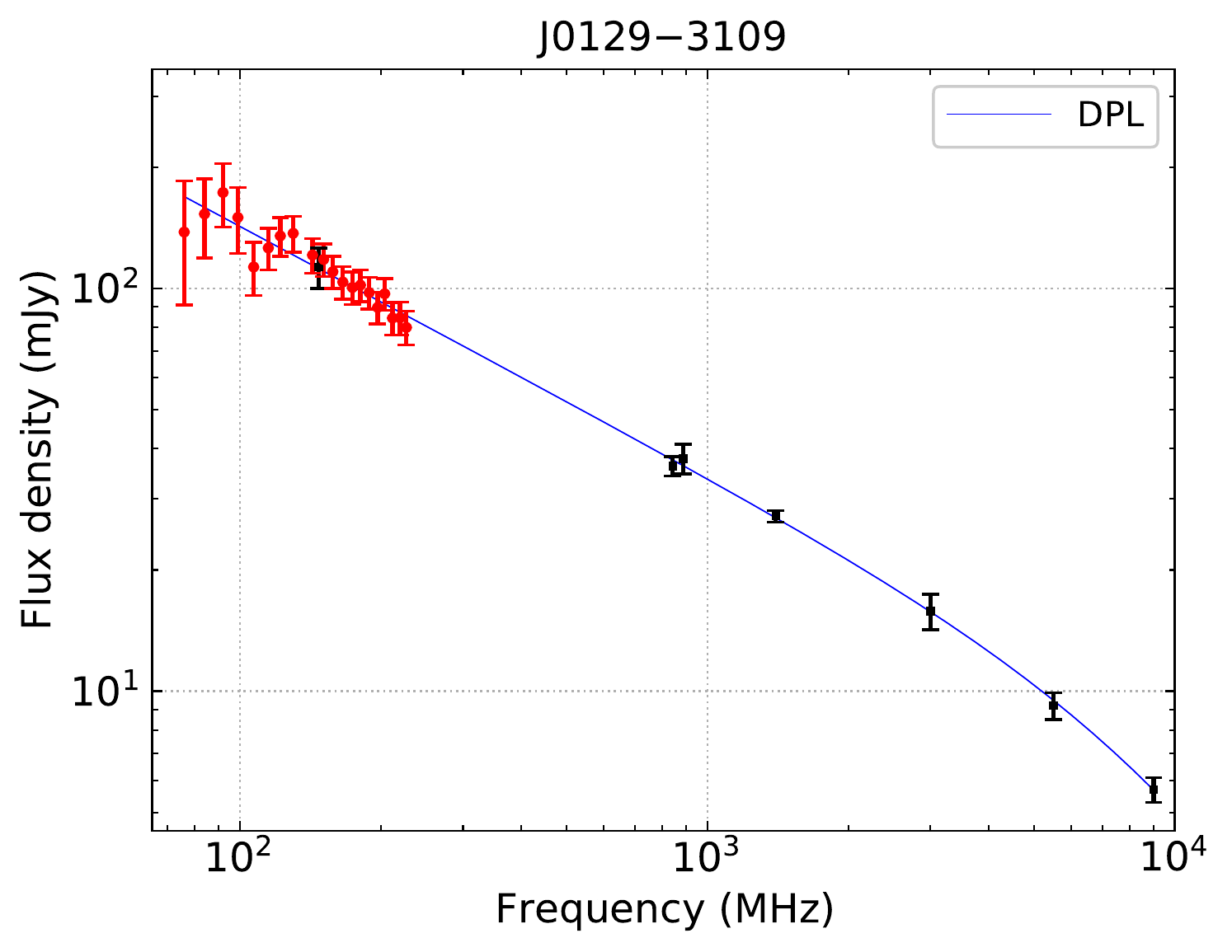}
\end{minipage}
\begin{minipage}{0.5\textwidth}
\vspace{0.2cm}
\includegraphics[width=7.5cm]{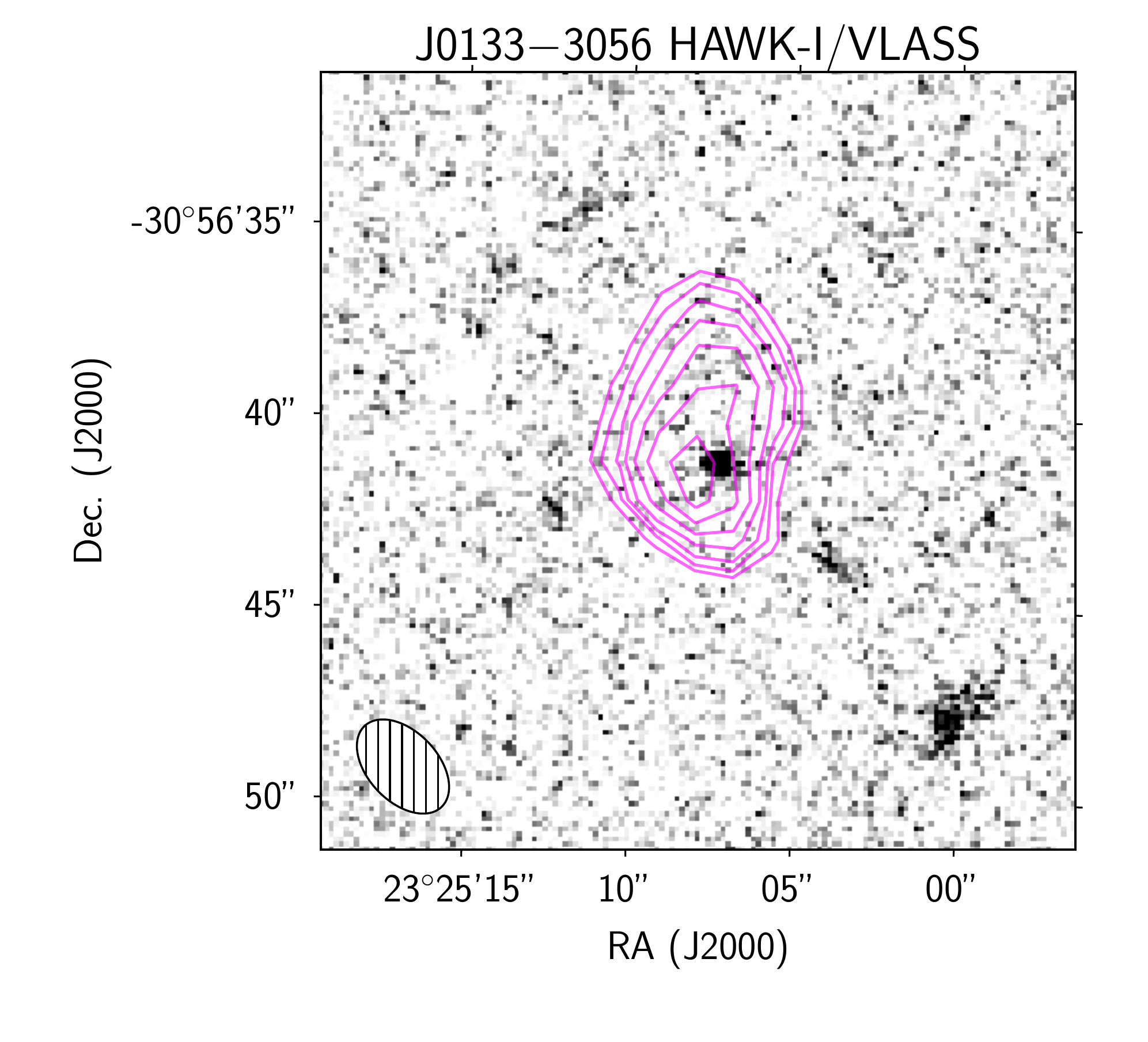}
\end{minipage}
\begin{minipage}{0.5\textwidth}
\includegraphics[width=8.5cm]{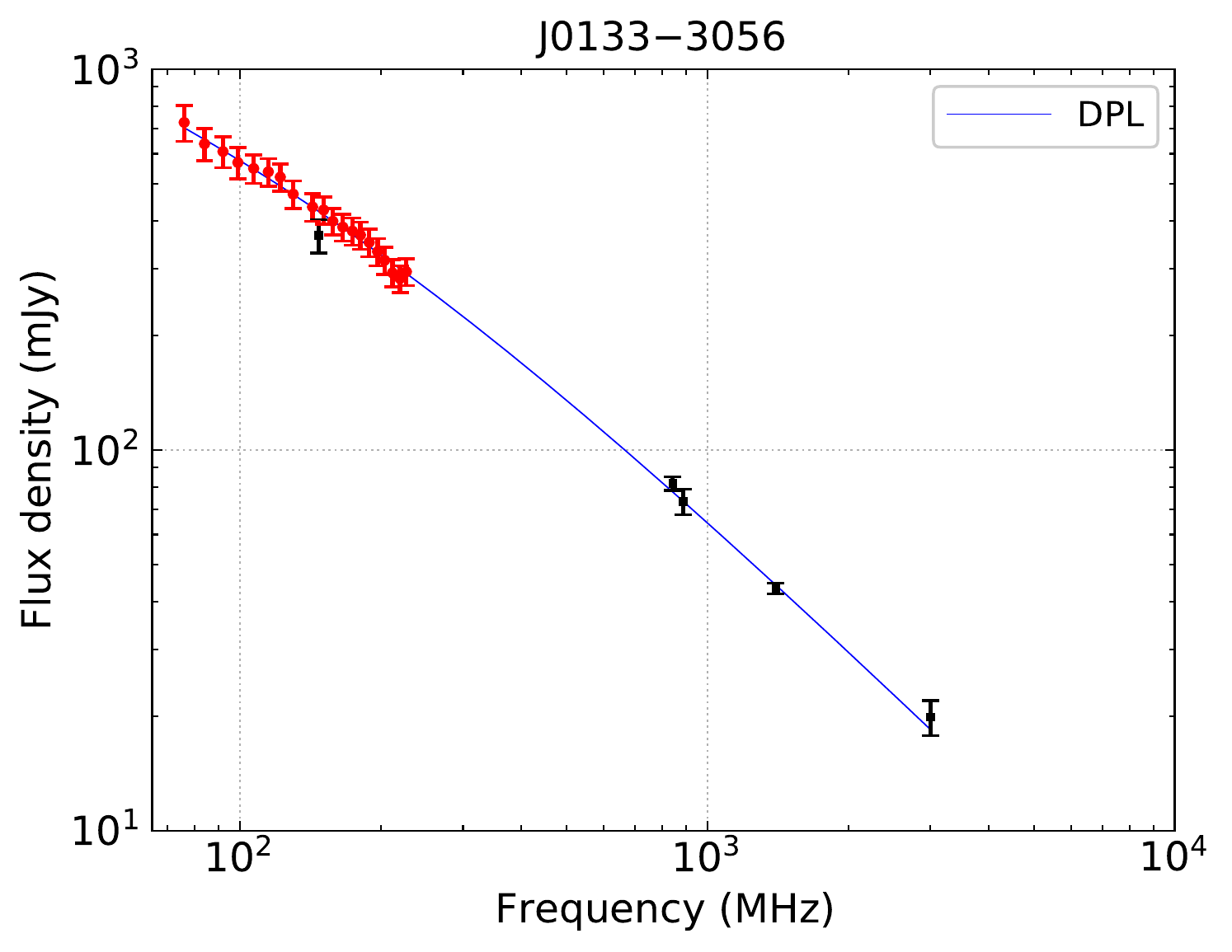}
\end{minipage}
\caption{{\em - continued.}}
\end{figure*}

\setcounter{figure}{1} 
\begin{figure*}
\begin{minipage}{0.5\textwidth}
\vspace{0.2cm}
\includegraphics[width=7.5cm]{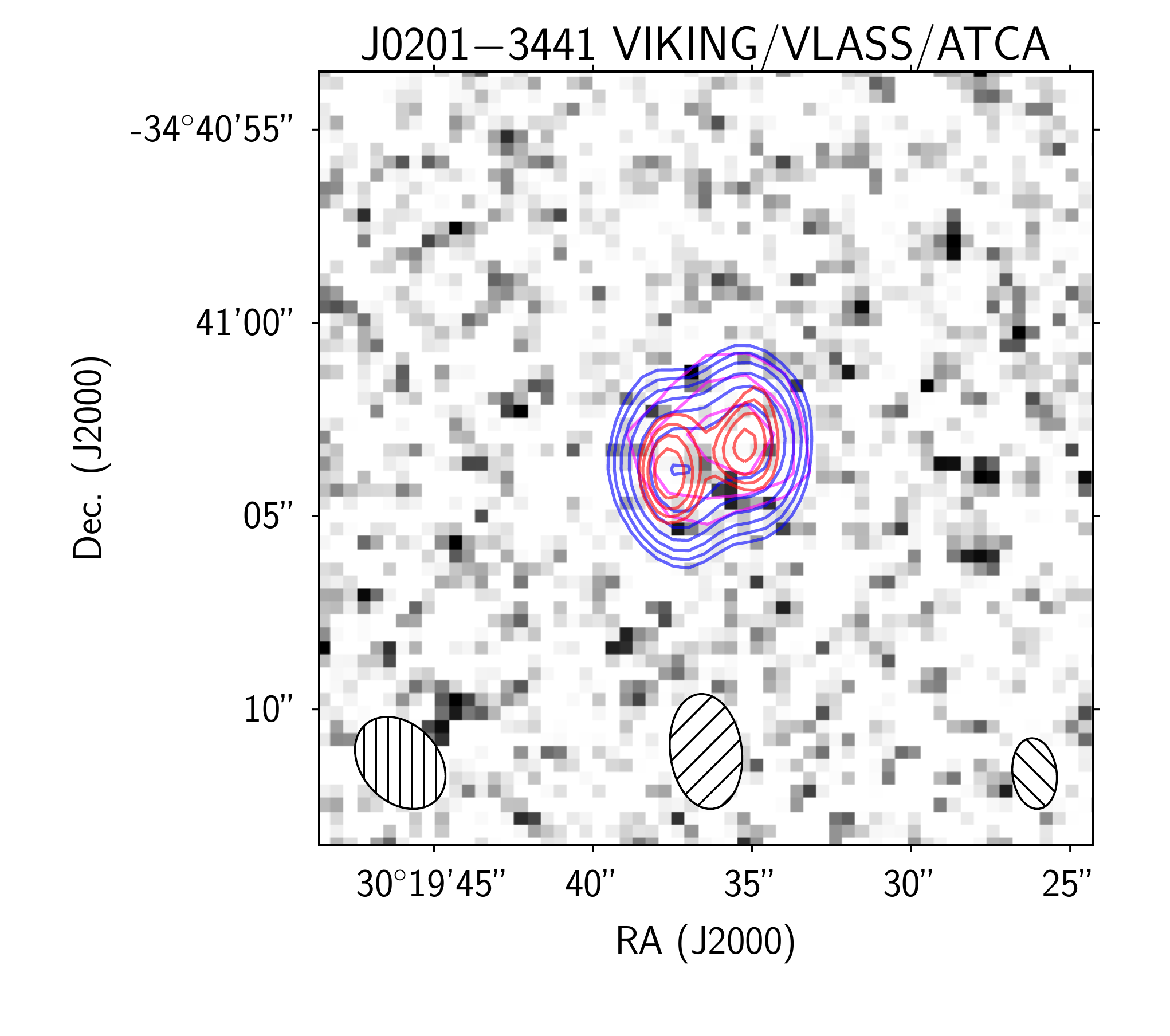}
\end{minipage}
\begin{minipage}{0.5\textwidth}
\includegraphics[width=8.5cm]{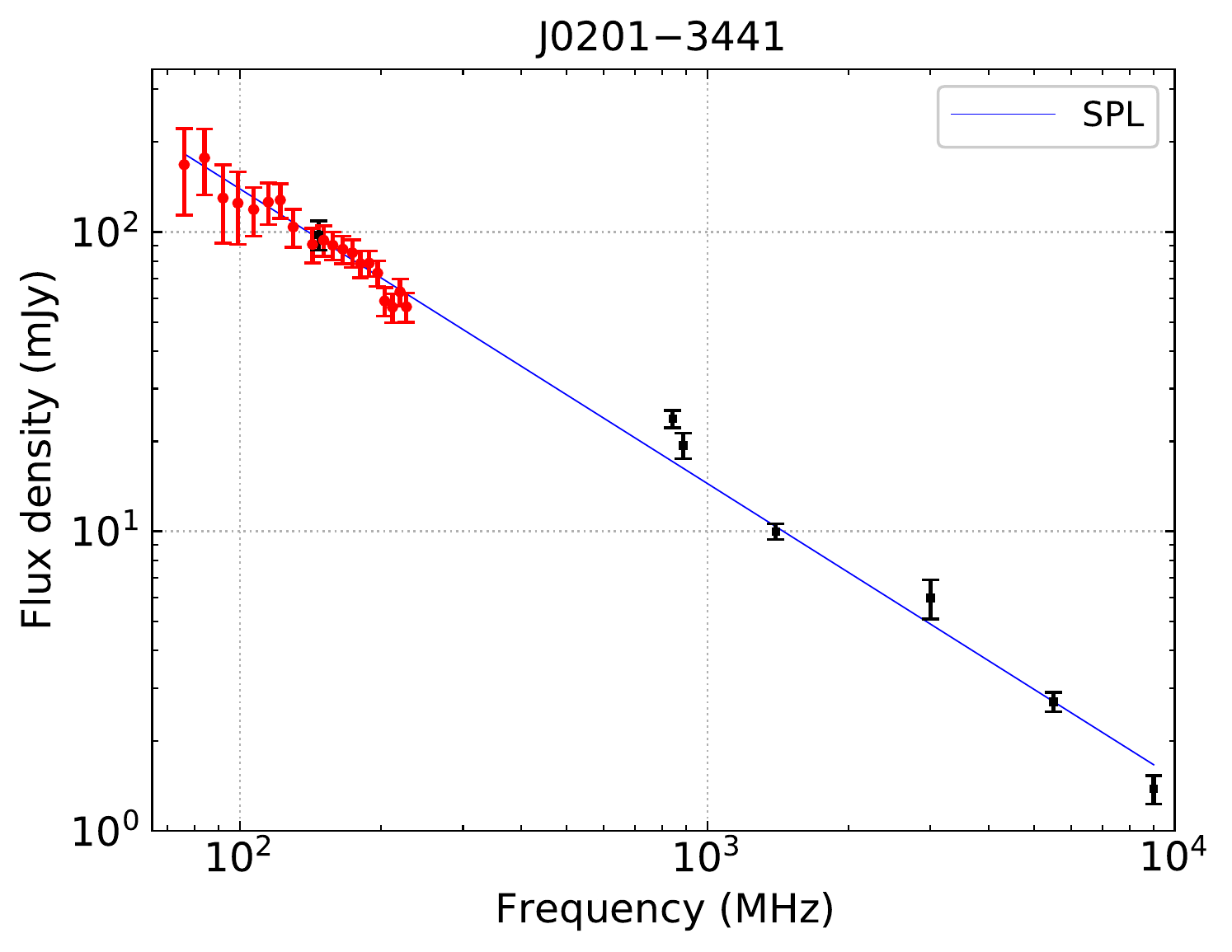}
\end{minipage}
\begin{minipage}{0.5\textwidth}
\vspace{0.2cm}
\includegraphics[width=7.5cm]{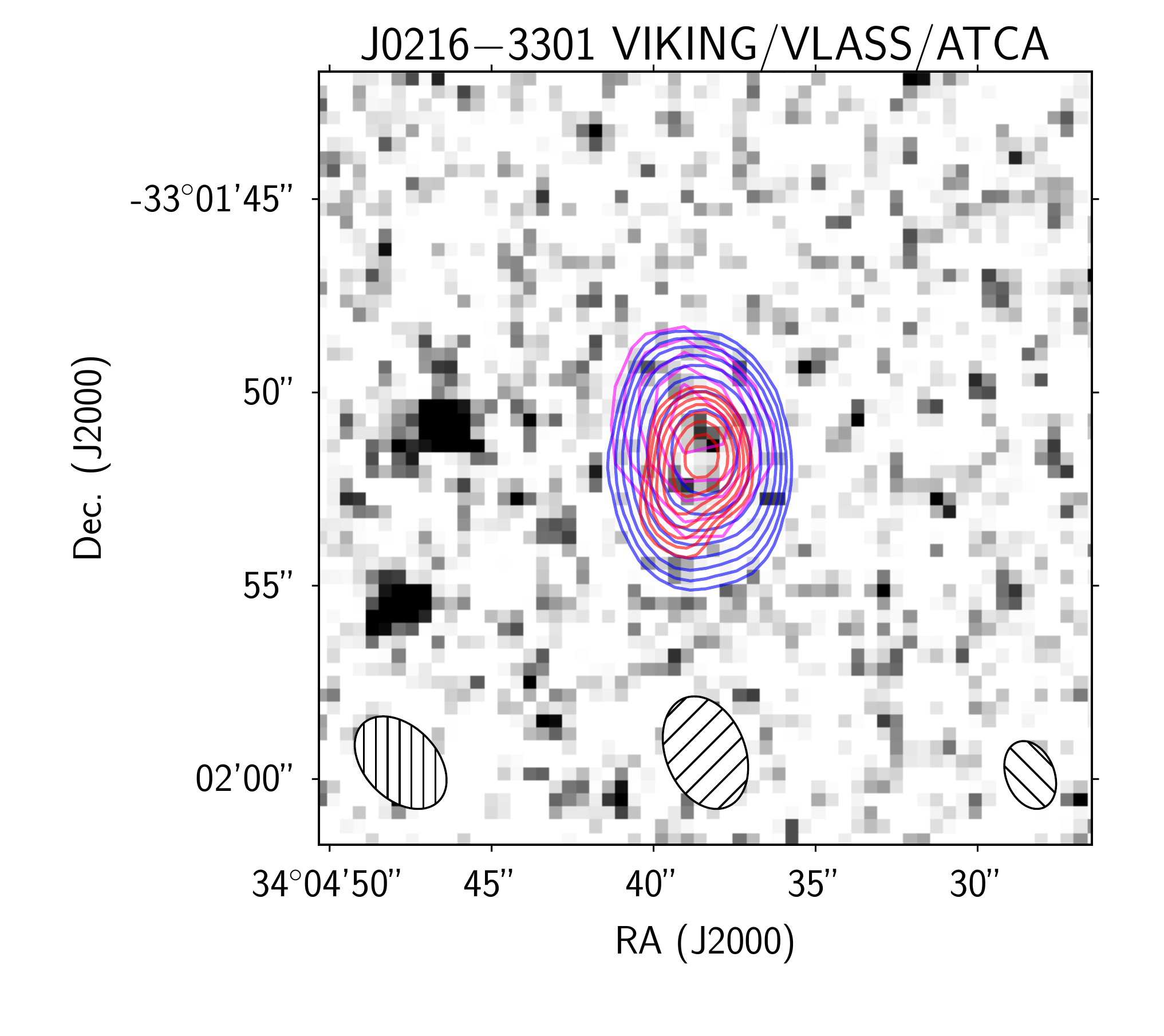}
\end{minipage}
\begin{minipage}{0.5\textwidth}
\includegraphics[width=8.5cm]{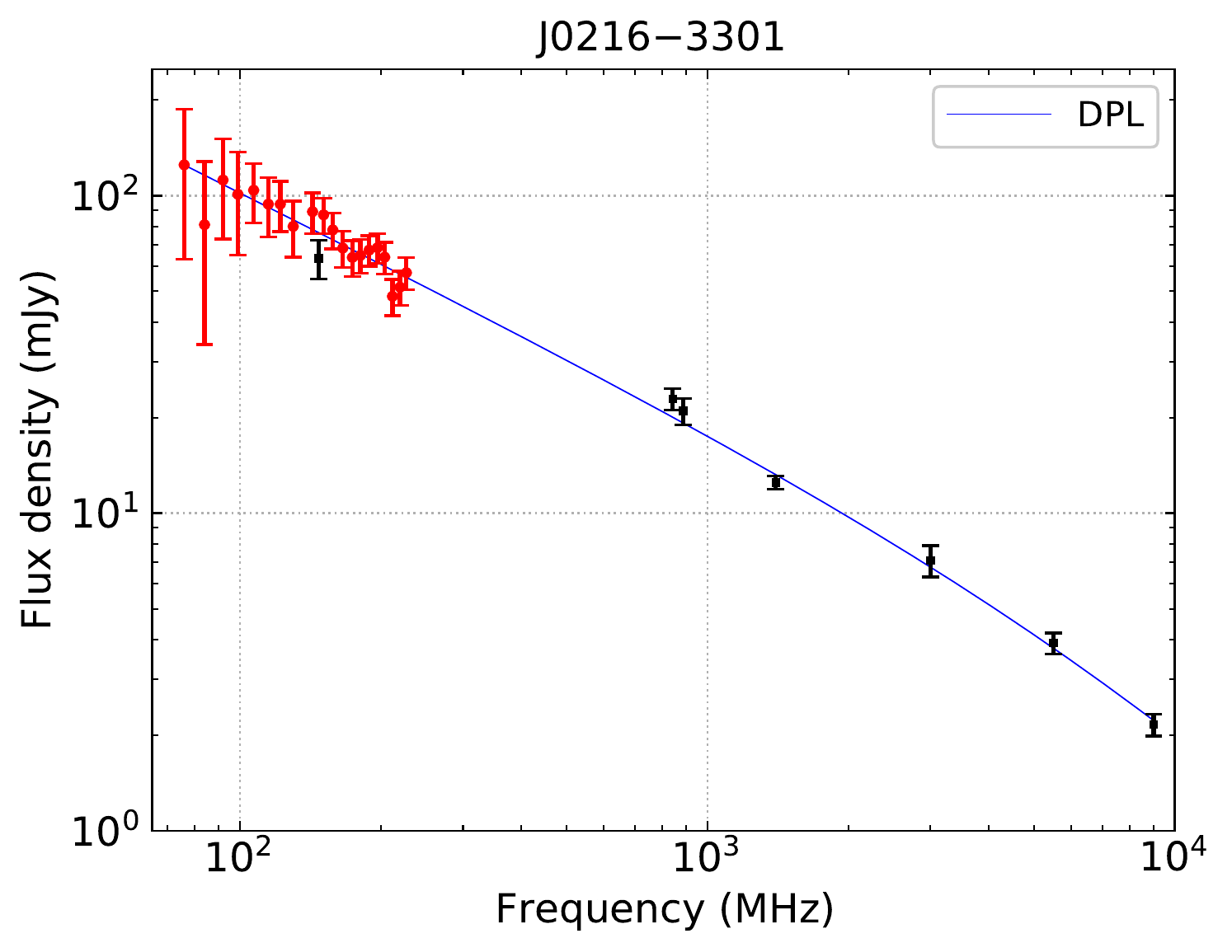}
\end{minipage}
\begin{minipage}{0.5\textwidth}
\vspace{0.2cm}
\includegraphics[width=7.5cm]{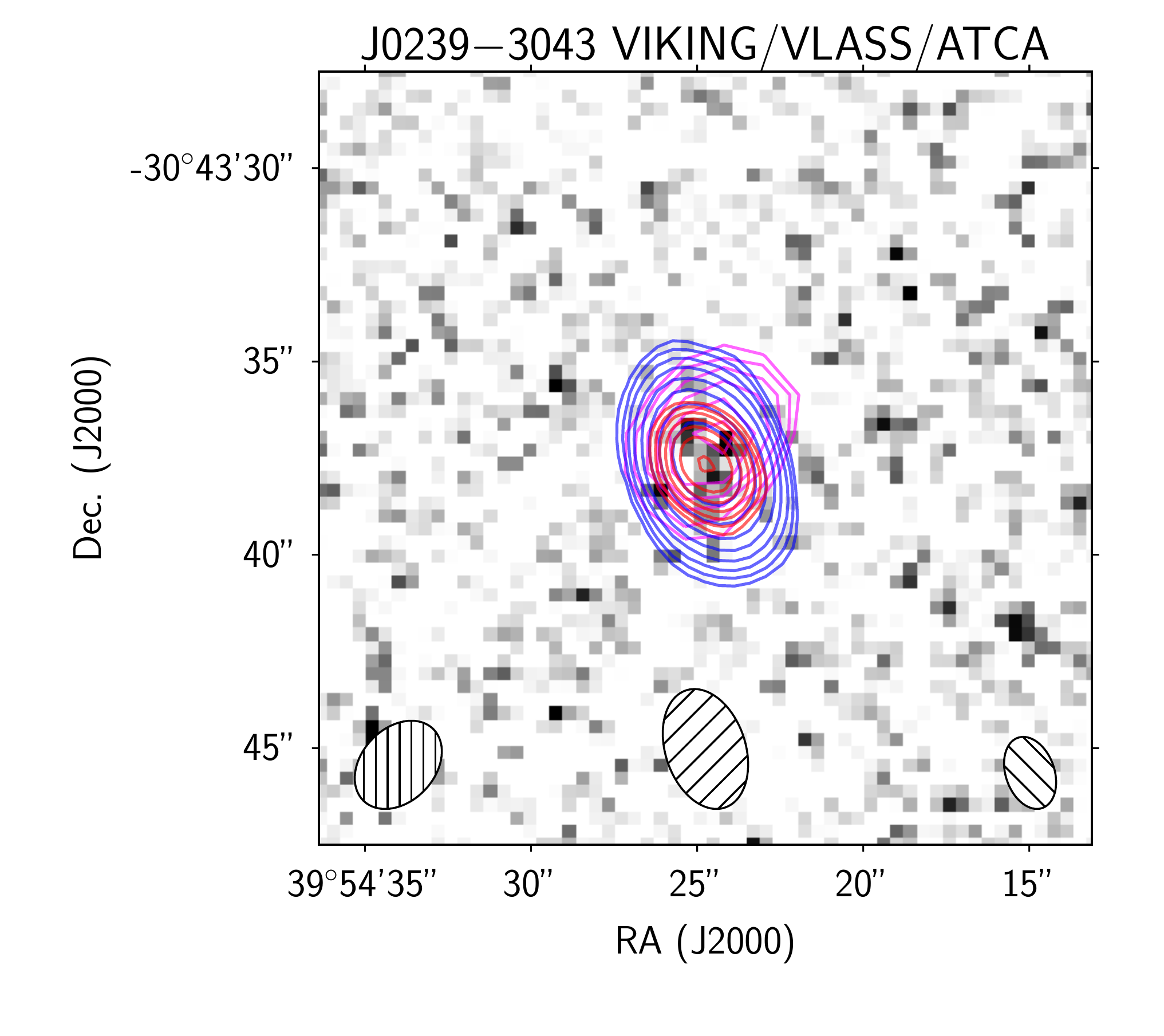}
\end{minipage}
\begin{minipage}{0.5\textwidth}
\includegraphics[width=8.5cm]{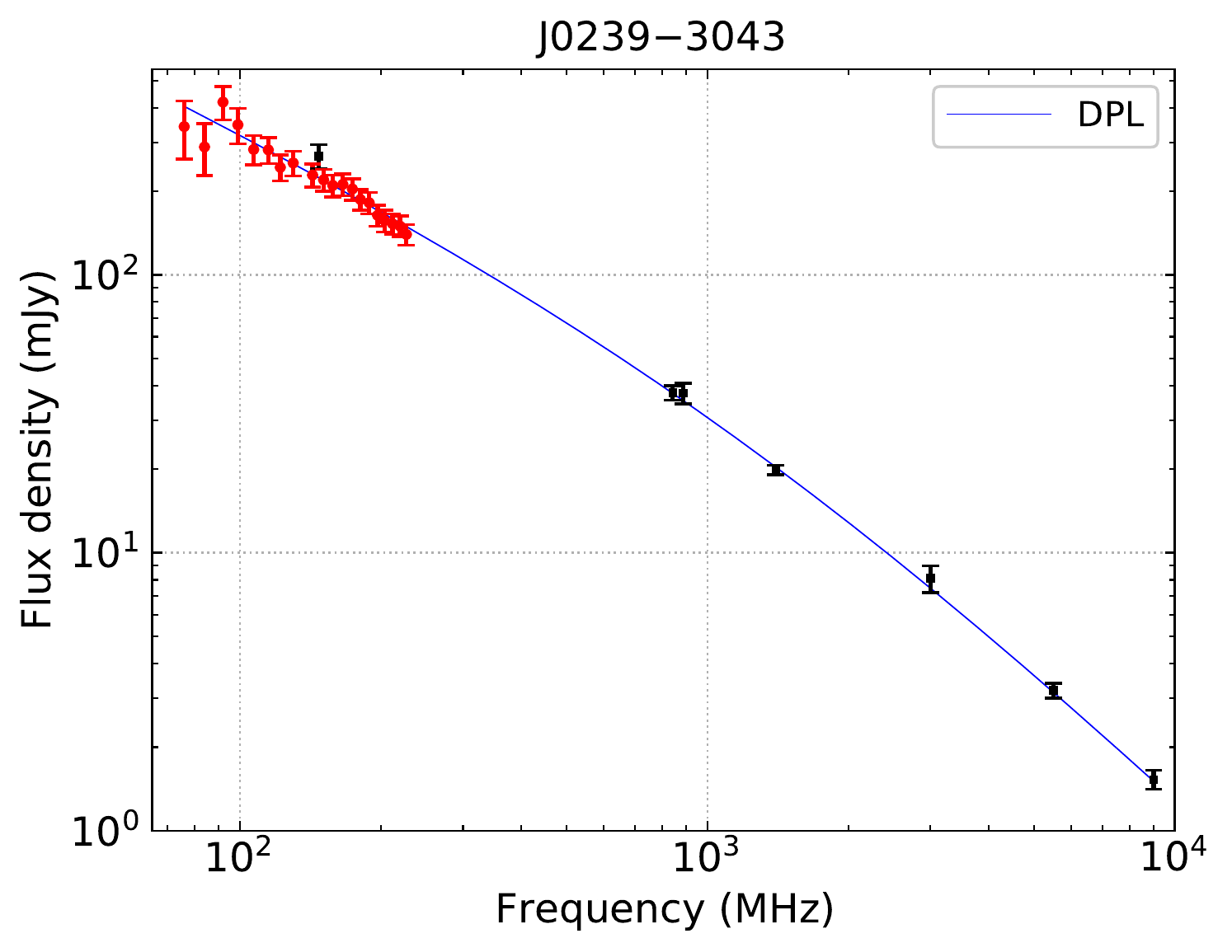}
\end{minipage}
\caption{{\em - continued.}}
\end{figure*}

\setcounter{figure}{1} 
\begin{figure*}
\begin{minipage}{0.5\textwidth}
\vspace{0.2cm}
\includegraphics[width=7.5cm]{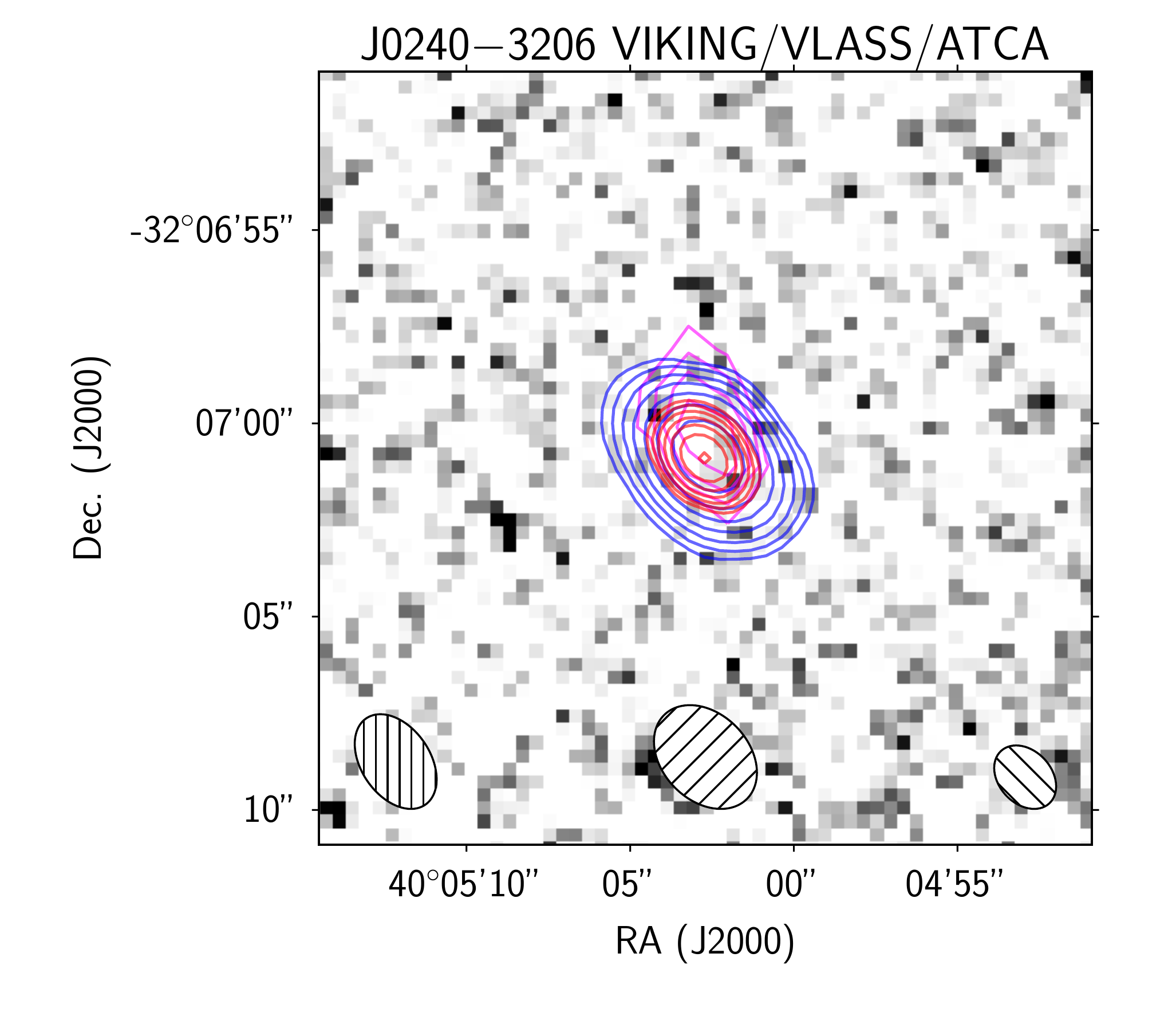}
\end{minipage}
\begin{minipage}{0.5\textwidth}
\includegraphics[width=8.5cm]{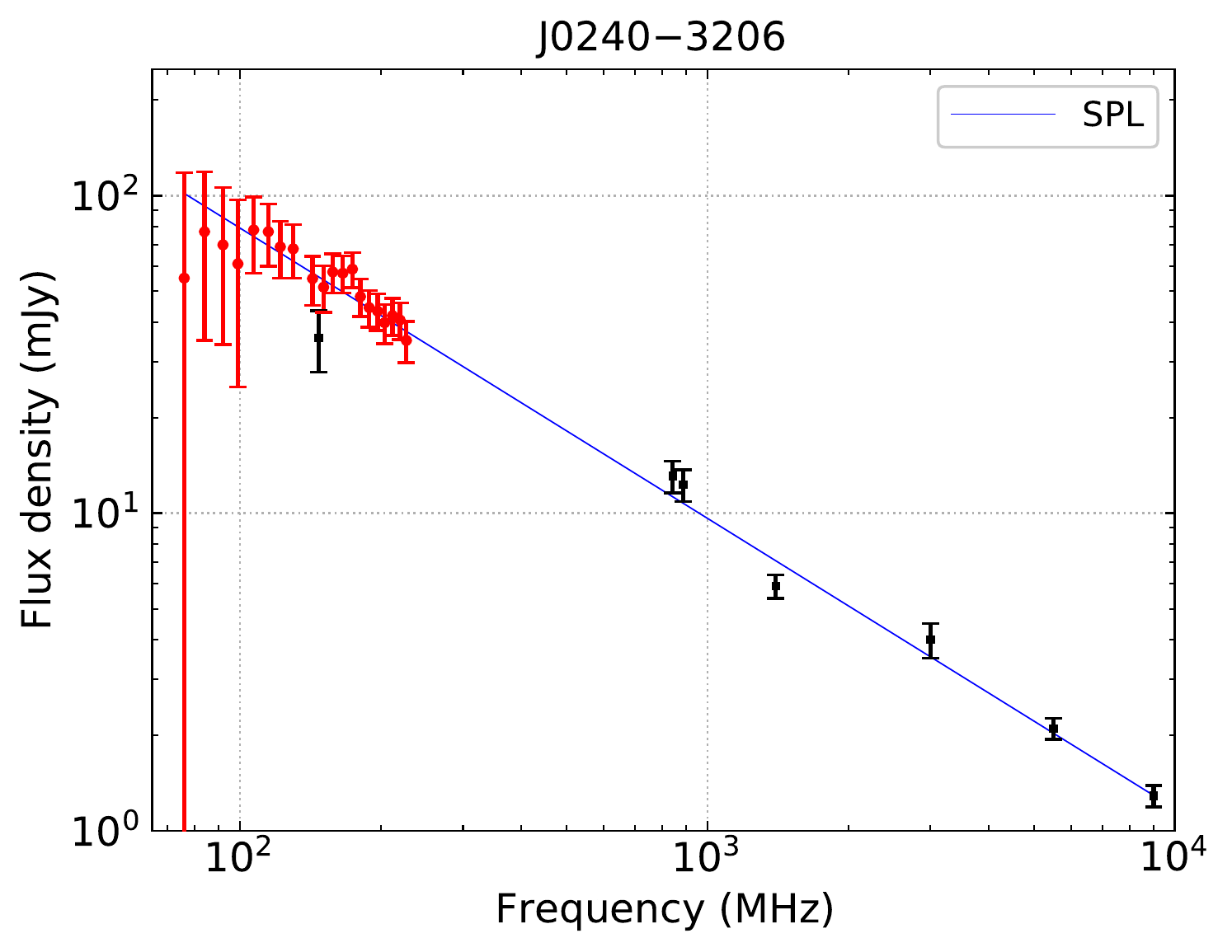}
\end{minipage}
\begin{minipage}{0.5\textwidth}
\vspace{0.2cm}
\includegraphics[width=7.5cm]{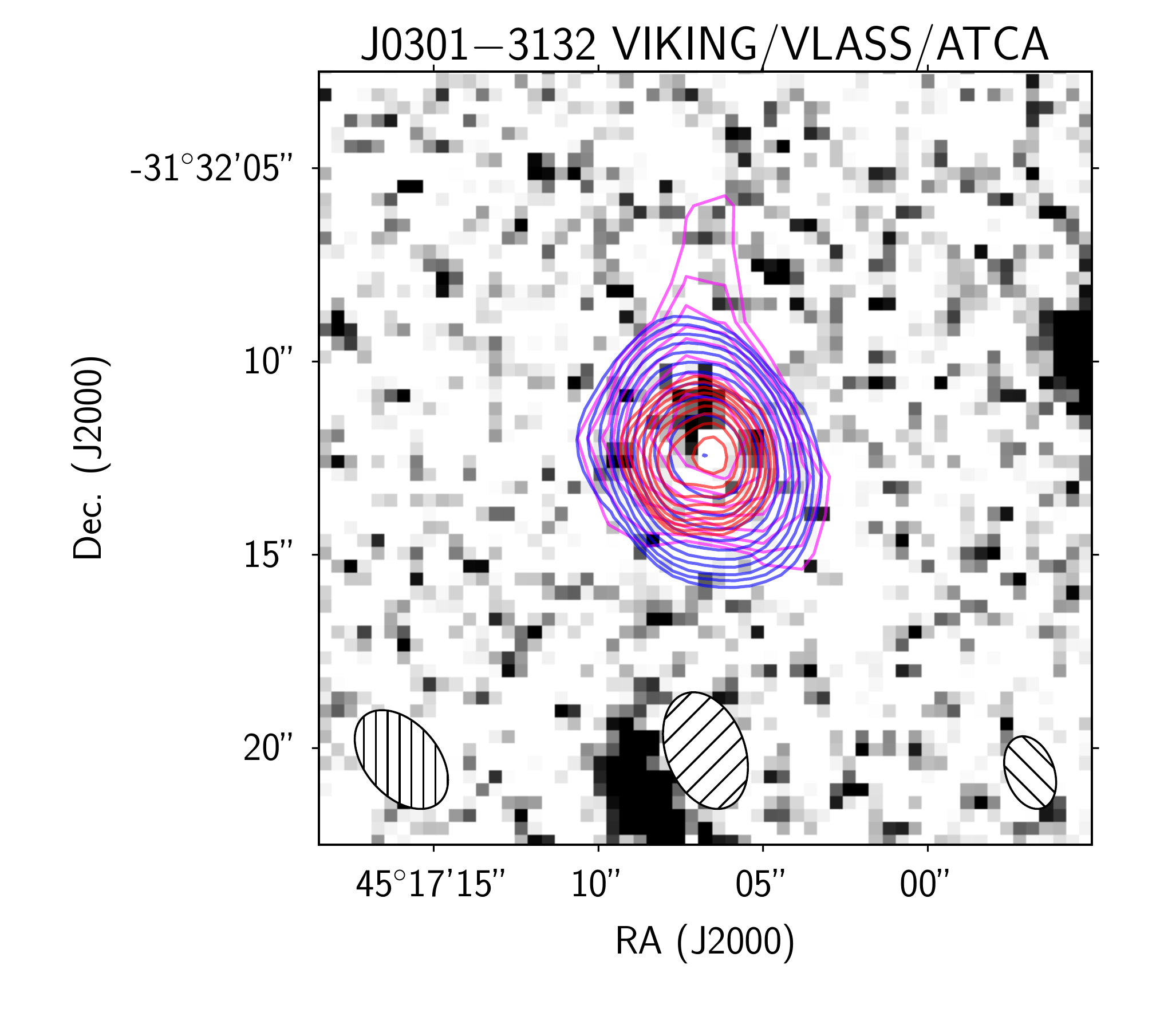}
\end{minipage}
\begin{minipage}{0.5\textwidth}
\includegraphics[width=8.5cm]{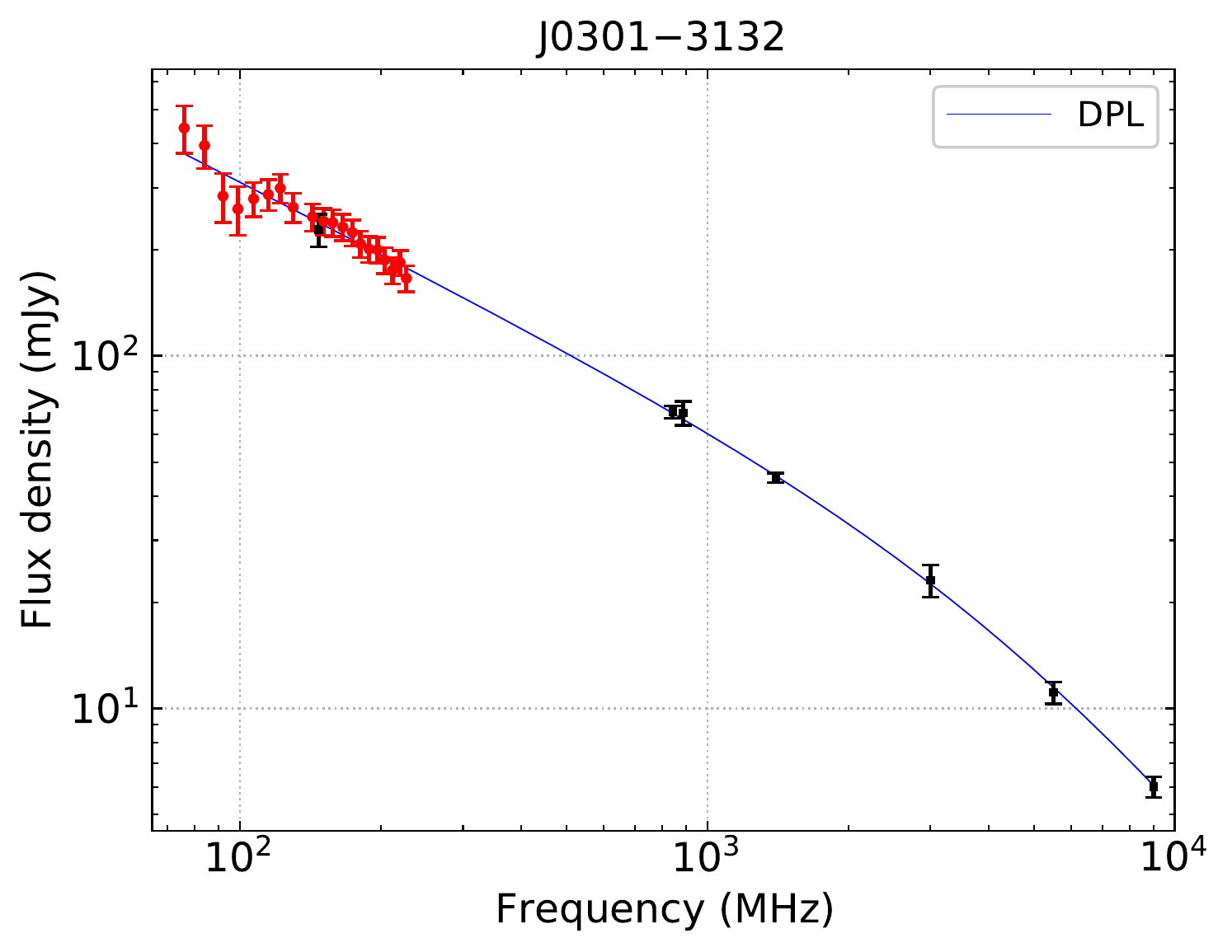}
\end{minipage}
\begin{minipage}{0.5\textwidth}
\vspace{0.2cm}
\includegraphics[width=7.5cm]{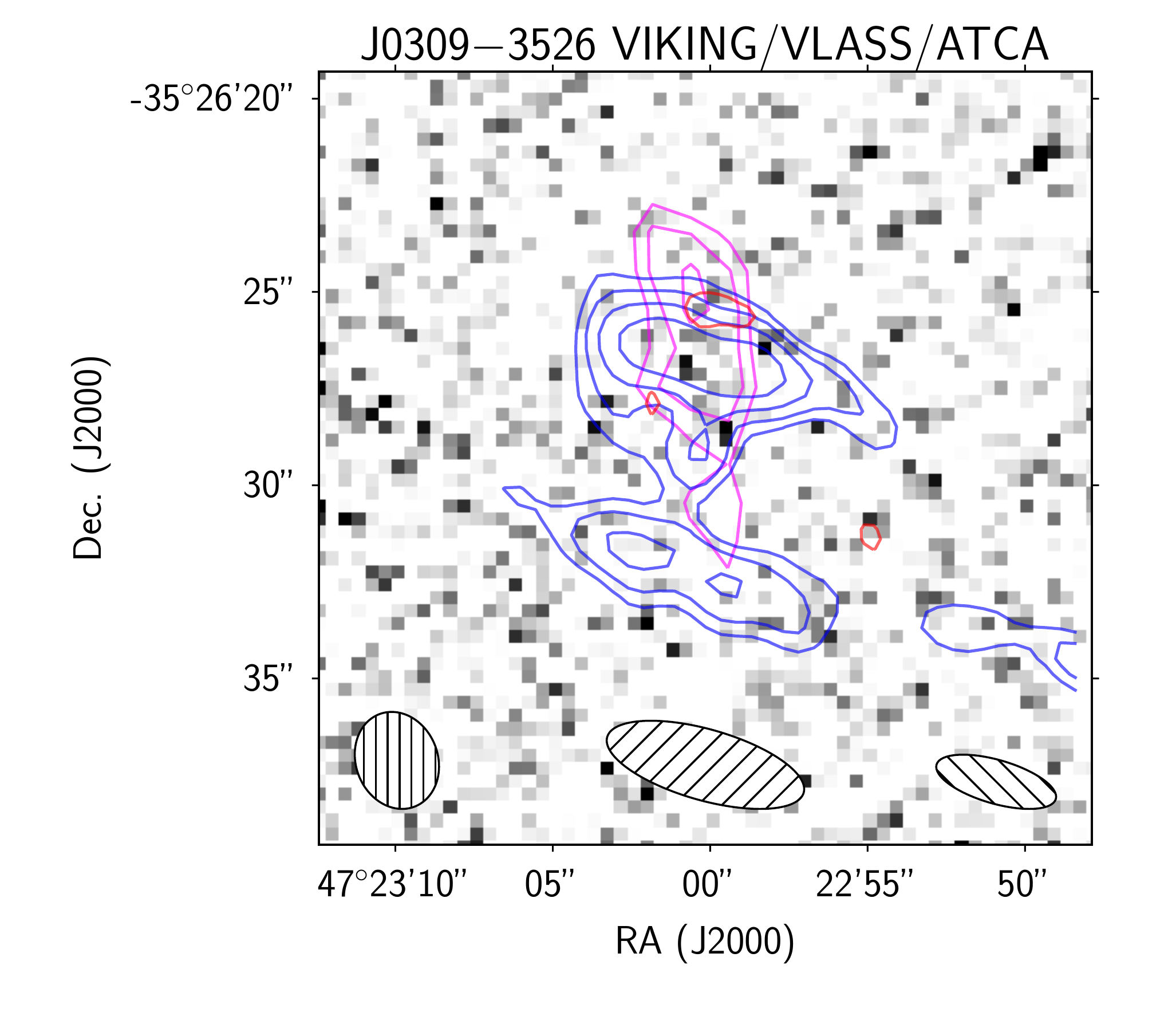}
\end{minipage}
\begin{minipage}{0.5\textwidth}
\includegraphics[width=8.5cm]{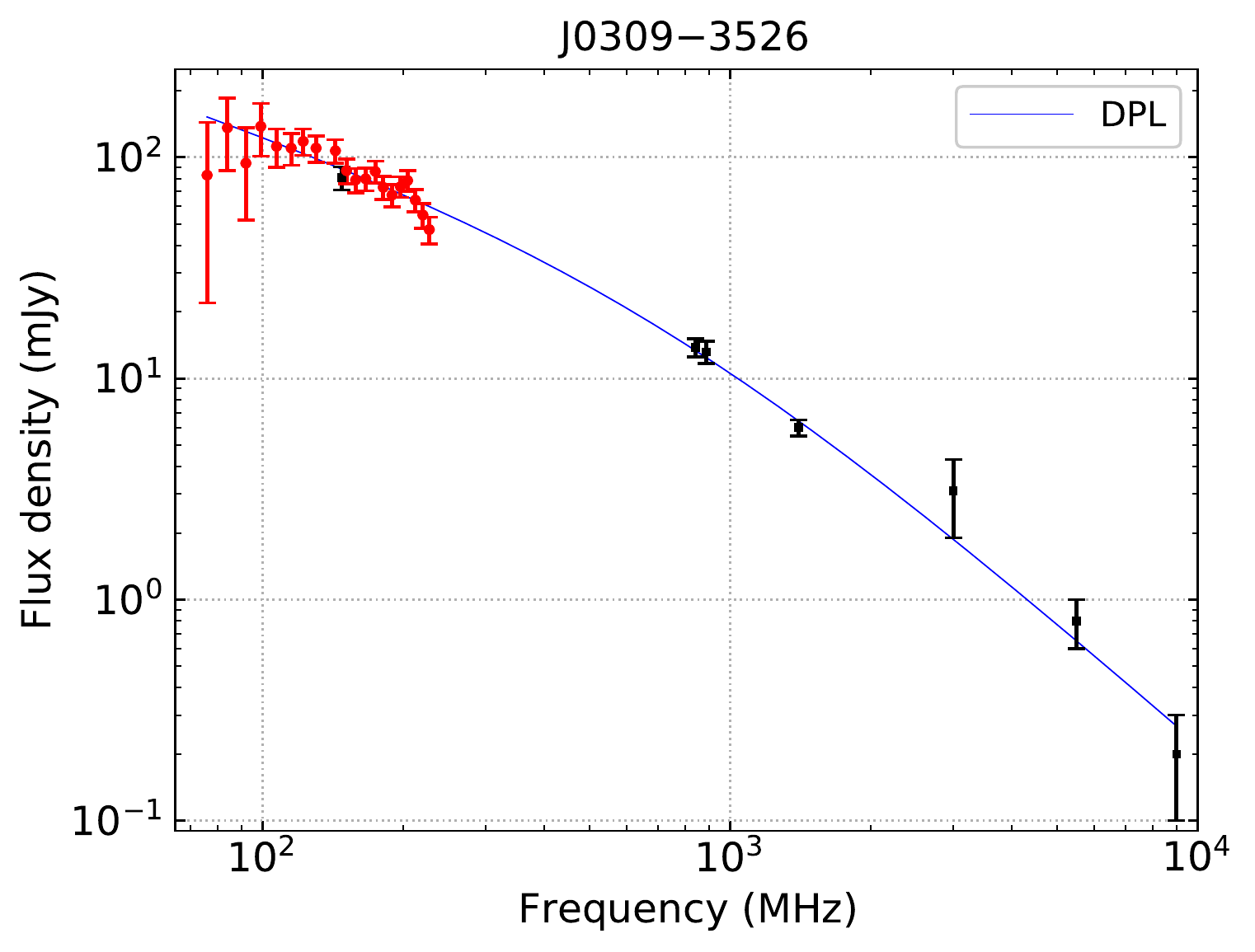}
\end{minipage}
\caption{{\em - continued.}}
\end{figure*}

\setcounter{figure}{1} 
\begin{figure*}
\begin{minipage}{0.5\textwidth}
\vspace{0.2cm}
\includegraphics[width=7.5cm]{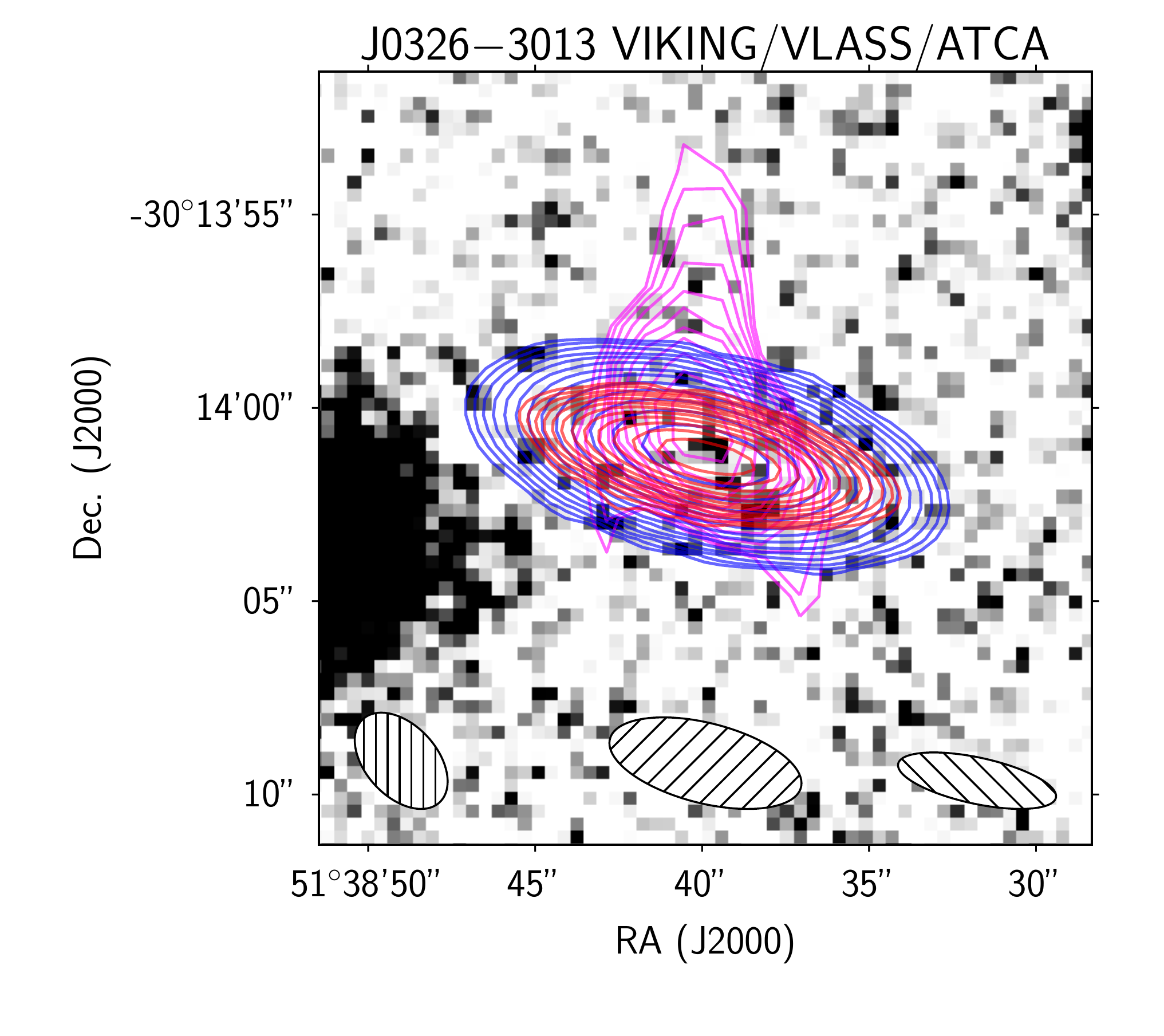}
\end{minipage}
\begin{minipage}{0.5\textwidth}
\includegraphics[width=8.5cm]{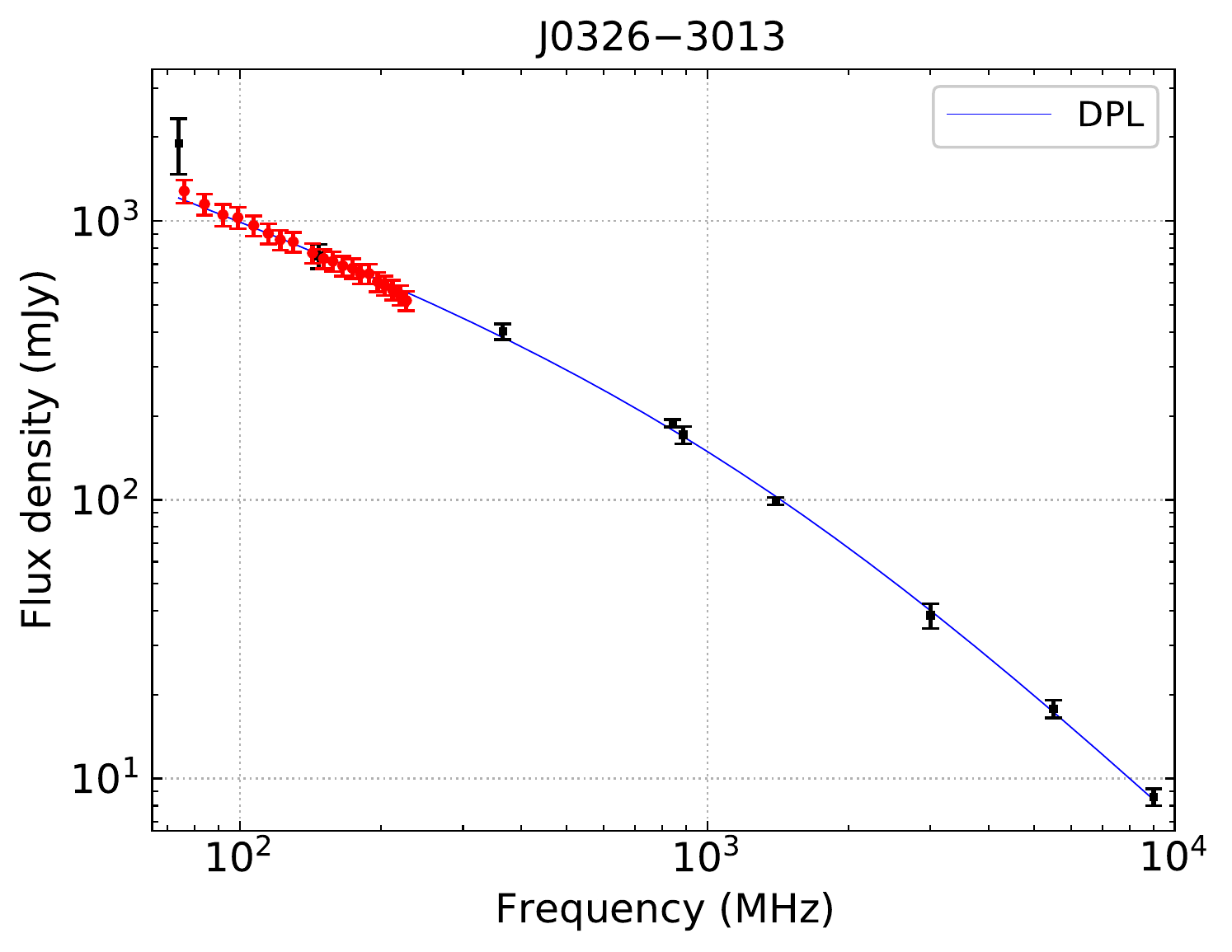}
\end{minipage}
\begin{minipage}{0.5\textwidth}
\vspace{0.2cm}
\includegraphics[width=7.5cm]{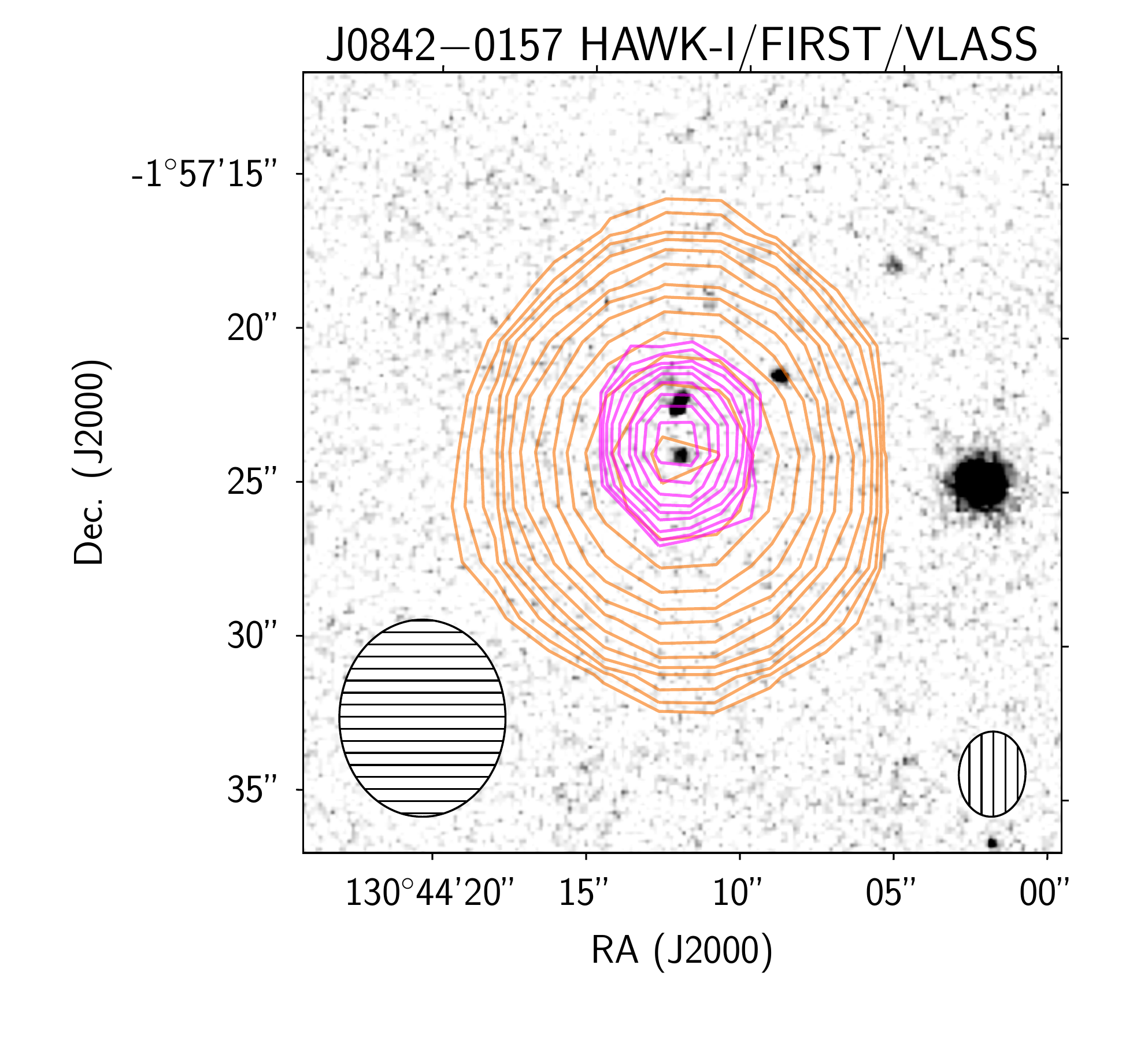}
\end{minipage}
\begin{minipage}{0.5\textwidth}
\includegraphics[width=8.5cm]{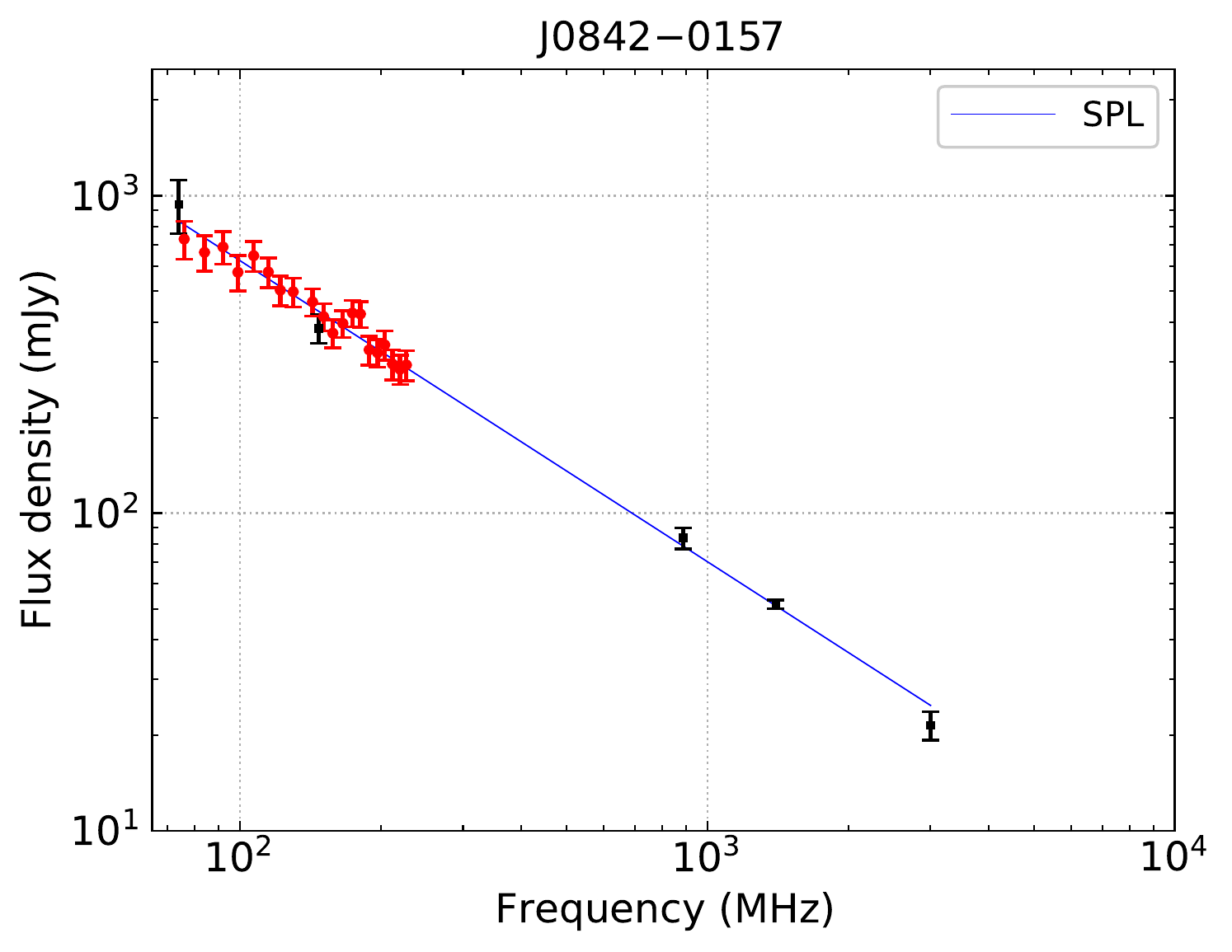}
\end{minipage}
\begin{minipage}{0.5\textwidth}
\vspace{0.2cm}
\includegraphics[width=7.5cm]{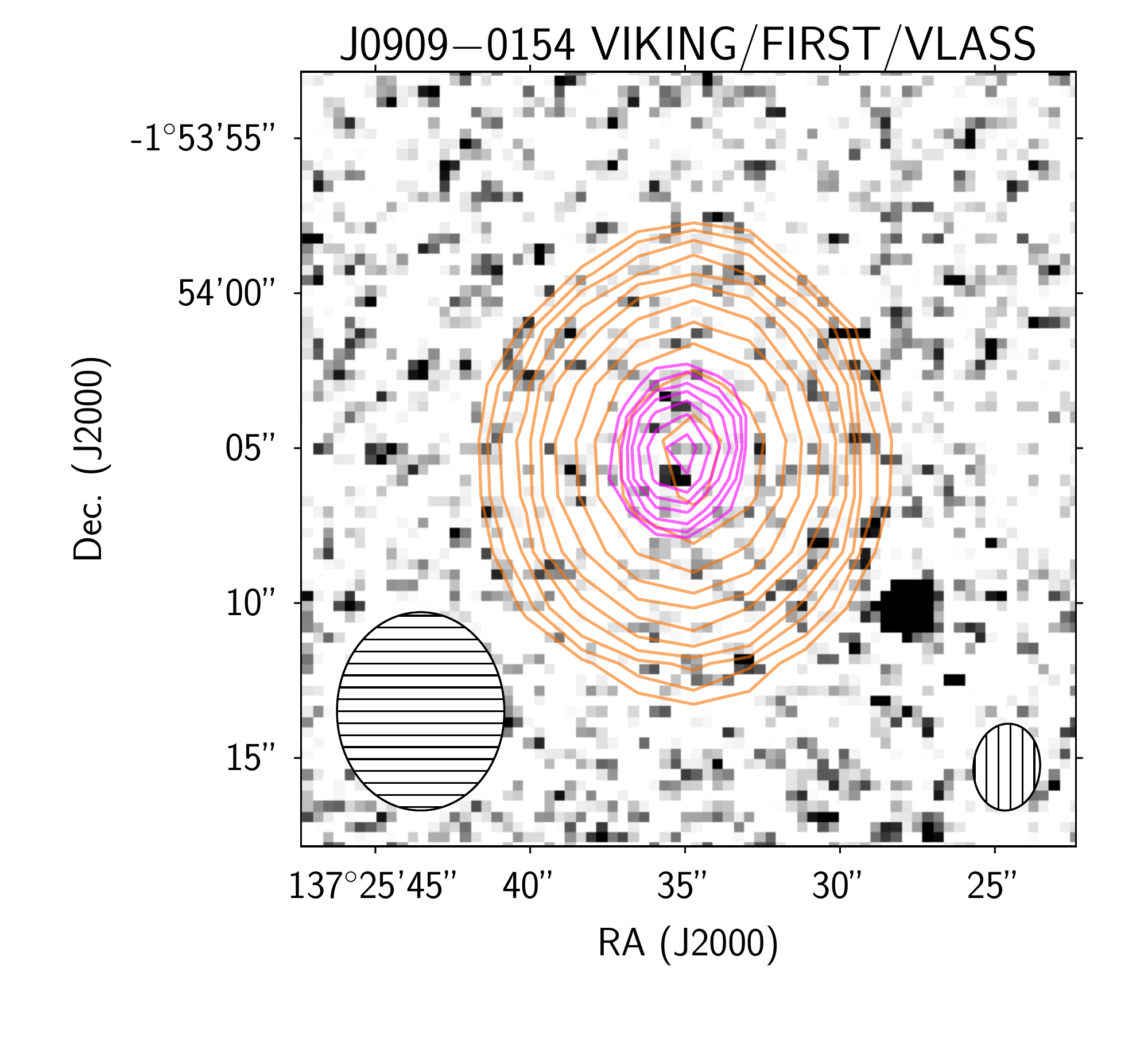}
\end{minipage}
\begin{minipage}{0.5\textwidth}
\includegraphics[width=8.5cm]{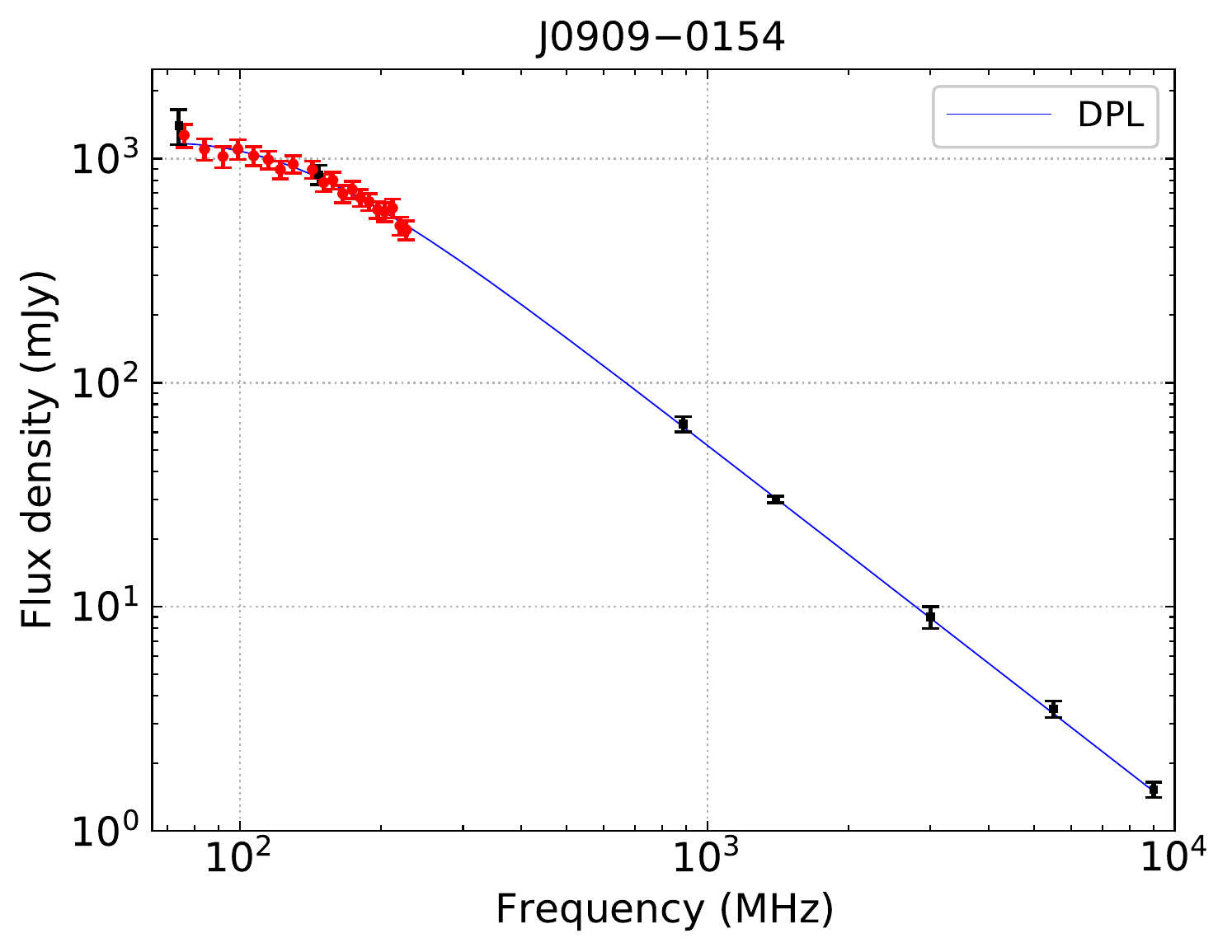}
\end{minipage}
\caption{{\em - continued.}}
\end{figure*}

\setcounter{figure}{1} 
\begin{figure*}
\begin{minipage}{0.5\textwidth}
\vspace{0.2cm}
\includegraphics[width=7.5cm]{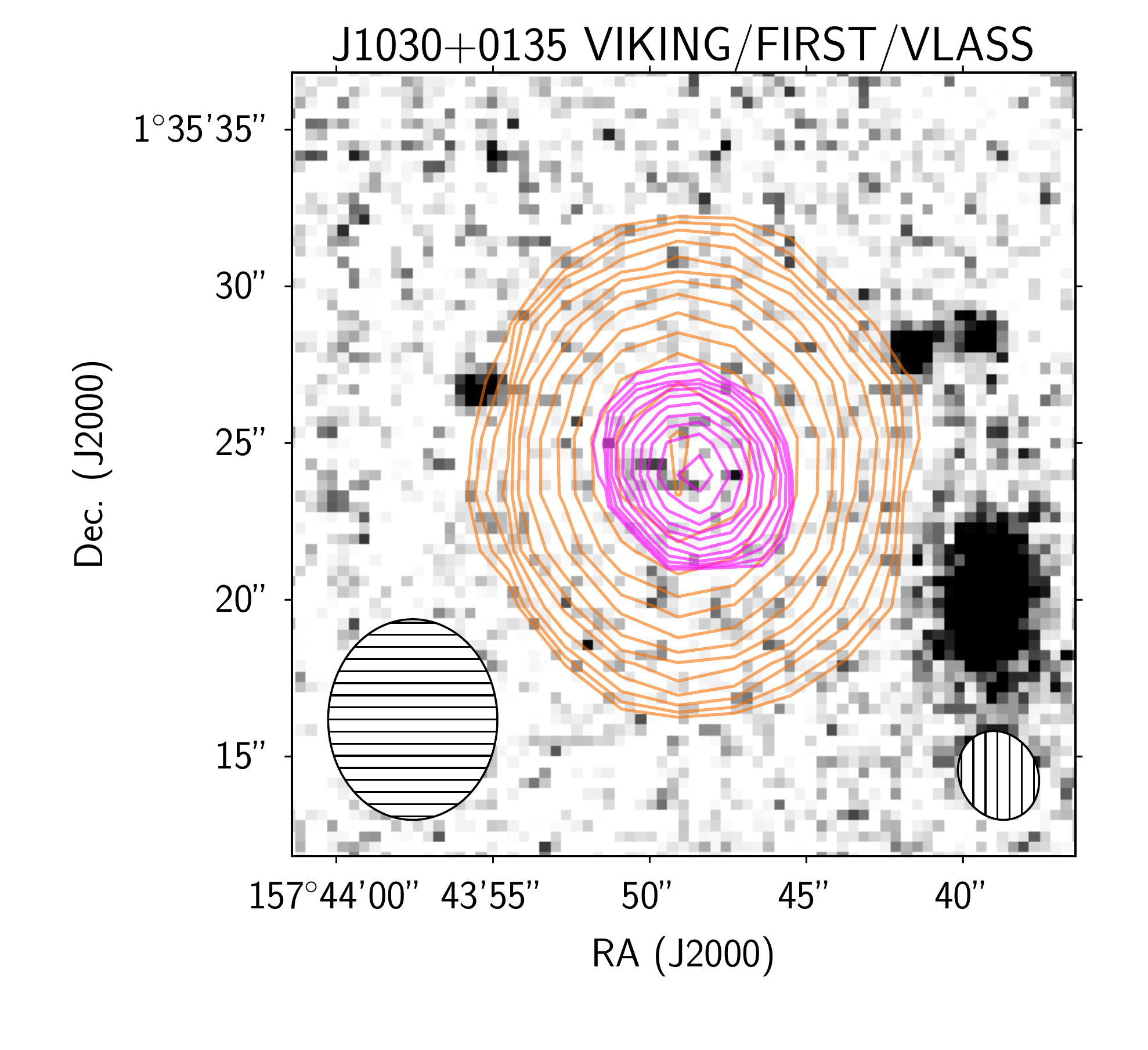}
\end{minipage}
\begin{minipage}{0.5\textwidth}
\includegraphics[width=8.5cm]{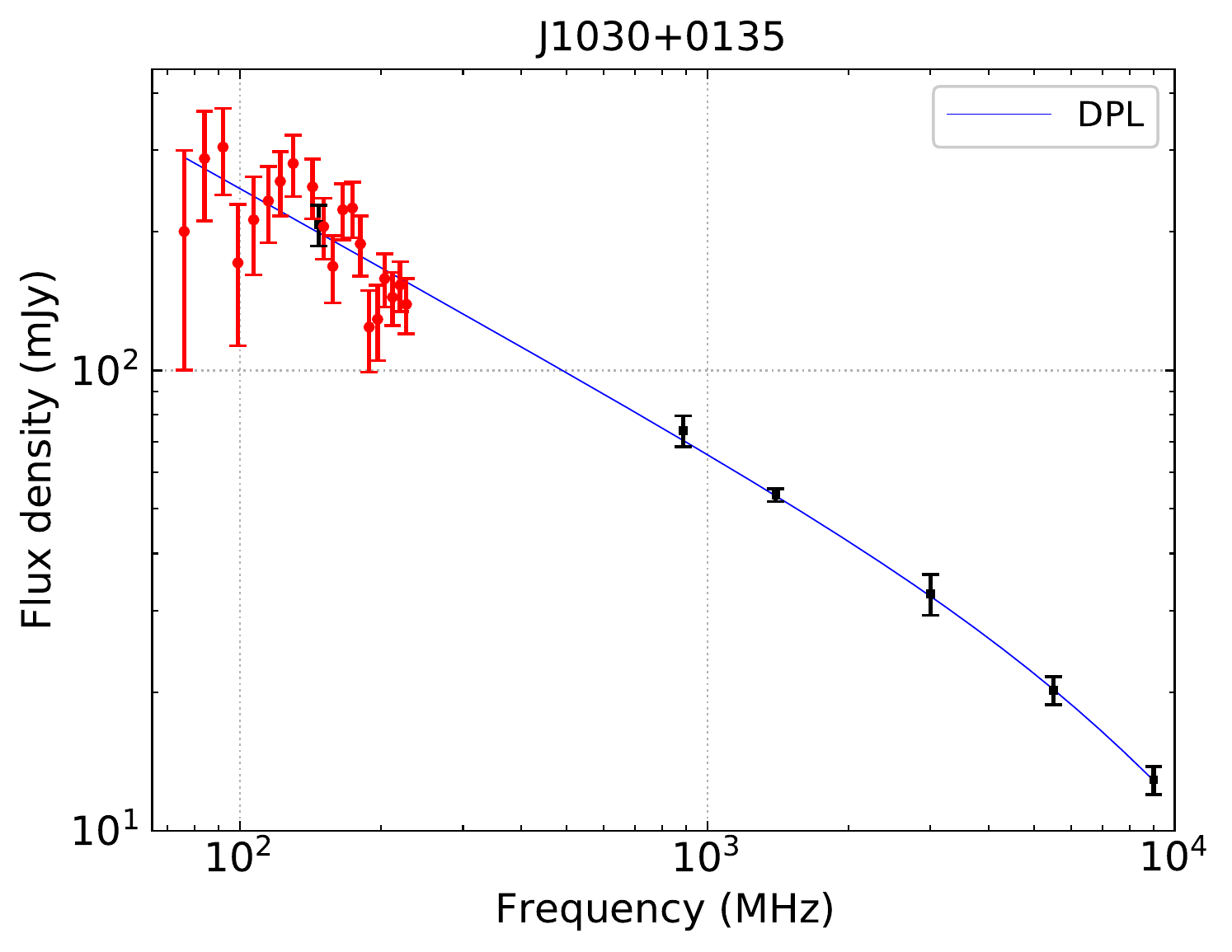}
\end{minipage}
\begin{minipage}{0.5\textwidth}
\vspace{0.2cm}
\includegraphics[width=7.5cm]{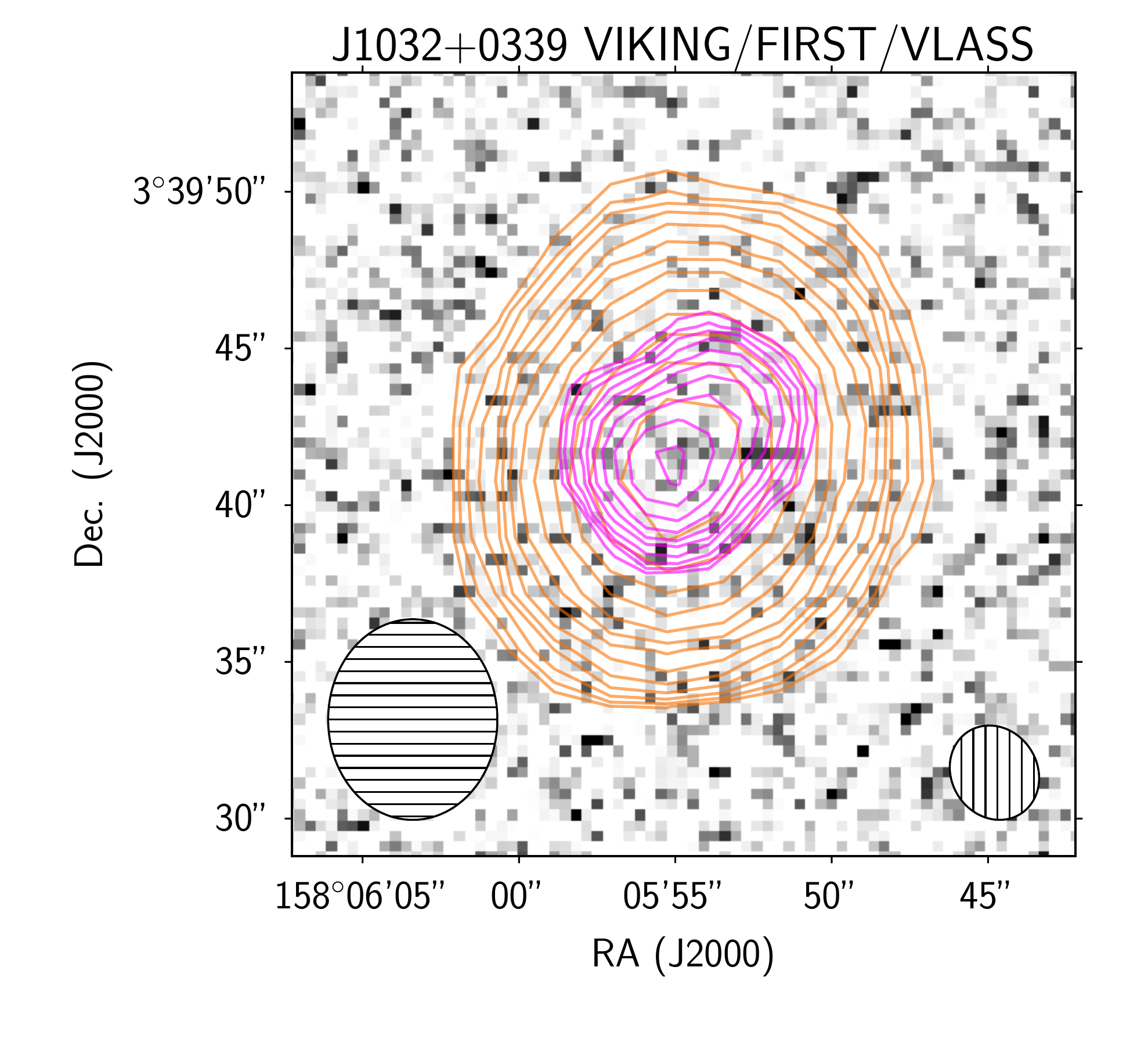}
\end{minipage}
\begin{minipage}{0.5\textwidth}
\includegraphics[width=8.5cm]{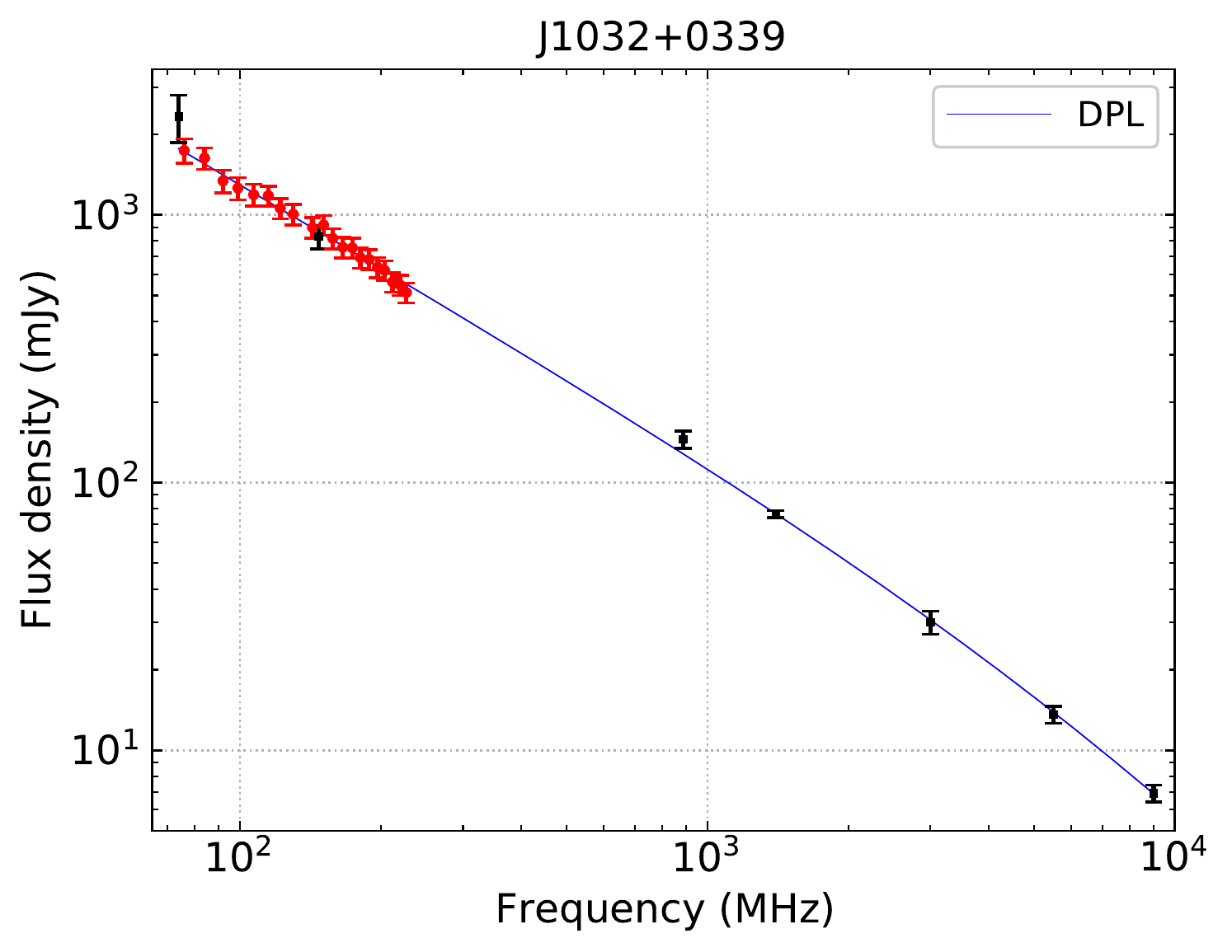}
\end{minipage}
\begin{minipage}{0.5\textwidth}
\vspace{0.2cm}
\includegraphics[width=7.5cm]{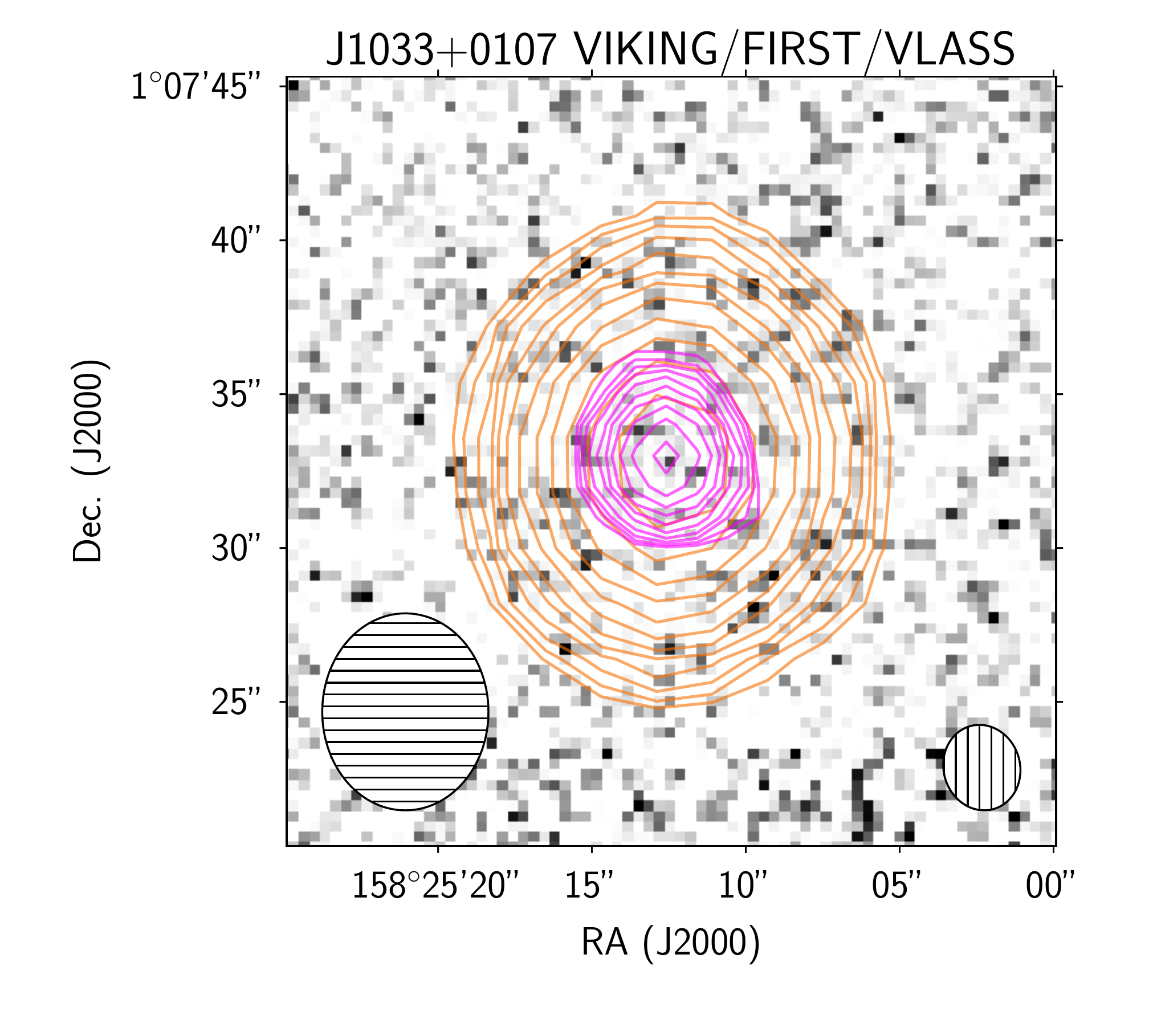}
\end{minipage}
\begin{minipage}{0.5\textwidth}
\includegraphics[width=8.5cm]{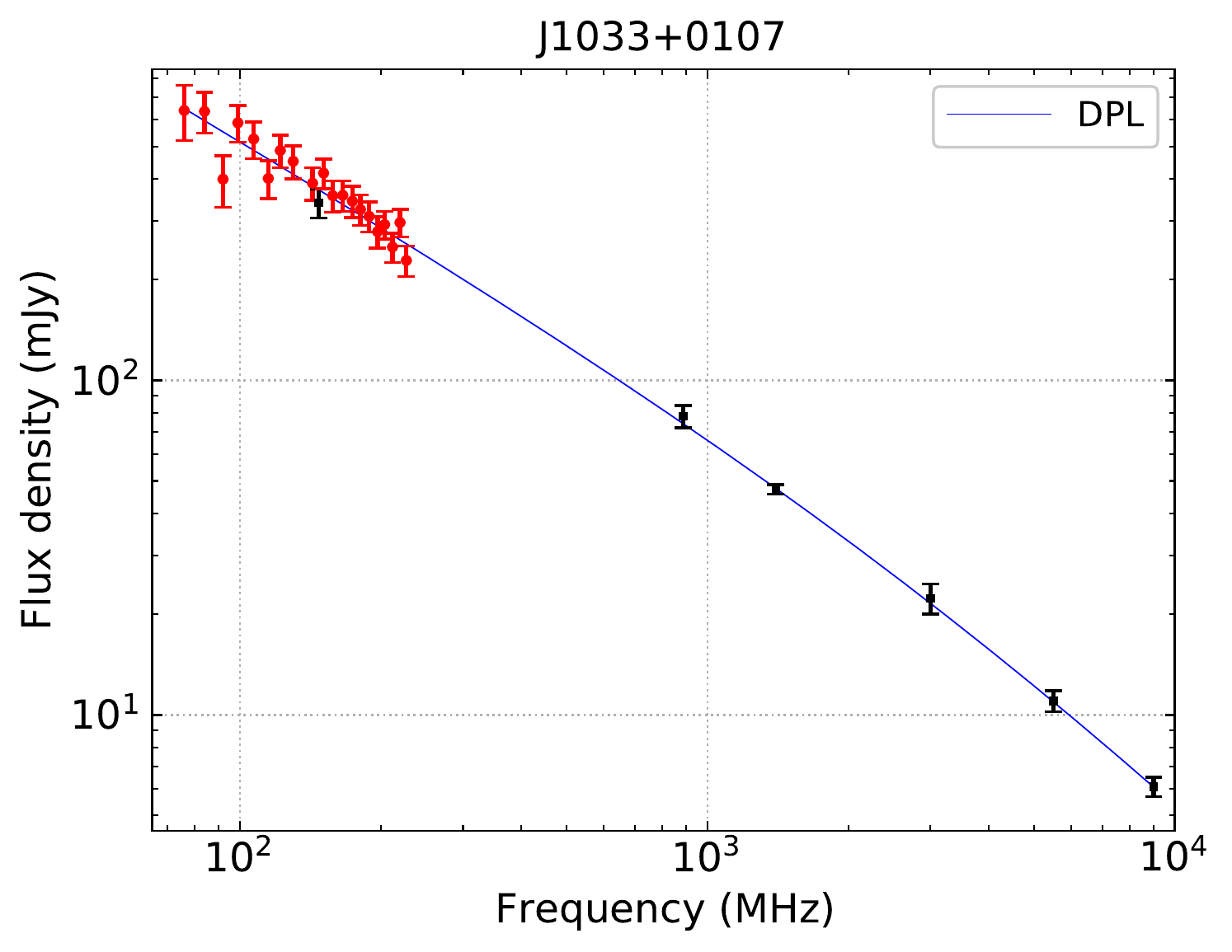}
\end{minipage}
\caption{{\em - continued.}}
\end{figure*}

\setcounter{figure}{1} 
\begin{figure*}
\begin{minipage}{0.5\textwidth}
\vspace{0.2cm}
\includegraphics[width=7.5cm]{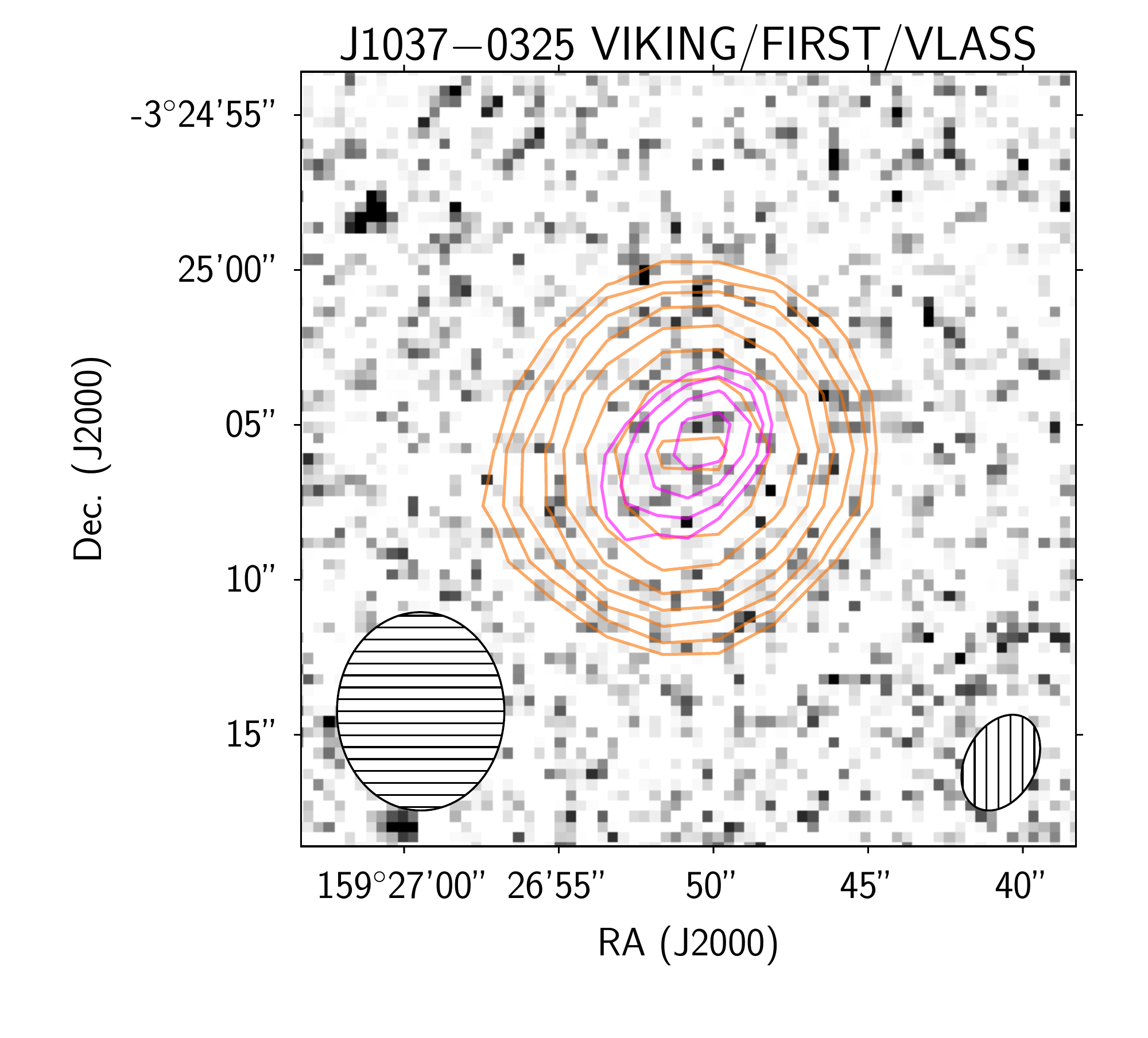}
\end{minipage}
\begin{minipage}{0.5\textwidth}
\includegraphics[width=8.5cm]{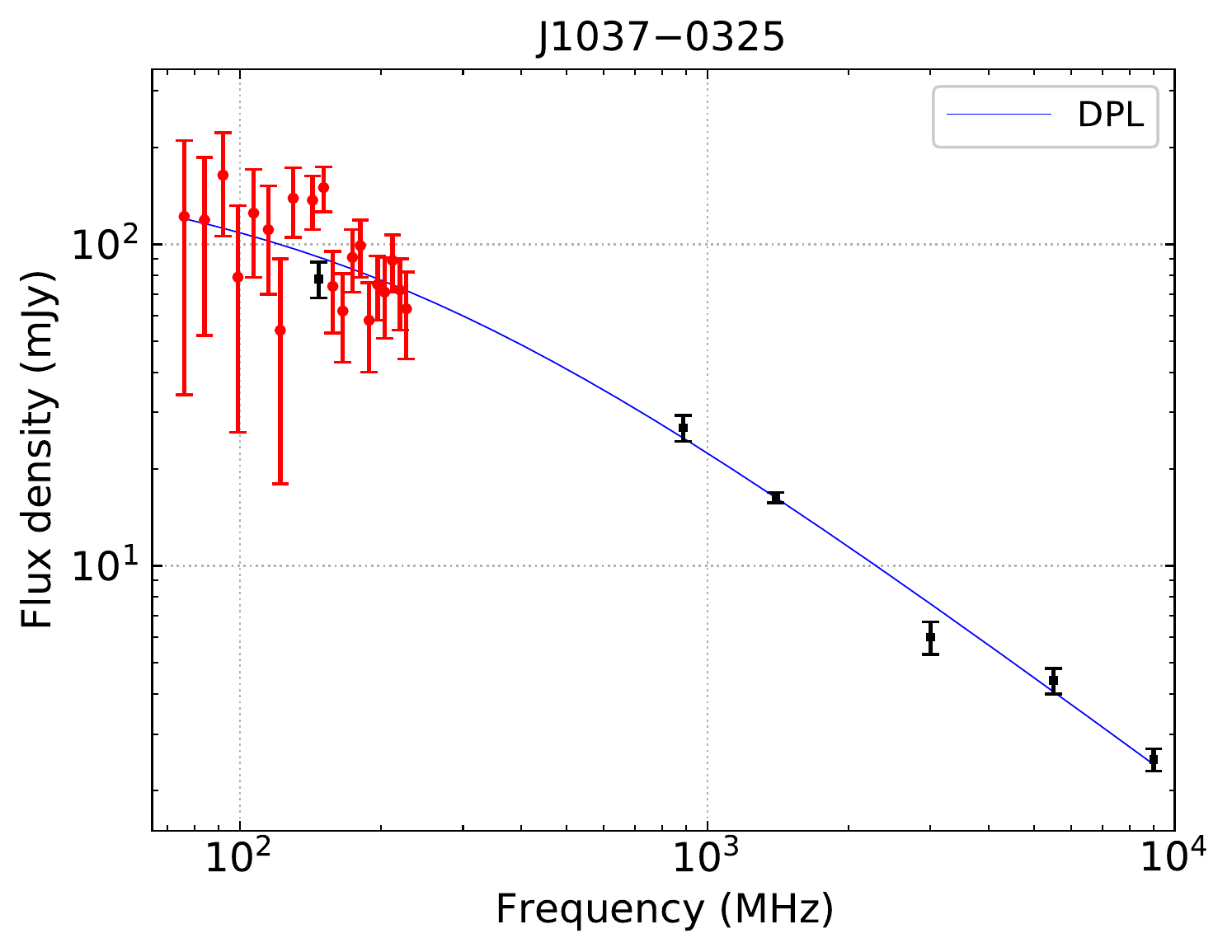}
\end{minipage}
\begin{minipage}{0.5\textwidth}
\vspace{0.2cm}
\includegraphics[width=7.5cm]{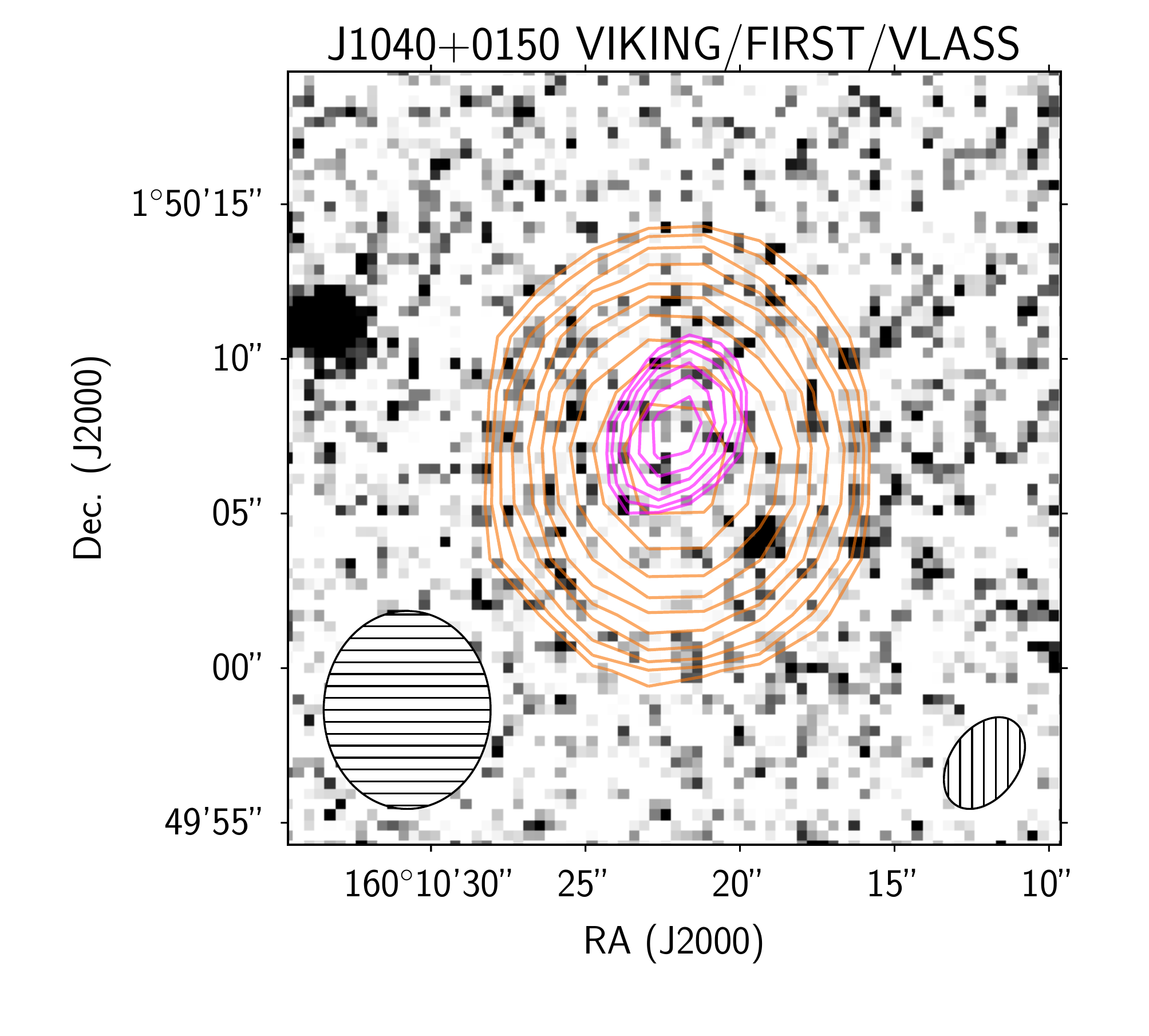}
\end{minipage}
\begin{minipage}{0.5\textwidth}
\includegraphics[width=8.5cm]{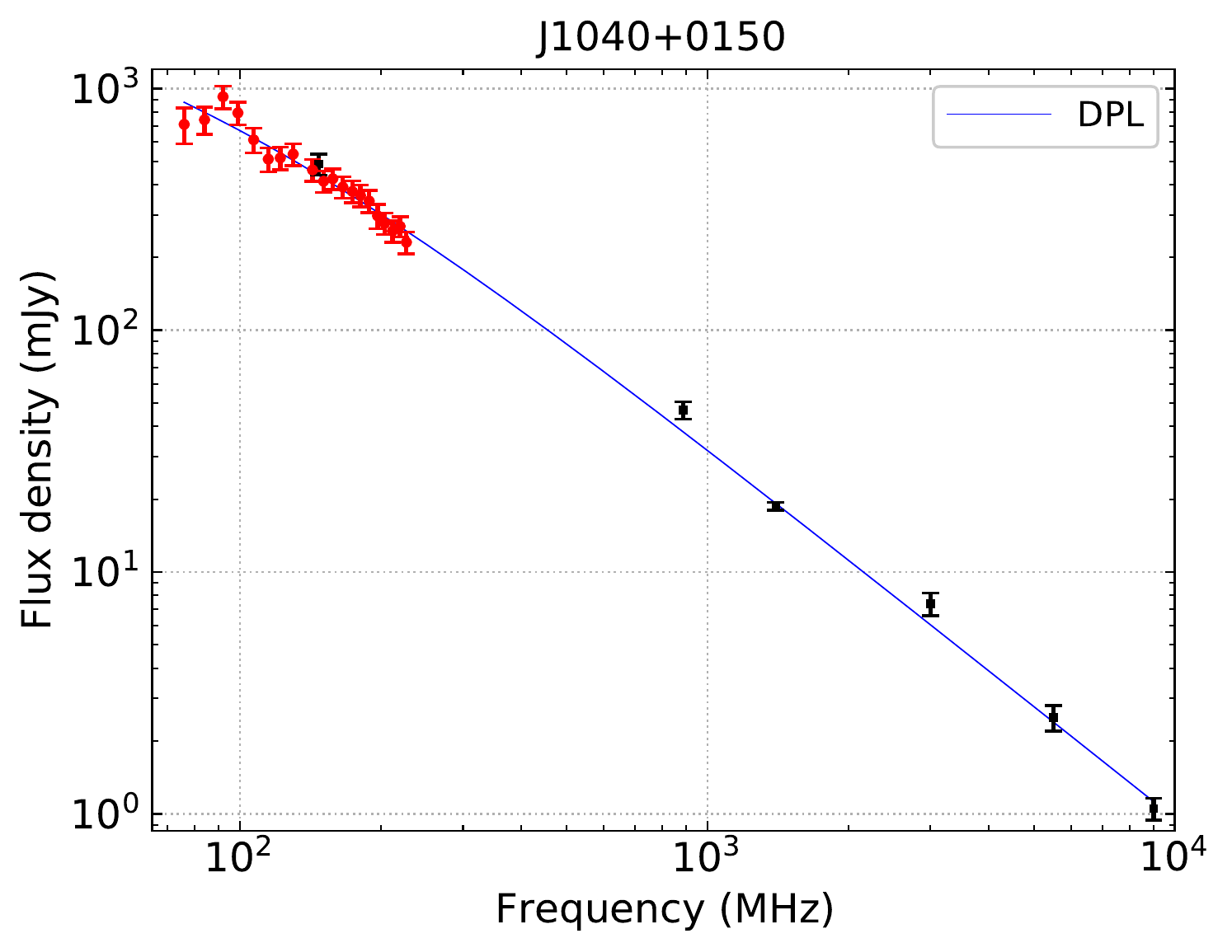}
\end{minipage}
\begin{minipage}{0.5\textwidth}
\vspace{0.2cm}
\includegraphics[width=7.5cm]{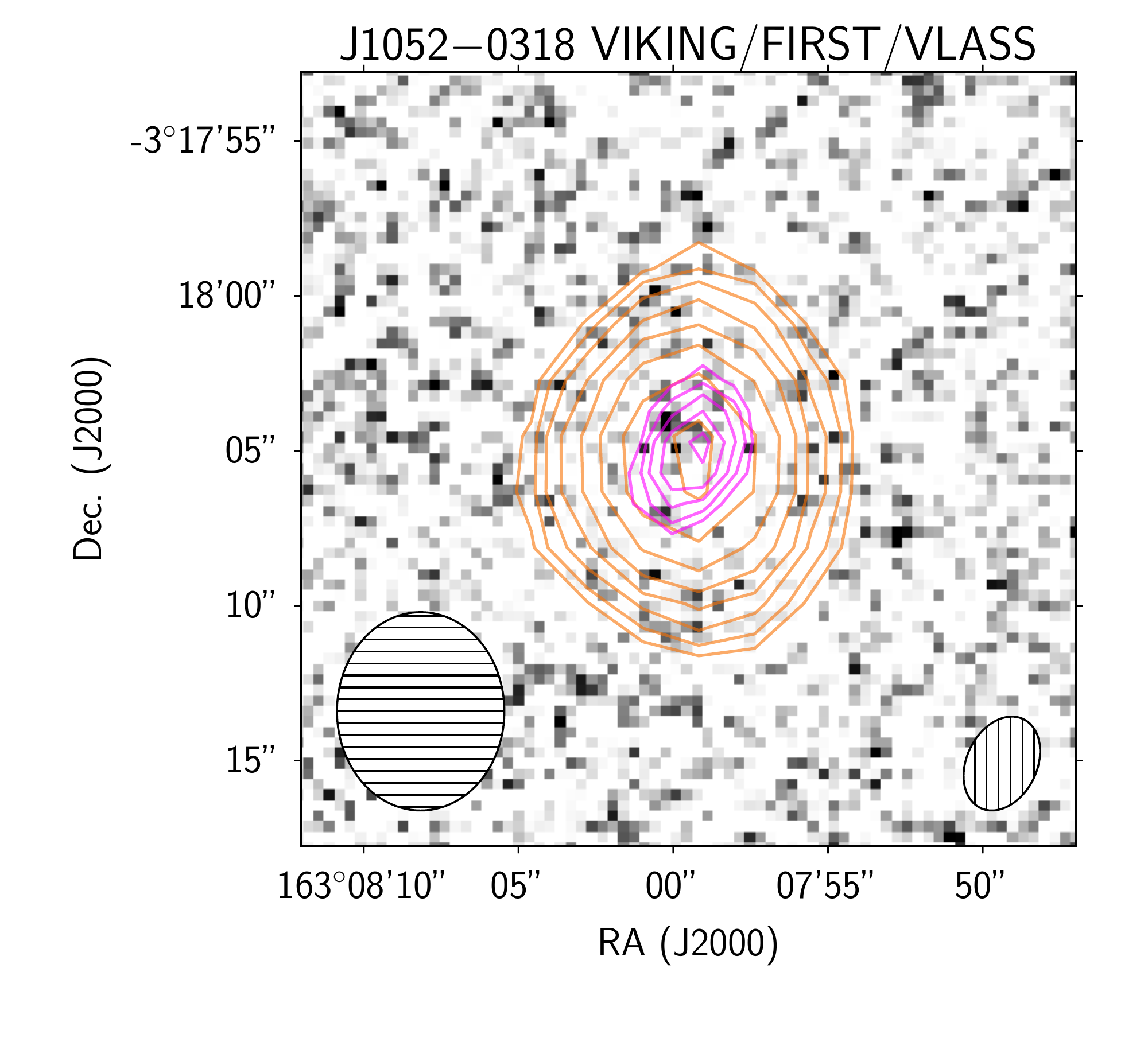}
\end{minipage}
\begin{minipage}{0.5\textwidth}
\includegraphics[width=8.5cm]{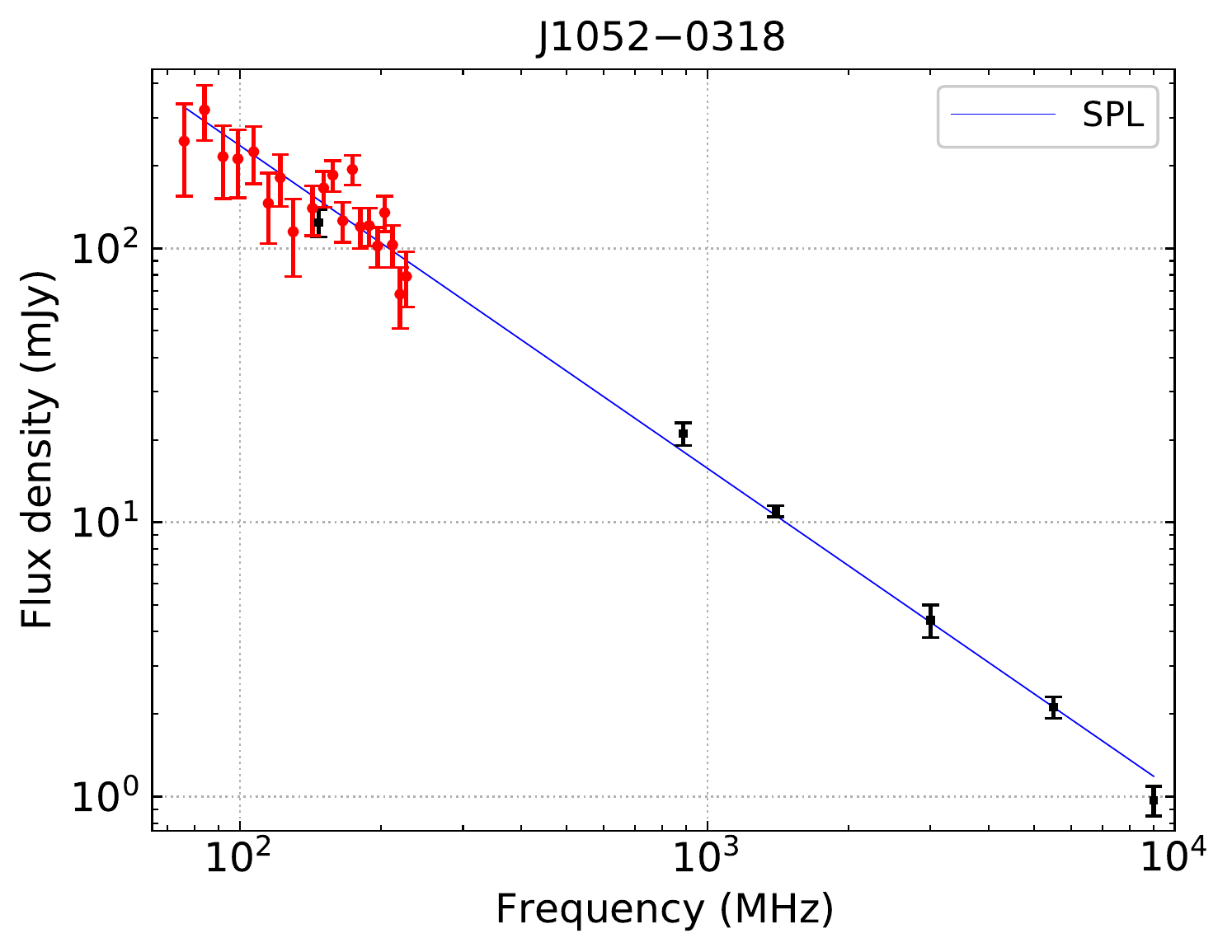}
\end{minipage}
\caption{{\em - continued.}}
\end{figure*}

\setcounter{figure}{1} 
\begin{figure*}
\begin{minipage}{0.5\textwidth}
\vspace{0.2cm}
\includegraphics[width=7.5cm]{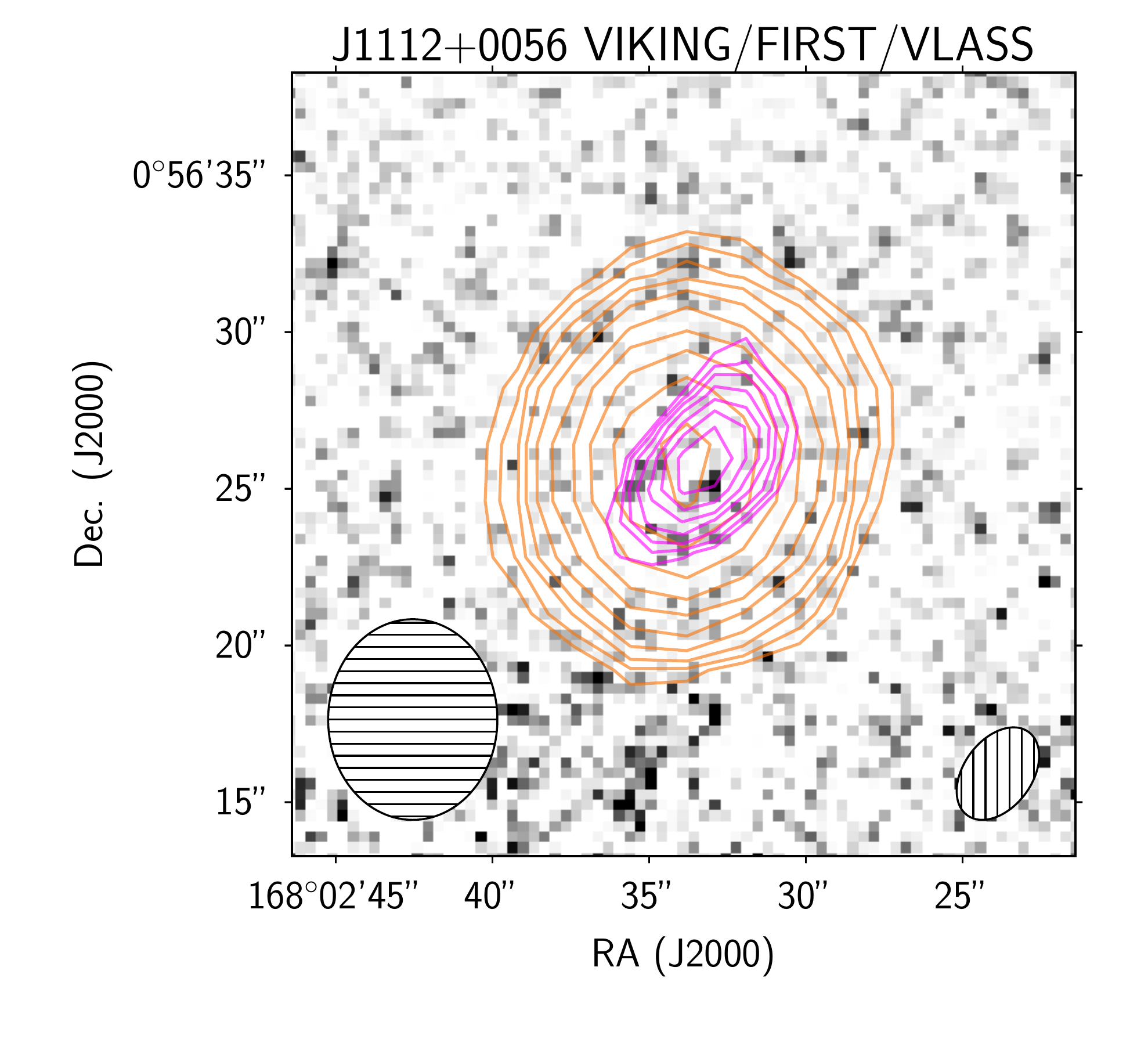}
\end{minipage}
\begin{minipage}{0.5\textwidth}
\includegraphics[width=8.5cm]{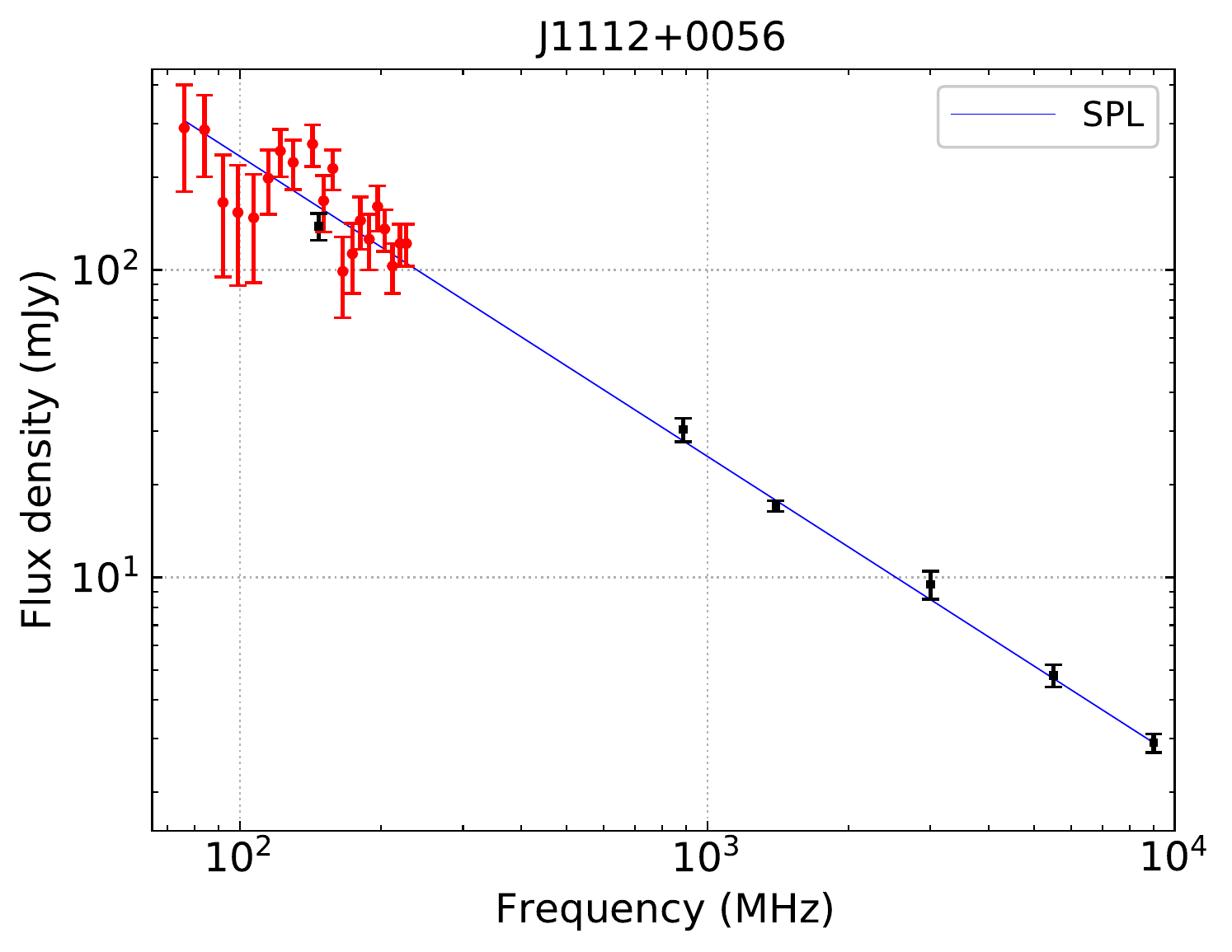}
\end{minipage}
\begin{minipage}{0.5\textwidth}
\vspace{0.2cm}
\includegraphics[width=7.5cm]{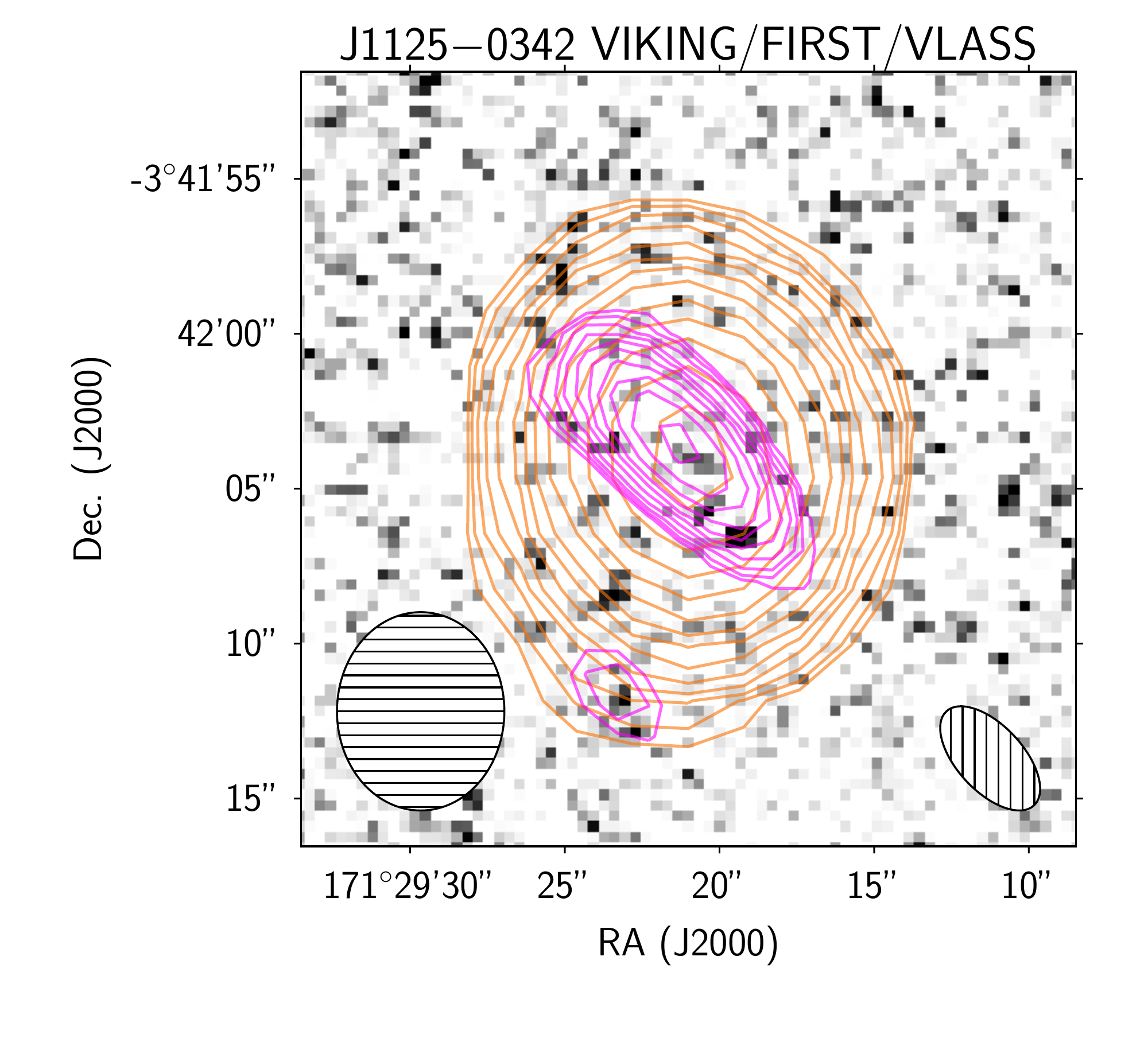}
\end{minipage}
\begin{minipage}{0.5\textwidth}
\includegraphics[width=8.5cm]{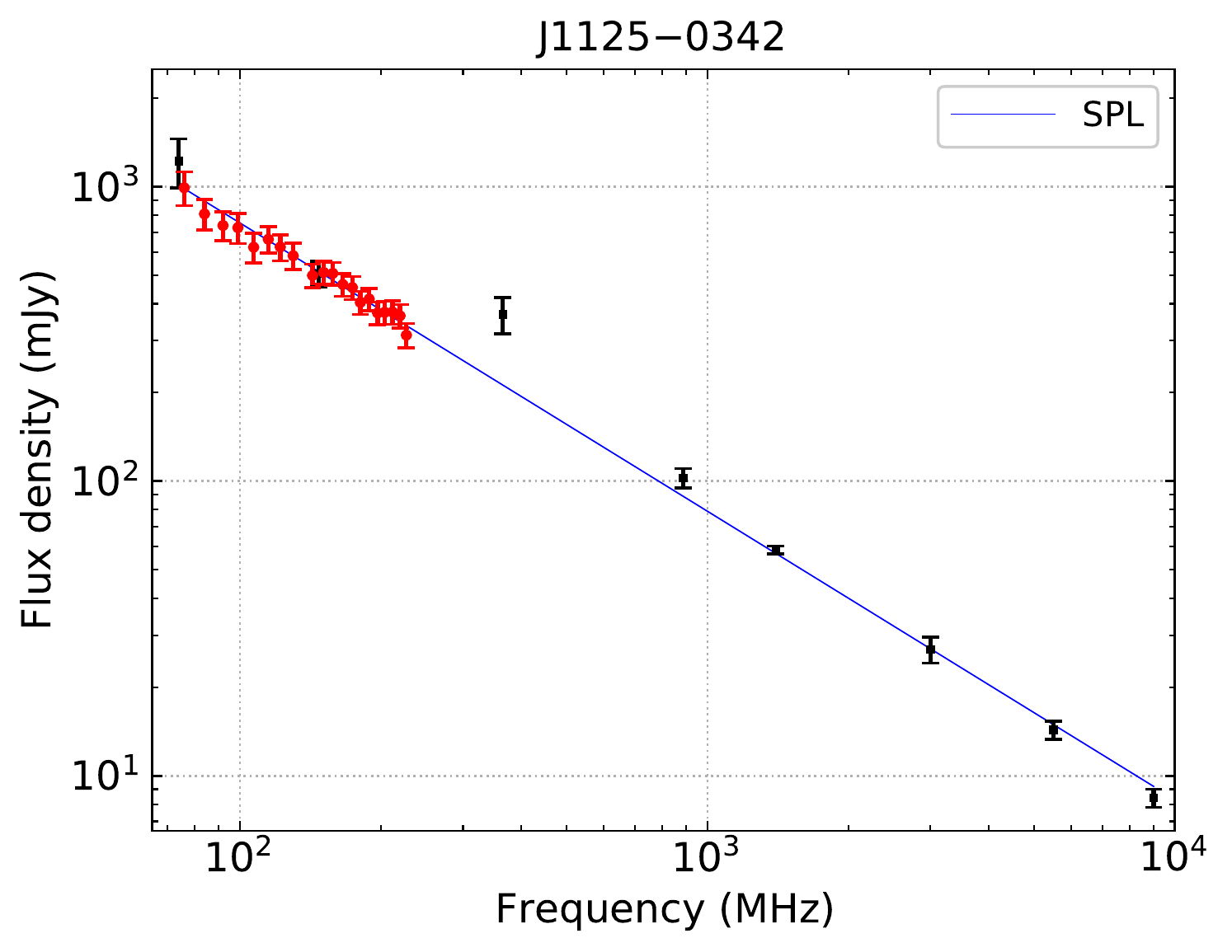}
\end{minipage}
\begin{minipage}{0.5\textwidth}
\vspace{0.2cm}
\includegraphics[width=7.5cm]{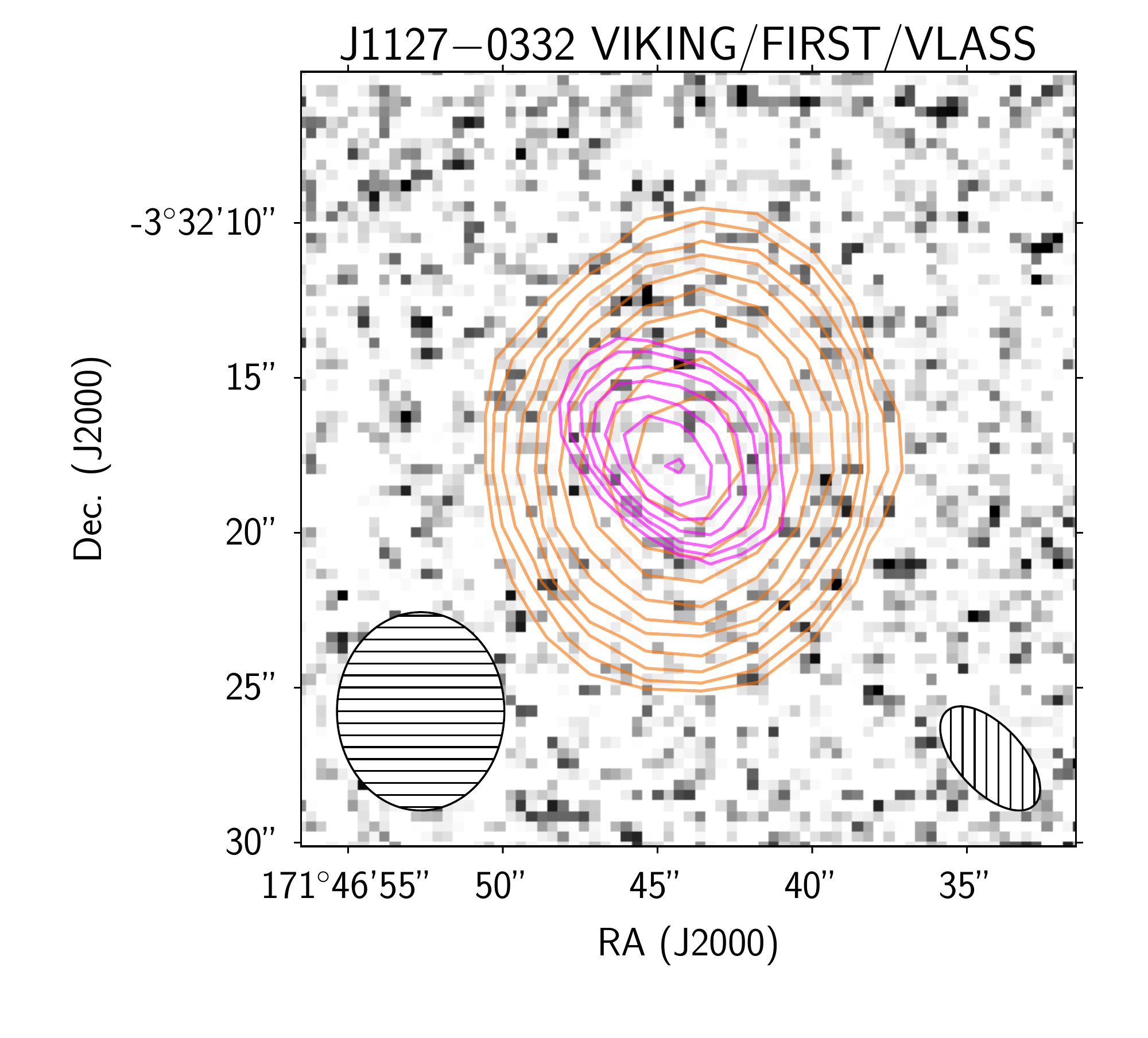}
\end{minipage}
\begin{minipage}{0.5\textwidth}
\includegraphics[width=8.5cm]{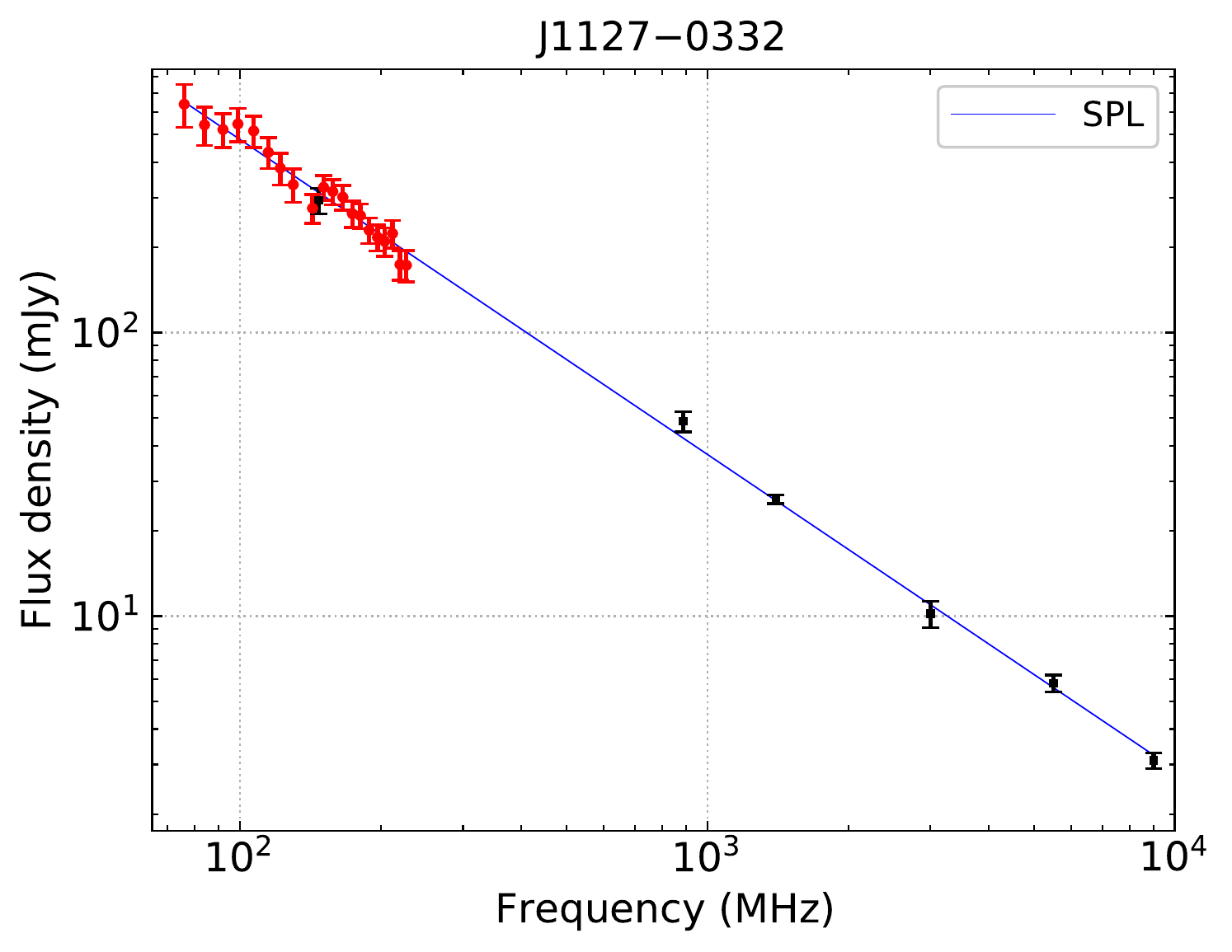}
\end{minipage}
\caption{{\em - continued.}}
\end{figure*}

\setcounter{figure}{1} 
\begin{figure*}
\begin{minipage}{0.5\textwidth}
\vspace{0.2cm}
\includegraphics[width=7.5cm]{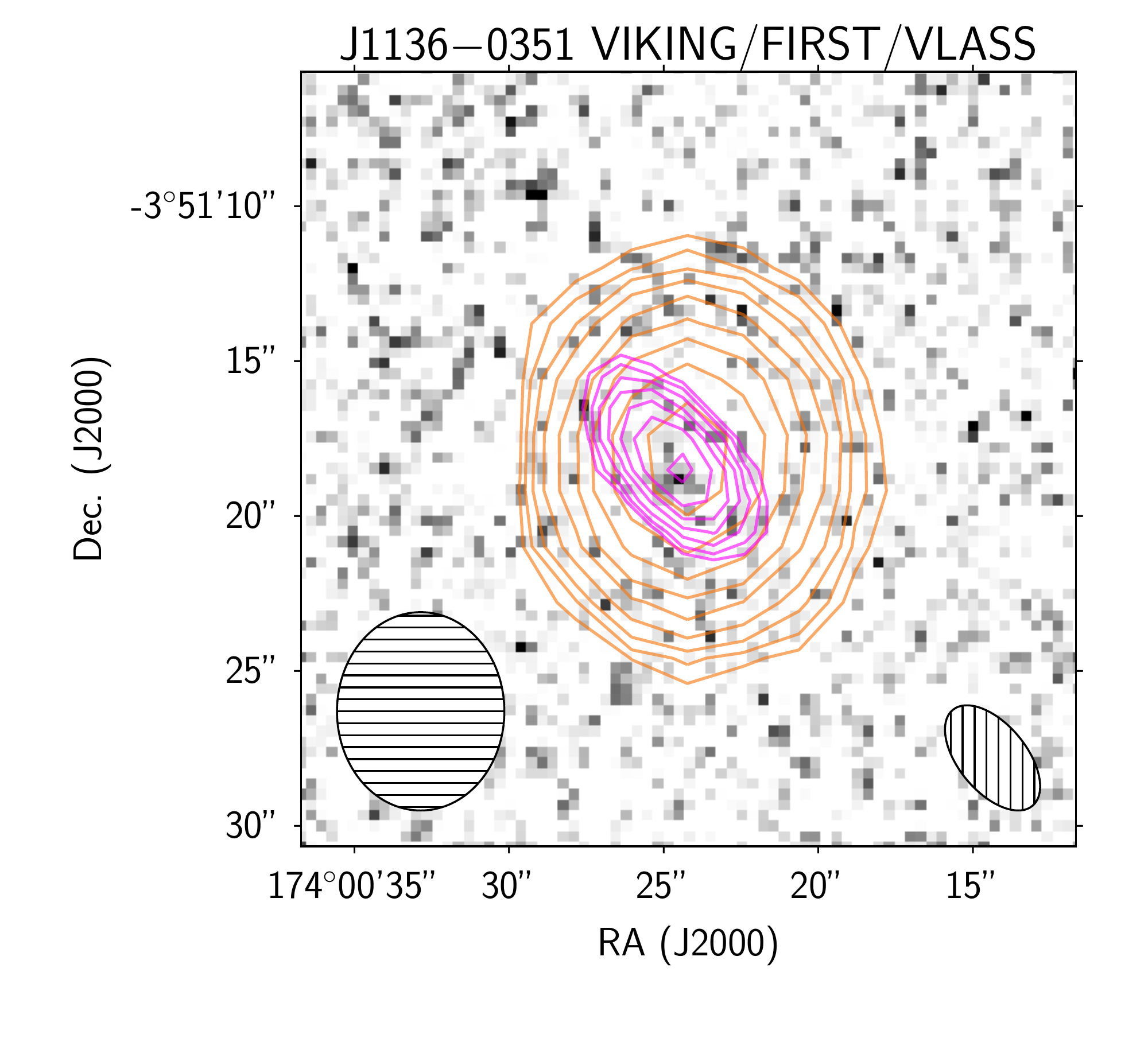}
\end{minipage}
\begin{minipage}{0.5\textwidth}
\includegraphics[width=8.5cm]{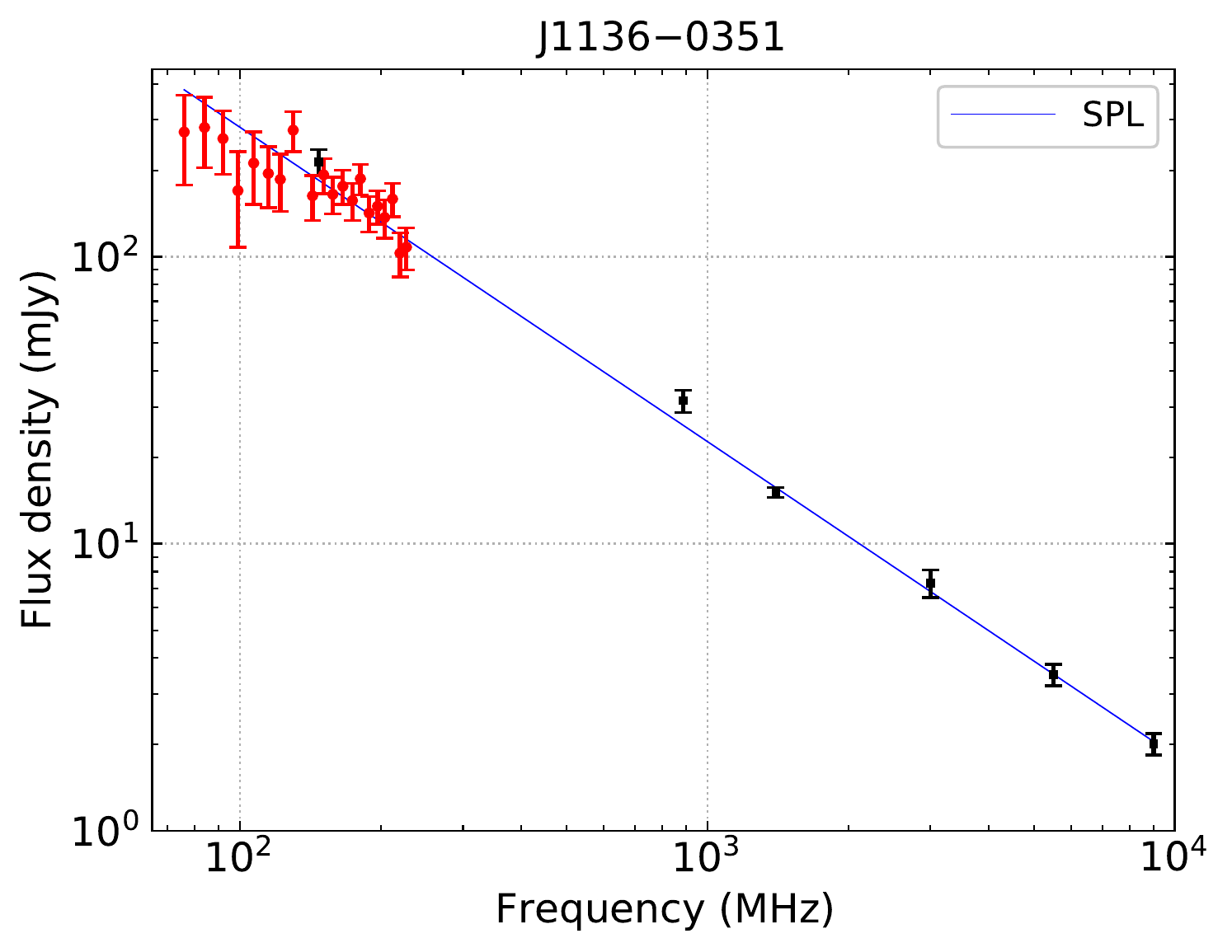}
\end{minipage}
\begin{minipage}{0.5\textwidth}
\vspace{0.2cm}
\includegraphics[width=7.5cm]{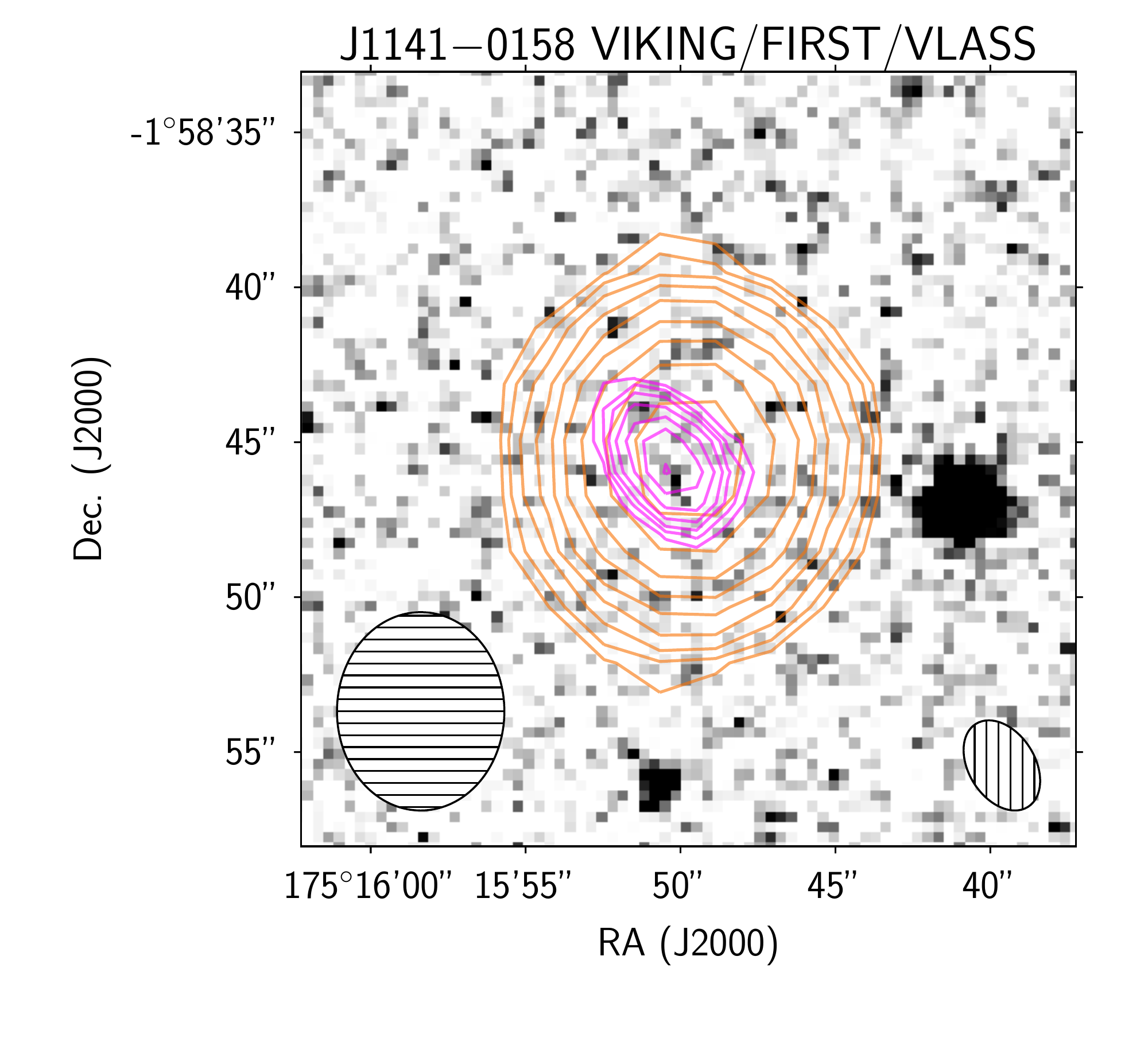}
\end{minipage}
\begin{minipage}{0.5\textwidth}
\includegraphics[width=8.5cm]{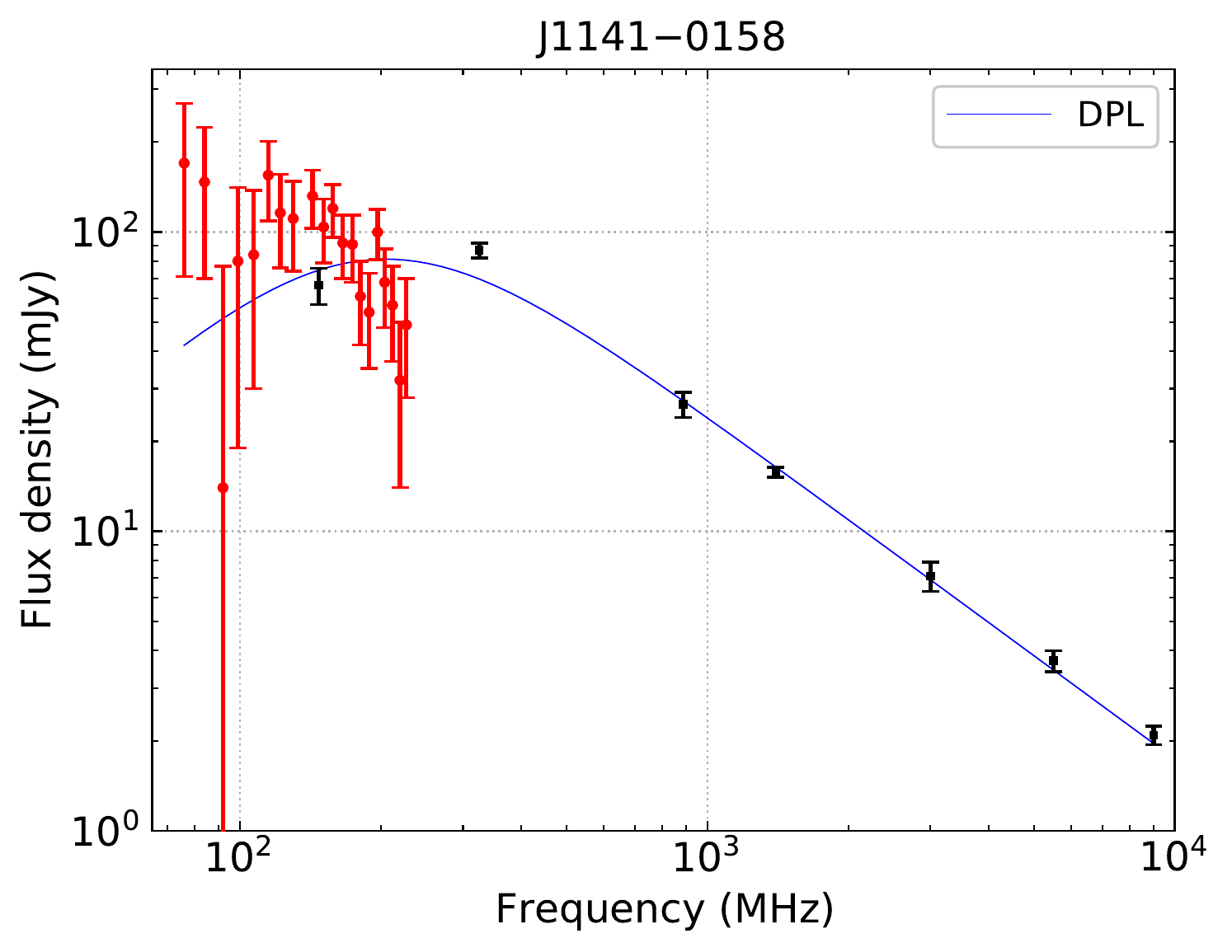}
\end{minipage}
\begin{minipage}{0.5\textwidth}
\vspace{0.2cm}
\includegraphics[width=7.5cm]{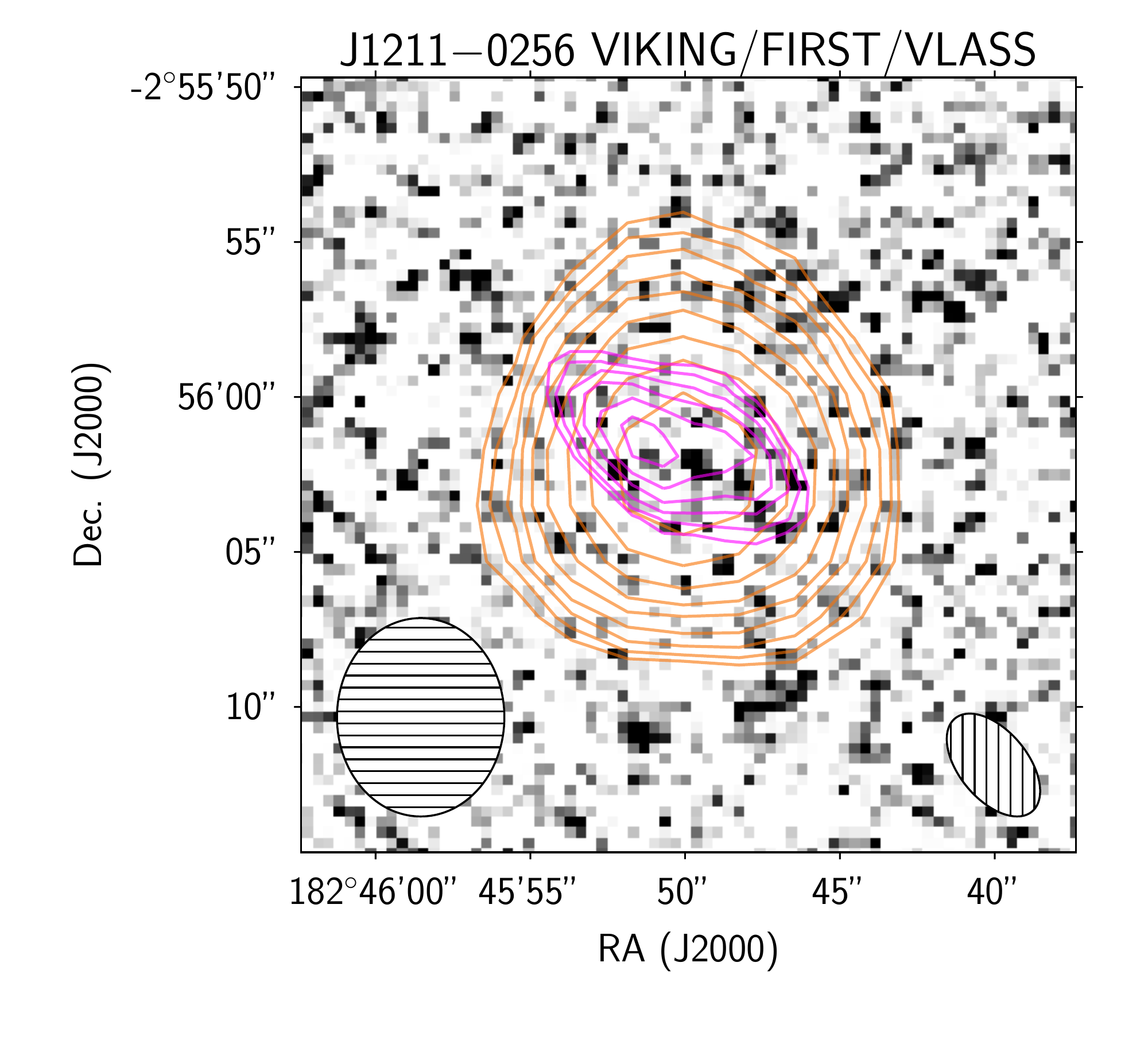}
\end{minipage}
\begin{minipage}{0.5\textwidth}
\includegraphics[width=8.5cm]{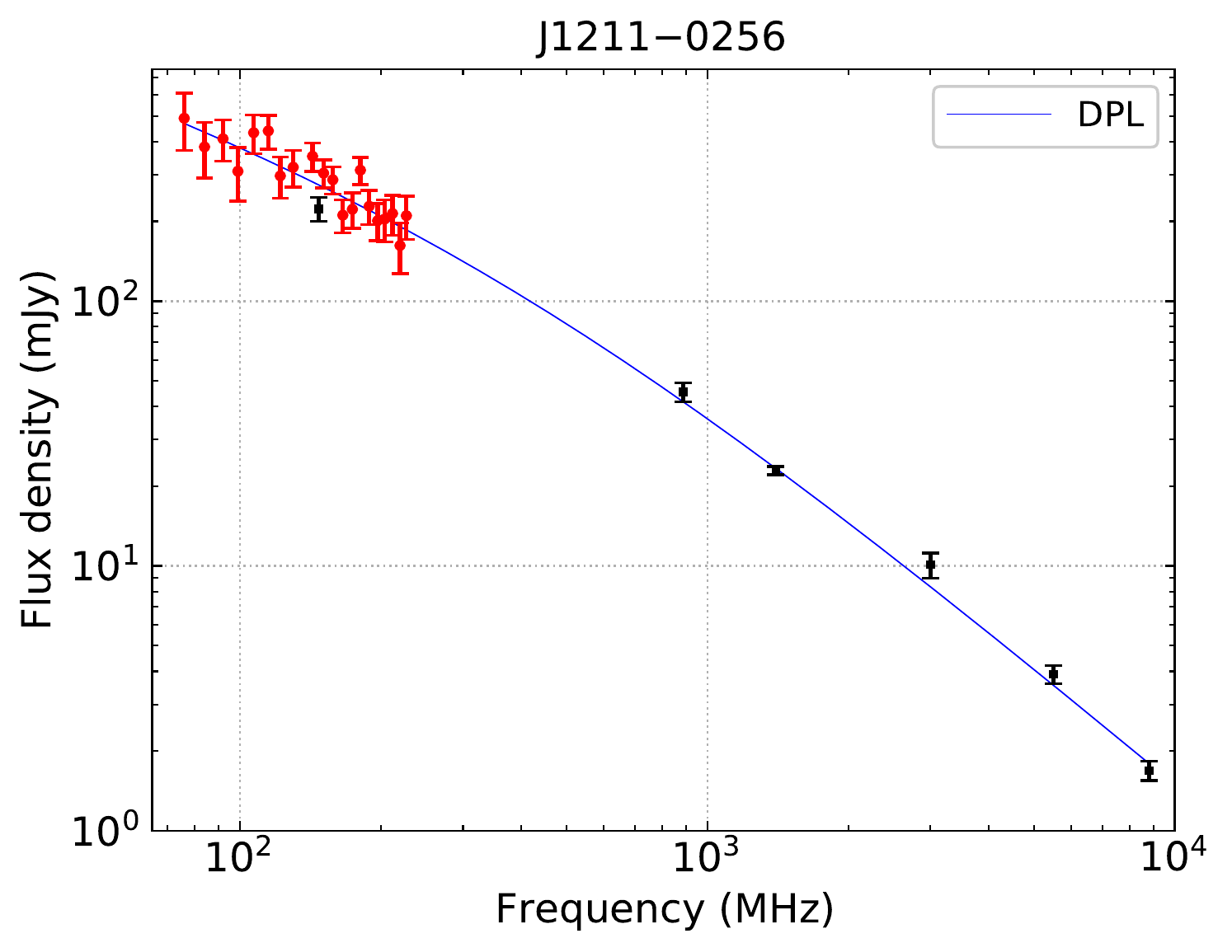}
\end{minipage}
\caption{{\em - continued.}}
\end{figure*}

\setcounter{figure}{1} 
\begin{figure*}
\begin{minipage}{0.5\textwidth}
\vspace{0.2cm}
\includegraphics[width=7.5cm]{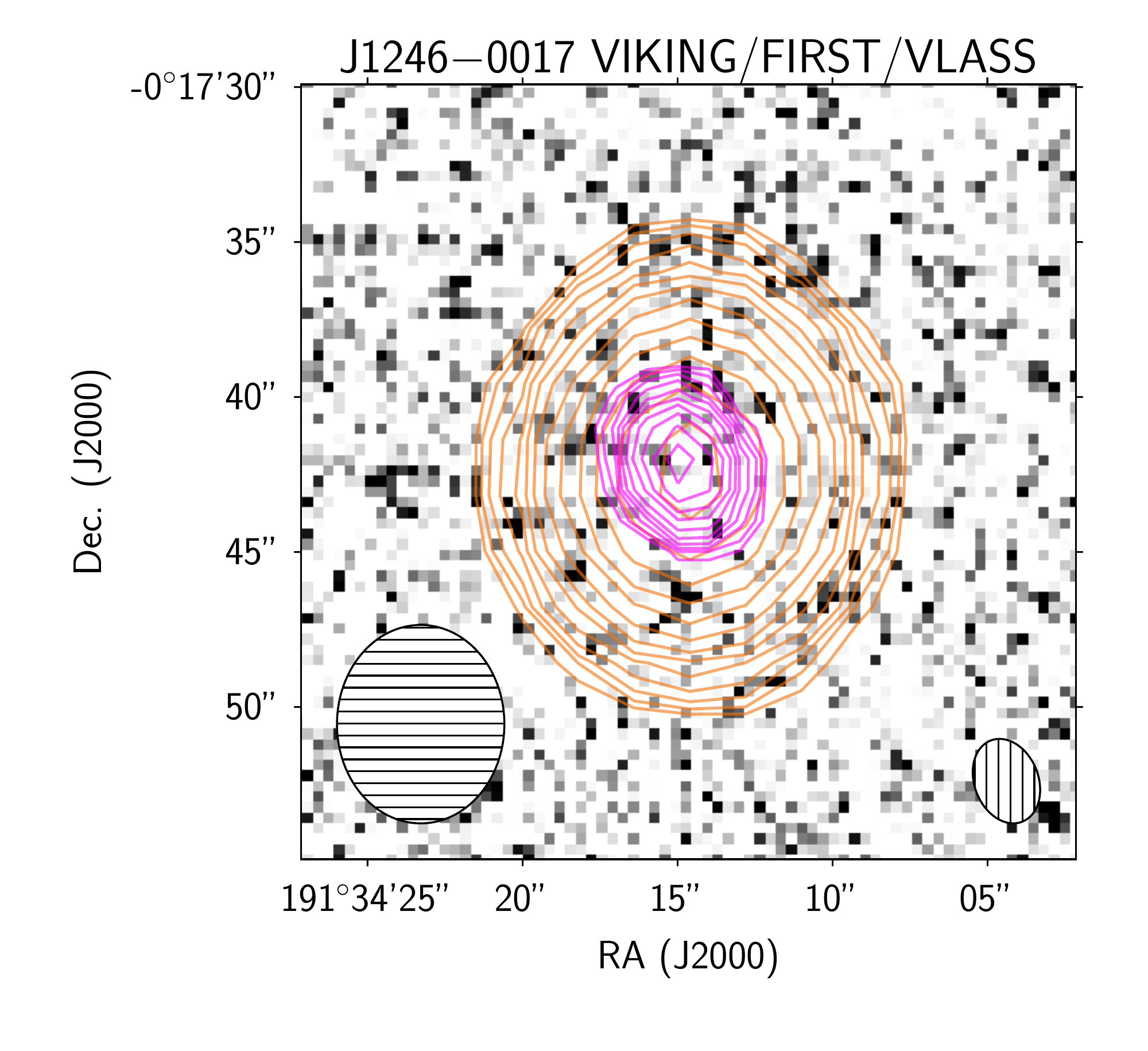}
\end{minipage}
\begin{minipage}{0.5\textwidth}
\includegraphics[width=8.5cm]{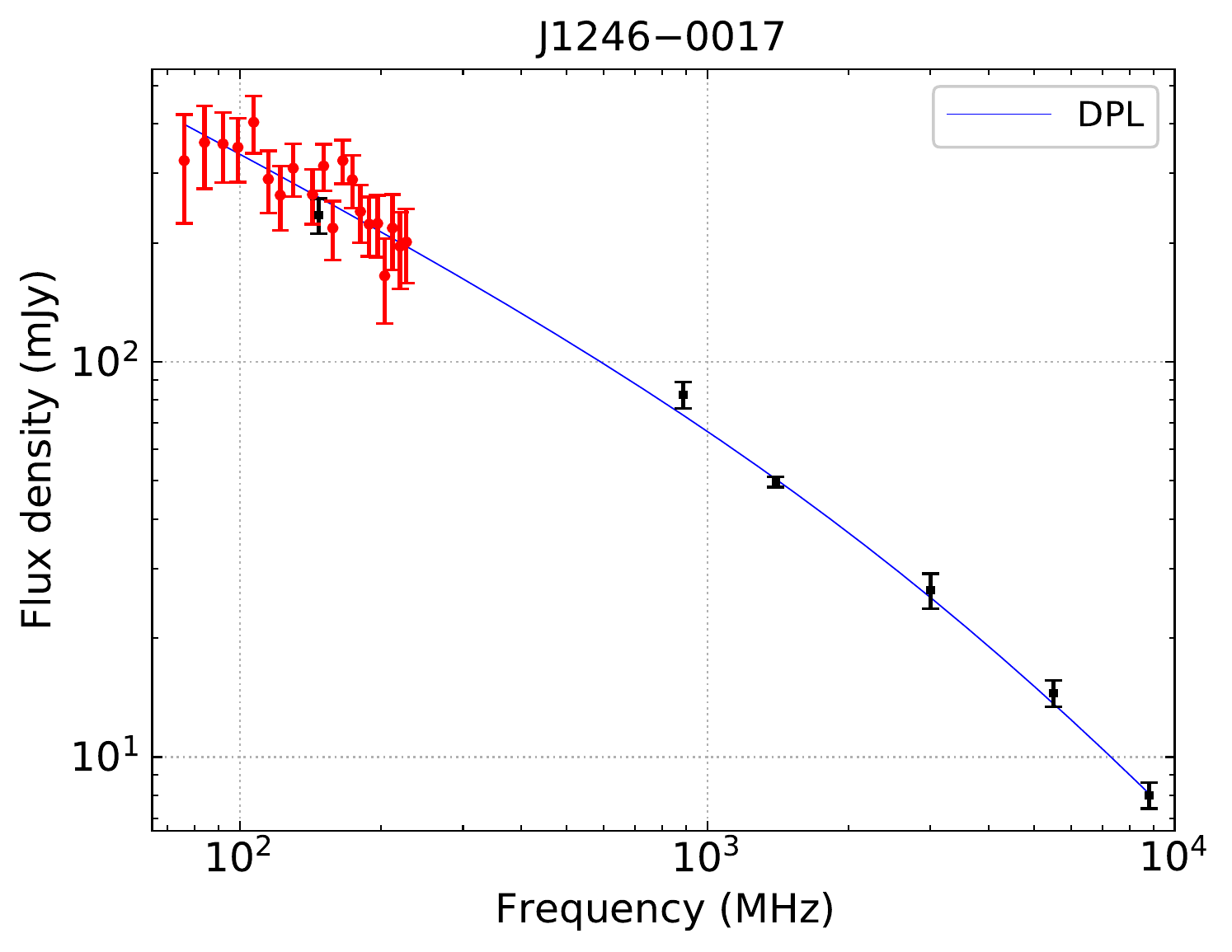}
\end{minipage}
\begin{minipage}{0.5\textwidth}
\vspace{0.2cm}
\includegraphics[width=7.5cm]{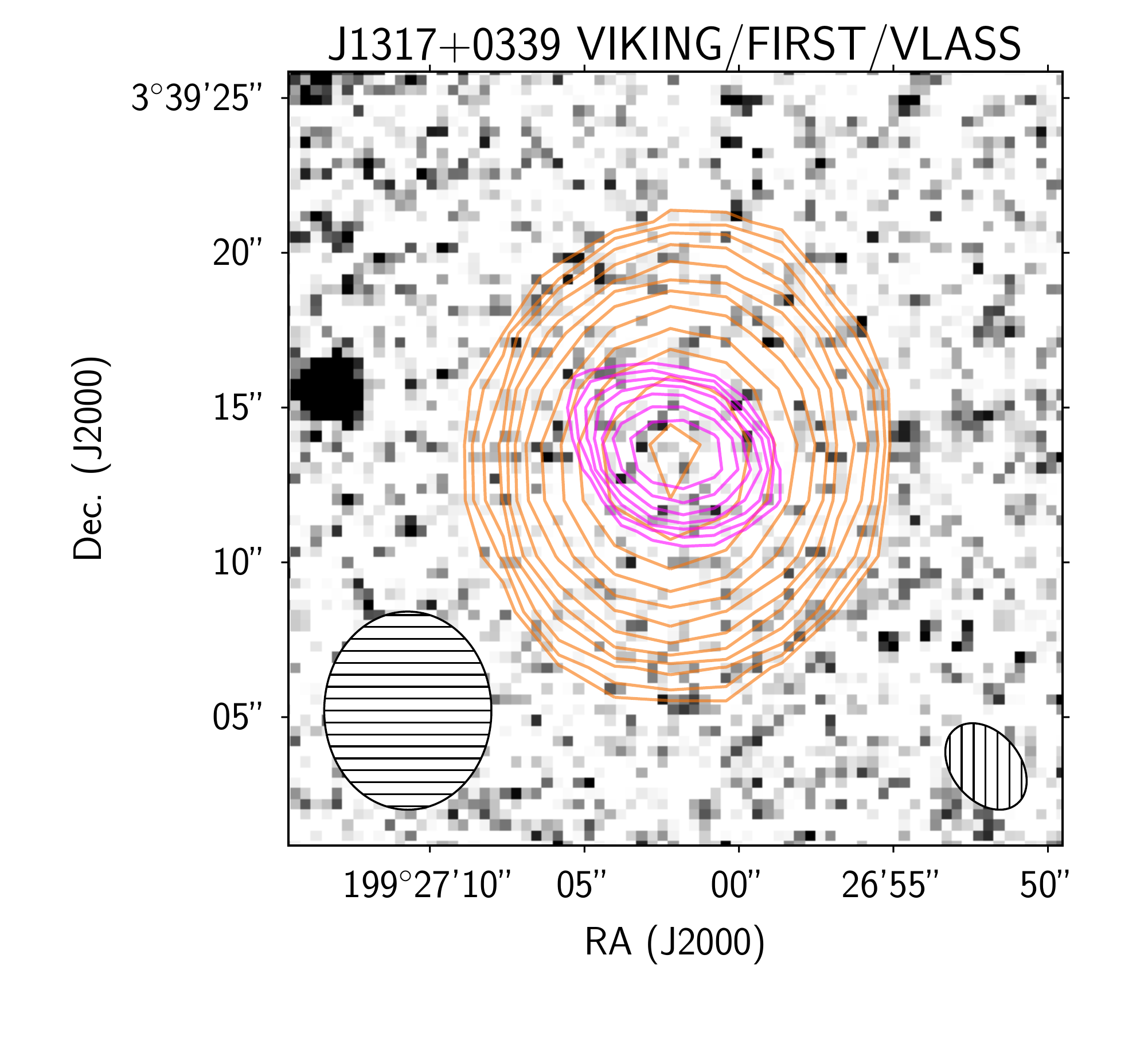}
\end{minipage}
\begin{minipage}{0.5\textwidth}
\includegraphics[width=8.5cm]{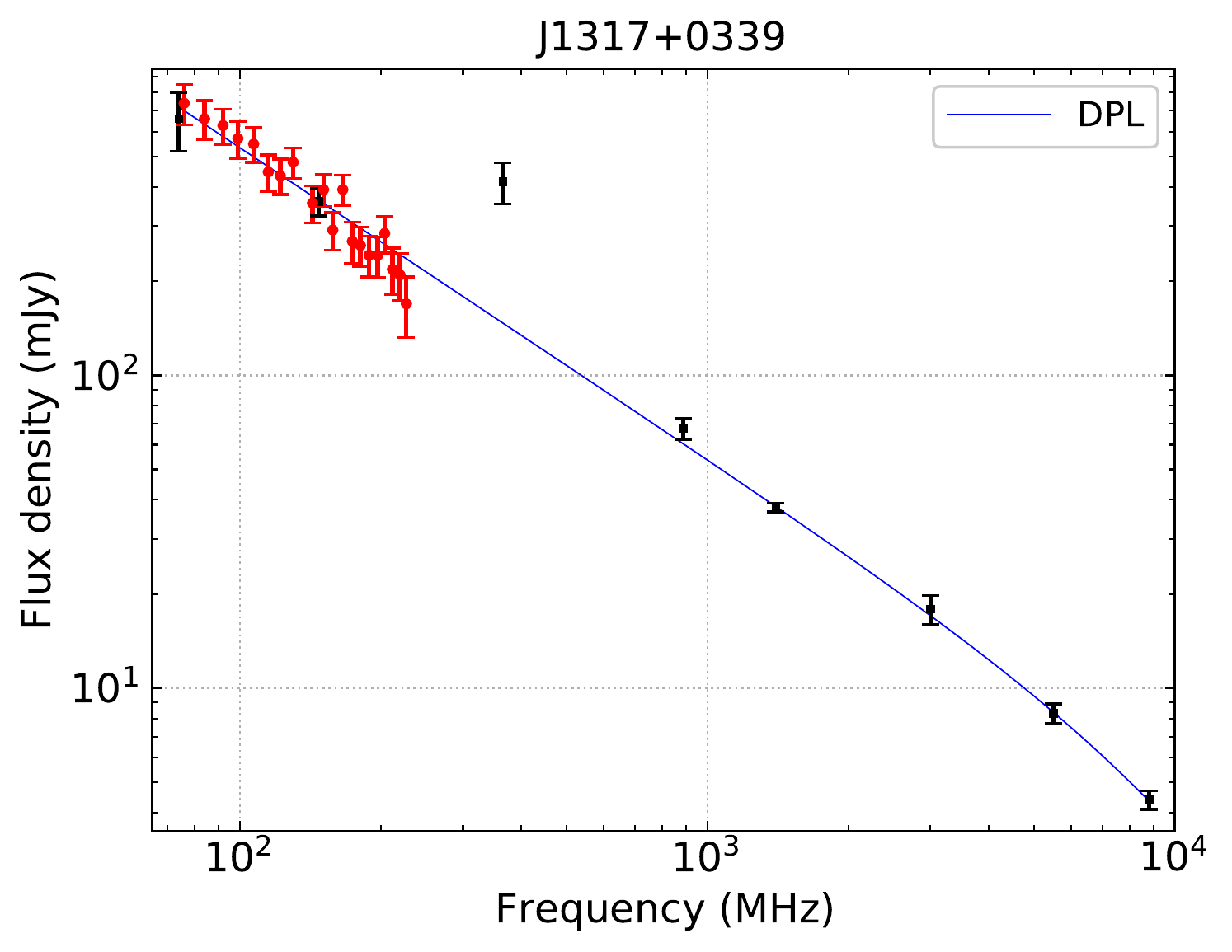}
\end{minipage}
\begin{minipage}{0.5\textwidth}
\vspace{0.2cm}
\includegraphics[width=7.5cm]{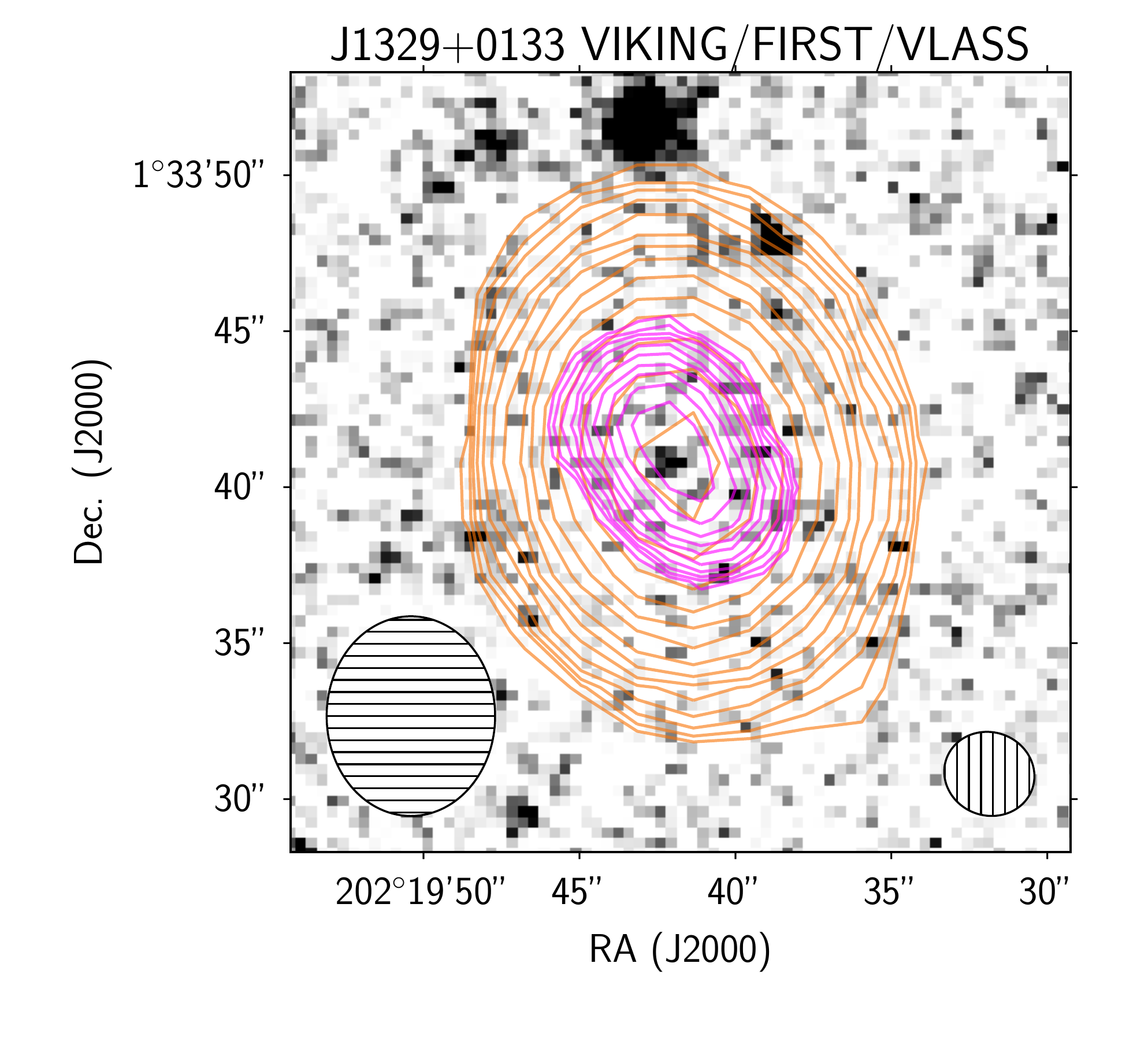}
\end{minipage}
\begin{minipage}{0.5\textwidth}
\includegraphics[width=8.5cm]{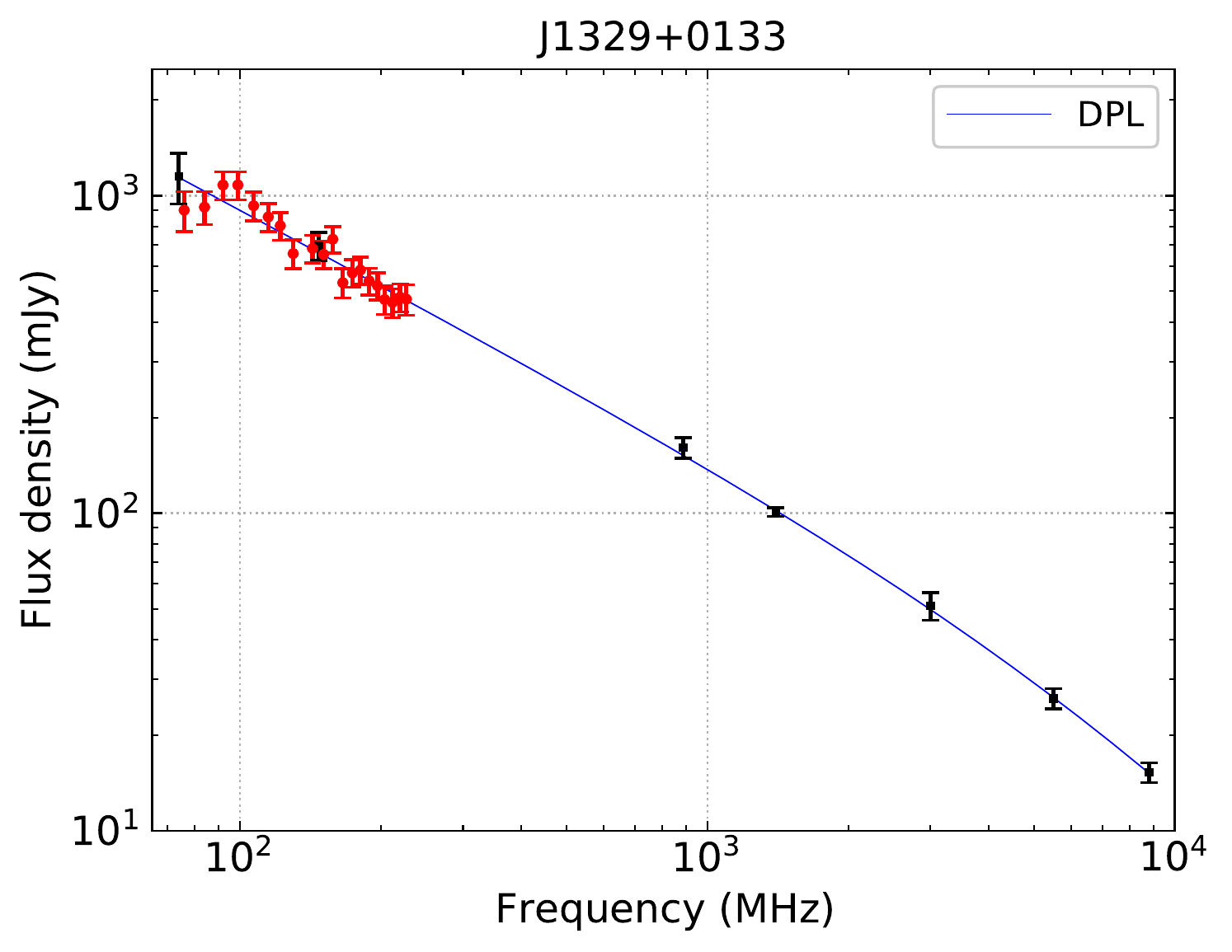}
\end{minipage}
\caption{{\em - continued.}}
\end{figure*}

\setcounter{figure}{1} 
\begin{figure*}
\begin{minipage}{0.5\textwidth}
\vspace{0.2cm}
\includegraphics[width=7.5cm]{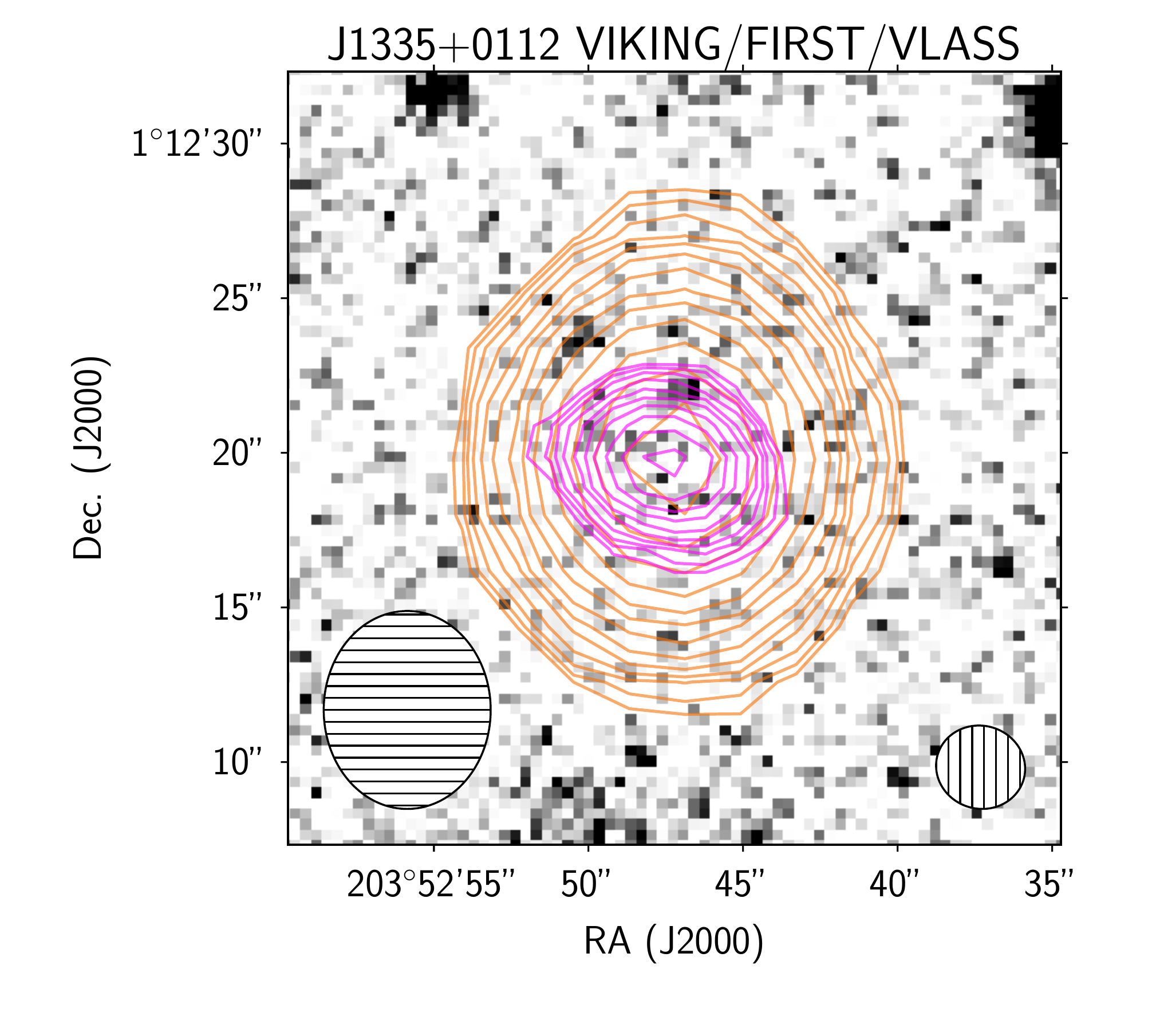}
\end{minipage}
\begin{minipage}{0.5\textwidth}
\includegraphics[width=8.5cm]{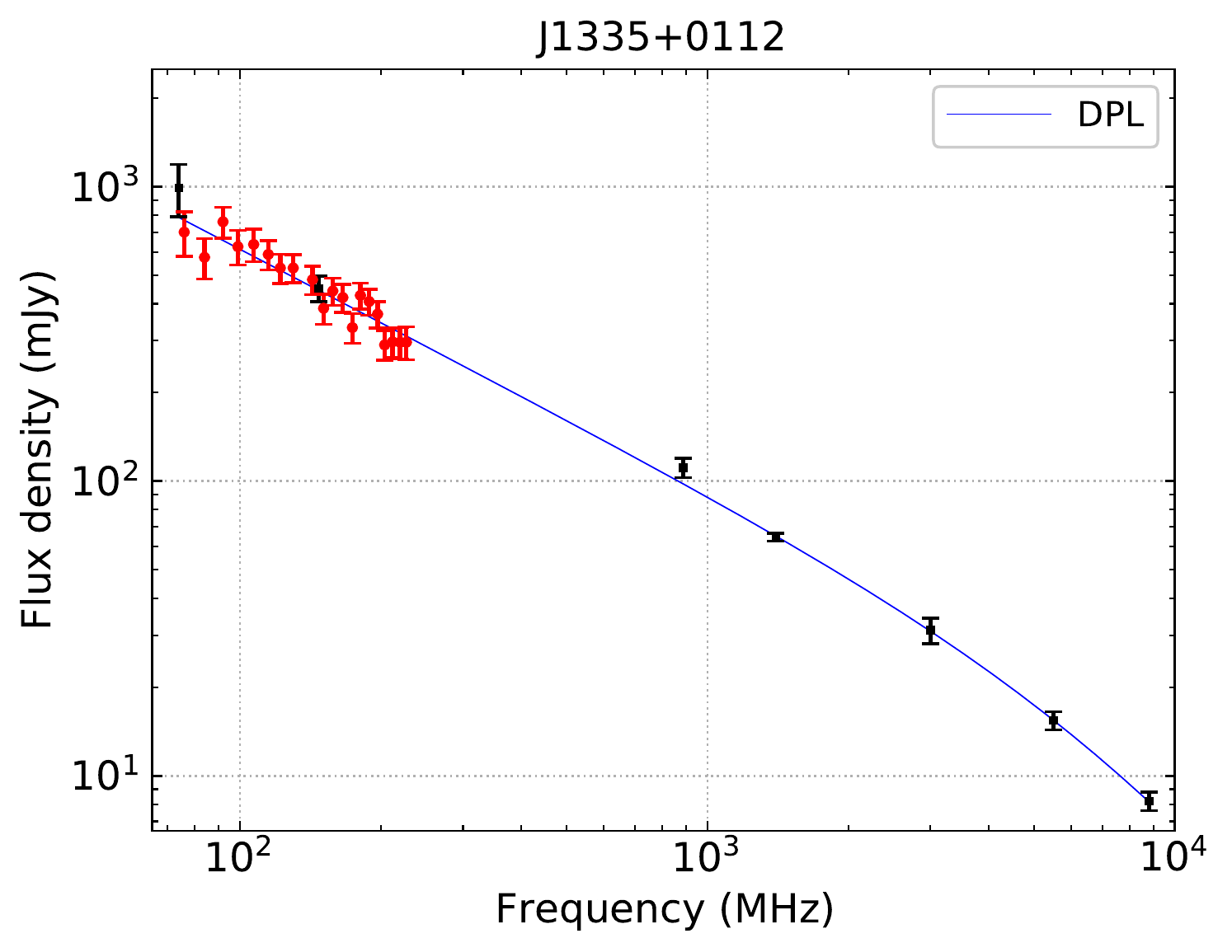}
\end{minipage}
\begin{minipage}{0.5\textwidth}
\vspace{0.2cm}
\includegraphics[width=7.5cm]{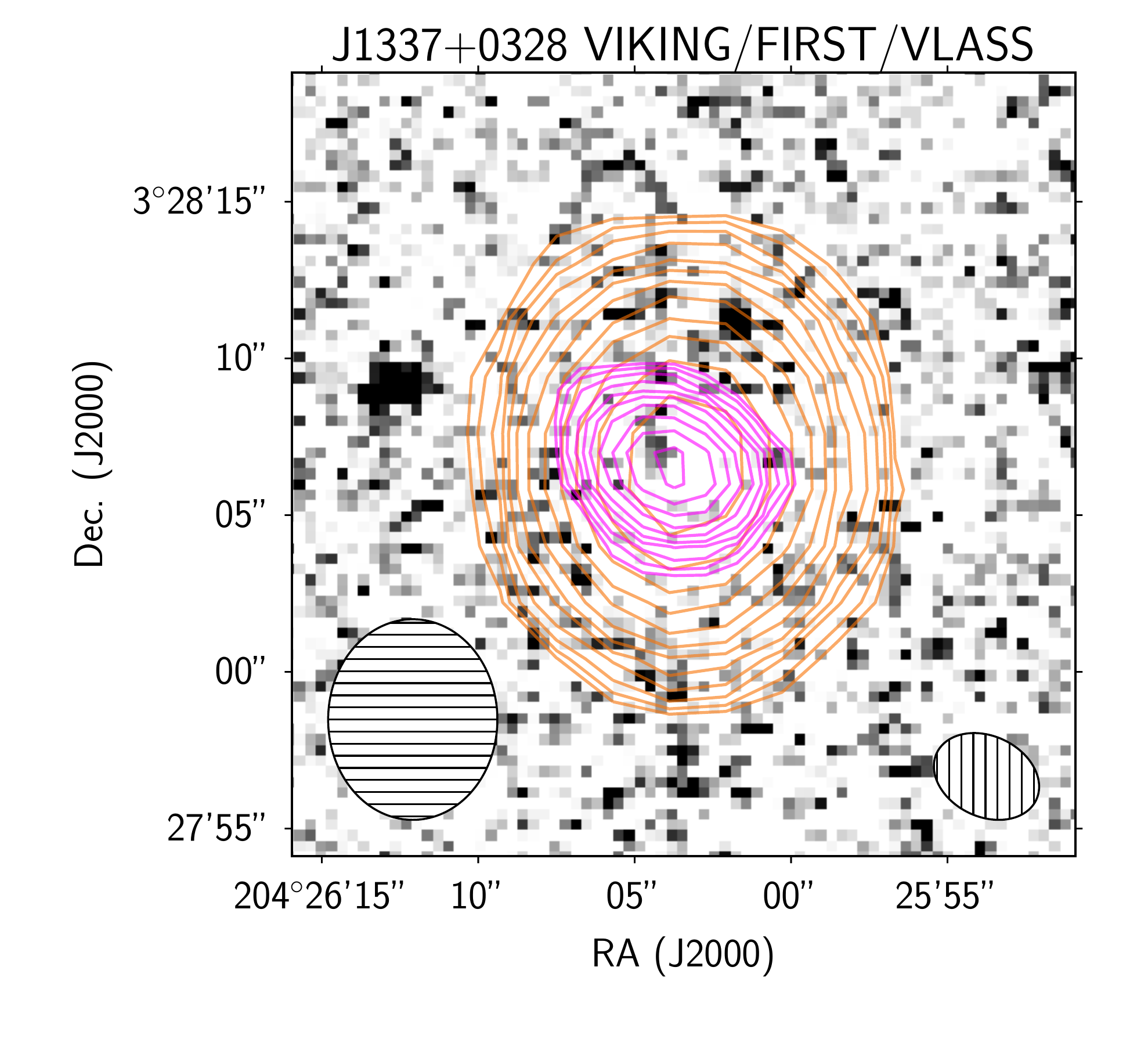}
\end{minipage}
\begin{minipage}{0.5\textwidth}
\includegraphics[width=8.5cm]{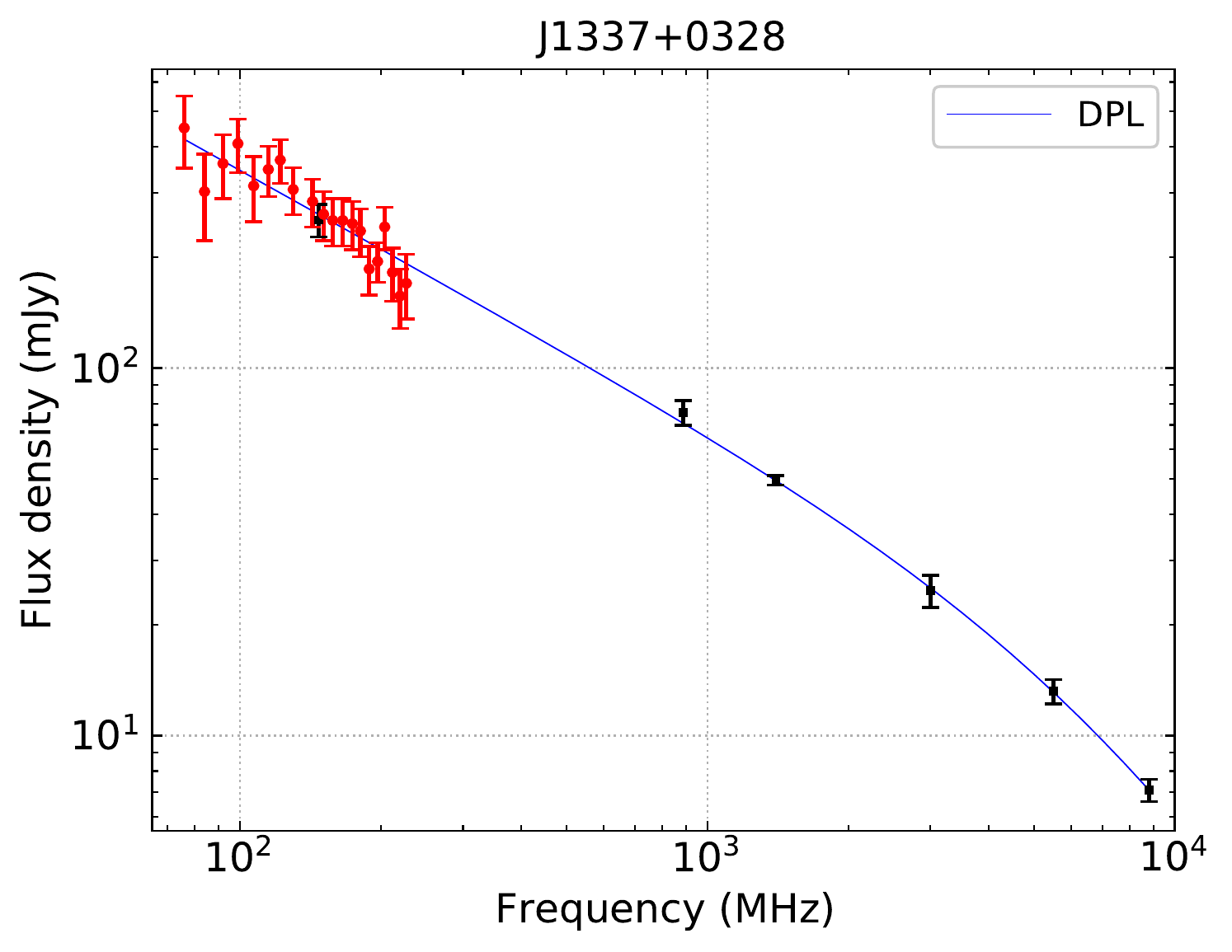}
\end{minipage}
\begin{minipage}{0.5\textwidth}
\vspace{0.2cm}
\includegraphics[width=7.5cm]{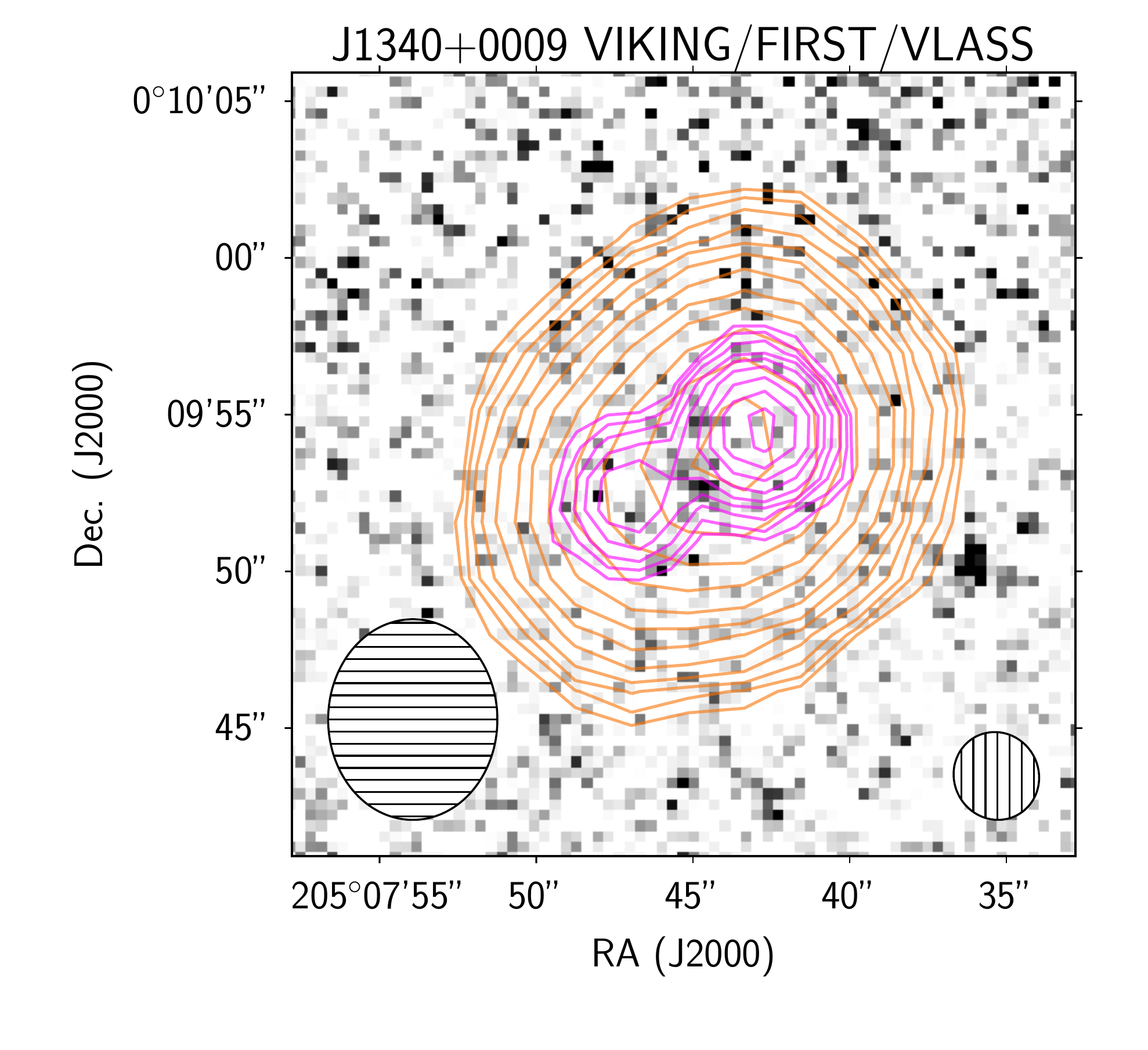}
\end{minipage}
\begin{minipage}{0.5\textwidth}
\includegraphics[width=8.5cm]{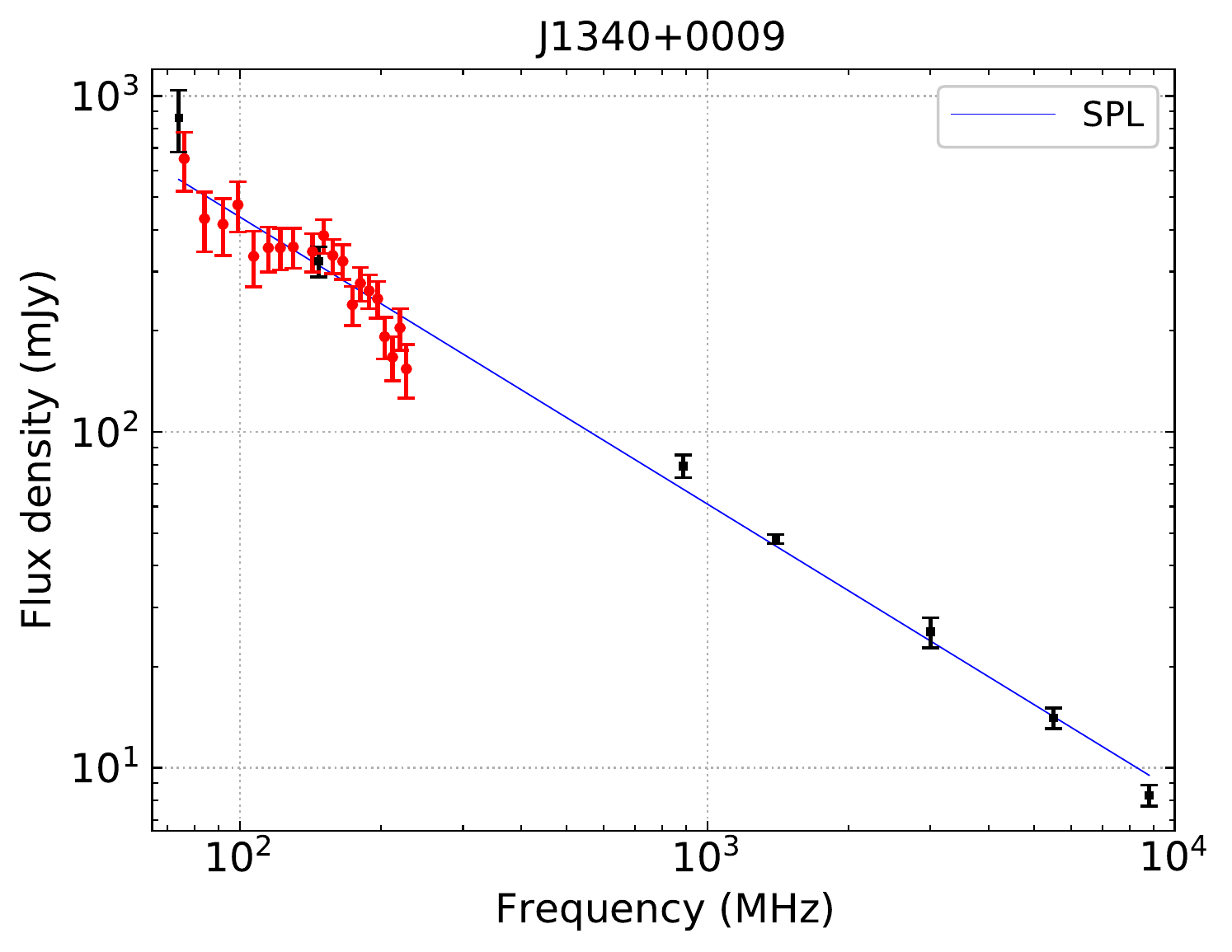}
\end{minipage}
\caption{{\em - continued.}}
\end{figure*}

\setcounter{figure}{1} 
\begin{figure*}
\begin{minipage}{0.5\textwidth}
\vspace{0.2cm}
\includegraphics[width=7.5cm]{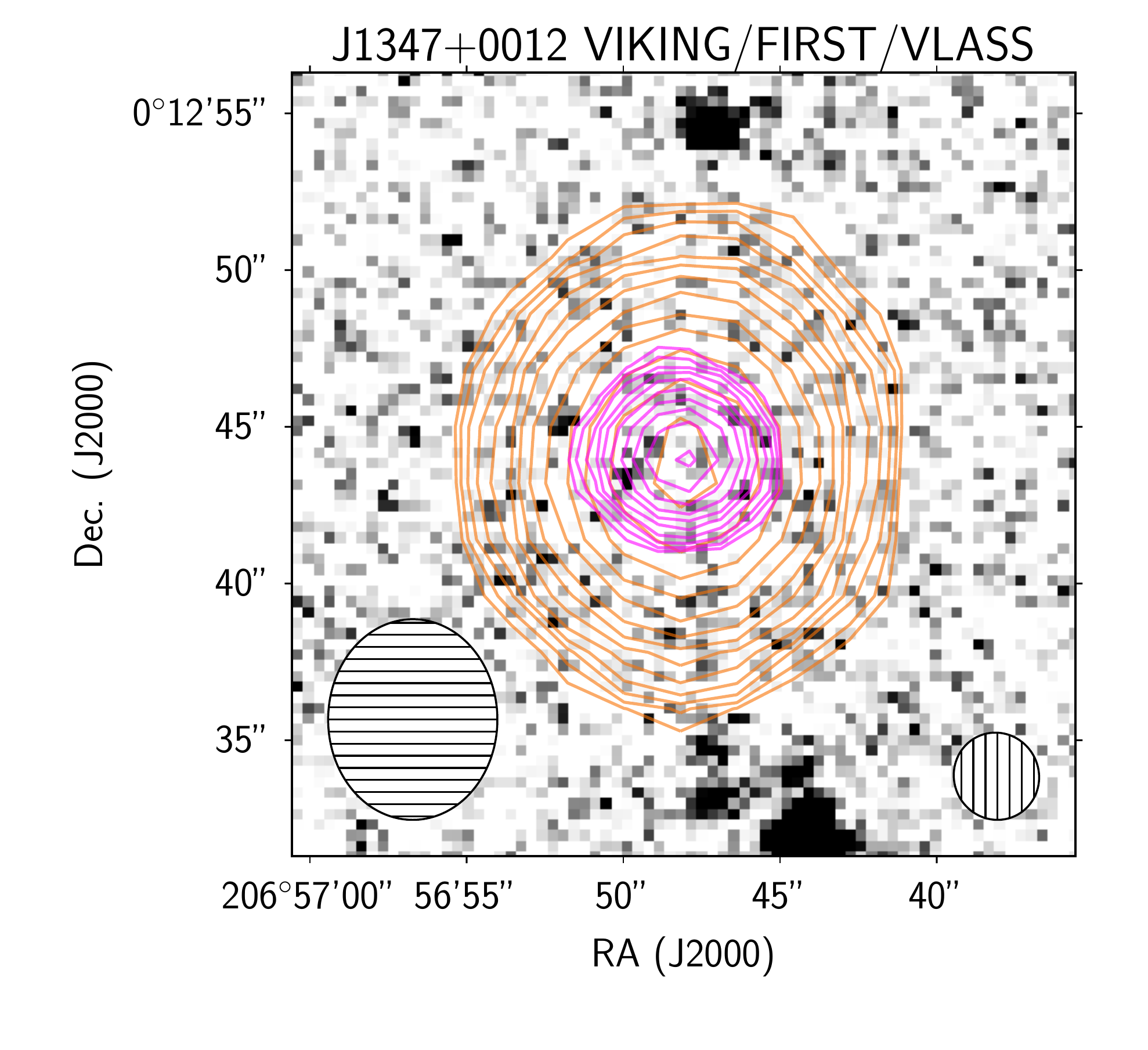}
\end{minipage}
\begin{minipage}{0.5\textwidth}
\includegraphics[width=8.5cm]{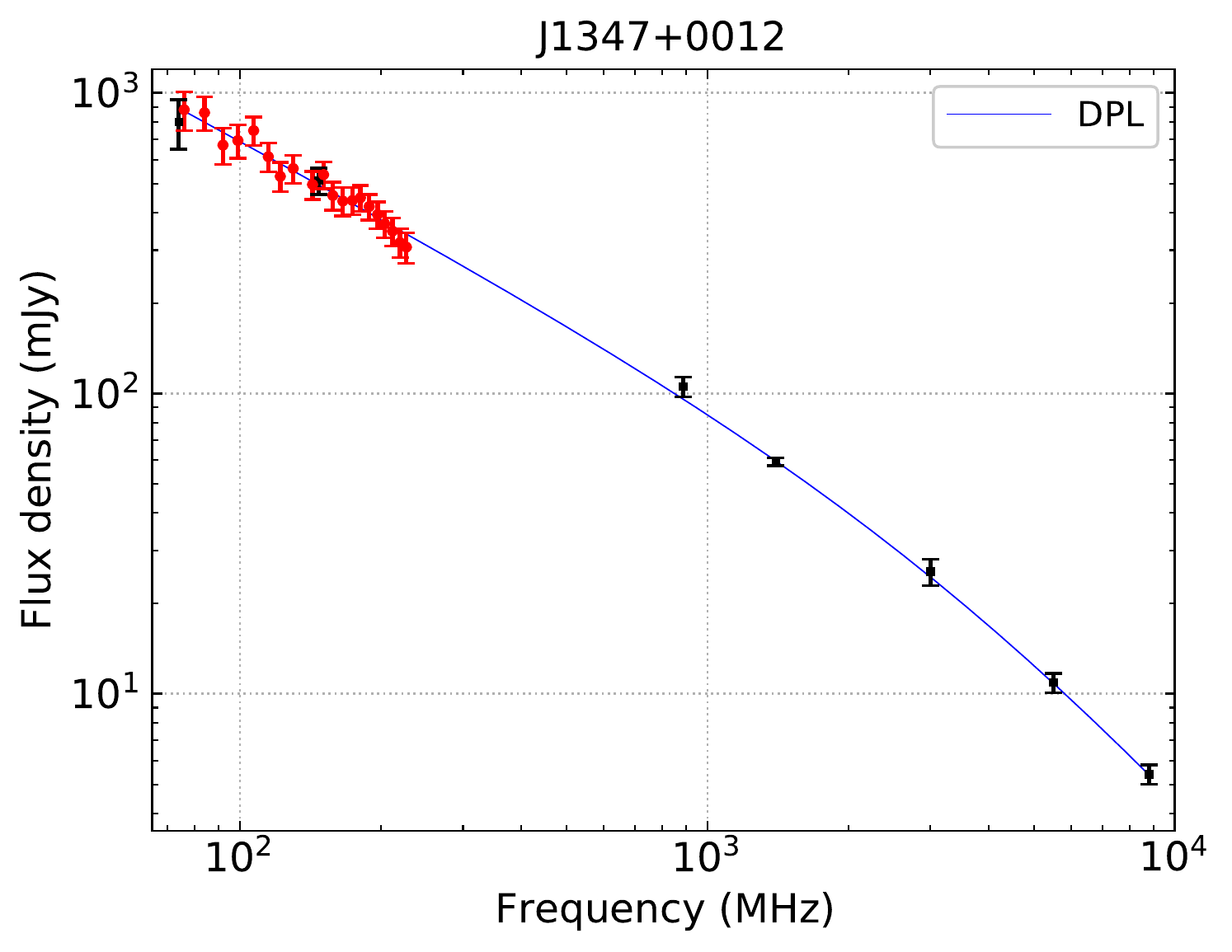}
\end{minipage}
\begin{minipage}{0.5\textwidth}
\vspace{0.2cm}
\includegraphics[width=7.5cm]{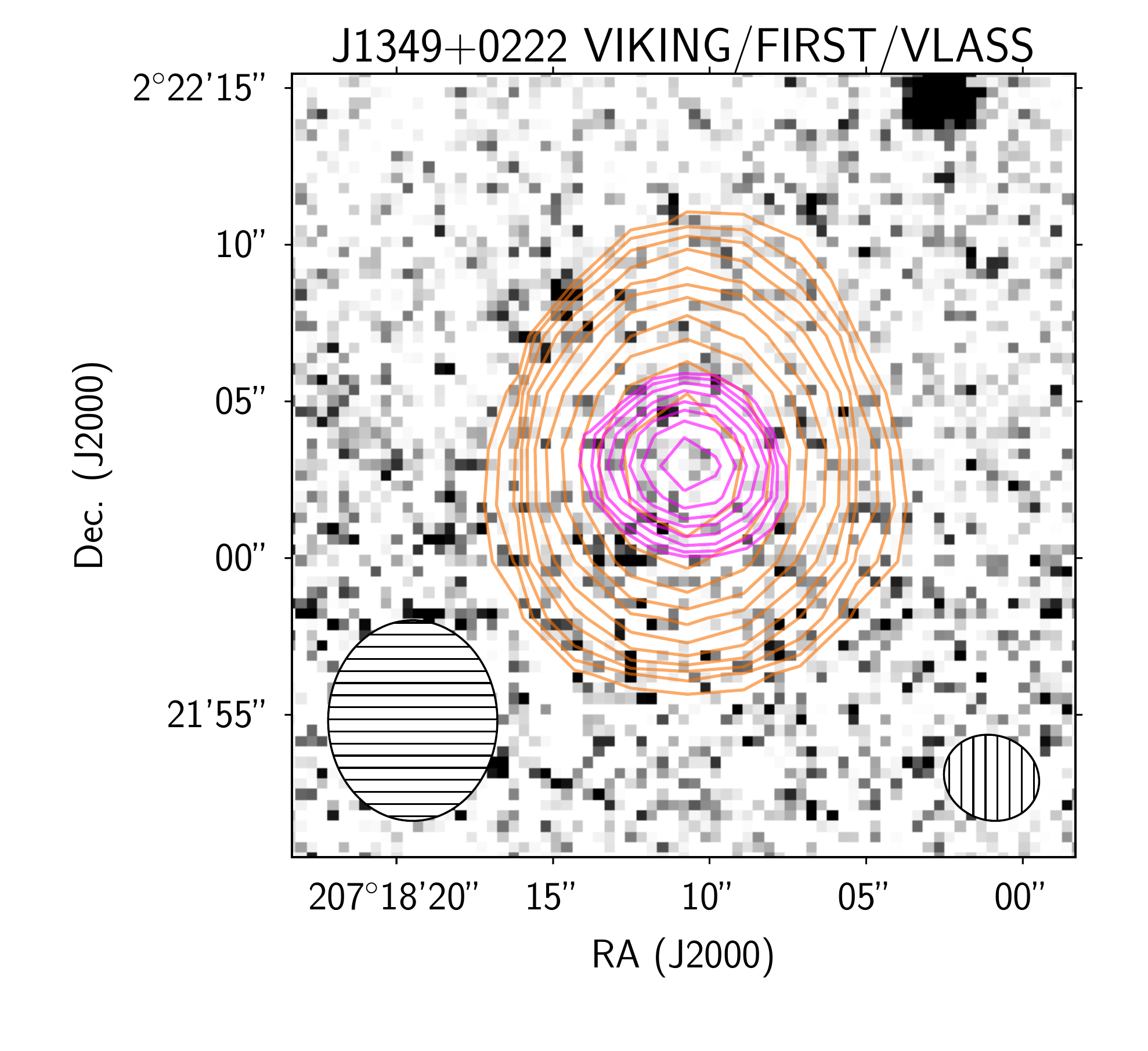}
\end{minipage}
\begin{minipage}{0.5\textwidth}
\includegraphics[width=8.5cm]{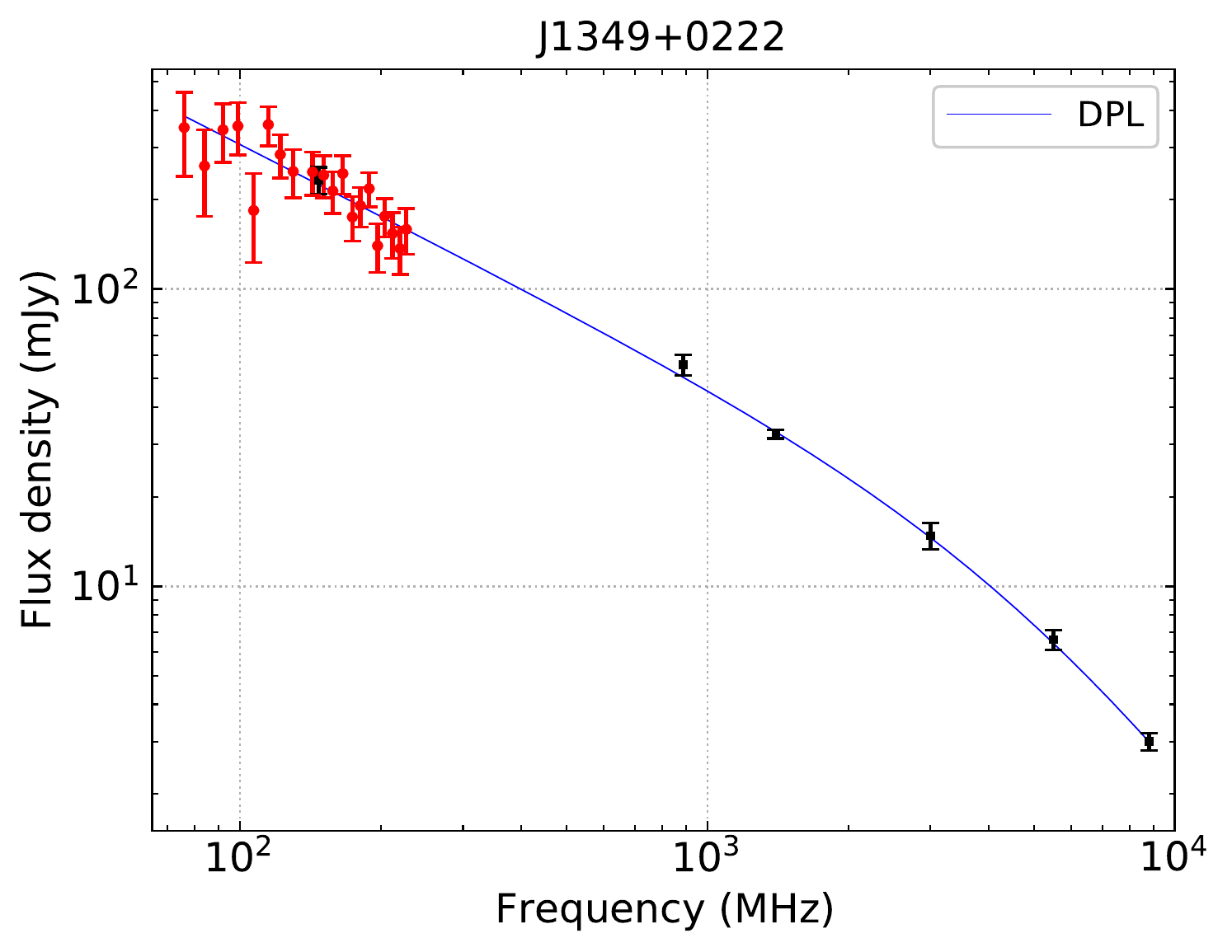}
\end{minipage}
\begin{minipage}{0.5\textwidth}
\vspace{0.2cm}
\includegraphics[width=7.5cm]{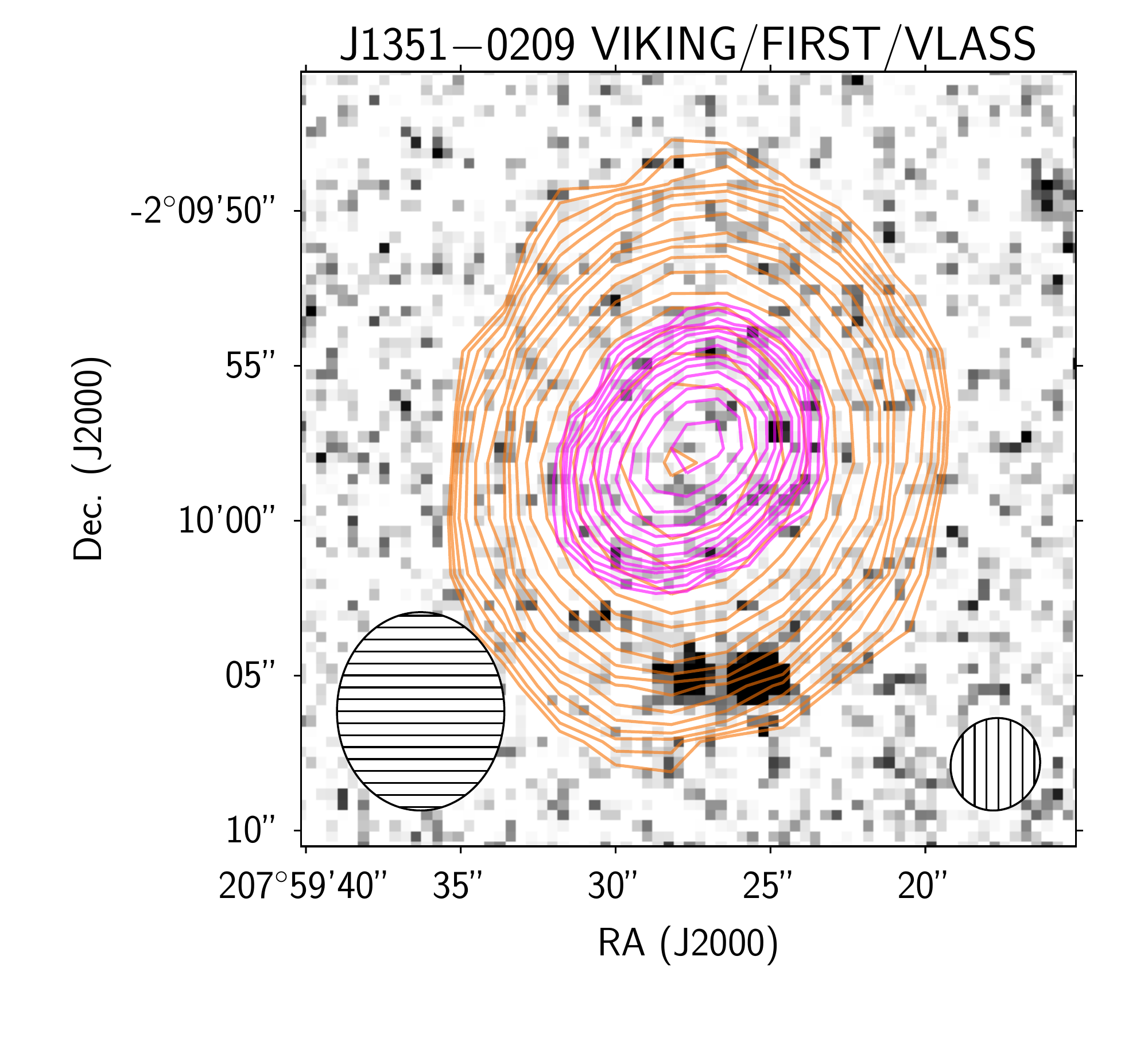}
\end{minipage}
\begin{minipage}{0.5\textwidth}
\includegraphics[width=8.5cm]{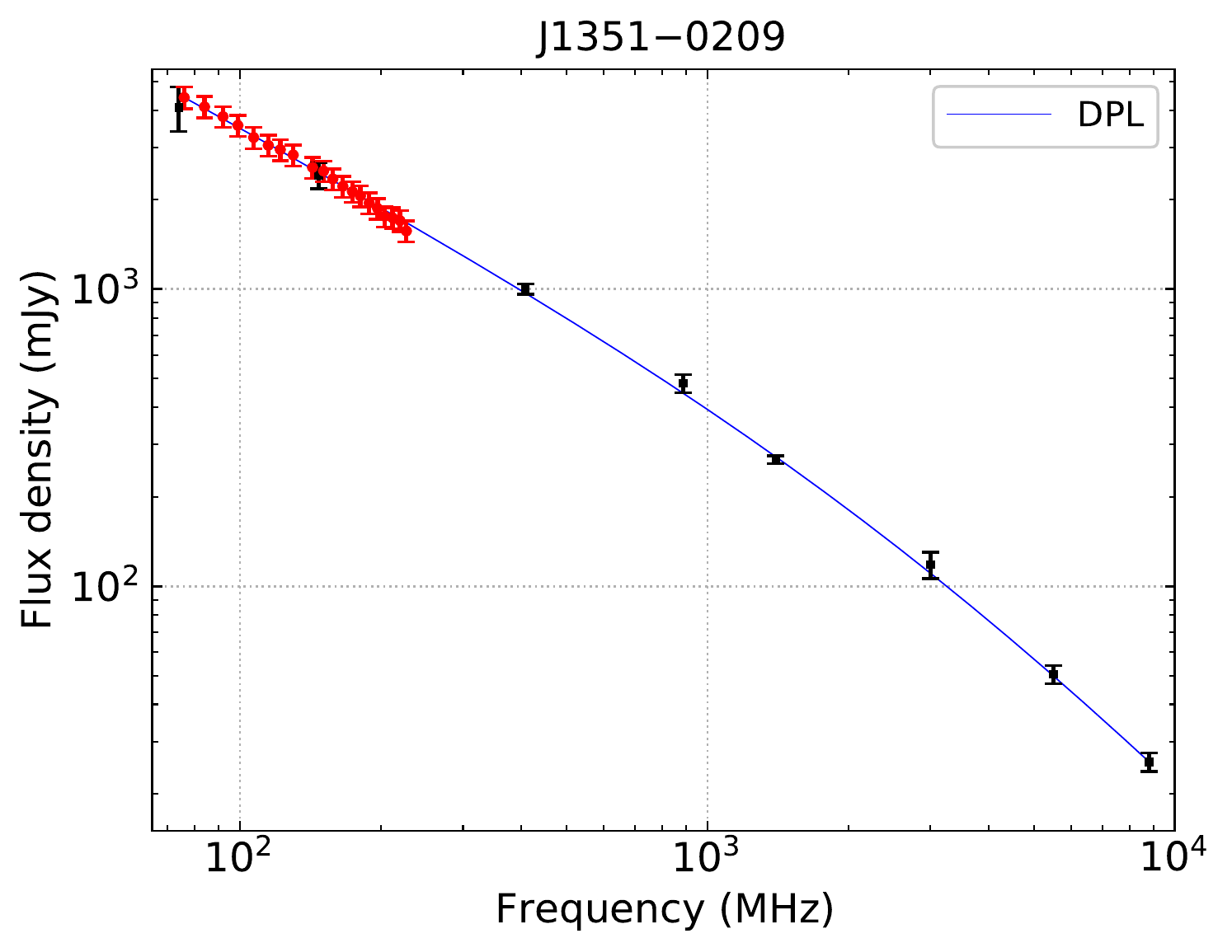}
\end{minipage}
\caption{{\em - continued.}}
\end{figure*}

\setcounter{figure}{1} 
\begin{figure*}
\begin{minipage}{0.5\textwidth}
\vspace{0.2cm}
\includegraphics[width=7.5cm]{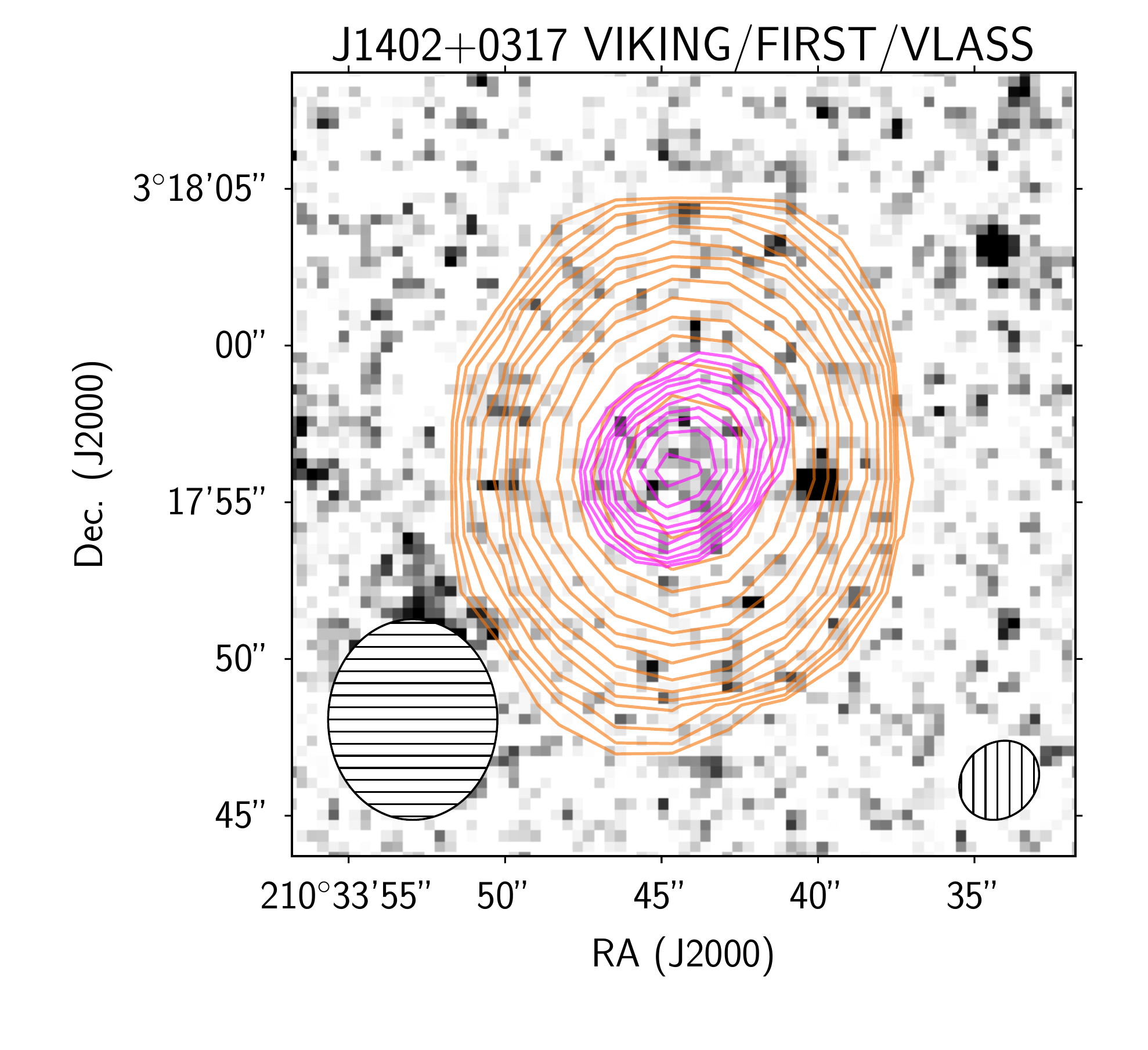}
\end{minipage}
\begin{minipage}{0.5\textwidth}
\includegraphics[width=8.5cm]{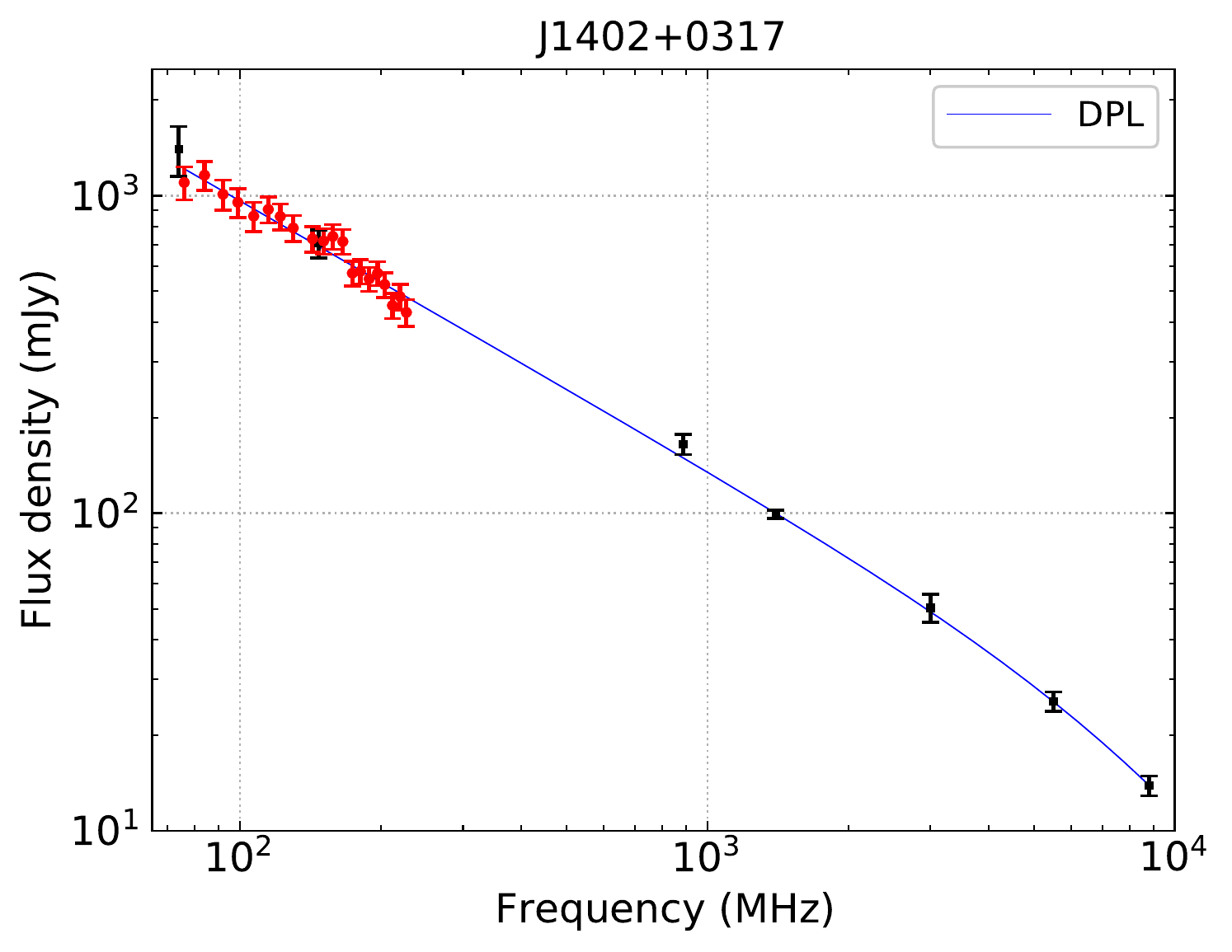}
\end{minipage}
\begin{minipage}{0.5\textwidth}
\vspace{0.2cm}
\includegraphics[width=7.5cm]{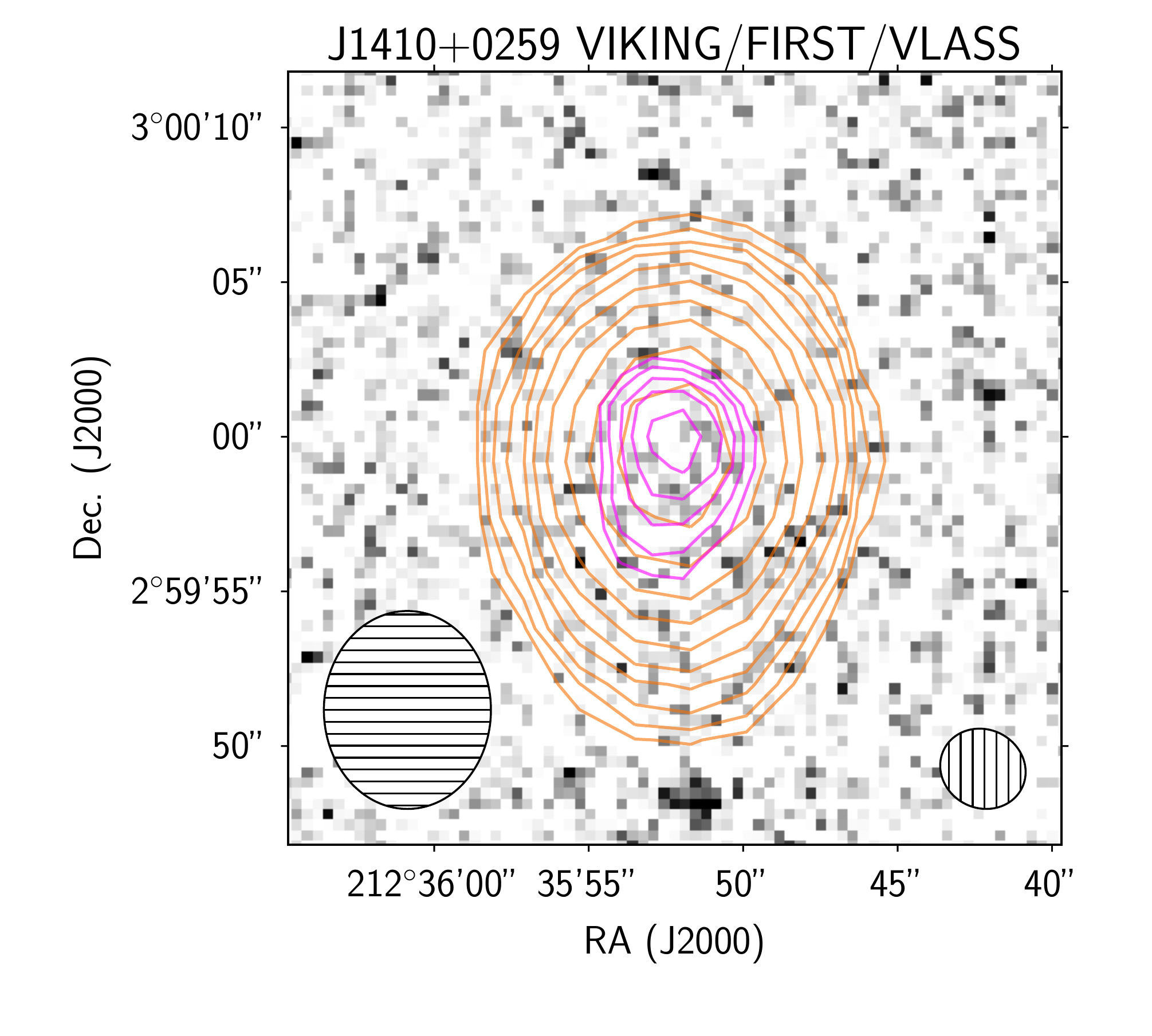}
\end{minipage}
\begin{minipage}{0.5\textwidth}
\includegraphics[width=8.5cm]{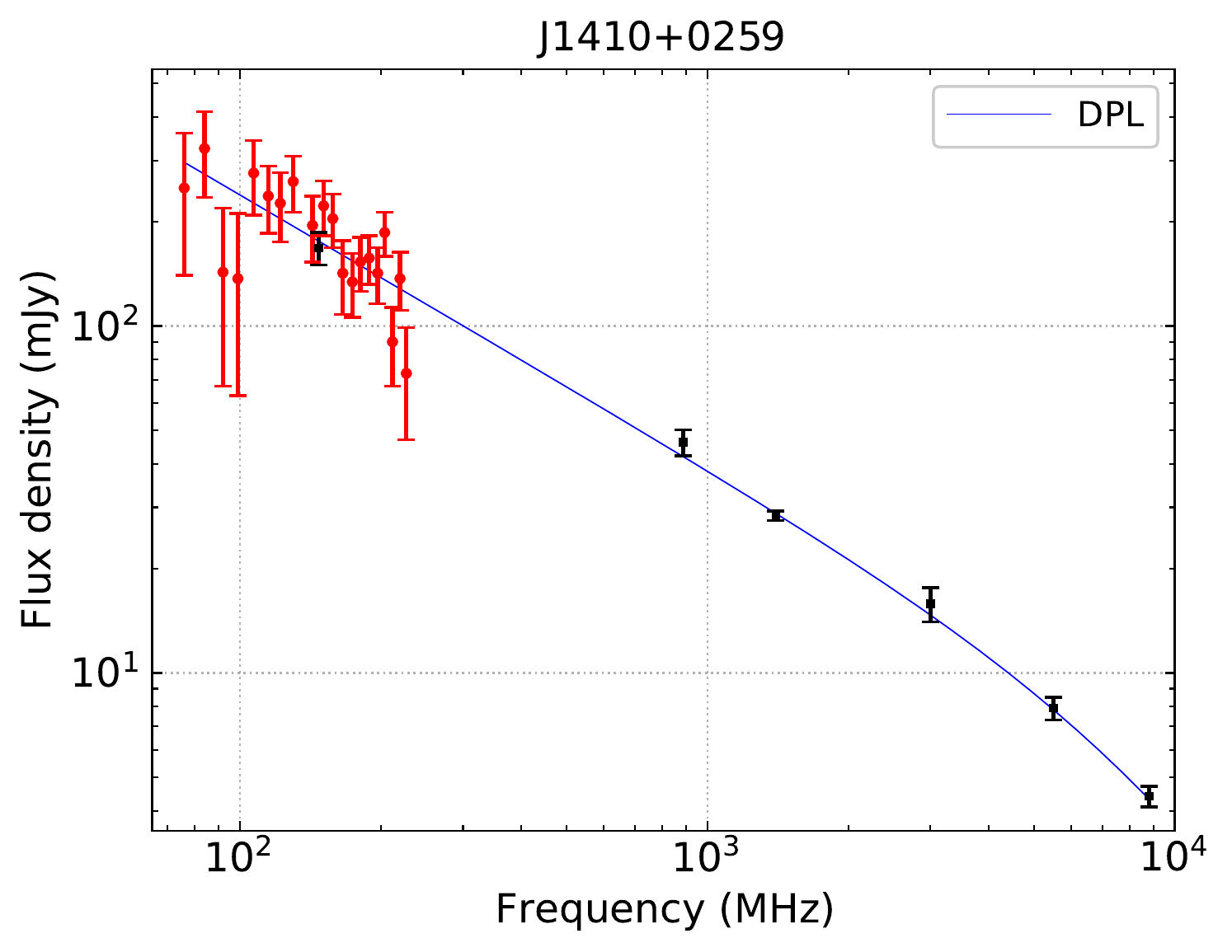}
\end{minipage}
\begin{minipage}{0.5\textwidth}
\vspace{0.2cm}
\includegraphics[width=7.5cm]{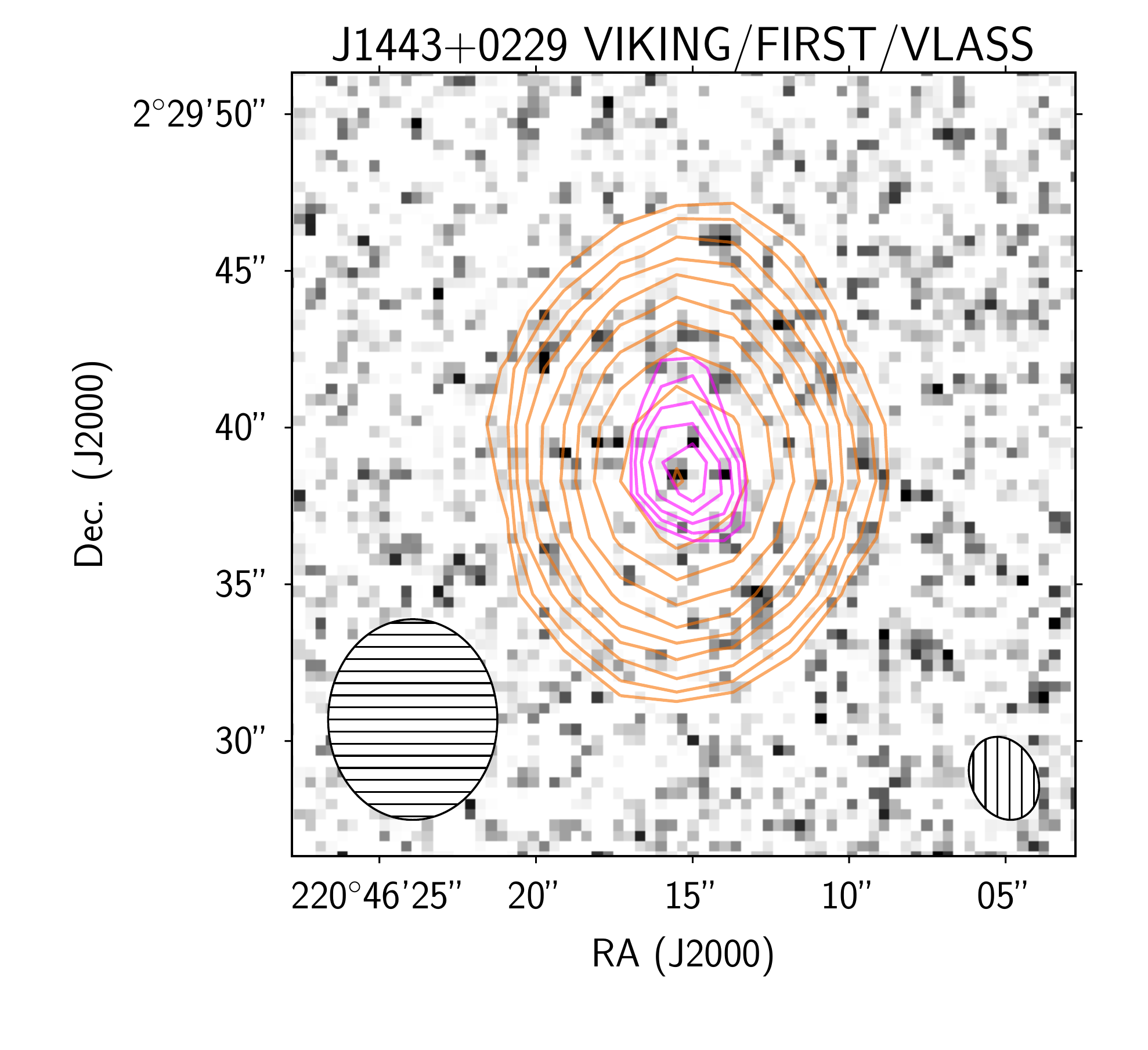}
\end{minipage}
\begin{minipage}{0.5\textwidth}
\includegraphics[width=8.5cm]{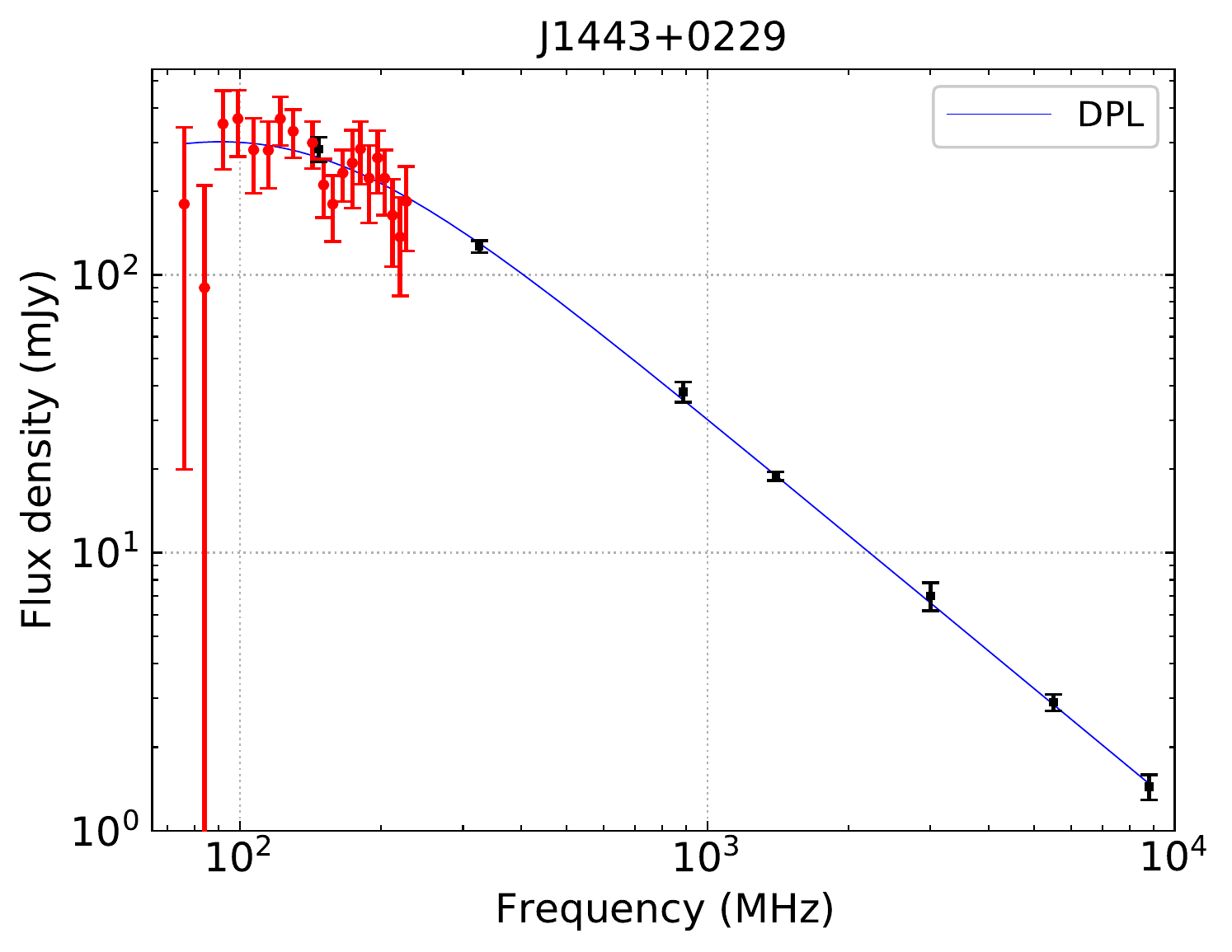}
\end{minipage}
\caption{{\em - continued.}}
\end{figure*}

\setcounter{figure}{1} 
\begin{figure*}
\begin{minipage}{0.5\textwidth}
\vspace{0.2cm}
\includegraphics[width=7.5cm]{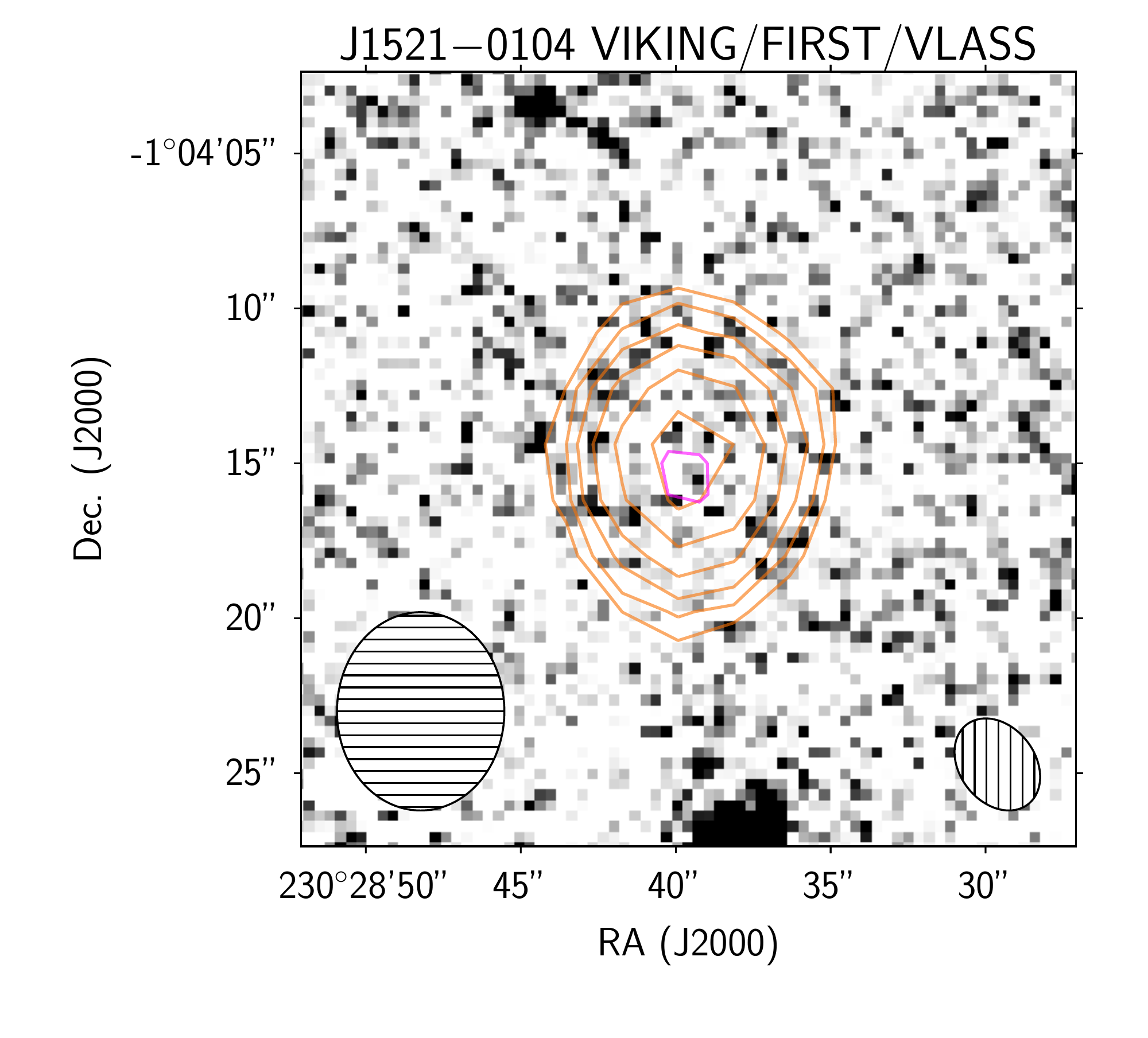}
\end{minipage}
\begin{minipage}{0.5\textwidth}
\includegraphics[width=8.5cm]{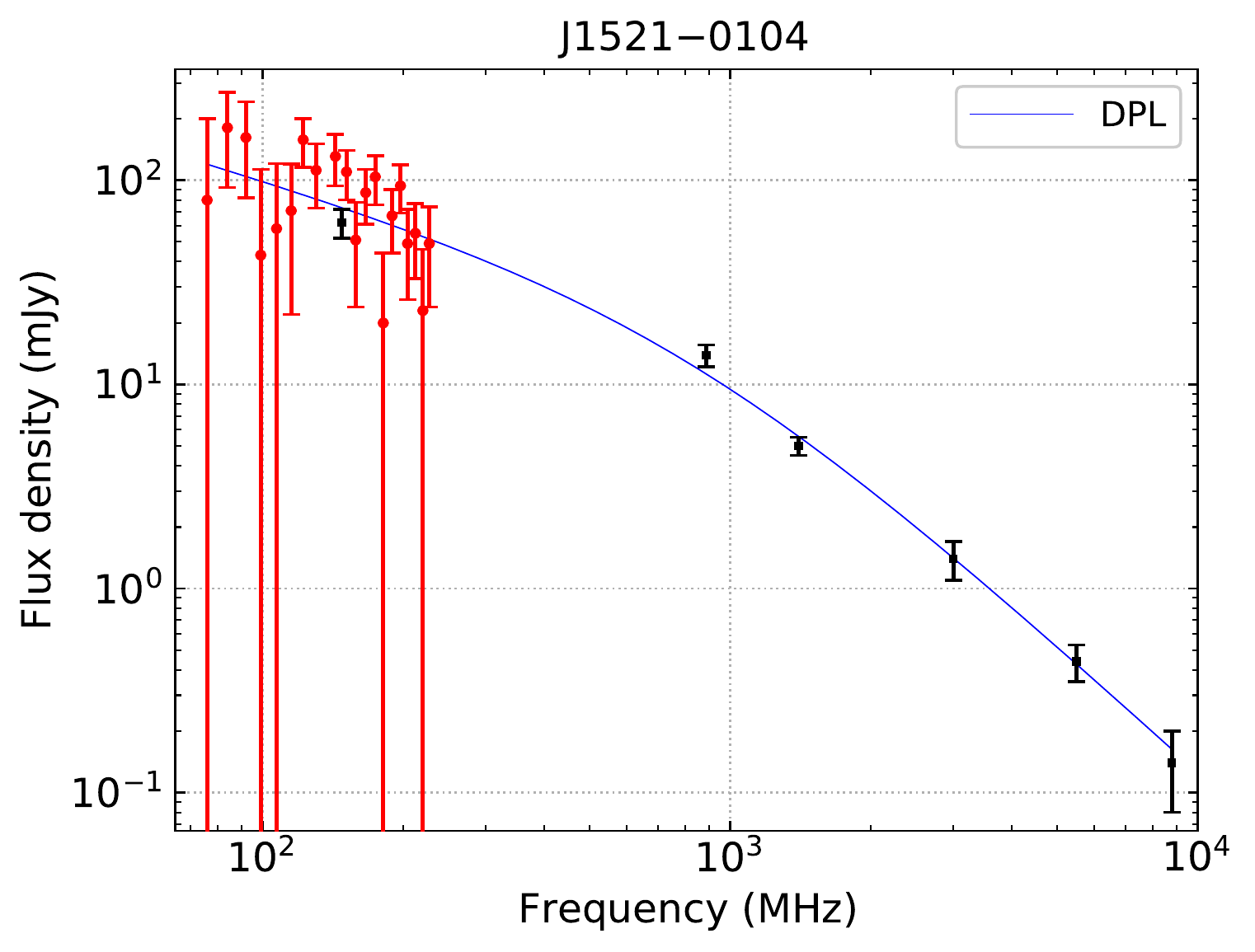}
\end{minipage}
\begin{minipage}{0.5\textwidth}
\vspace{0.2cm}
\includegraphics[width=7.5cm]{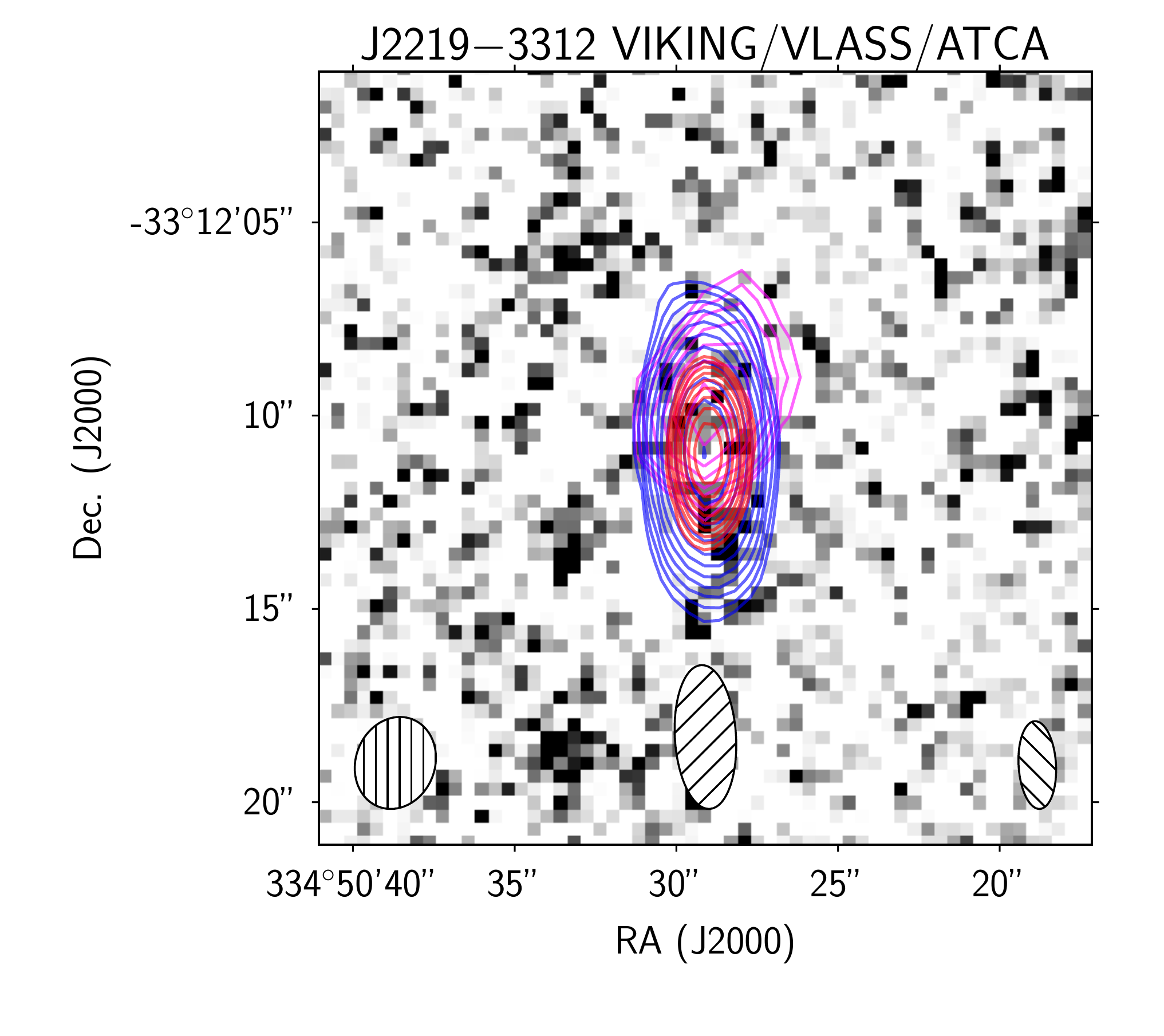}
\end{minipage}
\begin{minipage}{0.5\textwidth}
\includegraphics[width=8.5cm]{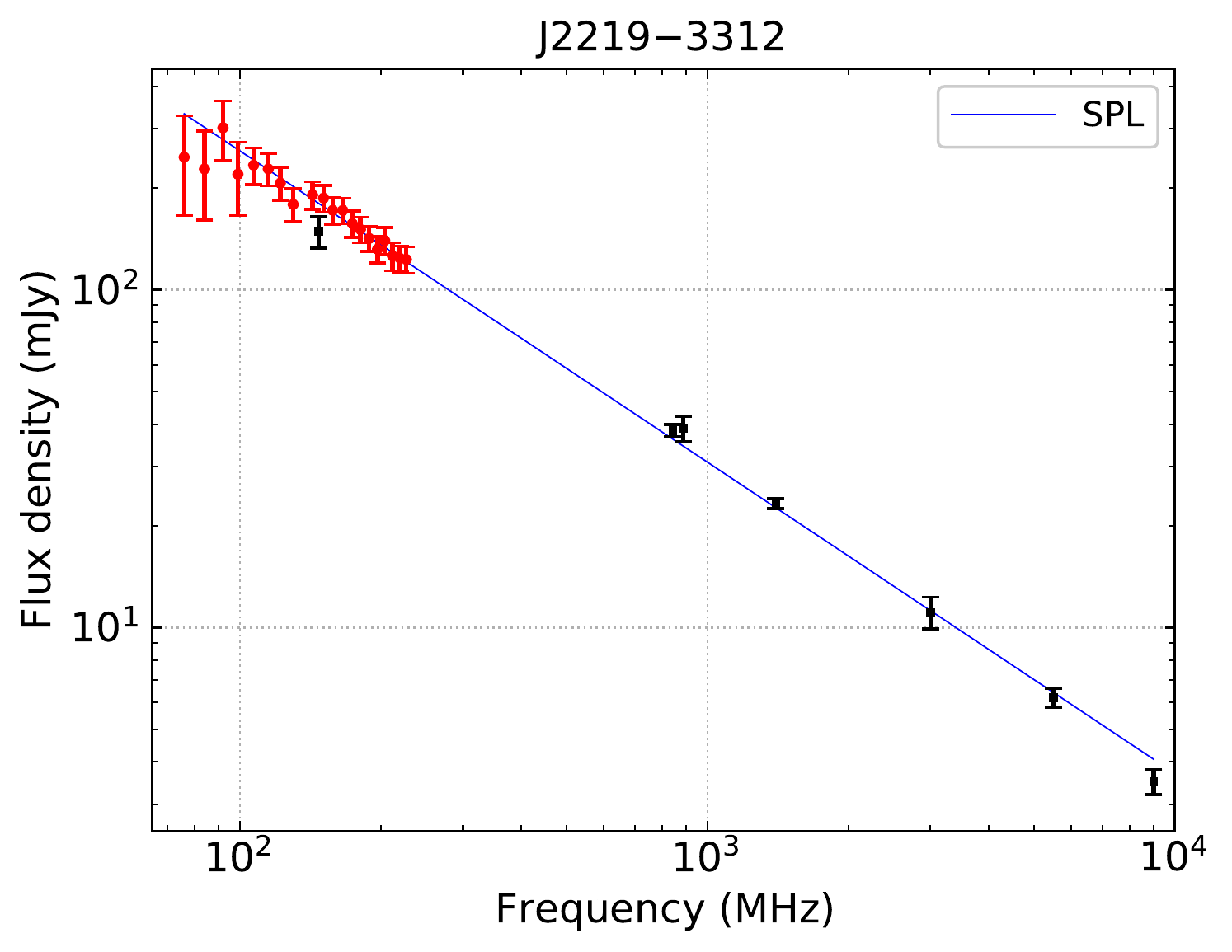}
\end{minipage}
\begin{minipage}{0.5\textwidth}
\vspace{0.2cm}
\includegraphics[width=7.5cm]{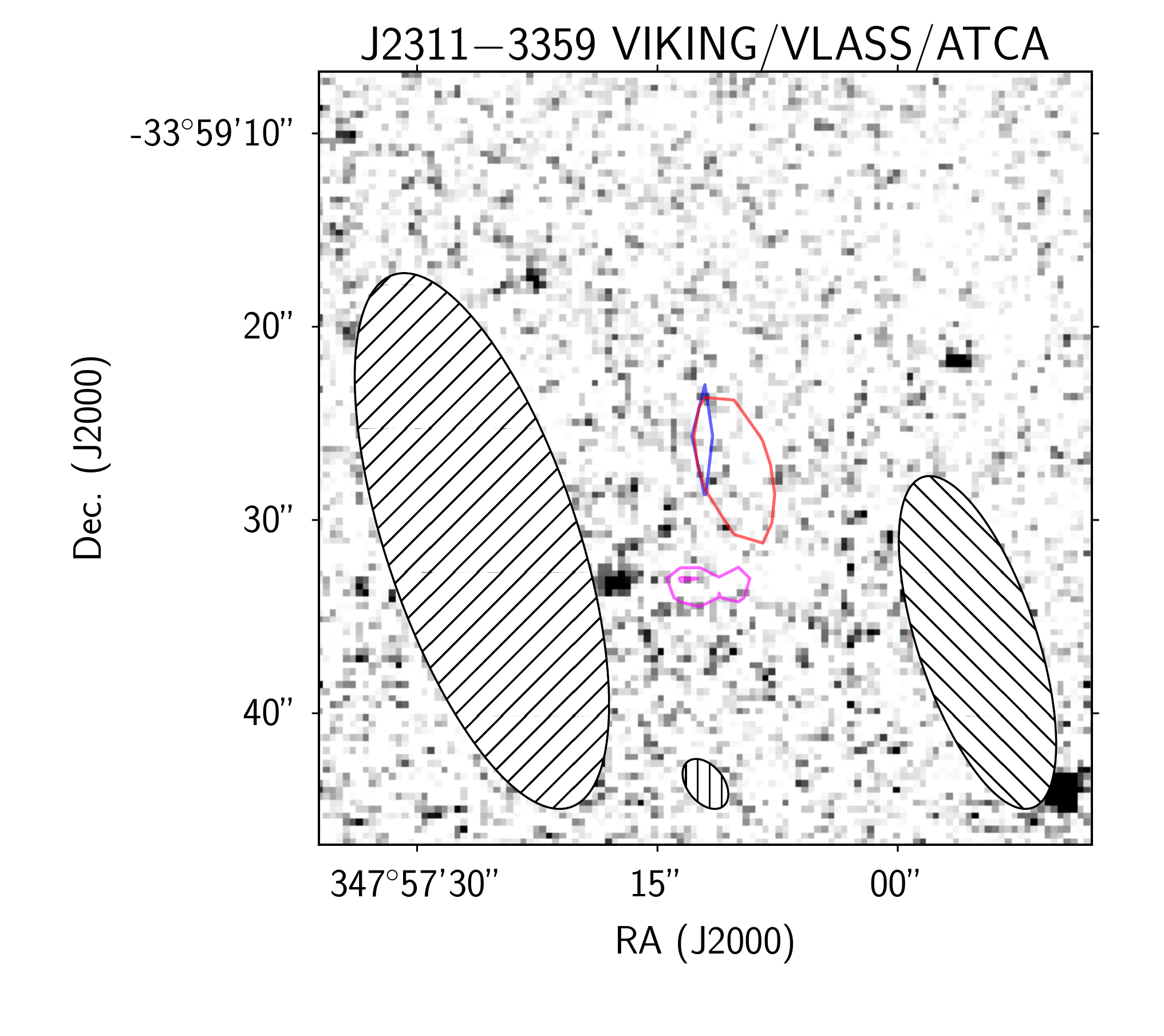}
\end{minipage}
\begin{minipage}{0.5\textwidth}
\includegraphics[width=8.5cm]{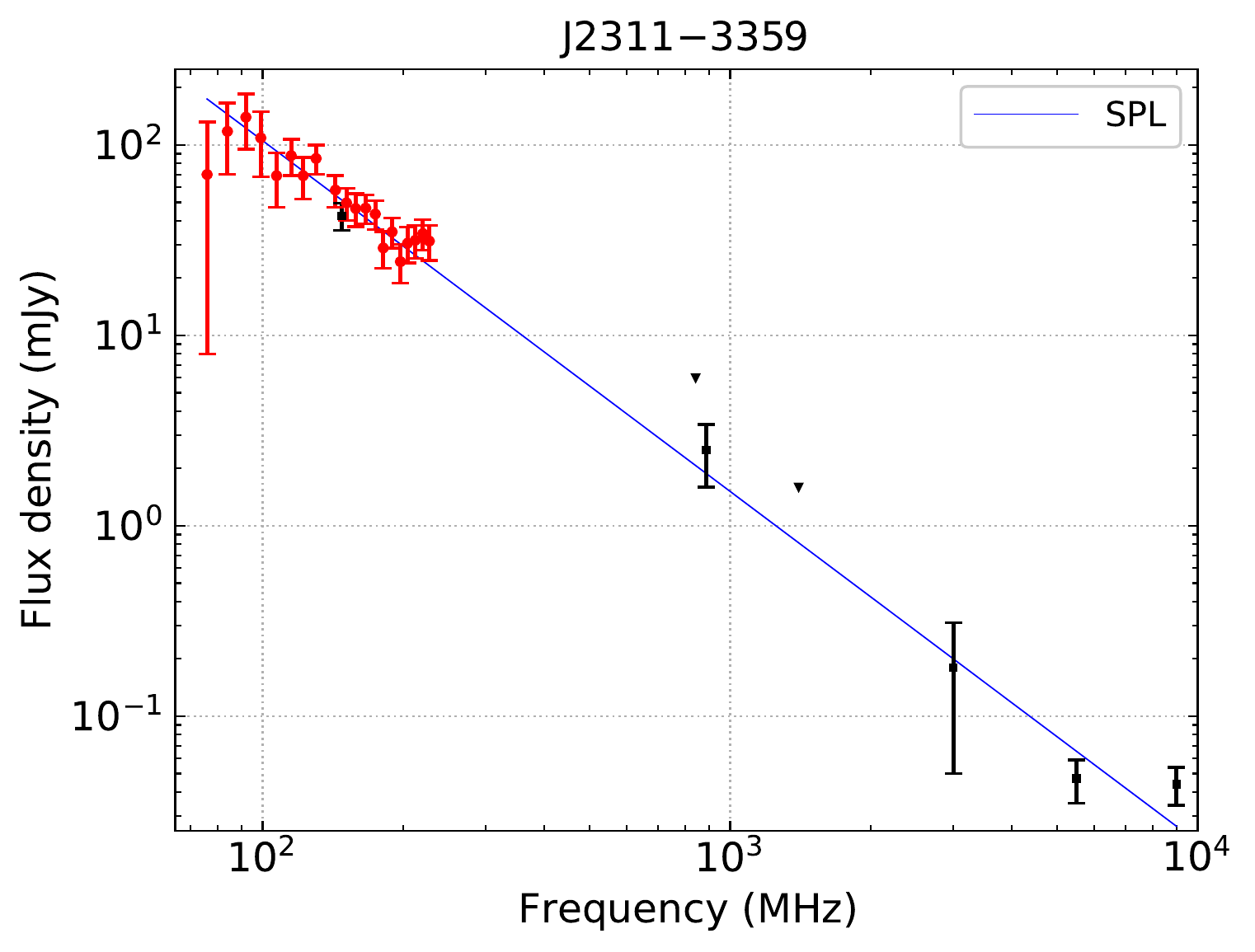}
\end{minipage}
\caption{{\em - continued.}}
\end{figure*}

\setcounter{figure}{1} 
\begin{figure*}
\begin{minipage}{0.5\textwidth}
\vspace{0.2cm}
\includegraphics[width=7.5cm]{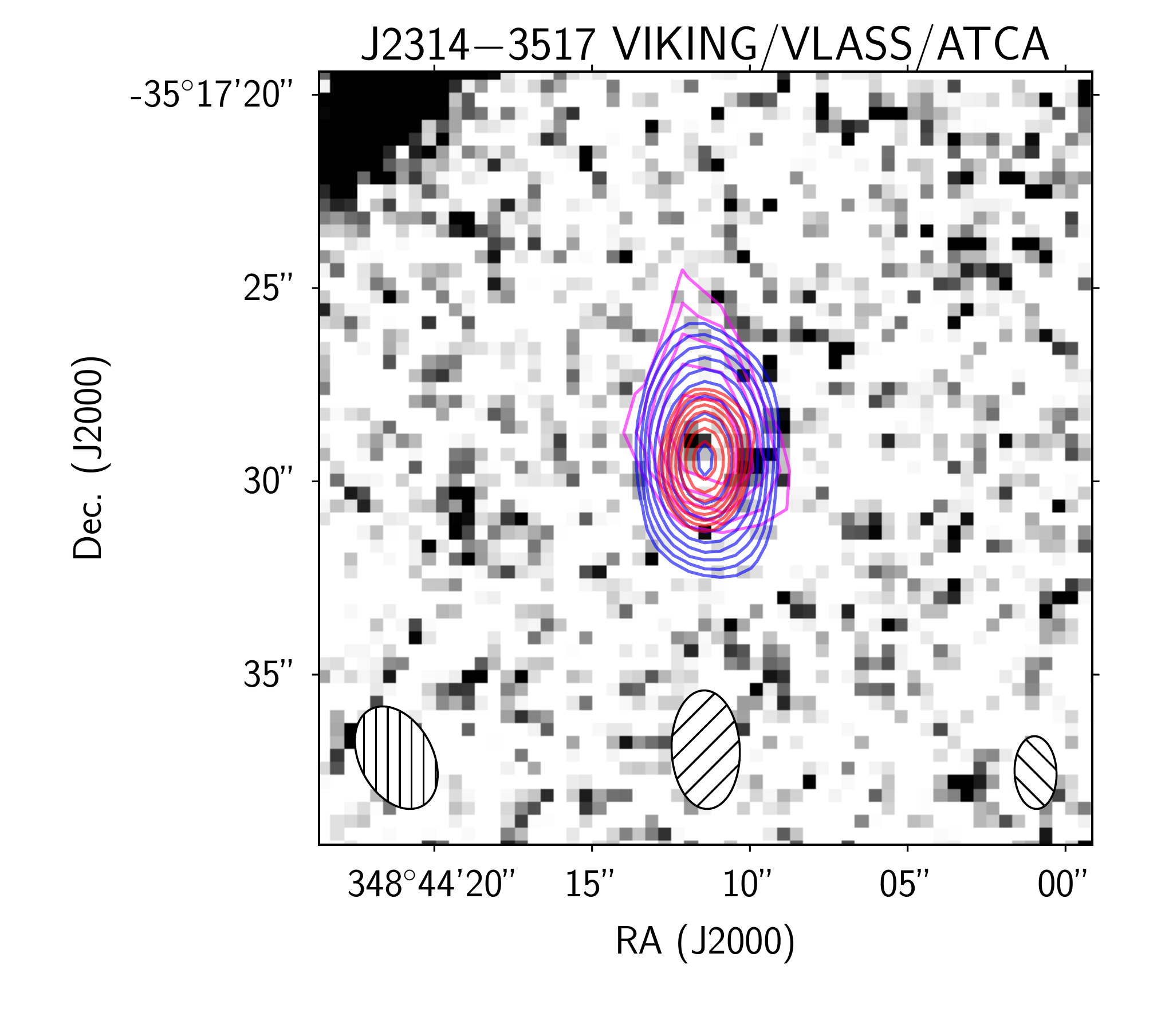}
\end{minipage}
\begin{minipage}{0.5\textwidth}
\includegraphics[width=8.5cm]{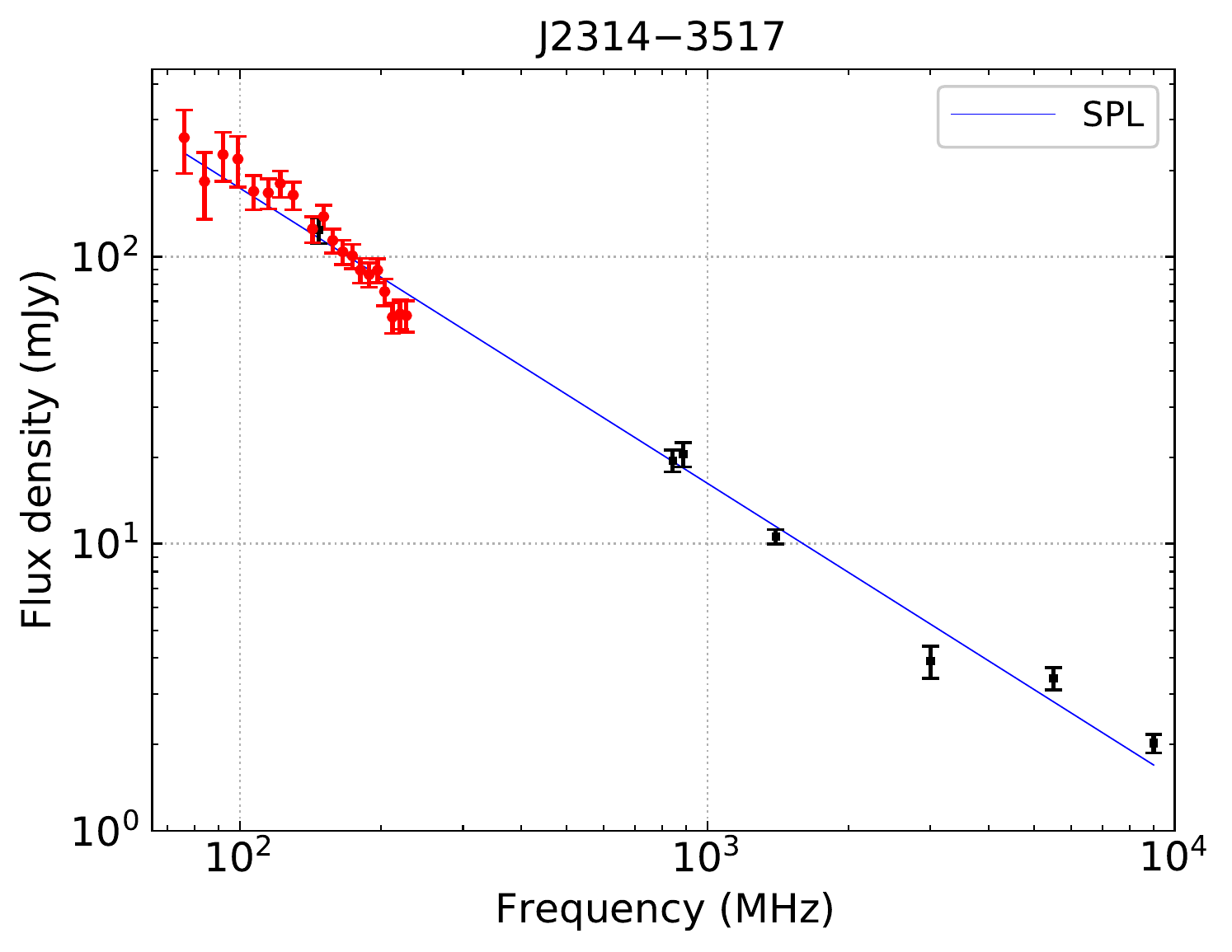}
\end{minipage}
\begin{minipage}{0.5\textwidth}
\vspace{0.2cm}
\includegraphics[width=7.5cm]{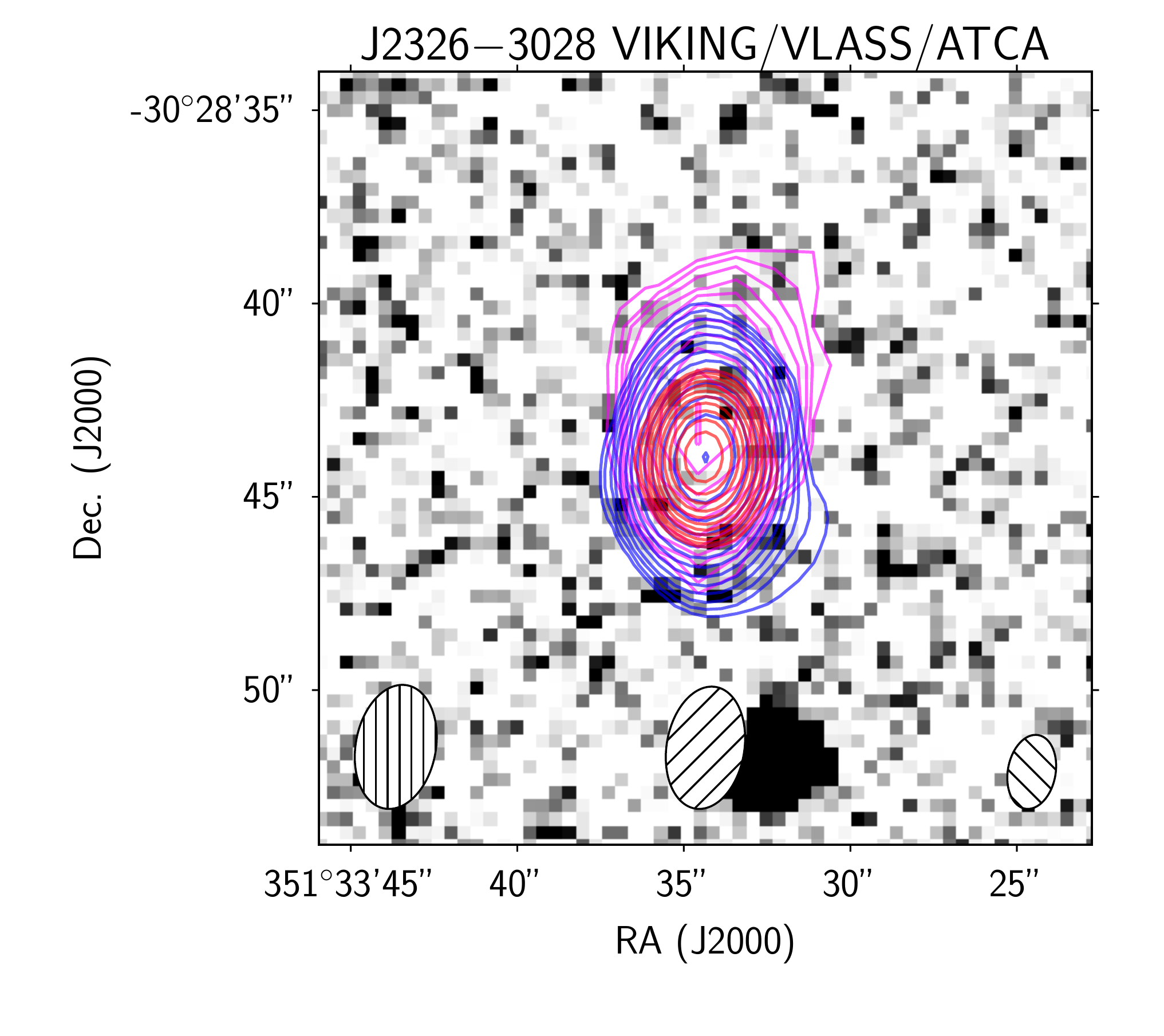}
\end{minipage}
\begin{minipage}{0.5\textwidth}
\includegraphics[width=8.5cm]{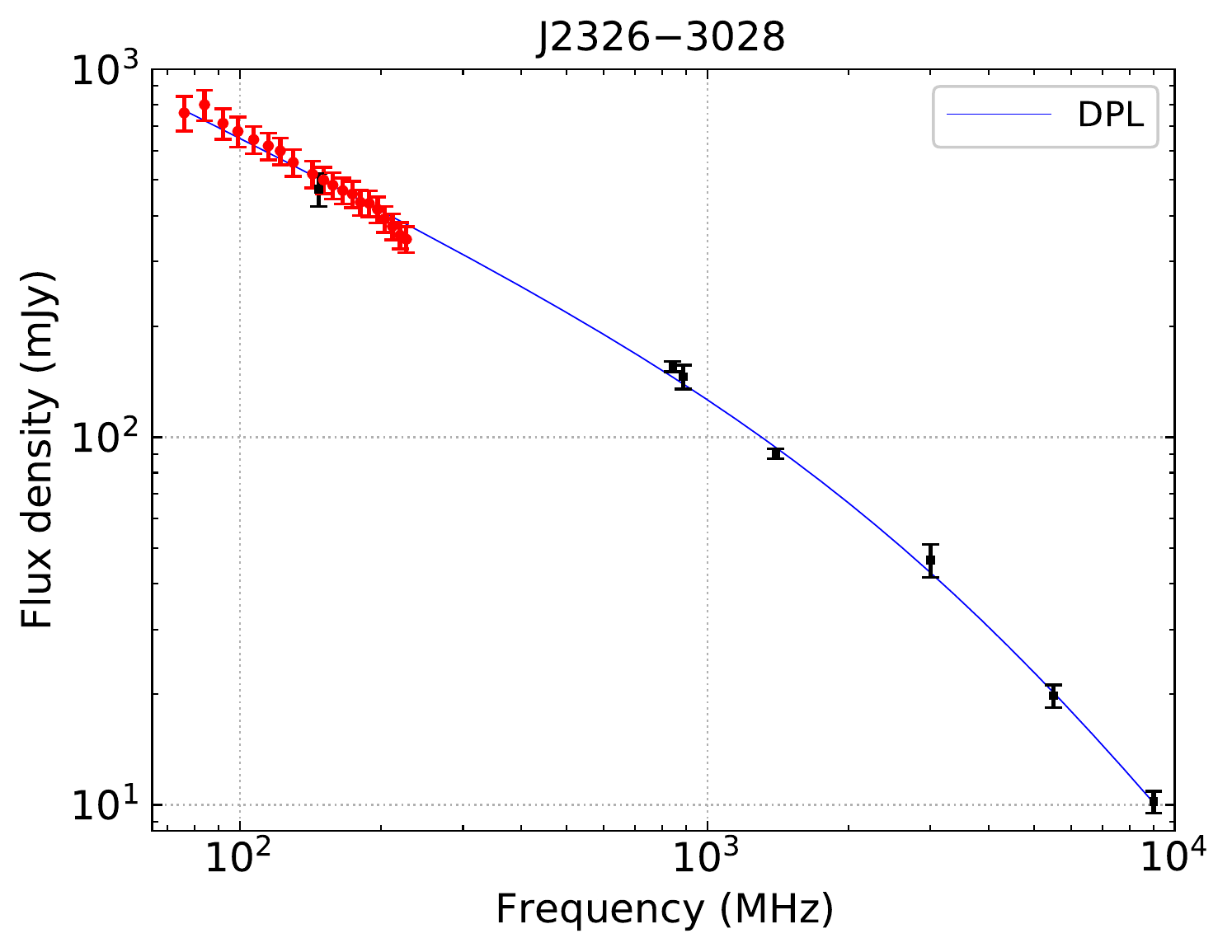}
\end{minipage}
\begin{minipage}{0.5\textwidth}
\vspace{0.2cm}
\includegraphics[width=7.5cm]{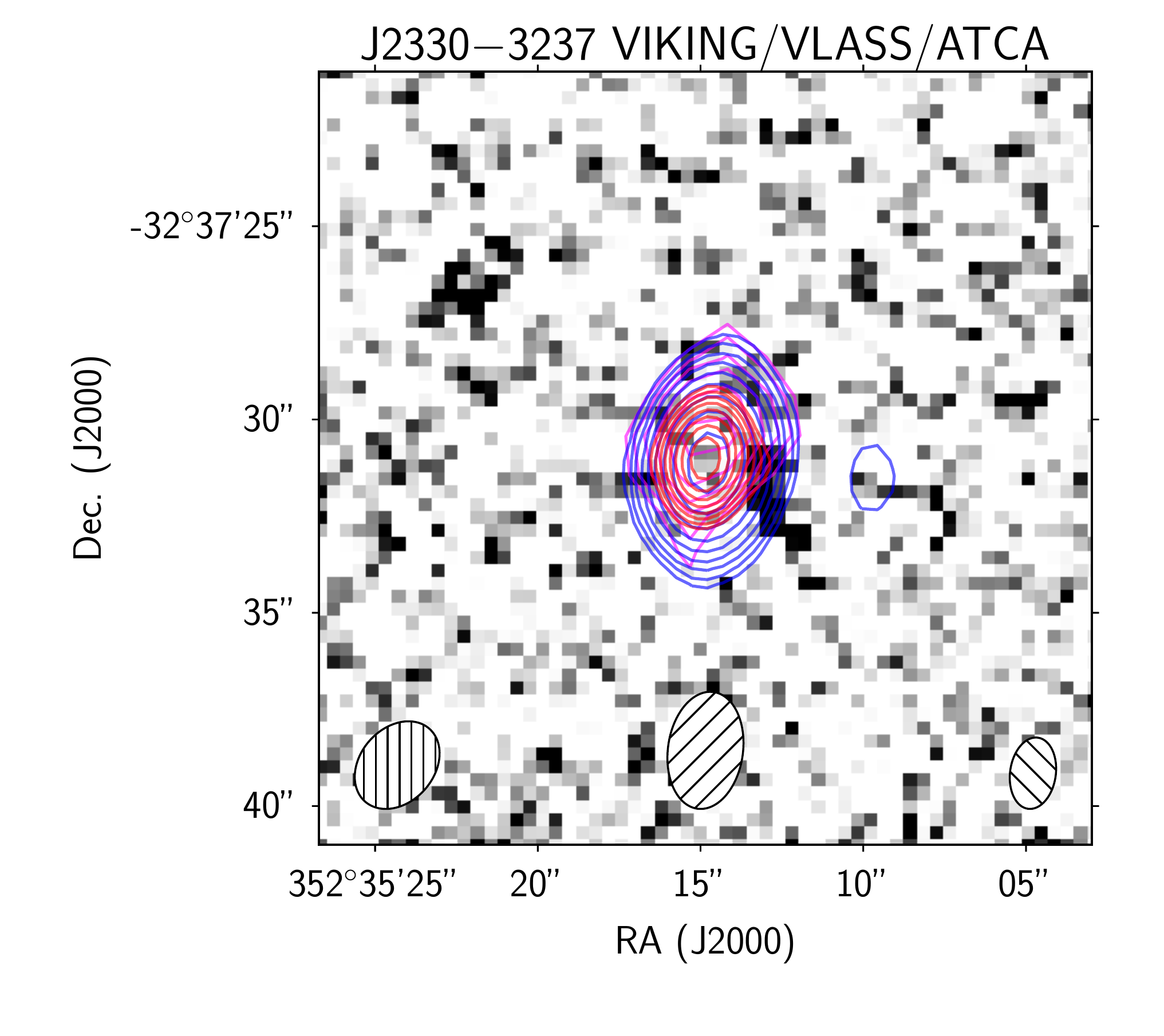}
\end{minipage}
\begin{minipage}{0.5\textwidth}
\includegraphics[width=8.5cm]{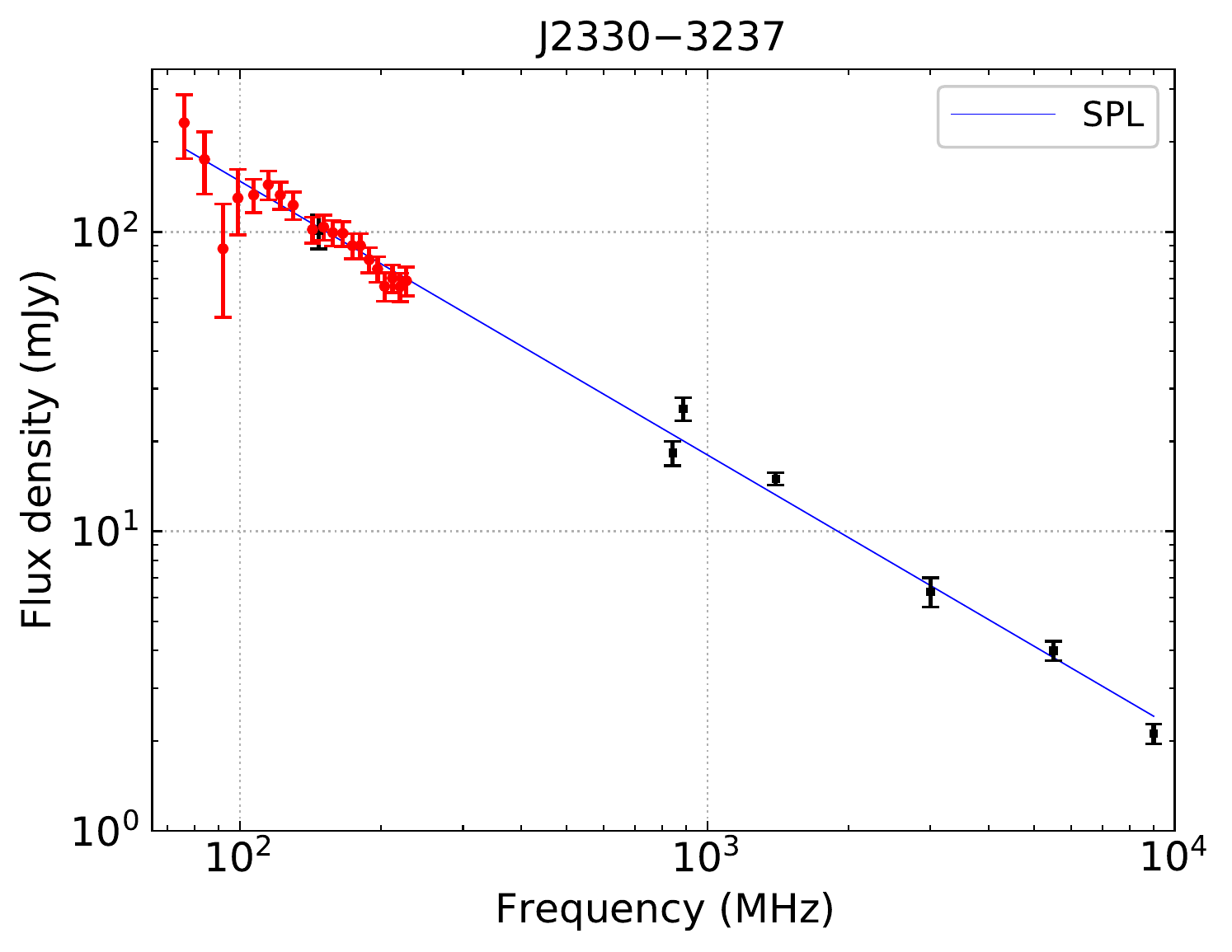}
\end{minipage}
\caption{{\em - continued.}}
\end{figure*}

\setcounter{figure}{1} 
\begin{figure*}
\begin{minipage}{0.5\textwidth}
\vspace{0.2cm}
\includegraphics[width=7.5cm]{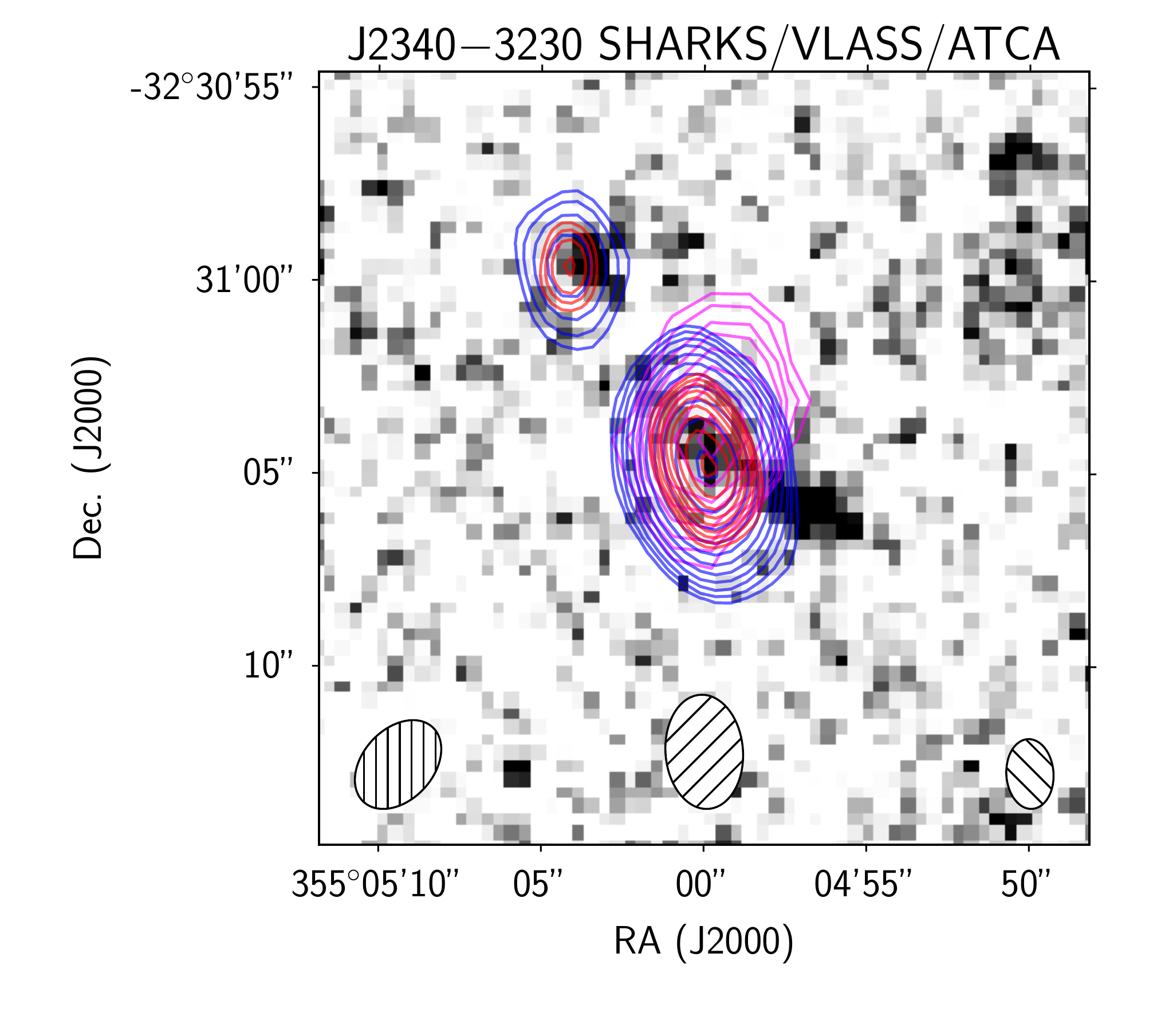}
\end{minipage}
\begin{minipage}{0.5\textwidth}
\includegraphics[width=8.5cm]{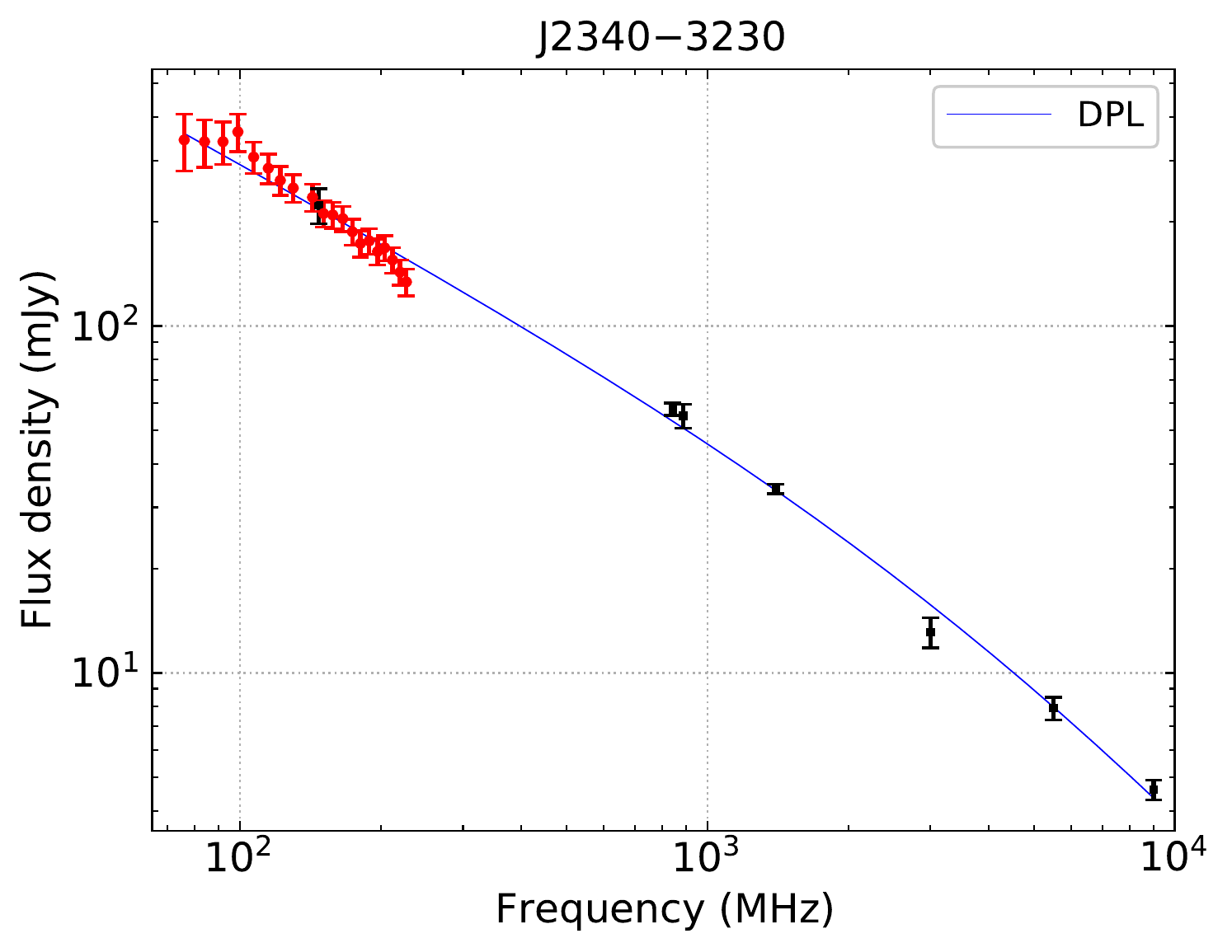}
\end{minipage}
\caption{{\em - continued.}}
\end{figure*}

\begin{table*}[!h]
\begin{minipage}{1.0\textwidth}
\centering
\caption{Summary of the $K_{\rm s}$-band images and lowest radio contour levels used in the overlay plots in Figure~\ref{fig:overlays_spectra}. The reported radio contour levels are usually $5\sigma$, but in a handful of cases they are either $3\sigma$ or $4\sigma$. We also report the $K_{\rm s}$-band host galaxy magnitudes (limits at the $5\sigma$ level); see Section~\ref{section: K-band} for further details.}
\small{
\begin{tabular}{ccrrrrr}
\hline\hline
\multicolumn{1}{c}{Source} & \multicolumn{1}{c}{$K_{\rm s}$-band} & \multicolumn{1}{c}{$K_{\rm s}$-band} & \multicolumn{1}{c}{FIRST} & \multicolumn{1}{c}{VLASS$^{\rm a}$} & \multicolumn{1}{c}{ATCA 5.5 GHz} & \multicolumn{1}{c}{ATCA 9 GHz} \\
 & \multicolumn{1}{c}{image} & magnitude &  \multicolumn{1}{c}{(mJy beam$^{-1}$)} & \multicolumn{1}{c}{(mJy beam$^{-1}$)} & \multicolumn{1}{c}{(mJy beam$^{-1}$)} & \multicolumn{1}{c}{(mJy beam$^{-1}$)} \\
\hline
J0002$-$3514 & VIKING & $>21.9$ & $\cdots$  &  0.70 &  0.19 &  0.14 \\ 
J0006$-$2946 & VIKING  & $>21.5$ & $\cdots$  &  0.74 &  0.19 &  0.16 \\
J0007$-$3040 & SHARKS & $>22.7$ & $\cdots$  &  0.71 &  0.19 &  0.13 \\
J0008$-$3007 & SHARKS & $>22.7$ & $\cdots$  &  0.74 &  0.20 &  0.16 \\
J0034$-$3112 & VIKING  & $>22.3$ & $\cdots$  &  0.74 &  0.19 &  0.17 \\
J0042$-$3515 & VIKING  & $>21.6$ & $\cdots$  &  0.70 &  0.20 &  0.20  \\
J0048$-$3540 & VIKING  & $>21.6$ & $\cdots$  &  0.74 &  0.16 &  0.16 \\
J0053$-$3256 & VIKING  & $>21.6$ & $\cdots$  &  0.66 &  0.19 &  0.17 \\
J0108$-$3501 & VIKING  & $>21.6$ & $\cdots$  &  0.82 &  0.20 &  0.15 \\
J0129$-$3109 & VIKING  & $>21.7$ & $\cdots$  &  0.79 &  0.20 &  0.15 \\
J0133$-$3056 & HAWK-I & $22.23 \pm 0.08$ & $\cdots$  &  0.78 & $\cdots$  & $\cdots$   \\
J0201$-$3441 & VIKING  & $>22.1$ & $\cdots$  &  0.75 &  0.18 &  0.17 \\
J0216$-$3301 & VIKING  & $>21.3$ & $\cdots$  &  0.74 &  0.19 &  0.14 \\
J0239$-$3043 & VIKING  & $>21.4$ & $\cdots$  &  0.72 &  0.18 &  0.14 \\
J0240$-$3206 & VIKING  & $>22.0$ & $\cdots$  &  0.61 &  0.18 &  0.15 \\
J0301$-$3132 & VIKING  & $>22.3$ & $\cdots$  &  0.76 &  0.27 &  0.20 \\
J0309$-$3526 & VIKING  & $>21.7$ & $\cdots$  &  0.37\rlap{$^{\rm b}$} &  0.10\rlap{$^{\rm b}$} &  0.078\rlap{$^{\rm b}$} \\
J0326$-$3013 & VIKING  & $>22.0$ & $\cdots$  &  0.61 &  0.20 &  0.14 \\
J0842$-$0157 & HAWK-I & $22.96 \pm 0.12$ &  0.70 &  0.76 & $\cdots$  & $\cdots$ \\
J0909$-$0154 & VIKING  & $>21.7$ & 0.79 &  0.72 & $\cdots$  & $\cdots$ \\
J1030$+$0135 & VIKING  & $>21.9$ &  0.75 &  0.72 & $\cdots$  & $\cdots$ \\ 	
J1032$+$0339 & VIKING  & $>21.5$ &  0.75 &  0.67 & $\cdots$  & $\cdots$ \\ 
J1033$+$0107 & VIKING  & $>21.8$ &  0.73 &  0.68 & $\cdots$  & $\cdots$ \\ 
J1037$-$0325 & VIKING  & $>21.6$ &  0.79 &  0.92 & $\cdots$  & $\cdots$ \\ 
J1040$+$0150 & VIKING  & $>21.9$ &  0.70 &  0.72 & $\cdots$  & $\cdots$ \\ 
J1052$-$0318 & VIKING  & $>21.6$ &  0.70 &  0.72 & $\cdots$  & $\cdots$ \\ 
J1112$+$0056 & VIKING  & $>21.7$ &  0.70 &  0.62 & $\cdots$  & $\cdots$ \\ 
J1125$-$0342 & VIKING  & $>21.4$ &  0.65 &  0.66 & $\cdots$  & $\cdots$ \\ 
J1127$-$0332 & VIKING  & $>21.7$ &  0.70 &  0.63 & $\cdots$  & $\cdots$ \\ 
J1136$-$0351 & VIKING  & $>21.8$ &  0.79 &  0.63 & $\cdots$  & $\cdots$ \\ 
J1141$-$0158 & VIKING  & $>21.9$ &  0.70 &  0.65 & $\cdots$  & $\cdots$ \\ 
J1211$-$0256 & VIKING  & $>21.6$ &  0.78 &  0.66 & $\cdots$  & $\cdots$ \\ 
J1246$-$0017 & VIKING  & $>21.8$ &  0.66 &  0.78 & $\cdots$  & $\cdots$ \\ 
J1317$+$0339 & VIKING  & $>21.3$ & 0.70 &  0.80 & $\cdots$  & $\cdots$ \\ 
J1329$+$0133 & VIKING  & $>21.5$ &  0.80 &  0.83 & $\cdots$  & $\cdots$ \\ 
J1335$+$0112 & VIKING  & $>21.2$ &  0.75 &  0.80 & $\cdots$  & $\cdots$  \\ 
J1337$+$0328 & VIKING  & $>21.5$ &  0.73 &  0.77 & $\cdots$  & $\cdots$ \\ 
J1340$+$0009 & VIKING  & $>21.4$ &  0.70 &  0.79 & $\cdots$  & $\cdots$ \\ 
J1347$+$0012 & VIKING  & $>21.7$ &  0.79 &  0.81 & $\cdots$  & $\cdots$ \\ 
J1349$+$0222 & VIKING  & $>21.8$ &  0.70 &  0.79 & $\cdots$  & $\cdots$ \\ 
J1351$-$0209 & VIKING  & $>21.2$ &  0.90 &  0.93 & $\cdots$  & $\cdots$ \\ 
J1402$+$0317 & VIKING  & $>21.5$ &  0.73 &  1.3 & $\cdots$  & $\cdots$ \\    
J1410$+$0259 & VIKING  & $>21.7$ &  0.68 &  1.3  &  $\cdots$ & $\cdots$ \\  
J1443$+$0229 & VIKING  & $>22.1$ &  0.67 &  0.80  & $\cdots$ & $\cdots$ \\  
J1521$-$0104 & VIKING  & $>22.2$ &  0.66 &  0.98  & $\cdots$ & $\cdots$ \\ 
J2219$-$3312 & VIKING  & $>22.3$ & $\cdots$  &  0.70 &  0.18 &  0.14 \\
J2311$-$3359 & VIKING  & $>21.9$ & $\cdots$ & 0.033\rlap{$^{\rm b}$} &  0.046\rlap{$^{\rm c}$} &  0.036\rlap{$^{\rm c}$} \\
J2314$-$3517 & VIKING  & $>21.6$ & $\cdots$  &  0.57 &  0.19 &  0.14 \\
J2326$-$3028 & VIKING  & $>22.2$  & $\cdots$  &  0.85 &  0.28 &  0.19 \\
J2330$-$3237 & VIKING  & $>21.6$ & $\cdots$  &  0.65 &  0.18 &  0.13 \\
J2340$-$3230 & SHARKS & $>22.7$ & $\cdots$  &  0.65 &  0.21 &  0.20 \\
\hline\hline
\multicolumn{7}{p{160mm}}{Notes. $^{\rm a}$Not corrected using the 1/0.87 scaling factor discussed in Section~\ref{section:overview available data}. $^{\rm b}$$3\sigma$ contour. $^{\rm c}$$4\sigma$ contour.}  \\
\end{tabular}
}
\label{table:overlay_contour_levels}
\end{minipage}
\end{table*}

\subsection{Radio spectral modelling}\label{section:SED modelling}

The radio data for each source in our sample spans a maximum frequency range of 74 MHz -- 9 GHz, with flux density measurements at up to 29 different frequencies. We were therefore able to explore the broadband radio spectral properties of our sample, which could then be compared with the GLEAM-only $\alpha$--$\beta$ fitting that was used as part of the sample selection (Section~\ref{section: alpha_beta selection}). For each source for which we present an overlay plot in Figure~\ref{fig:overlays_spectra}, we have also plotted the corresponding observed-frame broadband radio spectrum in the right-hand column of this figure. Overlaid on each spectrum is the preferred model: either a single or double power law. Radio spectra of particular interest are discussed in the notes on individual sources in Section~\ref{section:notes sources}, followed by further discussion in Sections~\ref{section:discussion spectra} and \ref{section:discussion spectra2}. 

For J0856$+$0223 and J0917$-$0012, radio spectra over a wider frequency range were presented in D20, \citet[][]{drouart21} and \citet[][]{seymour22}. We do not show spectra for these sources in Figure~\ref{fig:overlays_spectra}, but instead analyse these sources later in this paper in Section~\ref{section:discussion spectra2} and Figure~\ref{fig:all_sed}.  

We now describe the steps taken to construct the broadband radio spectra and carry out the spectral modelling.  

\subsubsection{Flux density scale corrections}\label{section: flux density scale corrections} 

Before we could model the broadband radio spectrum of each source, we first had to consider the fact that the radio data presented in Tables~\ref{table:fluxes} and \ref{table:extra_fluxes} were calibrated using a variety of flux density scales. Descriptions of these scales as well as relevant summaries and overviews can be found in the following references: \citet{wyllie69a,wyllie69b}, \citet{wills73}, \citet*{roger73}, \citet{baars77}, \citet{hunstead91}, \citet{reynolds94}, \citet{douglas96}, \citet{scaife12}, \citet{mauch13}, \citet{hurleywalker17}, \citet{hurleywalker17b} and \citet{perley17}. In particular, given that GLEAM is tied to the \citet[][]{baars77} flux density scale, we chose to rescale our other data sets, if required, to this scale. While this scale is known to become less accurate at low frequencies (e.g. discussion in \citealt[][]{rees90}, \citealt[][]{scaife12}, \citealt[][]{hurleywalker17} and \citealt[][]{perley17}), to first order this should not affect the spectral modelling presented below. 

As described above in Section~\ref{section:overview available data}, we used the rescaled TGSS flux densities from \citet[][]{hurleywalker17b}. Our ATCA data are also tied to the \citet[][]{baars77} scale, as are the RACS, FIRST and NVSS flux densities.\footnote{At 1.4 GHz, we used the NVSS flux densities only; these measurements are generally in excellent agreement with the FIRST measurements (see Section~\ref{section: flux comparison mid}).} It was not deemed necessary to rescale the SUMSS data given the agreement between SUMSS and RACS \citep[Section~\ref{section: flux comparison mid} and][]{hale21}; the consistency between the \citet[][]{baars77} and \citet[][]{perley17} scales at 3 GHz meant that a correction was not needed for VLASS either.    

The rescaled VLSSr, 325-MHz GMRT, TXS and MRC flux densities, as well as the multiplicative correction factors used, can be found in Table~\ref{table:extra_fluxes}. For VLSSr and TXS, these factors were directly available in the references for these surveys \citep[][]{douglas96,lane14}; we also increased the VLSSr calibration uncertainty by 5 per cent as recommended by \citet[][]{lane14} when rescaling the catalogued flux densities, which are tied to the \citet[][]{scaife12} flux density scale. For the GMRT data, the correction factor could be determined using the information in both \citet[][]{mauch13} and \citet[][]{perley17}, and for the MRC a correction factor is available in \citet[][]{baars77}. 

With all of the radio data on a consistent flux density scale, we could then fit each broadband spectrum. 

\subsubsection{Modelling the data}\label{section:modelling data} 

As in D20, we fitted a single, double and triple power law to each spectrum. We made use of the {\sc emcee} \citep[][]{foreman13a,foreman13b} and {\sc george} \citep[][]{ambikasaran15,foreman15} modules in {\sc python}, so as to carry out a Markov Chain Monte Carlo analysis of the parameter space. A single-power-law fit was defined as follows:
\begin{equation}\label{fit:spl}
S_{\nu} = N\left(\frac{\nu}{\nu_{0}}\right)^{\alpha},    
\end{equation}
where $N$ is a constant, $\alpha$ the spectral index across the observed-frequency range, and $\nu_{0}$ the reference frequency, which we chose to be 1000 MHz. Thus, $N$ is the fitted flux density at 1000 MHz for the single-power-law fit. For the smoothly varying double-power-law fit,    
\begin{equation}\label{fit:dpl}
S_{\nu} = N\left(\frac{\nu}{\nu_{0}}\right)^{\alpha_{\rm l}} \times \left[1 + \left(\frac{\nu}{\nu_{\rm b}}\right)^{\lvert\alpha_{\rm h} - \alpha_{\rm l}\rvert}\right]^{\sgn{(\alpha_{\rm h} - \alpha_{\rm l})}},
\end{equation}
where $\alpha_{\rm l}$ and $\alpha_{\rm h}$ are the spectral indices either side of the break frequency $\nu_{\rm b}$ and $\sgn{}$ the signum function. Lastly, the smoothly varying triple-power-law fit was of the form 
\begin{equation}\label{fit:tpl}
\begin{split}
S_{\nu} = N\left(\frac{\nu}{\nu_{0}}\right)^{\alpha_{\rm l}} \times \left[1 + \left(\frac{\nu}{\nu_{\rm bl}}\right)^{\lvert\alpha_{\rm m} - \alpha_{\rm l}\rvert}\right]^{\sgn{(\alpha_{\rm m} - \alpha_{\rm l})}} \\
\times \left[1 + \left(\frac{\nu}{\nu_{\rm bh}}\right)^{\lvert\alpha_{\rm h} - \alpha_{\rm m}\rvert}\right]^{\sgn{(\alpha_{\rm h} - \alpha_{\rm m})}},
\end{split}
\end{equation}
where $\alpha_{\rm l}$ and $\alpha_{\rm m}$ are the spectral indices either side of the lower break frequency $\nu_{\rm bl}$ and similarly for the spectral indices $\alpha_{\rm m}$ and $\alpha_{\rm h}$ as well as the higher break frequency $\nu_{\rm bh}$. 

The data were fitted with input units of MHz and mJy. We used non-informative priors for each of the parameters in Equations~\ref{fit:spl}--\ref{fit:tpl}; the priors are listed in Table~\ref{table:priors}. The priors were chosen such that we assumed a triple-power-law fit with a low-frequency turnover (that could result from synchrotron self-absorption and/or free--free absorption). For both the break frequency in the double-power-law fit and the higher break frequency in the triple-power-law fit, the prior was sufficiently general such that we could model spectral steepening at higher frequencies due to one or more energy loss mechanisms, or high-frequency spectral flattening due to a radio core component beginning to dominate over the lobe emission.

\begin{table}
 \centering
  \caption{Non-informative priors for each of the parameters in Equations~\ref{fit:spl}--\ref{fit:tpl}. Section~\ref{section:SED modelling} describes how we modelled the broadband radio spectra.}
  \begin{tabular}{lc}
  \hline\hline
Model & Range of priors\\
\hline
\multirow{2}{*}{Single power law} &  $0.01 < N < 5000$ \\
                                   & $-3 < \alpha < 3$ \\
\hline
\multirow{4}{*}{Double power law} & $0.01 < N < 50000$ \\
                                   & $-3 < \alpha_{\rm l} < 3$ \\
                                   & $0 < \nu_{\rm b} < 20000$ \\
                                   & $-3 < \alpha_{\rm h} < 3$ \\
\hline
\multirow{6}{*}{Triple power law} & $0.01 < N < 50000$ \\
                                   & $0 < \alpha_{\rm l} < 3$ \\
                                   & $0 < \nu_{\rm bl} < 500$ \\
                                   & $-3 < \alpha_{\rm m} < 3$ \\ 
                                   & $500 < \nu_{\rm bh} < 20000$ \\
                                   & $-3 < \alpha_{\rm h} < 3$ \\
\hline\hline
\end{tabular}
\label{table:priors}
\end{table}

When fitting the data, the uncertainty for each GLEAM flux density was calculated by combining the fitting and absolute calibration uncertainties, the latter being 8 per cent for all of the sources in our sample \citep[][]{hurleywalker17,franzen21}. Furthermore, it was necessary to take into account the correlations that exist between the GLEAM sub-bands, so as to avoid erroneous fits \citep[see discussion in][]{hurleywalker17}. To do so, we used a blocked Mat\'{e}rn covariance function for the GLEAM flux densities \citep[][]{rasmussen06}. 

The preferred model for each radio spectrum was determined using the sample-size-corrected Akaike Information Criterion \citep[AICc;][]{akaike74,burnham02}. In particular,
\begin{equation}\label{eqn:Akaike}
{\rm AICc} = \chi^2 + 2k + \frac{ 2k(k+1)}{n - k - 1},
\end{equation}
where $n$ and $k$ are the number of data points and free parameters, respectively, and $\chi^2$ is the standard goodness-of-fit statistic. For all possible realisations from the {\sc emcee} fitting, we determined the minimum value of AICc, $\min$(AICc), for each of the possible three models. We then selected the preferred model by using the standard convention of examining the difference
\begin{equation}\label{eqn:Akaike2}
\begin{split}
\Delta{\rm AICc} = \min({\rm AICc})_{i} - \\ \min\{\min(\rm{AICc})_{\rm SPL}, \min({\rm AICc})_{\rm DPL}, \min({\rm AICc})_{\rm TPL}\}, 
\end{split}
\end{equation}
where the subscripts SPL, DPL and TPL refer to the single-power-law, double-power-law, and triple-power-law fits, respectively, and the $ith$ model can be one of the three possibilities. Our preferred model first satisfied the condition $0 \leq \Delta{\rm AICc} < 4$ and secondly was the model with the fewest free parameters satisfying this condition. Note that this can mean, for example, that a double-power-law fit has the lowest AICc value, but the single power law was selected as the preferred fit because $\Delta{\rm AICc}$ is sufficiently small enough.        

The fitted model parameters are presented in Table~\ref{table:fitting_results}. For 34 sources, the preferred model is a double power law, with the remaining 17 sources described by a single power law. The triple power law is not the preferred model for any of the sources (but see \citealt[][]{drouart21} and \citealt[][]{seymour22}, where the radio emission from J0917$-$0012 was modelled with a triple power law). 

\begin{table*}
\setlength{\tabcolsep}{1.5pt}
\begin{minipage}{1.0\textwidth}
\centering
\caption{Fitted parameters for the broadband radio spectra. We give the preferred model type: SPL for a single power law and DPL for a double power law. The 16th, 50th and 84th percentiles are reported for the marginalised parameter distributions. We also list the parameter values corresponding to the best fit with the smallest value of AICc as well as the reduced chi-squared goodness-of-fit statistic for this model. Further information on the modelling can be found in Section~\ref{section:SED modelling}.}
\small{
{\setlength{\extrarowheight}{5pt}
\begin{tabular}{ccrrrrrrrrrrrr}
\hline\hline
& & \multicolumn{5}{c}{Marginalised distribution} & & \multicolumn{6}{c}{From model fit with min(AICc)}\\\cline{3-7}\cline{9-14}
\multicolumn{1}{c}{Source} & \multicolumn{1}{c}{Fit} & \multicolumn{1}{c}{$N$} & \multicolumn{1}{c}{$\alpha$} &  \multicolumn{1}{c}{$\alpha_{\rm l}$} &
\multicolumn{1}{c}{$\nu_{{\rm b}}$} & \multicolumn{1}{c}{$\alpha_{\rm h}$} & &
\multicolumn{1}{c}{$N$} & \multicolumn{1}{c}{$\alpha$} & \multicolumn{1}{c}{$\alpha_{\rm l}$} &
\multicolumn{1}{c}{$\nu_{{\rm b}}$} & \multicolumn{1}{c}{$\alpha_{\rm h}$} &
\multicolumn{1}{c}{$\chi^2_{\rm red}$} \\
& type & & & & \multicolumn{1}{c}{(GHz)} & & & & & & \multicolumn{1}{c}{(GHz)} & \\
\hline
J0002$-$3514	&	SPL	& $	18.31	_{-	0.52	}^{+	0.46	}$ & $	-1.064	_{-	0.019	}^{+	0.017	}$ & $\cdots$ & $\cdots$ & $\cdots$ & & $	18.35	$ & $	-1.066	$ & $\cdots$ & $\cdots$ & $\cdots$ & $	1.3	$ \\																
J0006$-$2946	&	DPL	& $	29.9	_{-	2.7	}^{+	11.3	}$ & $\cdots$ & $	-0.972	_{-	0.032	}^{+	0.048	}$ & $	14.9	_{-	6.5	}^{+	3.6	}$ & $	-1.94	_{-	0.78	}^{+	0.69	}$ & & $	26.9	$ & $\cdots$ & $	-0.982	$ & $	15.5	$ & $	-2.95	$ & $	1.0	$ \\
J0007$-$3040	&	DPL	& $	42.6	_{-	3.7	}^{+	5.8	}$ & $\cdots$ & $	-1.402	_{-	0.043	}^{+	0.056	}$ & $	2.14	_{-	0.52	}^{+	0.55	}$ & $	-2.80	_{-	0.14	}^{+	0.16	}$ & & $	44.2	$ & $\cdots$ & $	-1.385	$ & $	1.97	$ & $	-2.75	$ & $	0.7	$ \\
J0008$-$3007	&	DPL	& $	57	_{-	20	}^{+	42	}$ & $\cdots$ & $	-0.94	_{-	0.15	}^{+	0.19	}$ & $	0.75	_{-	0.43	}^{+	1.04	}$ & $	-1.95	_{-	0.23	}^{+	0.16	}$ & & $	82	$ & $\cdots$ & $	-0.81	$ & $	0.43	$ & $	-1.84	$ & $	0.4	$ \\
J0034$-$3112	&	SPL	& $	8.07	_{-	0.36	}^{+	0.36	}$ & $	-0.913	_{-	0.026	}^{+	0.027	}$ & $\cdots$ & $\cdots$ & $\cdots$ & & $	8.15	$ & $	-0.910	$ & $\cdots$ & $\cdots$ & $\cdots$ & $	0.5	$ \\																
J0042$-$3515	&	SPL	& $	10.34	_{-	0.41	}^{+	0.38	}$ & $	-1.006	_{-	0.025	}^{+	0.023	}$ & $\cdots$ & $\cdots$ & $\cdots$ & & $	10.35	$ & $	-1.010	$ & $\cdots$ & $\cdots$ & $\cdots$ & $	0.8	$ \\																
J0048$-$3540	&	SPL	& $	18.10	_{-	0.57	}^{+	0.61	}$ & $	-1.301	_{-	0.020	}^{+	0.017	}$ & $\cdots$ & $\cdots$ & $\cdots$ & & $	18.11	$ & $	-1.303	$ & $\cdots$ & $\cdots$ & $\cdots$ & $	0.3	$ \\																
J0053$-$3256	&	DPL	& $	27.0	_{-	2.5	}^{+	5.7	}$ & $\cdots$ & $	-1.203	_{-	0.045	}^{+	0.070	}$ & $	5.7	_{-	2.7	}^{+	3.0	}$ & $	-2.32	_{-	0.40	}^{+	0.34	}$ & & $	28.8	$ & $\cdots$ & $	-1.172	$ & $	4.5	$ & $	-2.25	$ & $	0.2	$ \\
J0108$-$3501	&	DPL	& $	84	_{-	11	}^{+	22	}$ & $\cdots$ & $	-0.729	_{-	0.061	}^{+	0.095	}$ & $	3.7	_{-	1.8	}^{+	2.0	}$ & $	-1.85	_{-	0.37	}^{+	0.24	}$ & & $	114	$ & $\cdots$ & $	-0.604	$ & $	1.5	$ & $	-1.58	$ & $	0.3	$ \\
J0129$-$3109	&	DPL	& $	35.3	_{-	2.1	}^{+	5.6	}$ & $\cdots$ & $	-0.608	_{-	0.033	}^{+	0.048	}$ & $	12.9	_{-	4.4	}^{+	4.4	}$ & $	-1.82	_{-	0.72	}^{+	0.59	}$ & & $	34.2	$ & $\cdots$ & $	-0.620	$ & $	13.4	$ & $	-2.16	$ & $	0.4	$ \\
J0133$-$3056	&	DPL	& $	170	_{-	75	}^{+	299	}$ & $\cdots$ & $	-0.66	_{-	0.18	}^{+	0.29	}$ & $	0.53	_{-	0.42	}^{+	1.99	}$ & $	-1.37	_{-	0.45	}^{+	0.19	}$ & & $	543	$ & $\cdots$ & $	-0.32	$ & $	0.11	$ & $	-1.21	$ & $	0.3	$ \\
J0201$-$3441	&	SPL	& $	14.39	_{-	0.44	}^{+	0.47	}$ & $	-0.982	_{-	0.020	}^{+	0.018	}$ & $\cdots$ & $\cdots$ & $\cdots$ & & $	14.46	$ & $	-0.984	$ & $\cdots$ & $\cdots$ & $\cdots$ & $	1.4	$ \\																
J0216$-$3301	&	DPL	& $	19.0	_{-	1.7	}^{+	3.8	}$ & $\cdots$ & $	-0.744	_{-	0.041	}^{+	0.050	}$ & $	12.9	_{-	5.9	}^{+	4.6	}$ & $	-1.73	_{-	0.73	}^{+	0.48	}$ & & $	19.4	$ & $\cdots$ & $	-0.728	$ & $	12.3	$ & $	-1.61	$ & $	0.7	$ \\
J0239$-$3043	&	DPL	& $	38.2	_{-	6.3	}^{+	20.6	}$ & $\cdots$ & $	-0.942	_{-	0.074	}^{+	0.160	}$ & $	4.0	_{-	2.9	}^{+	3.5	}$ & $	-1.90	_{-	0.41	}^{+	0.29	}$ & & $	57.2	$ & $\cdots$ & $	-0.795	$ & $	1.2	$ & $	-1.65	$ & $	0.5	$ \\
J0240$-$3206	&	SPL	& $	9.66	_{-	0.37	}^{+	0.35	}$ & $	-0.920	_{-	0.022	}^{+	0.023	}$ & $\cdots$ & $\cdots$ & $\cdots$ & & $	9.64	$ & $	-0.914	$ & $\cdots$ & $\cdots$ & $\cdots$ & $	0.9	$ \\																
J0301$-$3132	&	DPL	& $	64.0	_{-	4.3	}^{+	13.4	}$ & $\cdots$ & $	-0.693	_{-	0.037	}^{+	0.077	}$ & $	7.1	_{-	3.3	}^{+	2.0	}$ & $	-1.99	_{-	0.50	}^{+	0.43	}$ & & $	67.5	$ & $\cdots$ & $	-0.667	$ & $	6.1	$ & $	-1.84	$ & $	0.4	$ \\
J0309$-$3526	&	DPL	& $	15.4	_{-	3.8	}^{+	11.7	}$ & $\cdots$ & $	-0.94	_{-	0.14	}^{+	0.22	}$ & $	1.9	_{-	1.3	}^{+	2.6	}$ & $	-2.17	_{-	0.44	}^{+	0.38	}$ & & $	27.3	$ & $\cdots$ & $	-0.70	$ & $	0.7	$ & $	-1.87	$ & $	0.8	$ \\
J0326$-$3013	&	DPL	& $	217	_{-	37	}^{+	61	}$ & $\cdots$ & $	-0.684	_{-	0.080	}^{+	0.095	}$ & $	2.05	_{-	0.93	}^{+	1.41	}$ & $	-1.74	_{-	0.24	}^{+	0.16	}$ & & $	272	$ & $\cdots$ & $	-0.594	$ & $	1.20	$ & $	-1.61	$ & $	0.4	$ \\
J0842$-$0157	&	SPL	& $	70.3	_{-	2.2	}^{+	2.2	}$ & $	-0.948	_{-	0.019	}^{+	0.019	}$ & $\cdots$ & $\cdots$ & $\cdots$ & & $	70.4	$ & $	-0.948	$ & $\cdots$ & $\cdots$ & $\cdots$ & $	0.7	$ \\																
J0909$-$0154	&	DPL	& $	18\,000	_{-	11\,000	}^{+	18\,000	}$ & $\cdots$ & $	0.91	_{-	0.34	}^{+	0.27	}$ & $	0.0979	_{-	0.0064	}^{+	0.0090	}$ & $	-1.602	_{-	0.030	}^{+	0.025	}$ & & $	10\,000	$ & $\cdots$ & $	0.67	$ & $	0.1015	$ & $	-1.620	$ & $	0.3	$ \\
J1030$+$0135	&	DPL	& $	68.7	_{-	4.5	}^{+	10.9	}$ & $\cdots$ & $	-0.553	_{-	0.032	}^{+	0.049	}$ & $	13.8	_{-	5.0	}^{+	4.1	}$ & $	-1.76	_{-	0.74	}^{+	0.59	}$ & & $	67.4	$ & $\cdots$ & $	-0.567	$ & $	15.0	$ & $	-1.89	$ & $	1.0	$ \\
J1032$+$0339	&	DPL	& $	119.5	_{-	7.6	}^{+	18.8	}$ & $\cdots$ & $	-1.041	_{-	0.032	}^{+	0.050	}$ & $	11.5	_{-	4.8	}^{+	4.1	}$ & $	-2.09	_{-	0.50	}^{+	0.39	}$ & & $	124.9	$ & $\cdots$ & $	-1.021	$ & $	9.8	$ & $	-1.97	$ & $	0.3	$ \\
J1033$+$0107	&	DPL	& $	72.5	_{-	6.3	}^{+	19.9	}$ & $\cdots$ & $	-0.869	_{-	0.040	}^{+	0.078	}$ & $	12.2	_{-	7.8	}^{+	5.0	}$ & $	-1.79	_{-	0.60	}^{+	0.39	}$ & & $	86.1	$ & $\cdots$ & $	-0.802	$ & $	5.4	$ & $	-1.51	$ & $	0.8	$ \\
J1037$-$0325	&	DPL	& $	390	_{-	310	}^{+	1080	}$ & $\cdots$ & $	0.28	_{-	0.60	}^{+	0.60	}$ & $	0.132	_{-	0.041	}^{+	0.181	}$ & $	-1.025	_{-	0.102	}^{+	0.067	}$ & & $	150	$ & $\cdots$ & $	-0.05	$ & $	0.188	$ & $	-1.080	$ & $	1.3	$ \\
J1040$+$0150	&	DPL	& $	1210	_{-	860	}^{+	3270	}$ & $\cdots$ & $	-0.14	_{-	0.45	}^{+	0.45	}$ & $	0.073	_{-	0.017	}^{+	0.030	}$ & $	-1.525	_{-	0.058	}^{+	0.047	}$ & & $	750	$ & $\cdots$ & $	-0.33	$ & $	0.076	$ & $	-1.540	$ & $	1.1	$ \\
J1052$-$0318	&	SPL	& $	15.72	_{-	0.57	}^{+	0.59	}$ & $	-1.179	_{-	0.022	}^{+	0.024	}$ & $\cdots$ & $\cdots$ & $\cdots$ & & $	15.76	$ & $	-1.177	$ & $\cdots$ & $\cdots$ & $\cdots$ & $	1.4	$ \\																
J1112$+$0056	&	SPL	& $	24.86	_{-	0.74	}^{+	0.85	}$ & $	-0.982	_{-	0.018	}^{+	0.022	}$ & $\cdots$ & $\cdots$ & $\cdots$ & & $	24.77	$ & $	-0.976	$ & $\cdots$ & $\cdots$ & $\cdots$ & $	1.2	$ \\																
J1125$-$0342	&	SPL	& $	78.9	_{-	1.8	}^{+	2.0	}$ & $	-0.977	_{-	0.013	}^{+	0.015	}$ & $\cdots$ & $\cdots$ & $\cdots$ & & $	79.0	$ & $	-0.979	$ & $\cdots$ & $\cdots$ & $\cdots$ & $	0.8	$ \\																
J1127$-$0332	&	SPL	& $	37.14	_{-	0.92	}^{+	1.03	}$ & $	-1.111	_{-	0.016	}^{+	0.015	}$ & $\cdots$ & $\cdots$ & $\cdots$ & & $	37.21	$ & $	-1.111	$ & $\cdots$ & $\cdots$ & $\cdots$ & $	0.6	$ \\																
J1136$-$0351	&	SPL	& $	22.68	_{-	0.72	}^{+	0.69	}$ & $	-1.089	_{-	0.020	}^{+	0.019	}$ & $\cdots$ & $\cdots$ & $\cdots$ & & $	22.70	$ & $	-1.095	$ & $\cdots$ & $\cdots$ & $\cdots$ & $	1.1	$ \\																
J1141$-$0158	&	DPL	& $	16\,000	_{-	10\,000	}^{+	15\,000	}$ & $\cdots$ & $	2.61	_{-	0.56	}^{+	0.28	}$ & $	0.175	_{-	0.018	}^{+	0.016	}$ & $	-1.130	_{-	0.045	}^{+	0.045	}$ & & $	1500	$ & $\cdots$ & $	1.36	$ & $	0.191	$ & $	-1.147	$ & $	2.2	$ \\
J1211$-$0256	&	DPL	& $	60	_{-	24	}^{+	138	}$ & $\cdots$ & $	-0.86	_{-	0.19	}^{+	0.43	}$ & $	1.7	_{-	1.4	}^{+	6.4	}$ & $	-1.70	_{-	0.69	}^{+	0.24	}$ & & $	138	$ & $\cdots$ & $	-0.56	$ & $	0.3	$ & $	-1.50	$ & $	1.3	$ \\
J1246$-$0017	&	DPL	& $	74.6	_{-	8.3	}^{+	60.8	}$ & $\cdots$ & $	-0.678	_{-	0.053	}^{+	0.189	}$ & $	8.4	_{-	7.5	}^{+	4.6	}$ & $	-1.60	_{-	0.69	}^{+	0.42	}$ & & $	87.4	$ & $\cdots$ & $	-0.603	$ & $	4.0	$ & $	-1.44	$ & $	0.8	$ \\
\hline\hline
\end{tabular}
}}
\label{table:fitting_results}
\end{minipage}
\end{table*}

\setcounter{table}{6} 
\begin{table*}[t]
\setlength{\tabcolsep}{1.5pt}
\begin{minipage}{1.0\textwidth}
\centering
\caption{{\em - continued.}}
\small{
{\setlength{\extrarowheight}{5pt}
\begin{tabular}{ccrrrrrrrrrrrr}
\hline\hline
& & \multicolumn{5}{c}{Marginalised distribution} & & \multicolumn{6}{c}{From model fit with min(AICc)} \\\cline{3-7}\cline{9-14}
\multicolumn{1}{c}{Source} & \multicolumn{1}{c}{Fit} & \multicolumn{1}{c}{$N$} & \multicolumn{1}{c}{$\alpha$} &  \multicolumn{1}{c}{$\alpha_{\rm l}$} &
\multicolumn{1}{c}{$\nu_{{\rm b}}$} & \multicolumn{1}{c}{$\alpha_{\rm h}$} & &
\multicolumn{1}{c}{$N$} & \multicolumn{1}{c}{$\alpha$} & \multicolumn{1}{c}{$\alpha_{\rm l}$} &
\multicolumn{1}{c}{$\nu_{{\rm b}}$} & \multicolumn{1}{c}{$\alpha_{\rm h}$} &
\multicolumn{1}{c}{$\chi^2_{\rm red}$} \\
& type & & & & \multicolumn{1}{c}{(GHz)} & & & & & & \multicolumn{1}{c}{(GHz)} & \\
\hline
J1317$+$0339	&	DPL	& $	56.9	_{-	3.6	}^{+	11.3	}$ & $\cdots$ & $	-0.966	_{-	0.032	}^{+	0.039	}$ & $	14.3	_{-	4.9	}^{+	3.8	}$ & $	-2.12	_{-	0.55	}^{+	0.66	}$ & & $	54.1	$ & $\cdots$ & $	-0.994	$ & $	14.3	$ & $	-2.76	$ & $	1.6	$ \\
J1329$+$0133	&	DPL	& $	147	_{-	11	}^{+	22	}$ & $\cdots$ & $	-0.793	_{-	0.034	}^{+	0.046	}$ & $	12.1	_{-	4.9	}^{+	4.7	}$ & $	-1.83	_{-	0.68	}^{+	0.40	}$ & & $	150	$ & $\cdots$ & $	-0.781	$ & $	10.9	$ & $	-1.76	$ & $	0.7	$ \\
J1335$+$0112	&	DPL	& $	92.2	_{-	5.9	}^{+	14.5	}$ & $\cdots$ & $	-0.823	_{-	0.034	}^{+	0.061	}$ & $	10.0	_{-	4.0	}^{+	3.9	}$ & $	-2.03	_{-	0.58	}^{+	0.41	}$ & & $	92.2	$ & $\cdots$ & $	-0.823	$ & $	9.6	$ & $	-2.18	$ & $	0.8	$ \\
J1337$+$0328	&	DPL	& $	67.2	_{-	3.9	}^{+	9.8	}$ & $\cdots$ & $	-0.713	_{-	0.038	}^{+	0.055	}$ & $	8.8	_{-	3.0	}^{+	2.6	}$ & $	-2.05	_{-	0.58	}^{+	0.45	}$ & & $	68.0	$ & $\cdots$ & $	-0.707	$ & $	8.4	$ & $	-2.08	$ & $	0.5	$ \\
J1340$+$0009	&	SPL	& $	61.3	_{-	2.0	}^{+	3.0	}$ & $	-0.858	_{-	0.034	}^{+	0.021	}$ & $\cdots$ & $\cdots$ & $\cdots$ & & $	61.0	$ & $	-0.854	$ & $\cdots$ & $\cdots$ & $\cdots$ & $	1.5	$ \\																
J1347$+$0012	&	DPL	& $	92.8	_{-	7.8	}^{+	16.9	}$ & $\cdots$ & $	-0.878	_{-	0.044	}^{+	0.061	}$ & $	6.2	_{-	2.6	}^{+	2.1	}$ & $	-2.09	_{-	0.46	}^{+	0.34	}$ & & $	103.3	$ & $\cdots$ & $	-0.833	$ & $	4.3	$ & $	-1.89	$ & $	0.4	$ \\
J1349$+$0222	&	DPL	& $	48.6	_{-	4.1	}^{+	8.5	}$ & $\cdots$ & $	-0.803	_{-	0.049	}^{+	0.080	}$ & $	5.9	_{-	2.4	}^{+	1.7	}$ & $	-2.22	_{-	0.46	}^{+	0.40	}$ & & $	49.2	$ & $\cdots$ & $	-0.797	$ & $	5.6	$ & $	-2.23	$ & $	0.7	$ \\
J1351$-$0209	&	DPL	& $	449	_{-	47	}^{+	101	}$ & $\cdots$ & $	-0.902	_{-	0.048	}^{+	0.074	}$ & $	6.1	_{-	3.2	}^{+	3.1	}$ & $	-1.93	_{-	0.41	}^{+	0.30	}$ & & $	530	$ & $\cdots$ & $	-0.837	$ & $	3.3	$ & $	-1.72	$ & $	0.2	$ \\
J1402$+$0317	&	DPL	& $	142.3	_{-	7.8	}^{+	21.1	}$ & $\cdots$ & $	-0.832	_{-	0.026	}^{+	0.041	}$ & $	13.0	_{-	4.2	}^{+	3.3	}$ & $	-1.99	_{-	0.61	}^{+	0.52	}$ & & $	137.3	$ & $\cdots$ & $	-0.846	$ & $	12.7	$ & $	-2.39	$ & $	0.6	$ \\
J1410$+$0259	&	DPL	& $	40.6	_{-	2.8	}^{+	8.1	}$ & $\cdots$ & $	-0.773	_{-	0.040	}^{+	0.066	}$ & $	11.9	_{-	5.2	}^{+	4.8	}$ & $	-1.88	_{-	0.69	}^{+	0.49	}$ & & $	38.9	$ & $\cdots$ & $	-0.788	$ & $	11.9	$ & $	-2.35	$ & $	1.1	$ \\
J1443$+$0229	&	DPL	& $	14\,000	_{-	11\,000	}^{+	24\,000	}$ & $\cdots$ & $	1.44	_{-	0.61	}^{+	0.46	}$ & $	0.112	_{-	0.013	}^{+	0.015	}$ & $	-1.355	_{-	0.044	}^{+	0.035	}$ & & $	2800	$ & $\cdots$ & $	0.75	$ & $	0.121	$ & $	-1.392	$ & $	0.6	$ \\
J1521$-$0104	&	DPL	& $	14.2	_{-	4.8	}^{+	33.2	}$ & $\cdots$ & $	-0.91	_{-	0.21	}^{+	0.52	}$ & $	1.4	_{-	1.0	}^{+	2.1	}$ & $	-2.27	_{-	0.45	}^{+	0.35	}$ & & $	22.4	$ & $\cdots$ & $	-0.66	$ & $	0.8	$ & $	-2.11	$ & $	1.1	$ \\
J2219$-$3312	&	SPL	& $	30.98	_{-	0.71	}^{+	0.69	}$ & $	-0.919	_{-	0.015	}^{+	0.014	}$ & $\cdots$ & $\cdots$ & $\cdots$ & & $	30.88	$ & $	-0.922	$ & $\cdots$ & $\cdots$ & $\cdots$ & $	0.7	$ \\																
J2311$-$3359	&	SPL	& $	1.53	_{-	0.17	}^{+	0.18	}$ & $	-1.837	_{-	0.062	}^{+	0.055	}$ & $\cdots$ & $\cdots$ & $\cdots$ & & $	1.51	$ & $	-1.842	$ & $\cdots$ & $\cdots$ & $\cdots$ & $	1.0	$ \\																
J2314$-$3517	&	SPL	& $	16.23	_{-	0.70	}^{+	0.75	}$ & $	-1.030	_{-	0.038	}^{+	0.044	}$ & $\cdots$ & $\cdots$ & $\cdots$ & & $	16.24	$ & $	-1.028	$ & $\cdots$ & $\cdots$ & $\cdots$ & $	1.9	$ \\																
J2326$-$3028	&	DPL	& $	138	_{-	11	}^{+	22	}$ & $\cdots$ & $	-0.682	_{-	0.041	}^{+	0.061	}$ & $	5.1	_{-	1.9	}^{+	1.5	}$ & $	-2.01	_{-	0.42	}^{+	0.31	}$ & & $	158	$ & $\cdots$ & $	-0.622	$ & $	3.4	$ & $	-1.75	$ & $	0.6	$ \\
J2330$-$3237	&	SPL	& $	17.93	_{-	0.53	}^{+	0.52	}$ & $	-0.913	_{-	0.017	}^{+	0.017	}$ & $\cdots$ & $\cdots$ & $\cdots$ & & $	18.02	$ & $	-0.914	$ & $\cdots$ & $\cdots$ & $\cdots$ & $	1.2	$ \\																
J2340$-$3230	&	DPL	& $	49.5	_{-	4.2	}^{+	10.5	}$ & $\cdots$ & $	-0.799	_{-	0.042	}^{+	0.062	}$ & $	9.8	_{-	4.7	}^{+	5.3	}$ & $	-1.81	_{-	0.54	}^{+	0.36	}$ & & $	56.2	$ & $\cdots$ & $	-0.728	$ & $	5.3	$ & $	-1.61	$ & $	0.9	$ \\
\hline\hline
\end{tabular}
}}
\end{minipage}
\end{table*}

\subsubsection{Possible effects of source blending}\label{section:blending} 

We also investigated whether the GLEAM $\alpha$--$\beta$ selection and broadband spectral fitting could be affected by source blending in GLEAM. While we used a selection criterion of a single NVSS match within 50\arcsec\:of the GLEAM position (Table~\ref{table:sample selection}) so as to preferentially select isolated, compact sources, the GLEAM synthesised beam half width at half maximum (HWHM) generally extends beyond 50\arcsec, especially so at the lower end of the GLEAM band. There are also some cases where relatively faint, unassociated sources are visible within 50\arcsec\:in the RACS, VLASS and/or ATCA radio maps at the various frequencies (Figure~\ref{fig:overlays_spectra}; also see Section~\ref{section:notes sources}). These sources are not visible in NVSS due to either source blending in this survey, or they are too faint to have been detected.    

Having inspected the RACS, VLASS and ATCA radio maps with the GLEAM synthesised beam FWHMs across the full frequency range overlaid, in general we are confident that, in the vast majority of cases, any source blending in GLEAM has not affected the accuracy of both the GLEAM $\alpha$--$\beta$ selection and broadband spectral fitting. Contributions from unassociated sources should be contained within the GLEAM flux density uncertainties. Similarly, we also considered whether blending is affecting the reliability of our flux density measurements at other frequencies. However, given the discussion and analysis in Section~\ref{section:notes sources} of relevant cases of interest, this is unlikely to be a significant effect. 

\subsection{Notes on individual sources}\label{section:notes sources}  

In this section, we discuss the overlay plots and/or radio spectra of a number of sources in the sample. 

{\bf J0007$-$3040:} This source is of particular interest, with $K_{\rm s} > 22.7$ from SHARKS, a relatively compact radio morphology with LAS $=3\farcs0$ in both the ATCA and VLASS images, and a double-power-law spectrum with best-fitting spectral indices of $\alpha_{\rm l} = -1.385$ and $\alpha_{\rm h} = -2.75$. While $\alpha_{\rm l}$ indicates a USS spectrum at frequencies below $\nu_{\rm b} = 1.97$ GHz, $\alpha_{\rm h}$ is exceptionally steep and relatively well constrained. This source therefore appears to be a promising HzRG candidate; possible scenarios for explaining the properties of the radio spectrum are discussed in Section~\ref{section:discussion spectra}. Another possibility is that this source may be an as of yet undetected pulsar, but this seems less likely given the LAS and the fact that there are hints of an incipient double radio morphology for this source. 

We also note that we checked whether emission potentially being resolved out in the ATCA maps could explain at least some of the observed spectral curvature. Adjusting the robust weighting parameter at 5.5 GHz to $-$2 (i.e. close to uniform weighting) gave an image with a very similar resolution to the 9-GHz map generated with a robust weighting parameter of 0.5. The measured 5.5-GHz flux density was not significantly different to the value reported in Table~\ref{table:fluxes}. Additionally, a 5.5-GHz map generated with a robust weighting parameter of 2 (i.e. close to natural weighting) also gave a very similar flux density to the value in Table~\ref{table:fluxes}.

{\bf J0133$-$3056:} This source is an incipient double in VLASS with LAS $= 4\farcs3$. The likely host galaxy is seen in the HAWK-I $K_{\rm s}$-band image close to the peak of the radio emission, with magnitude $K_{\rm s} = 22.23 \pm 0.08$.

{\bf J0309$-$3526:} The ATCA 5.5-GHz and VLASS data both suggest a multi-component morphology, perhaps a triple source, although as can be seen in the corresponding panel in Figure~\ref{fig:overlays_spectra}, the morphologies are not fully consistent between the two frequencies. There is clear extension at 5.5 GHz in the direction of the synthesised beam, almost orthogonal to what appears to be the main axis of the radio emission. Given the VLASS morphology and the fact that we could not phase self-calibrate the 5.5-GHz data due to low S/N, it seems most plausible that the extension at 5.5 GHz in the direction of the synthesised beam is spurious. The spectral index between 5.5 and 9 GHz is extremely steep: $\alpha^{9000}_{5500} = -2.8 \pm 1.1$. There is evidence that the accuracy of $\alpha^{9000}_{5500}$ is affected by diffuse emission being resolved out; this can be seen, for example, by decreasing the robust weighting parameter from 0.5 to $-$2 (i.e. uniform weighting in the latter case) and reimaging the 5.5-GHz data. Higher S/N data at mid/high frequencies in array configurations with sufficient low-surface-brightness sensitivity are needed for this source. 

{\bf J0842$-$0157:} The FIRST and VLASS morphologies are very compact: LAS $=1\farcs7$ and $1\farcs1$, respectively. We take the HAWK-I $K_{\rm s}$-band source closest to the radio centroid as the host galaxy identification; the magnitude of the host galaxy is $K_{\rm s} = 22.96 \pm 0.12$.

{\bf J0909$-$0154, J1141$-$0158 and J1443$+$0229:} The best-fitting model for J0909$-$0154 has significantly flattened at the bottom end of the GLEAM band and would be expected to begin to turn over at frequencies $\lesssim 70$ MHz. J1141$-$0158 and J1443$+$0229 are the only sources in the sample where the best fit peaks and turns over (within the GLEAM band at 205 and 91 MHz, respectively); higher S/N data at the lower GLEAM frequencies are needed to confirm the turnover in each case, however.  

{\bf J1112$+$0056:} There is a second NVSS source 51\farcs3 from the GLEAM position (beyond the region shown in the panel for this source in Figure~\ref{fig:overlays_spectra}) that has a 1.4-GHz flux density that is about 40 per cent of the NVSS flux density of the HzRG candidate (NVSS J111212$+$005519 with $S_{1400} = 6.7 \pm 0.5$ mJy; cf. $S_{1400} = 17.1 \pm 0.7$ mJy for the HzRG candidate). Both of these sources have very similar two-point spectral indices $\alpha^{1400}_{887.5}$; if this spectral similarity is also the case at low frequencies, the accuracy of the GLEAM $\alpha$--$\beta$ fitting should not be affected significantly. The GLEAM/TGSS flux density ratio in Table~\ref{table:fluxes} is $1.27 \pm 0.15$; this tentatively suggests that there could be an excess in GLEAM due to source blending, albeit not statistically significant at e.g. the $3\sigma$ level.   

{\bf J1125$-$0342 and J1317$+$0339:} These are the USS-selected sources TN~J1125$-$0342 and TN~J1317$+$0339, respectively, from \citet[][]{debreuck00}. Their redshifts remain unknown and the VIKING non-detections are the deepest constraints on their $K_{\rm s}$-band magnitudes: $>21.4$ (J1125$-$0342) and $> 21.3$ (J1317$+$0339).    

For J1125$-$0342, two sources separated by 8\farcs3 are visible in the VLASS Epoch 1 image. The 3-GHz flux density of the southern source is $\approx 1.4$ mJy, a factor of $\sim$ 20 fainter than the much brighter source to the north. J1125$-$0342 could therefore be a very asymmetric double, or the fainter source to the south is unrelated. In the case of the former scenario, the angular size would then suggest that the source is too extended to be at a very high redshift. For the purpose of analysis in this paper, we have assumed the latter scenario. Deeper $K_{\rm s}$-band imaging and higher-resolution radio imaging at e.g. 5.5 and 9 GHz are needed. 

For J1317$+$0339, it can be seen in Figure~\ref{fig:overlays_spectra} that the 365-MHz flux density point is a clear outlier; we did not include this data point when fitting the radio spectrum. The $\alpha_{\rm l}$ value in Table~\ref{table:fitting_results} implies that the spectrum is not as steep as suggested by the two-point spectral index between the TXS and NVSS surveys. This is the case for J1125$-$0342 as well, although not to the same extent as for J1317$+$0339. We included the 365-MHz data point when fitting the radio spectrum of J1125$-$0342.    

{\bf J1329$+$0133:} There is a hint of a host galaxy identification in the VIKING $K_{\rm s}$-band image, but the brightest pixel value is only at the $3.5\sigma$ level (which we do not consider to be a secure detection). 

{\bf J1335$+$0112:} As previously discussed in Section~\ref{section: HzRG sample}, this source has a detection in AllWISE: the 3.37-\textmu m W1-band magnitude is $20.033 \pm 0.130$ \citep[][]{cutri14}. The source is not detected in any of the other three longer-wavelength {\it WISE} bands. The $K_{\rm s}$-band magnitude is $> 21.2$ (Table~\ref{table:overlay_contour_levels}); the $\gtrsim 1.2$ mag break between 2.15 and 3.37 \textmu m might potentially indicate a redshifted 4000~\AA\:break (i.e. $z \gtrsim 4.4$). The WISE W2-band (4.62-\textmu m) magnitude is $> 19.064$ ($5\sigma$). Additional follow-up and analysis is needed.    

{\bf J1340$+$0009:} There is a hint of a host galaxy identification in the VIKING $H$-band image, but the brightest pixel value is only at the $4.2\sigma$ level (which we do not consider to be a secure detection).

{\bf J1351$-$0209:} This is the brightest radio source in our sample (by a factor of about three at 151 MHz) with $S_{151} = 2.44$ Jy. It is also catalogued as the Parkes source PKS B1349$-$019 \citep[][]{wright90}. The PKS 2.7- and 5-GHz flux densities from \citet[][]{wright90} are 130 and 50 mJy, respectively, consistent with the VLASS and ATCA 5.5-GHz data. Given this consistency at very similar frequencies as well as the fact that uncertainties are not reported for these Parkes measurements, we chose not to use the Parkes flux densities in the radio spectrum modelling. 

Additionally, \citet[][]{downes86} reported 1.5- and 4.9-GHz flux densities from their high-resolution VLA data (250 and 60 mJy, respectively); the former is consistent with the NVSS flux density and the latter with both the Parkes 5-GHz and ATCA 5.5-GHz measurements. Again, there is no significant advantage in including these data in our radio spectrum modelling, particularly as it is unclear what the flux density uncertainties are in these cases as well. The radio morphology in the 4.9-GHz VLA map shows two components; the angular extent ($\sim 2\farcs2$; position angle $\sim -45\degr$) is reasonably consistent with the FIRST and VLASS LAS measurement (2\farcs8).  

\citet[][]{dunlop89} did not detect the host galaxy in optical imaging; the reported $B$- and $R$-band magnitude limits are $\gtrsim 24$. The source is listed in \citet[][]{dunlop90} as a candidate high-redshift object, but with $K \approx 19.75$; this is most likely a misidentification or erroneous catalogue entry given that our overlay plot shows no evidence of a $K_{\rm s}$-band identification to a much greater depth ($ K_{\rm s} > 21.2$).   

{\bf J1521$-$0104:} There is a hint of extension in RACS about 20\arcsec to the north-west (beyond the region shown in the panel for this source in Figure~\ref{fig:overlays_spectra}), albeit at a marginal level ($\sim 3.9\sigma$). The extension is coincident with the near-infrared source 2MASS J15215337$-$0104034, where $K_{\rm s} = 14.449 \pm 0.033$ \citep[][]{cutri03}. This 2MASS source is 22\farcs2 from our HzRG candidate. The possible radio/near-infrared association might then suggest that the HzRG candidate shown in Figure~\ref{fig:overlays_spectra} is part of a larger, head--tail source. On the other hand, there is no evidence of similar extension in our other radio data sets. Given that the extended 887.5-MHz radio emission and in turn the radio/near-infrared association is tentative, we regard the source shown in Figure~\ref{fig:overlays_spectra} as an HzRG candidate with the requisite compact radio morphology. A deeper $K_{\rm s}$-band image would allow us to determine if a host galaxy is associated with this radio source.

\citet[][]{thyagarajan11} classified J1521$-$0104 as variable at 1.4 GHz based on analysis of the three separate snapshot observations taken for this source as part of FIRST. These authors reported a minimum variability time-scale of 8 days.  The catalogued FIRST and NVSS flux densities in Table~\ref{table:fluxes} are consistent on a longer time-scale (2.1 yr), however.\footnote{Using the information available in \citet[][]{ofek11}; also see the VLA Data Archive at \url{https://science.nrao.edu/facilities/vla/archive/index}.} Short-time-scale variability from a scintillating radio core would rule out that J1521$-$0104 is a component in a larger radio source. Alternatively, the variability may have been due to a scintillating lobe hotspot; this alone would not provide conclusive evidence of the true angular extent of this radio source.      

{\bf J2219$-$3312:} While this source passed step 8 in Table~\ref{table:sample selection}, there is an AllWISE source 1\farcs2 from the ATCA position. The ATCA centroid is slightly further to the south-south-east than in TGSS (offset 1\farcs3), and there is also evidence in Figure~\ref{fig:overlays_spectra} of a small offset between the ATCA and VLASS centroids. The AllWISE source has W1 and  W2 magnitudes of $20.089 \pm 0.161$ and $20.188 \pm 0.347$, respectively \citep[][]{cutri14}, but is not detected in the longer-wavelength {\it WISE} bands. The $K_{\rm s}$-band limit is $> 22.3$ (Table~\ref{table:overlay_contour_levels}); the $\gtrsim 2.2$ mag break between $K_{\rm s}$-band and W1 is larger than for J1335$+$0112 discussed above. We regard the AllWISE source as a tentative host galaxy identification; this association needs to be confirmed with follow-up work, particularly deep $K_{\rm s}$-band imaging.     

{\bf J2311$-$3359:} This source, not selected from our $\alpha$--$\beta$ criteria (as discussed in Section~\ref{section: HzRG sample}), has an extremely steep spectrum well described by a single power law with $\alpha = -1.842$. Therefore, J2311$-$3359 is faint in our higher-frequency images. The source is unresolved in the ATCA data, but the low S/N does not allow an accurate LAS determination. This source is also unresolved in RACS, but we measured an LAS of 5\farcs6 in an 887.5-MHz ASKAP early science image of the GAMA-23 field \citep[see][]{seymour20}. Such an LAS value falls outside of our LAS $\leq 5\arcsec$ criterion, but needs to be confirmed with additional radio data. Note in Figure~\ref{fig:overlays_spectra} that the VLASS contours are offset from the ATCA contours, but the former are from a line-like artefact in the map.    

{\bf J2330$-$3237:} We interpret this source as an asymmetric double with LAS 4\farcs2, where the components have significantly different flux densities (the western lobe being much fainter). Using the available high-resolution radio data, the flux densities of the brighter eastern lobe are $6.3 \pm 0.7$, $3.74 \pm 0.27$ and $2.11 \pm 0.16$ mJy at 3, 5.5 and 9 GHz, respectively. Similarly, the values for the significantly fainter component to the west are $<0.45$, $0.22 \pm 0.06$ and $<0.078$ mJy, respectively (upper limits at the $3\sigma$ level). Given that the western lobe is detected at 5.5 GHz only, we verified that this source was not spuriously created as a result of our phase-only self-calibration step. The two point spectral indices are $\alpha^{5500}_{3000} = -0.86 \pm 0.22$ and $\alpha^{9000}_{5500} = -1.16 \pm 0.21$ for the eastern lobe; similarly these values are $\alpha^{5500}_{3000} \gtrsim -1.2 $  and $\alpha^{9000}_{5500} \lesssim -2.1$ for the western lobe. The significant spectral steepening of the western lobe at higher frequencies could at least be partly due to flux density possibly being resolved out in the higher-resolution 9-GHz map. Further evidence in favour of a double-lobed morphology is the hint of a $K_{\rm s}$-band identification between the two components (brightest pixel value $= 3.7\sigma$); there is similar marginal evidence in $H$-band (brightest pixel value $= 4.4\sigma$).   

{\bf J2340$-$3230:} Two radio sources are visible, with the north-eastern source having a possible marginal (brightest pixel value $= 4.6\sigma$) $K_{\rm s}$-band detection in SHARKS. The radio flux densities of this source are $\sim$~0.75, $0.79 \pm 0.11$ and $0.58 \pm 0.07$ mJy at 3, 5.5 and 9~GHz, respectively. Therefore, there is tentative evidence that this source is turning over at GHz frequencies. For the purpose of analysis in this paper, we have assumed that the HzRG candidate is the south-western source (which is e.g. an order of magnitude brighter at 5.5~GHz) and that the north-eastern source is an unrelated source that is nearby in projection. Alternatively, this could be a lower-redshift source with a one-sided jet (assuming that one of the two sources is the radio core), where the LAS is 6\farcs2. A deeper $K_{\rm s}$-band image is needed to identify a possible near-infrared counterpart coincident with the south-western source; there is tentative evidence of a SHARKS $K_{\rm s}$-band detection (brightest pixel value $= 3.8\sigma$).  

\section{Discussion}\label{section:discussion}

\subsection{Properties of the broadband radio spectra}\label{section:discussion spectra} 

As was summarised in Section~\ref{section:modelling data}, 34 out of 51 sources ($\approx 70$ per cent) have broadband spectra that can be best modelled with a double power law, with the remaining sources having a single power law as the preferred model (Table~\ref{table:fitting_results}). While our GLEAM selection technique fits for both spectral steepness and curvature, the advantage of broadband spectral modelling is the much longer `lever arm'. Given that curvature is part of our selection process as well as our wide frequency coverage, we find a slightly larger fraction of sources with curved broadband spectra than, for example, \citet[][]{saxena18a}, where 10 out of 17 of the sources in their USS-selected sample have spectra that are flatter between 370 and 147.5 MHz compared with between 1400 and 370 MHz (also see \citealt[][]{debreuck00} and \citealt[][]{bornancini07} for evidence of low-frequency flattening in other USS-selected samples). The fraction of single-power-law broadband-spectrum sources in our sample is also far smaller than, for example, in the SUMSS--NVSS USS sample, where 33 out of 37 sources were found to have single-power-law spectra, with the remaining four sources flattening rather than steepening with increasing frequency, albeit with modelling between 843 MHz and 18 GHz and no low-frequency coverage \citep[][]{klamer06}.    

Of the 17 sources with single-power-law broadband spectra, the median spectral index is $-0.984$. Only two of the sources would traditionally be classified as USS: J0048$-$3540 with $\alpha=-1.303$ and J2311$-$3359 with $\alpha=-1.842$. J2311$-$3359 also has a USS spectral index from the GLEAM $\alpha$--$\beta$ fitting (as remarked in Section~\ref{section: HzRG sample}; $\alpha = -1.58 \pm 0.19$), whereas as for J0048$-$3540 the GLEAM-only spectral index is just below a typical USS cutoff ($\alpha = -1.27 \pm 0.04$). On the other hand, J2314$-$3517, also noted earlier in Section~\ref{section: HzRG sample} as a source with a USS spectral index and significant curvature from the GLEAM-only fitting ($\alpha = -1.46 \pm 0.08$ and $\beta = -2.03 \pm 0.65$), has a flatter single-power-law broadband spectrum ($\alpha=-1.028$). This is clearly apparent in Figure~\ref{fig:overlays_spectra}, demonstrating the value of wide frequency coverage in radio spectral modelling.  

For nine sources best fitted with a single power law, $\min$(AICc) occurs for a double-power-law fit rather than a single power law: J0002$-$3514, J0042$-$3515, J0201$-$3441, J1052$-$0318, J1125$-$0342, J1136$-$0351, J1340$+$0009, J2219$-$3312 and J2330$-$3237. However, the simpler single-power-law fit still satisfies our selection condition $0 \leq \Delta{\rm AICc} < 4$ (Section~\ref{section:modelling data}). As can be seen in Figure~\ref{fig:overlays_spectra}, there are hints of curvature for these sources (note in particular that J1340$+$0009 has the most curvature in GLEAM from the subset of sources with broadband single-power-law fits; $\beta = -2.59 \pm 0.81$), and coverage over a wider frequency range would be useful to further explore the broadband spectral properties. 

In the case of the double-power-law fits, the best-fitting break frequencies range from the bottom of the GLEAM band ($\nu_{\rm b} = 76$ MHz for J1040$+$0150) to frequencies above our highest-frequency data point ($\nu_{\rm b} = 15.5$ GHz for J0006$-$2946). This latter behaviour is possible given the smoothly varying nature of the model in Equation~\ref{fit:dpl} and the corresponding `transition region' around the break where the spectral index gradually changes. While a direct comparison between the GLEAM-only curved fits and the broadband double-power-law fits (i.e. between Equations~\ref{eqn:curved} and \ref{fit:dpl}) is not possible given the different frequency coverage and number of fitted parameters, when the best-fitting break frequency in Table~\ref{table:fitting_results} is above the GLEAM band (i.e. above 227 MHz), the spectral index at 151 MHz is systematically flatter in the broadband double-power-law fits compared with the GLEAM-only curved fits (i.e. $\alpha_{\rm l}$ versus $\alpha$(GLEAM); median difference $\approx$ 0.15).  

As was discussed in Section~\ref{section:notes sources}, J1141$-$0158 and J1443$+$0029 have spectra that turn over in the GLEAM band and J0909$-$0154 flattens significantly at GLEAM frequencies as well. Otherwise, as for the single-power-law sources discussed above, the remaining sources with double-power-law fits are expected to exhibit spectral turnovers significantly below the GLEAM band. While not turning over, many of the double-power-law sources in our sample have spectra that steepen by $\Delta \alpha \sim -1$ across the break, but we note that $\alpha_{\rm h}$ is often not well constrained. Indeed, the interested reader should inspect the 16th, 50th and 84th percentiles in Table~\ref{table:fitting_results} to assess how well a particular parameter is constrained. 

If not due to synchrotron self-absorption and/or free--free absorption, a change in spectral index of $\Delta \alpha \sim -1$ suggests that some sources in our sample, if at high redshift, could be exhibiting energy losses resulting from inverse-Compton scattering (which would steepen the spectrum by $\Delta\alpha = -0.5$ for active sources; e.g. see \citealt[][]{klamer06} for an overview of energy loss mechanisms in the context of HzRG radio spectra) along with another mechanism that steepens the spectrum further (or is instead the main cause for the significant spectral steepening). Note that $\Delta \alpha \sim -1$ can also be seen, for example, in Figure~1 in \citealt[][]{miley08} for the HzRG 4C~23.56 at $z=2.48$. As previously discussed in Section~\ref{section:notes sources}, J0007$-$3040 is a particularly interesting curved source that is USS at low frequencies ($\alpha_{\rm l} = -1.385$; this is also apparent in the GLEAM $\alpha$--$\beta$ fitting results where $\alpha = -1.51 \pm 0.02$) and exceptionally steep at higher frequencies ($\alpha_{\rm h} = -2.75$). 

Apart from spectral steepening expected for active sources from inverse-Compton losses, steepening beyond $\Delta\alpha = -0.5$ at high frequencies could occur if the jets have switched off, or, alternatively, in the case of either jet- or lobe-dominated emission, if the jet power is intermittent (e.g. modelling by \citealt[][]{hardcastle18}, \citealt[][]{turner18b} and \citealt[][]{shabala20}; also see references therein). Another potential scenario is that if the radio emission is lobe dominated, then the high-energy end of the electron energy distribution (where the Lorentz factor $\gamma \gg 10^5$) may have been redshifted into the observed frame below $\sim$ 20 GHz; the spectrum would then deviate from a single power law and steepen because there are relatively few high-energy electrons that radiate at high rest-frame frequencies. A further possibility is that the lobe magnetic field strength has dropped very rapidly; an abrupt change in the magnetic field strength could occur, for example, when the jet leaves the host galaxy \citep[e.g.][]{shabala17}. \citet[][]{turner18a} found that the bulk of the synchrotron emission at 1 GHz for low-redshift sources results from the most recent 5--15 per cent of the radio source evolution; freshly injected electrons could have a lower emissivity and hence the spectrum would steepen beyond $\Delta\alpha = -0.5$ at high frequencies.    

While flux density being resolved out at high angular resolution should be an effect that is generally minimised given the compact nature of our targets, we cannot fully rule out that artificial curvature is apparent in at least some of the spectra shown in Figure~\ref{fig:overlays_spectra}, particularly given the significant overall improvement in angular resolution with increasing frequency in our data sets. We addressed this topic for J0007$-$3040 and J0309$-$3526 in Section~\ref{section:notes sources}. More generally, for the EQU sources, the comparison between the FIRST and NVSS flux densities carried out in Section~\ref{section: flux comparison mid} gives us confidence that resolution effects are not widespread in the broadband spectra of these sources, as does the observation that the VLASS flux densities in Figure~\ref{fig:overlays_spectra} are not significantly underestimated compared to the lower-resolution measurements either side of the VLASS data point. For the SGP sources, the significance of this potential issue is more challenging to assess, as a comparison between FIRST and NVSS was not possible. However, one test that we carried out was to adjust the robust weighting parameter and reimage the 5.5-GHz ATCA data for these sources. We found that only in the case of J0309$-$3526 was the flux density significantly affected by the change in weighting (as previously discussed in Section~\ref{section:notes sources}). A further test for the SGP sources would be to obtain matched-resolution ATCA observations with more compact array configurations than 6A \citep[e.g. the radio spectral fitting in][]{klamer06}.   

As with the single-power-law sources, we can also assess whether any of the sources best modelled by a double power law would be classified as USS sources. However, this very much depends on the frequency at which the analysis is done. As a simple approach, let us first consider the two-point spectral index $\alpha^{1400}_{147.5}$, as was used in \citet[][]{saxena18a}. The median spectral index is $-0.90$, and only four of the 34 sources with a double-power-law fit have $\alpha^{1400}_{147.5} \leq -1.3$: J0007$-$3040, J0008$-$3007, J0909$-$0154 and J1040$+$0150. If we instead consider the two-point spectral index $\alpha^{1400}_{887.5}$, similar to $\alpha^{1400}_{843}$ used in \citet[][]{debreuck04}, then the median spectral index steepens to $-1.19$, and the above four sources as well as J0053$-$3256, J0239$-$3043, J0309$-$3526, J1032$+$0339, J1211$-$0256, J1443$+$0229 and J1521$-$0104 would be classified as USS (i.e. 11 out of 34 sources). The fractional increase is consistent with the steepening in the double-power-law fits with increasing frequency. Both parts of the above exercise further emphasise that our sample contains (far) fewer USS HzRG candidates than previous investigations in the literature.   

For the equatorial sources in our sample, lower-frequency observations using the LOFAR low-band antennas (frequency range 30--80 MHz) could be used to search for and model a low-frequency spectral turnover; some sources also fall within the planned sky coverage of LoLSS ($\delta > 0\degr$). Additionally, higher-frequency observations in e.g. the ATCA 12-mm band would help to refine the modelling of the spectral curvature, particularly for the ten sources with fitted break frequencies above our frequency coverage. Note that the sources in our sample are too faint to have been detected in the Australia Telescope 20-GHz Survey \citep[AT20G; flux density limit $S_{19\,904} = 40$ mJy;][]{murphy10}.   

\subsection{MHz-peaked-spectrum sources at high redshift}\label{section:discussion spectra2} 

\begin{figure*}
\vspace{-0.5cm}
\begin{minipage}{0.5\textwidth}
\includegraphics[width=9.5cm]{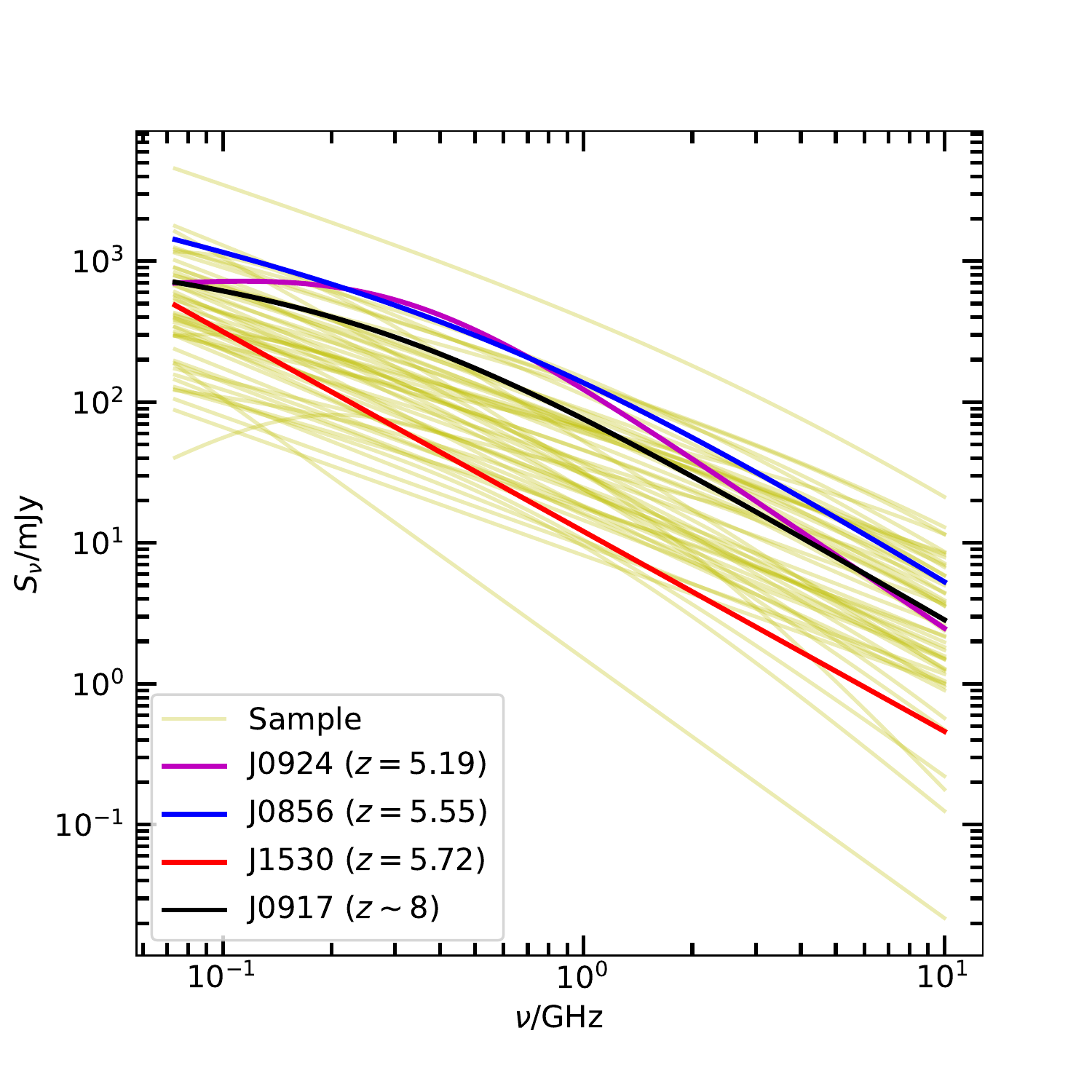}
\end{minipage}%
\begin{minipage}{0.5\textwidth}
\includegraphics[width=9.5cm]{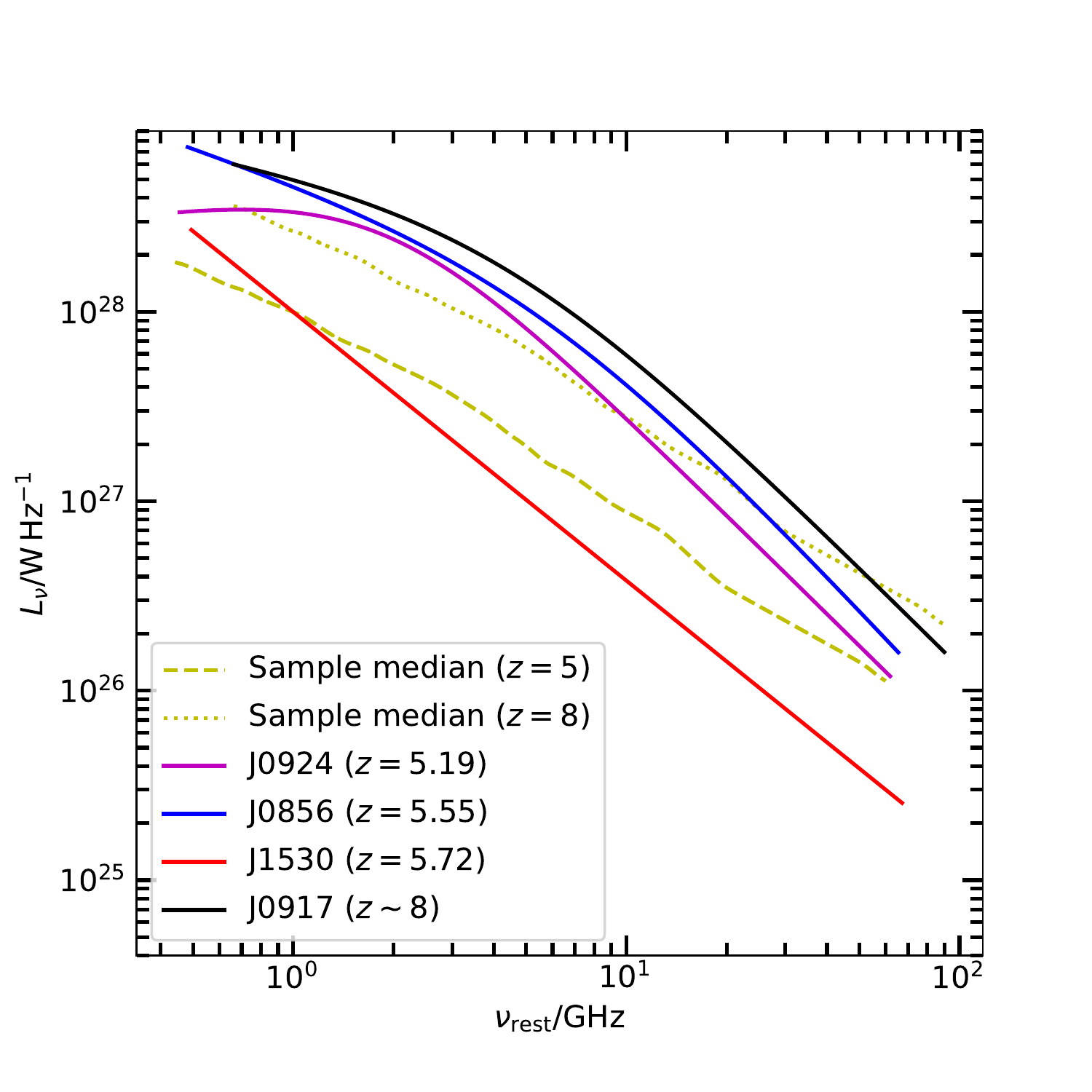}
\end{minipage}
\caption{{\em Left:} Compilation of the best-fitting observed-frame radio spectra of the 51 new HzRG candidates in our sample. We also show the best-fitting spectra for J0856$+$0223 and J0917$-$0012 as well as for the HzRGs J0924$-$2201 and J1530$+$1049. For consistency, the models for J0856$+$0223, J0917$-$0012 and J0924$-$2201 (double power law), as well as for J1530$+$1049 (single power law), were determined with the same fitting code and over the same frequency range as the models computed in Section~\ref{section:SED modelling} for the 51 new HzRG candidates. {\em Right:} The corresponding rest-frame spectra for J0856$+$0223, J0917$-$0012 (assumed to be at $z=8$), J0924$-$2201 and J1530$+$1049. We also plot the median rest-frame spectra of the 51 new HzRG candidates at two fiducial redshifts: $z=5$ and $z=8$. Further details of the analysis can be found in Section~\ref{section:discussion spectra2}.} 
\label{fig:all_sed}
\end{figure*}

\begin{table*}
\begin{minipage}{1.0\textwidth}
\centering
\caption{Fitted parameters for the observed-frame broadband radio spectra of J0856$+$0223, J0917$-$0012, J0924$-$2201 and J1530$+$1049 plotted in the left panel of Figure~\ref{fig:all_sed}. The columns are the same as in Table~\ref{table:fitting_results}, although we show only a subset of the fitting results in this case.}
\begin{tabular}{ccrrrrrr}
\hline\hline
& & \multicolumn{6}{c}{From model fit with min(AICc)}\\\cline{3-8}
\multicolumn{1}{c}{Source} & \multicolumn{1}{c}{Fit} &
\multicolumn{1}{c}{$N$} & \multicolumn{1}{c}{$\alpha$} & \multicolumn{1}{c}{$\alpha_{\rm l}$} &
\multicolumn{1}{c}{$\nu_{{\rm b}}$} & \multicolumn{1}{c}{$\alpha_{\rm h}$} &
\multicolumn{1}{c}{$\chi^2_{\rm red}$} \\
& & & & & \multicolumn{1}{c}{(GHz)} & & \\
\hline
J0856$+$0223	&	DPL	&  $	370	$ & $\cdots$ & $	-0.55	$ & $0.585$ & $-1.586$ & $	3.4	$ \\
J0917$-$0012	&	DPL	&  $	460	$ & $\cdots$ & $	-0.23	$ & $0.277$ & $-1.497$ & $	3.0	$ \\
J0924$-$2201	&	DPL	&  $	1400	$ & $\cdots$ & $	0.24	$ & $0.306$ & $-1.733$ & $	0.8	$ \\
J1530$+$1049	&	SPL	&  $	12.05	$ & $	-1.418	$ & $\cdots$ & $\cdots$ & $\cdots$ & $	1.4	$ \\
\hline\hline
\end{tabular}
\label{table:fitting_results2}
\end{minipage}
\end{table*}

Peaked-spectrum compact radio sources have been the subject of considerable study \citep[review by][]{odea21}. An anticorrelation between the rest-frame turnover frequency and projected linear size has been observed for these sources (\citealt[][]{odea97}; also see the recent compilation in Figure~5 of \citealt[][]{wolowska21}). If such a relation continues to hold at the highest redshifts beyond $z \sim 5$, then selecting compact sources with observed-frame turnovers at MHz frequencies could be an efficient method for finding very distant HzRGs (\citealt*[][]{falcke04}; \citealt[][]{coppejans15,coppejans16a,coppejans16b,coppejans17}; \citealt[][]{callingham17}; \citealt*[][]{keim19}). As we have established in previous sections, our HzRG candidate sample comprises compact radio sources that peak at MHz frequencies: either within the GLEAM band or, for the vast majority of sources, at frequencies below the GLEAM coverage. For the latter case, we now briefly consider what this may allow us to infer about their redshifts. We use the relation found by \citet[][]{orienti14} for the anticorrelation between the rest-frame turnover frequency ($\nu_{\rm p}$ in GHz) and the largest linear size (LLS in kpc), such that  
\begin{equation}\label{equation_peak_LLS} 
\log(\nu_{\rm p}) = -0.21 - 0.59\log({\rm LLS}).  
\end{equation}
From Equation~\ref{equation_peak_LLS}, for a given LLS there will be an expected rest-frame turnover frequency whose equivalent frequency in the observed frame must be significantly below $\sim$ 70 MHz, or else we would see clear evidence of the start of a turnover in our broadband radio spectra. One can recast this exercise in the observed frame to find the minimum redshift required for a source of a given angular size such that the observed-frame turnover is well below GLEAM, assuming that Equation~\ref{equation_peak_LLS} holds. We do not exhaustively consider all possibilities here, but, for example, a source with LAS $=3\arcsec$ at $z \gtrsim 0.7$ would be expected to turn over at observed-frame frequencies $\lesssim 60$ MHz. For a similar scenario, LAS $=1\arcsec$ would correspond to $z \gtrsim 1.9$, LAS $=0\farcs5$ to $z \gtrsim 3.8$, and LAS $=0\farcs2$ to $z \gtrsim 10.6$. In practice, however, there is observed scatter about the anticorrelation, and it remains unclear if the relation holds at very high redshift. Also, a maximum LLS will become apparent at very high redshift due to significant inverse-Compton losses; this maximum LLS decreases with increasing redshift \citep[e.g.][]{saxena17}. Nonetheless, it is intriguing that the radio spectral properties may hint at high redshifts for some sources in our sample.      

In Figure~\ref{fig:all_sed}, we present a compilation of the best-fitting radio spectra of our 51 new HzRG candidates. For comparison, we also plot the best-fitting spectra for J0924$-$2201 ($z=5.19$; double power law), J0856$+$0223 ($z=5.55$; double power law), J1530$+$1049 ($z=5.72$; single power law) and J0917$-$0012 (henceforth assumed to be at $z \sim 8$; see Section~\ref{section:intro}; double power law). For all four of these additional sources, the best-fitting spectra (parameters presented in Table~\ref{table:fitting_results2}) were determined using the fitting code described in Section~\ref{section:SED modelling} and using the input flux density catalogues over the same frequency range as described in Section~\ref{section:overview available data}.\footnote{We therefore used data from 74 MHz to 9 GHz for J0856$+$0223, 76 MHz to 9 GHz for J0917$-$0012, and 76 MHz to 3 GHz for J1530$+$1049. For J0924$-$2201, we used data from 76 MHz to 3 GHz plus an extra flux density point at 4.85 GHz from \citet[][]{debreuck00}.} J1530$+$1049 is similar to a number of sources in our sample in the sense that it has significant curvature in GLEAM (Figure~\ref{fig:alpha_beta}), yet the longer lever arm of the broadband modelling results in a single power law being the preferred fit (also see discussion in Section~\ref{section:discussion spectra}).   

 We show radio spectra in both the observed frame and rest frame (i.e. luminosity as a function of rest-frame frequency in the latter case) in Figure~\ref{fig:all_sed}. In the observed frame (left panel of Figure~\ref{fig:all_sed}), the mix of single- and double-power-law fits span about 2 dex in flux density at low frequencies and about 3 dex at high frequencies. J1351$-$0209, the very bright source discussed in Section~\ref{section:notes sources}, can be seen at the top of the panel. The USS-selected source J2311$-$3359, also discussed previously in this paper, is seen at the bottom of the panel with by far the lowest high-frequency flux density. The left panel of Figure~\ref{fig:all_sed} further demonstrates that nearly all of our sources do not peak in the observed-frequency range, but that some of these sources begin to flatten at low frequency. As discussed in Section~\ref{section:discussion spectra}, many of the sources with a single power law as the preferred fit show indications that they could have curvature at low and/or high frequency. In summary, in the left panel of Figure~\ref{fig:all_sed}, broadly speaking our new HzRG candidates have radio spectra that are (i) very similar to or partly consistent with the spectra of J0856$+$0223 and J0917$-$0012, or (ii) flatter analogues of J1530$+$1049 (apart from the outlier J2311$-$3359). However, there are very few sources with spectra similar to J0924$-$2201 (J1141$-$0158 and J1443$+$0229 only; Section~\ref{section:notes sources}).  

Examining the rest-frame spectra (right panel of Figure~\ref{fig:all_sed}), we first note that we have plotted median rest-frame spectra for the 51 new HzRG candidates assuming two fiducial redshifts: $z=5$ and $z=8$. We see that J0924$-$2201, J0856$+$0223 and J0917$-$0012 are significantly more luminous than J1530$+$1049 over most of the frequency range. From the median radio spectra of the sample at the fiducial redshifts of $z=5$ and $z=8$, assuming a significant yield of sources within this redshift range, we would expect to find distant radio galaxies amongst the most powerful known at these redshifts, but also less luminous radio sources such as J1530$+$1049, particularly at lower frequencies.   

The modelling of J0924$-$2201 implies that it peaks at 113 MHz (699 MHz) in the observed (rest) frame. The observed-frame peak is broadly consistent with the modelling done by \citet[][]{callingham17} for this source with GLEAM, TGSS and NVSS only; these authors calculated a peak frequency in the observed frame of $160 \pm 30$ MHz. Referring to the discussion earlier in this section, redshifting a J0924$-$2201-like object such that the turnover frequency was at 60 MHz instead would require an extreme redshift: $z=10.7$. Our selection method may therefore be unlikely to find the most youthful, compact radio galaxies at very high redshift if they have gigahertz-peaked-spectrum (GPS) spectra like their lower-redshift counterparts (e.g. J1606$+$3124 with a tentative redshift of $z=4.56$; \citealt[][]{an22} and references therein). Instead we may have better prospects of finding systems similar to J0856$+$0223, J0917$-$0012 and J1530$+$1049 that also turn over at frequencies below GLEAM.  

\subsection{Further radio angular size constraints from interplanetary scintillation}\label{section:discussion ips} 

In Section~\ref{section: flux comparison low}, we investigated potential low-frequency variability of sources in our sample, not finding any statistically significant evidence of variability from a comparison of the TGSS 147.5-MHz and GLEAM fitted 151-MHz flux densities. However, assuming a sufficiently well-sampled data set, another possible mechanism for low-frequency variability is interplanetary scintillation \citep*[IPS;][]{clarke64,hewish64}, which is an ongoing key science driver for the MWA \citep[][]{morgan18,morgan19,chhetri18a,chhetri18b}. Given that IPS is a powerful technique for identifying compact radio sources, it has the potential to be an effective tool for finding HzRGs (\citealt[][]{sadler19}). Indeed, as discussed in \citet[][]{drouart21}, MWA IPS observations place additional constraints on the compact radio morphology of J0917$-$0012.

In the latest MWA IPS survey at 162 MHz (Morgan et al. in prep.), 13 sources in our sample have multiple detections, including J0856$+$0223 and J0917$-$0012 from our pilot study. These sources are listed in Table~\ref{table:ips}. \citet[][]{chhetri18a} defined a parameter called
the normalised scintillation index (NSI) that indicates how much of the source flux density scintillates. Our sources with IPS detections have median NSI values that range from $\lesssim 0.24$ to 0.81; interestingly, many of the measurements are clustered around NSI $\sim 0.5$. Sources with an NSI $= 1$, i.e. 100 per cent of their flux density scintillates, are compact at the $ \leq 0\farcs3$ scale, whereas sources with NSI $< 1$ can arise due to different scenarios, as explained in Figure~5 of \citet[][]{morgan19}. As an illustration, we take the source J1125$-$0342 that has NSI $= 0.55$. This NSI can arise from (i) a slightly resolved Gaussian (here approximately twice the size of the Fresnel diameter: $\sim$ 0\farcs6), (ii) a point source, with approximately half of the total flux density embedded in an extended component (that in theory can be extended on angular scales up to the synthesised beam size of the MWA), or (iii) a two-component compact source where the components have angular separation $> 0\farcs3$ and where one of the components is partially resolved at the 0\farcs3 scale or embedded in emission on this scale (this third scenario can result in NSI $< 0.7$). For scenario (i), the conversion from NSI to linear size is provided in Figure~6 of \citet[][]{chhetri18a}. For scenarios (ii) and (iii), we cannot directly place an upper limit on the angular size. 

For the source with the largest NSI in Table~\ref{table:ips}, J1030$+$0135, the expected IPS angular size constraint agrees well with the LAS of 0\farcs5 from the VLASS quick-look image (Table~\ref{table:fluxes}), although the LAS from FIRST is larger (1\farcs8). From Tables~\ref{table:fluxes}, \ref{table:overlay_contour_levels} and \ref{table:fitting_results} as well as Figure~\ref{fig:overlays_spectra}, this is a source with a relatively deep limit from VIKING ($K_{\rm s} > 21.9$), not USS at low and mid frequencies, and with a spectral turnover below the GLEAM band. Assuming an LAS $=0\farcs5$, the previous discussion concerning MHz-peaked-spectrum sources in Section~\ref{section:discussion spectra2} might possibly suggest a redshift $z \gtrsim 3.8$. LOFAR low-band data would be particularly useful for this source. 

We will fully analyse the IPS properties of our sample as part of a future publication. However, we can already conclude that our search method selects very strongly for sources with sub-arcsecond structure.

\begin{table}
 \centering
  \caption{IPS properties of 13 sources in our sample (from Morgan et al. in prep.). For each NSI measurement, we report the median $\pm$ median absolute deviation (apart from the upper limit for J1127$-$0332). See Section~\ref{section:discussion ips} for further information.}
  \begin{tabular}{cc}
  \hline\hline
   \multicolumn{1}{c}{Source} & \multicolumn{1}{c}{NSI} \\ 
\hline
J0842$-$0157 & $0.41 \pm 0.09$ \\
J0856$+$0223 & $0.52 \pm 0.06$ \\
J0909$-$0154 & $0.60 \pm 0.06$ \\
J0917$-$0012 & $0.48 \pm 0.04$ \\
J1030$+$0135 & $0.81 \pm 0.14$ \\
J1032$+$0339 & $0.58 \pm 0.08$ \\
J1033$+$0107 & $0.42 \pm 0.11$ \\
J1040$+$0150 & $0.65 \pm 0.09$ \\
J1052$-$0318 & $0.64 \pm 0.14$ \\
J1112$+$0056 & $0.41 \pm 0.16$ \\
J1125$-$0342 & $0.55 \pm 0.08$ \\
J1127$-$0332 & $\lesssim 0.24$ \\
J1136$-$0351 & $0.38 \pm 0.25$ \\
\hline\hline
\end{tabular}
\label{table:ips}
\end{table}

\subsection{How many HzRGs in our sample could be within the EoR?}

By compiling the sample presented in this paper, our ongoing primary objective is to find a significant number of HzRGs within the EoR, i.e. with $z \gtrsim 6.5$. From the SKA Design Study (SKADS) Simulated Skies project \citep[S$^3$;][]{wilman08}, there are ten radio sources at $z>6.5$ over 400 deg$^{2}$ of S$^3$ with $S_{151} \geq 40$ mJy. However, this number does not include the high-redshift declines in the AGN radio luminosity functions as recommended by \citet[][]{wilman08}, which, if factored in, leaves just one source meeting this flux density selection criterion. Note that these declines in the radio luminosity functions are extrapolated from lower redshift and therefore are rather uncertain. 

Scaling the above predictions to the footprint of VIKING ($\approx$ 1200 deg$^2$), we would expect to find $\sim$ 3--30 radio sources at $z>6.5$ with $S_{151} \geq 40$ mJy. This scaled range is naturally quite uncertain as it is based upon small number statistics as well as extrapolated models; moreover, no radio sources are yet confirmed at $z>6.5$ with $S_{151} \geq 40$ mJy. Our selection method is neither complete nor likely 100 per cent efficient, but even at a success rate of 25 per cent as in D20 (and with the success rate potentially being 50 per cent if J0917$-$0012 is confirmed as being at $z \sim 8$), we might still expect up to $\sim$ 8 sources in our sample to be at very high redshift. Even finding a handful of AGN with powerful radio emission at such an early cosmic epoch would be extremely valuable in improving our understanding of massive galaxy formation and evolution in the early Universe.

\section{Conclusions and future work}\label{section:conclusions}

In this paper, we have extended the successful \citet[][]{drouart20} pilot study of GLEAM-selected HzRGs. By searching a sky area twenty times larger than in the pilot project, our new sample of GLEAM- and VIKING-selected HzRG candidates is an order of magnitude larger. We have compiled a multi-wavelength data set that will form the basis for ongoing investigations of the properties of this sample. Our main conclusions are as follows. 
\begin{enumerate}
\item Applying a refined selection technique over a sky area of approximately 1200 deg$^{2}$, we defined a sample of 53 sources: 51 new HzRG candidates as well as the $z=5.55$ (J0856$+$0223) and candidate $z \sim 8$ (J0917$-$0012) powerful radio galaxies from the pilot study. These sources were selected on the basis of their fitted 151-MHz flux densities ($\geq 40$ mJy), curved low-frequency radio spectra ($\alpha \leq -0.7 \cap \beta \leq -0.2$), compact radio morphology (LAS $\leq 5\arcsec$) and faintness in $K_{\rm s}$-band ($K_{\rm s} > 21.2$; $5\sigma$). 

\item Of the new HzRG candidates in the sample, two sources, J0133$-$3056 and J0842$-$0157, have $K_{\rm s}$-band host galaxy detections from deep HAWK-I imaging. J0842$-$0157, with $K_{\rm s} = 22.96$, has a similar magnitude to J0856$+$0223 ($K_{\rm s} = 23.2$), J0917$-$0012 ($K_{\rm s} = 23.01$) and the HzRG J0924$-$2201 at $z=5.19$ ($K_{\rm s} = 23.2$). Given the near-infrared and radio properties, J0842$-$0157 is an excellent target for follow-up investigations. J0007$-$3040 (a target of particular interest), J0008$-$3007 and J2340$-$3230 have $K_{\rm s} > 22.7$ ($5\sigma$) from SHARKS, and there are also an additional nine sources with $K_{\rm s} > 22.0$ ($5\sigma$) from VIKING.
    
\item While our technique selects sources with curved low-frequency radio spectra in GLEAM, broadband modelling over the frequency range 74/76 MHz to 8.8/9 GHz revealed that about 30 per cent of the sample are best modelled with a single power law, with the remaining 70 per cent best modelled with a double power law. Unlike traditional searches for HzRGs, our sample has a low fraction of USS sources below 1400 MHz ($\approx 10$--$25$ per cent). 

\item The vast majority of sources in our sample have inferred low-frequency spectral turnovers below $\sim 70$ MHz, i.e. below GLEAM. Analogous to low-redshift sources with spectral peaks at frequencies of hundreds of MHz, our sample may contain high-redshift analogues of young, compact radio galaxies. However, our selection method will miss the more youthful and compact GPS sources, unless in the unlikely case that they are at extreme redshift. 
\item The sample that we have compiled will be a valuable resource for identifying more radio-loud AGN in the early Universe. There could be up to $\sim$ 30 HzRGs at $z > 6.5$ with $S_{151} \geq 40$ mJy in the VIKING survey region from which we selected our HzRG candidates.  
\end{enumerate}

We will follow up this work by obtaining further multi-wavelength data to enable host galaxy detections and spectroscopic redshift determinations. We will also carry out additional modelling using the broadband radio spectra and radio morphological properties so as to obtain redshift constraints (\citealt[][]{turner20,turner21}; see \citealt[][]{seymour22} for such analysis for J0917$-$0012). This follow-up will be the subject of a future publication.    

\begin{acknowledgements}

We acknowledge the Noongar people as the traditional owners and custodians of Wadjak boodjar, the land on which the majority of this work was completed. We thank the referee for their review of this paper. 

JMA acknowledges financial support from the Science and Technology Foundation (FCT, Portugal) through research grants PTDC/FIS-AST/29245/2017, UIDB/04434/2020 and UIDP/04434/2020. NHW is supported by an Australian Research Council Future Fellowship (project number FT190100231) funded by the Australian Government. GN acknowledges funding support from the Natural Sciences and Engineering Research Council (NSERC) of Canada through a Discovery Grant and Discovery Accelerator Supplement, and from the Canadian Space Agency through grant 18JWST-GTO1. The work of DS was carried out at the Jet Propulsion Laboratory, California Institute of Technology, under a contract with NASA.

This scientific work makes use of the Murchison Radio-astronomy Observatory, operated by CSIRO. We acknowledge the Wajarri Yamatji people as the traditional owners of the Observatory site. Support for the operation of the MWA is provided by the Australian Government (NCRIS), under a contract to Curtin University administered by Astronomy Australia Limited. We acknowledge the Pawsey Supercomputing Centre which is supported by the Western Australian and Australian Governments.

GAMA is a joint European-Australasian project based around a spectroscopic campaign using the Anglo-Australian Telescope. The GAMA input catalogue is based on data taken from the Sloan Digital Sky Survey and the UKIRT Infrared Deep Sky Survey. Complementary imaging of the GAMA regions is being obtained by a number of independent survey programmes including {\it GALEX} MIS, VST KiDS, VISTA VIKING, {\it WISE}, {\it Herschel}-ATLAS, GMRT and ASKAP providing UV to radio coverage. GAMA is funded by the STFC (UK), the ARC (Australia), the AAO, and the participating institutions. The GAMA website is \url{http://www.gama-survey.org/}. Based on observations collected at the European Organisation for Astronomical Research in the Southern Hemisphere under ESO programmes IDs 177.A-3016, 177.A-3017, 177.A-3018 and 179.A-2004. This work was supported by resources provided by the Pawsey Supercomputing Centre with funding from the Australian Government and the Government of Western Australia.

The Australia Telescope Compact Array is part of the Australia Telescope National Facility which is funded by the Australian Government for operation as a National Facility managed by CSIRO. We acknowledge the Gomeroi people as the traditional owners of the Observatory site. We are grateful to Australia Telescope National Facility staff for granting us permission to observe project CX437 as well as swiftly fixing some system issues that occurred during our C3377 observations. 

Based on observations collected at the European Organisation for Astronomical Research in the Southern Hemisphere under ESO programme 0104.A-0599(A).

For the creation of the data used in this work, the SHARKS team at the Instituto de Astrof\'{i}sica de Canarias (ACR, HD and CMG) has been financially supported by the Spanish Ministry of Science, Innovation and Universities (MICIU) under grant AYA2017-84061-P, co-financed by FEDER (European Regional Development Funds), by the Spanish Space Research Program ``Participation in the NISP instrument and preparation for the science of EUCLID'' (ESP2017-84272-C2-1-R) and by the ACIISI, Consejer\'{i}a de Econom\'{i}a, Conocimiento y Empleo del Gobierno de Canarias and the European Regional Development Fund (ERDF) under grant with reference PROID2020010107. Based on data products created from observations collected at the European Organisation for Astronomical Research in the Southern Hemisphere under ESO programme 198.A-2006. We thank the support of the Wide-Field Astronomy Unit for testing and parallelising the mosaic process and preparing the releases. The work of the Wide-Field Astronomy Unit is funded by the UK Science and Technology Facilities Council through grant ST/T002956/1.

The Australian SKA Pathfinder is part of the Australia Telescope National Facility which is managed by CSIRO. Operation of ASKAP is funded by the Australian Government with support from the National Collaborative Research Infrastructure Strategy. ASKAP uses the resources of the Pawsey Supercomputing Centre. Establishment of ASKAP, the Murchison Radio-astronomy Observatory and the Pawsey Supercomputing Centre are initiatives of the Australian Government, with support from the Government of Western Australia and the Science and Industry Endowment Fund. We acknowledge the Wajarri Yamatji people as the traditional owners of the Observatory site. This paper includes archived data obtained through the CSIRO ASKAP Science Data Archive, CASDA \url{(http://data.csiro.au)}. 

The National Radio Astronomy Observatory is a facility of the National Science Foundation operated under cooperative agreement by Associated Universities, Inc.

AllWISE makes use of data from {\it WISE}, which is a joint project of the University of California, Los Angeles, and the Jet Propulsion Laboratory/California Institute of Technology, and NEOWISE, which is a project of the Jet Propulsion Laboratory/California Institute of Technology. {\it WISE} and NEOWISE are funded by the National Aeronautics and Space Administration.

The Hyper Suprime-Cam (HSC) collaboration includes the astronomical communities of Japan and Taiwan, and Princeton University. The HSC instrumentation and software were developed by the National Astronomical Observatory of Japan (NAOJ), the Kavli Institute for the Physics and Mathematics of the Universe (Kavli IPMU), the University of Tokyo, the High Energy Accelerator Research Organization (KEK), the Academia Sinica Institute for Astronomy and Astrophysics in Taiwan (ASIAA), and Princeton University. Funding was contributed by the FIRST program from the Japanese Cabinet Office, the Ministry of Education, Culture, Sports, Science and Technology (MEXT), the Japan Society for the Promotion of Science (JSPS), Japan Science and Technology Agency (JST), the Toray Science Foundation, NAOJ, Kavli IPMU, KEK, ASIAA, and Princeton University. 

This paper makes use of software developed for the Large Synoptic Survey Telescope. We thank the LSST Project for making their code available as free software at \url {http://dm.lsst.org}. 

This paper is based in part on data collected at the Subaru Telescope and retrieved from the HSC data archive system, which is operated by the Subaru Telescope and Astronomy Data Center (ADC) at National Astronomical Observatory of Japan. Data analysis was in part carried out with the cooperation of Center for Computational Astrophysics (CfCA), National Astronomical Observatory of Japan. The Subaru Telescope is honoured and grateful for the opportunity of observing the Universe from Maunakea, which has the cultural, historical and natural significance in Hawaii. 

The Pan-STARRS1 Surveys (PS1) and the PS1 public science archive have been made possible through contributions by the Institute for Astronomy, the University of Hawaii, the Pan-STARRS Project Office, the Max Planck Society and its participating institutes, the Max Planck Institute for Astronomy, Heidelberg, and the Max Planck Institute for Extraterrestrial Physics, Garching, The Johns Hopkins University, Durham University, the University of Edinburgh, the Queen's University Belfast, the Harvard-Smithsonian Center for Astrophysics, the Las Cumbres Observatory Global Telescope Network Incorporated, the National Central University of Taiwan, the Space Telescope Science Institute, the National Aeronautics and Space Administration under grant No. NNX08AR22G issued through the Planetary Science Division of the NASA Science Mission Directorate, the National Science Foundation grant No. AST-1238877, the University of Maryland, Eotvos Lorand University (ELTE), the Los Alamos National Laboratory, and the Gordon and Betty Moore Foundation.

This research has made use of the NASA/IPAC Extragalactic Database (NED), which is operated by the Jet Propulsion Laboratory, California Institute of Technology, under contract with the National Aeronautics and Space Administration.

This research has made use of the CIRADA cutout service at URL \url{cutouts.cirada.ca}, operated by the Canadian Initiative for Radio Astronomy Data Analysis (CIRADA). CIRADA is funded by a grant from the Canada Foundation for Innovation 2017 Innovation Fund (Project 35999), as well as by the Provinces of Ontario, British Columbia, Alberta, Manitoba and Quebec, in collaboration with the National Research Council of Canada, the US National Radio Astronomy Observatory and Australia’s Commonwealth Scientific and Industrial Research Organisation.

This publication makes use of data products from the Two Micron All Sky Survey, which is a joint project of the University of Massachusetts and the Infrared Processing and Analysis Center/California Institute of Technology, funded by the National Aeronautics and Space Administration and the National Science Foundation.

This research has made use of the VizieR catalogue access tool, CDS, Strasbourg, France (DOI: 10.26093/cds/vizier). The original description of the VizieR service was published in A\&AS 143, 23 \citep*[][]{ochsenbein00}. This research has made use of the SIMBAD database, operated at CDS, Strasbourg, France \citep[][]{wenger00}. This research has made use of NASA's Astrophysics Data System Bibliographic Services. This project also made use of {\sc astropy} \citep[][]{astropy13,astropy18}, {\sc kern} \citep[][]{molenaar18}, {\sc kvis} \citep[][]{gooch95}, {\sc matplotlib} \citep[][]{hunter07}, {\sc numpy} \citep[][]{oliphant06}, {\sc scipy} \citep[][]{virtanen20}, {\sc topcat} \citep[][]{taylor05} and Ned Wright's Javascript Cosmology Calculator \citep[\url{http://www.astro.ucla.edu/~wright/CosmoCalc.html};][]{wright06}.

\end{acknowledgements}

\bibliographystyle{pasa-mnras}
\bibliography{references}

\end{document}